\documentclass[11pt]{article}

\usepackage{color,xspace,wrapfig,fullpage}
\newcommand{\EMS}[1]{\textcolor{red}{#1}} 

\newcommand{\etal}{\emph{et al.}\xspace}
\newcommand{\eg}{\emph{e.g.,}\xspace}
\newcommand{\Eg}{\emph{E.g.,}\xspace}
\newcommand{\ie}{\emph{i.e.,}\xspace}
\newcommand{\Ie}{\emph{I.e.,}\xspace}

\newcommand{\remove}[1]{}

\usepackage{multirow,booktabs,amsmath,amsfonts,cite,hyperref,float,epsfig,graphicx,stfloats,url,bm,booktabs,bbding,amssymb,diagbox}
\usepackage[hyphenbreaks]{breakurl}
\usepackage{pifont}
\usepackage{footnote}
\usepackage{verbatim}
\usepackage{subfig}
\usepackage{tikz}
\usepackage{tabularx}
\usepackage{colortbl}
\usepackage{xcolor}
\usepackage{inputenc}

\makesavenoteenv{table}

\begin{document}

\title{Evaluation of Open-source Tools for Differential Privacy}
	

\author{Shiliang Zhang\footnote{Chalmers University of Technology, Gothenburg, Sweden. Email: \texttt{shiliang@chalmers.se}, \texttt{anton.hagermalm@gmail.com}, \texttt{sanjinslavnic@gmail.com}, \texttt{elad@chalmers.se}, and \texttt{magnus.almgren@chalmers.se}.} \and Anton Hagermalm$^\ast$ \and Sanjin Slavnic$^\ast$ \and Elad M.\ Schiller$^\ast$ \and Magnus Almgren$^\ast$}

%

\maketitle

\newcommand{\tabincell}[2]{\begin{tabular}{@{}#1@{}}#2\end{tabular}}
	
\begin{abstract}	
Differential privacy (DP) defines privacy protection by promising quantified indistinguishability between individuals that consent to share their privacy-sensitive information and the ones that do not. DP aims to deliver this promise by including well-crafted elements of random noise in the published data and thus there is an inherent trade-off between the degree of privacy protection and the ability to utilize the protected data. Currently, several open-source tools were proposed for DP provision. To the best of our knowledge, there is no comprehensive study for comparing these open-source tools with respect to their ability to balance DP's inherent trade-off as well as the use of system resources. This work proposes an open-source evaluation framework for privacy protection solutions and offers evaluation for OpenDP Smartnoise, Google DP, PyTorch Opacus, Tensorflow Privacy, and Diffprivlib. In addition to studying their ability to balance the above trade-off, we consider discrete and continuous attributes by quantifying their performance under different data sizes. Our results reveal several patterns that developers should have in mind when selecting tools under different application needs and criteria. This evaluation survey can be the basis for an improved selection of open-source DP tools and quicker adaptation of DP.
\end{abstract}



\section{Introduction}

Privacy relates to one's ability to decide on the manner, context, and timing that one's personal information is managed by others. Data privacy protection is important for guaranteeing human dignity, safety, and self-determination as well as for guarding proprietary rights and economic interests. Despite extensive academic research and legal recognition that data privacy protection received, \eg General Data Protection Regulation (GDPR)~\cite{DBLP:journals/cr/SchererK18}, Health Insurance Portability and Accountability Act (HIPAA)~\cite{DBLP:journals/wicomm/MbonihankuyeNN19}, and California Consumer Privacy Act (CCPA)~\cite{goldman2020introduction}, the practice often does not use quantified methods for protecting data privacy. Differential privacy (DP) is a prominent proposal for quantifying privacy protection. It promises quantified indistinguishability between individuals that consent to share their privacy-sensitive information and the ones that do not. DP aims at delivering this promise by including well-crafted elements of random noise in the published data and thus there is an inherent trade-off between the degree of privacy protection and the ability to utilize the protected data. Recently several open-source tools were proposed for DP provision. To the best of our knowledge, there is no comprehensive study for comparing these open-source tools with respect to their ability to balance DP's inherent trade-offs as well as the use of system resources. This work proposes an open-source evaluation framework for DP tools and services. Using this framework, we offer evaluation results for OpenDP Smartnoise, Google DP, PyTorch Opacus, Tensorflow Privacy, and Diffprivlib. We consider both discrete and continuous attributes by quantifying their performance under different data sizes. Our results reveal several patterns that users should have in mind when selecting DP tools.

\remove{
	
	Differential privacy~\cite{DBLP:conf/tamc/Dwork08} (DP) has been proposed as a formal privacy guarantee for data. It reinforces a more substantial privacy basis for privacy regulations such as GDPR~\cite{DBLP:journals/cr/SchererK18} or HIPAA~\cite{DBLP:journals/wicomm/MbonihankuyeNN19} than existing approaches like anonymization, and is regarded as the most viable privacy protection technique for real-world data analysis~\cite{DBLP:journals/comsur/HassanRC20}. Nevertheless, a gap exists between fundamental and applied research on the topic of DP~\cite{DBLP:journals/corr/abs-1812-02890}. The academic literature around DP is less intuitive, with concepts that do not necessarily map well to the vernacular used in industry. Therefore, there lacks a bridge between the academic DP landscape and the practical application of DP in consumer technology. Particularly, there is an absence of knowledge over what software tools can be leveraged in privacy application development with DP principles, and to what level of performance can researchers and developers expect from the existing tools~\cite{DBLP:phd/basesearch/Zhang21b}. This disadvantage hinders further prosperity of DP in real-world applications and consequently induces practical obstacles to protect individual privacy from a software \& application perspective.
	
} 


\subsection{Differential privacy}

Dwork proposed differential privacy (DP)~\cite{DBLP:conf/tamc/Dwork08} as a means for guaranteeing that all individuals will be exposed to essentially the same risk of jeopardizing their privacy. This is done by quantifying the probability of his or her privacy-sensitive information being included in a DP analysis. \Ie~it ensures that personal private information cannot be revealed from the analysis results, regardless of the adversary's computational power or access to any additional information that may exist, or will ever exist.
As detailed in~\cite{DBLP:journals/jpc/DworkMNS16}, a randomized mechanism $M$ is said to provide $(\epsilon, \delta$)-DP protection for all events $S$, $S\subseteq \mbox{range}(M)$, given two datasets $D$ and $D'$, such that $\mbox{\large$\Pr[M(D) \in S] \leq e^\epsilon \cdot Pr[M(D') \in S] + \delta$}$, where $\mbox{range}(M)$ denotes the output range with a given input, and $\Pr[\;]$ denotes probability distribution. If $\delta = 0$, the randomized mechanism $M$ gives $\epsilon$-DP by its strictest definition. The parameter $\epsilon$ refers to the privacy budget, which controls the level of privacy guarantee achieved by mechanism $M$.
In other words, DP guarantees that the addition or removal of information related to a single individual in a dataset essentially does not affect the result of any analysis or query and limits the risk of privacy disclosing associated with providing or refraining from providing (privacy-sensitive) information to a dataset.


\subsection{A brief review of open-source tools with DP services}
\label{sec:rev}
We briefly review the most relevant open-source DP services.



\subsubsection{OpenDP Smartnoise}\label{architecutre_smartnoise}

OpenDP Smartnoise~\cite{hardvard_privacy} has its roots in Harvard University Privacy Tools Project. This project gained experience in building and deploying PSI~\cite{DBLP:conf/csfw/HsuGHKNPR14}, a system developed to share and explore privacy-sensitive datasets with privacy protections of DP, and ultimately contributed to their efforts toward Smartnoise. OpenDP Smartnoise has also incorporated insights from other DP tools, such as PinQ~\cite{DBLP:journals/cacm/McSherry10}, $\epsilon$ktelo~\cite{DBLP:conf/sigmod/ZhangMKHMM18}, PrivateSQL~\cite{DBLP:conf/cidr/KotsogiannisTMM19}, Fuzz~\cite{DBLP:conf/uss/HaeberlenPN11}, and LightDP~\cite{DBLP:conf/popl/ZhangK17}. While most of the tools like PSI and PinQ are research prototypes, OpenDP Smartnoise is now putting efforts into further developing DP concepts into production-ready tools. With those features, OpenDP Smartnoise has obtained significant popularity by the developer community, with more than two hundred stars in its open-source software repository.



\subsubsection{Google DP}\label{architecture_google-dp}

Google released an open-sourced version of its DP library that empowers some of its core products~\cite{google_blog_dp}. Available in Java and Go, this library captures years of Google's developer experience and offers practitioners and organizations potential benefits from their implementation, with a fairly low entrance level of expertise in DP.


\subsubsection{Opacus}\label{subsection_Opacus}

PyTorch Opacus~\cite{open_mind} is Facebook's DP library for ML services, built on top of PyTorch. It is developed in collaboration with Facebook AI Research, the PyTorch team, and OpenMined, an open-source community dedicated to developing privacy techniques for ML and AI. The service from Opacus targets both ML practitioners and professional DP researchers with its general and specific features.

\subsubsection{Tensorflow Privacy}\label{subsection_Tensorflow}

TensorFlow Privacy~\cite{DBLP:journals/corr/abs-2010-09063} (TFP) is an ML framework developed and released by Google, initially inspired by the work of Abadi \etal~\cite{DBLP:conf/ccs/AbadiCGMMT016}, who implemented a similar optimizer for TensorFlow and a privacy cost tracker. It emerges and adapts DP mechanisms to TensorFlow to allow users to leverage differential privacy in the training of ML models. Furthermore, TFP is configurable, and developers can define their own ML models, with which developers can implement their operators in their applications. With its flexibility and DP services, TFP has become an open DP tool that is leveraged and contributed by a large developer community.

\subsubsection{Diffprivlib}\label{sys_arch_diffprivlib}

Developed by the industrial giant IBM, Diffprivlib~\cite{DBLP:journals/corr/abs-1907-02444} empowers differential privacy in machine learning tasks, including classification, regression, clustering, dimensionality reduction, and data regularization. It is supposed to be a general-purpose tool for conducting experiments, investigations, and application developments with differential privacy. With its detailed product manual,\footnote{\url{https://diffprivlib.readthedocs.io/en/latest/}} practitioners of different levels can easily find what they need during their interaction with Diffprivlib. As a result, Diffprivlib's open repository~\cite{diffpriv_gh} has gained considerable attention amongst developers.


\subsubsection{Chorus}\label{subsection_chorus}

Chorus~\cite{DBLP:journals/corr/abs-1809-07750,DBLP:conf/eurosp/JohnsonNHS20} utilizes a cooperative architecture to achieve DP statistical queries. It leverages industrial-grade database management systems (DBMS) for data processing tasks and even queries that need to be modified or, in some cases, entirely rewritten. This architecture has three primary components, namely rewriting, analysis and post-processing. The rewriting component is used to modify queries to perform functions like clipping, the analysis component to analyze queries to determine different properties such as required noise to satisfy differential privacy, and the post-processing component to process the result of the queries. An example from Chorus is the implementation of a summation mechanism with clipping. The rewriting component can modify the original query so that the DBMS executes the clipping and the summation, leaving the rest of the summation mechanism to the analysis and the post-processing component.

What separates Chorus from previous work is that it is DBMS-independent. Unlike an integrated approach, Chorus does not require modifying or affecting the database or changing to a purpose-built database engine. Therefore, Chorus can leverage DMBS to ensure scalability when working on datasets that hold large amounts of data.

Added safeguards can be necessary when deploying Chorus to minimize the chance of a malicious actor acquiring sensitive data. For example, in the case of Chorus' deployment on Uber, privacy-sensitive data was only available through a centralized query interface, which was protected along with the privacy budget account and the DMBS from tampering.

While the early repository of Chorus archived by Uber is deprecated,\footnote{\url{https://github.com/uber-archive/sql-differential-privacy}} a new version emerged that is maintained and active among the open-source community.\footnote{\url{https://github.com/uvm-plaid/chorus}} The new repository of Chorus has not gained much attention since its relatively recent release. Nevertheless, Chorus showed strength in its early version, and thus, more concrete results and popularity among developers can be expected for its current version. In this work, we focus on the evaluation of DP tools that have already gained significant popularity, see Section~\ref{tools_datasets}.

\subsection{Evaluation approach}
\label{tools_datasets}
Several open-source tools are available for applying DP in statistical queries~\cite{DBLP:journals/popets/WilsonZLDSG20} and machine learning~\cite{DBLP:journals/corr/abs-1907-02444}. In this paper, we refer to statistical queries as the retrieval of features and aggregated information from datasets, \eg sum, count, and average. Machine learning tasks deal with the construction of models based on training data from a given dataset, say, for linear regression. Those two kinds of tasks, \ie statistical query or machine learning, are prevailing tools for application development and thus we select them to serve as evaluation test cases.




Table~\ref{table: tool list} presents the selected DP tools for evaluation. Our choice was based on the support that each of these tolls received from comparably larger communities, technology companies, and research institutions as well as their wide acceptance in the open-source community. We also required a sufficiently long history of being free from known bugs as well as developer-friendly documentation.
Our evaluation criteria for the studied DP tools focus on data utility and system overhead, \ie running time and required memory. The considered tools are evaluated within two task domains, \ie~statistical query and machine learning. Since tools of different categories follow different evaluating procedures, we analyze the evaluation results within every single domain. During the evaluation, we adopt a diversity of settings to investigate how different tools' performances vary under different conditions. We use the open-source data of United States Health Reform Monitoring Survey data~\cite{holahan2018health} for the experiments in statistical queries, and the UCI Parkinson dataset~\cite{DBLP:journals/tbe/TsanasLMR10} for machine learning experiments.


\begin{table}[htp]
	\centering
	\renewcommand\arraystretch{1.6}
	\resizebox{0.8\textwidth}{!}{
		\begin{tabular}{|l|l|l|}
			\cline{1-3}
			Name & Domain & Origin \\
			\cline{1-3}
			\href{https://github.com/IBM/differential-privacy-library}{Diffprivlib}~\cite{DBLP:journals/corr/abs-1907-02444} (v0.5.0) & Machine learning & IBM\\ 
			\cline{1-3}
			\href{https://github.com/google/differential-privacy}{Google Differential Privacy}~\cite{DBLP:journals/popets/WilsonZLDSG20} (v1.1.0) & Statistical query & Google\\ 
			\cline{1-3}
			\href{https://github.com/opendp/smartnoise-core}{OpenDP SmartNoise}~\cite{DBLP:journals/corr/abs-2011-05537} (v0.2.0) & Statistical query & Microsoft~\&~Harvard\\ 
			\cline{1-3}
			\href{https://github.com/pytorch/opacus}{PyTorch Opacus}~\cite{DBLP:journals/corr/abs-2105-05381} (v0.14.0) & Machine learning & Facebook\\ 
			\cline{1-3}
			\href{https://github.com/tensorflow/privacy}{TensorFlow Privacy}~\cite{DBLP:journals/corr/abs-2010-09063} (v0.6.2) & Machine learning & Google\\
			\cline{1-3}
		\end{tabular}
	}
	\caption{\label{table: tool list}DP tools with publicly available open-source repositories}
\end{table}


\subsection{Related work}

This section reviews existing works regarding performance evaluation of privacy tools that apply DP. While there have been numerous studies around DP, none, to the best of our knowledge, provide a comparative study between the various open-source tools that can be applied in practice, nor did they offer sufficient insight into how to apply and configure DP tools in privacy-preservation. 



\subsubsection{Statistical Queries}
Statistical queries in this work refer to the analysis of data to extract statistical features, \eg~the operation of \texttt{SUM}, \texttt{AVERAGE}, \texttt{COUNT}, and \texttt{HISTOGRAM}. One DP query engine that has been integrated into the industry is Flex~\cite{10.1145/3187009.3177733}. The open-source library for Flex is deprecated.\footnote{\url{https://github.com/uber-archive/sql-differential-privacy}} However, their evaluation remains relevant to provide clues on evaluating privacy tools.

Johnson~\etal~\cite{10.1145/3187009.3177733} evaluated Flex using an SQL-compatible interface, which makes it convenient to put the interface in front of any already deployed SQL-compatible database and, in turn, lowers the bar of adoption. Their evaluation uses a large set of real-world queries run by Uber's data engineers in production, which gives insights into how Flex would perform in the industry. However, this evaluation merely benchmarks one $\epsilon$ value ($\epsilon=0.1$) and presents only one single value of the additional overhead of 4.86ms corresponding to $0.03$ \% of the average execution time of their non-privacy protected queries. Furthermore, the 4.86ms overhead does not include the pre-collection of \textit{frequent join} attributes, which has to be updated each time the underlying data is updated. Such evaluating procedure induces the risk that the actual overhead might be more significant than presented by Johnson~\etal Besides, it is not clear whether overhead changes when settings differ, \eg in dataset sizes, privacy configurations or queries, etc.



The DP tools of Smartnoise~\cite{DBLP:journals/corr/abs-2011-05537} and Google DP~\cite{DBLP:journals/popets/WilsonZLDSG20} are integrated into private statistical-query services. Smartnoise results from years of cumulative experience building and deploying privacy tools for research and has recently become a tool in Microsoft's privacy ecosystem. However, even though the open-source software community provides transparency and demonstrates examples, no comparative research has been done on the tool's query engine.

Google DP has been evaluated alongside Flex~\cite{10.1145/3187009.3177733} and PinQ, which is a research project on Privacy Integrated Queries (PinQ). PinQ provides a programming language and execution platform in which all expressible programs satisfy DP~\cite{DBLP:journals/cacm/McSherry10, DBLP:journals/popets/WilsonZLDSG20}. The evaluation of Google DP included 1,000,000 runs with various aggregate functions, with a fixed $\epsilon$ value of $0.1$, where a benchmark TPC-H dataset was used. While in the comparison, Flex and PinQ were run only 10,000 times due to performance concerns. This evaluation points out that compared with Google DP, Flex and PinQ cannot enforce contribution bounds for databases where one single user can contribute multiple samples, leading to query results that are not DP. Furthermore, because Flex or PinQ assumes that the underlying database is associated with at most one record per user, their performance comparison with Google DP that supports the contribution of multiple samples can be problematic. The evaluation of Google DP merely considered a single value of privacy budget. Our extensive evaluation uses broad evaluation settings, \ie different privacy budgets and dataset sizes.

\subsubsection{Machine Learning}\label{related_work_machine_learning}

DP machine learning (ML) has gained attention at companies like Google, Facebook, and IBM. Investigations have been done on the performances of DP stochastic gradient descent (SGD)~\cite{DBLP:conf/ccs/AbadiCGMMT016}, which is a dominant algorithm for private training of ML models. Nevertheless, DP-SGD can increase the training time significantly compared to non-private SGD~\cite{DBLP:journals/corr/abs-2102-03013}.
In a recent study, Subramani \etal~\cite{DBLP:journals/corr/abs-2010-09063} reduced the run-time overhead when executing DP-SGD. They implemented the functionality in the open-source library of Tensorflow Privacy by exploiting language primitives~\cite{DBLP:journals/corr/abs-2102-03013}. Microsoft has shown support in the DP-ML field by implementing Opacus and demonstrating the impact of epsilon and dataset size on DP-ML~\cite{MicrosoftAzureWhitePapers}. To the best of our knowledge, we are the first to extensively compare these DP ML tools.


Beyond improvements regarding performance, there is also work to improve the privacy guarantees of DP-ML~\cite{DBLP:conf/uss/Jayaraman019} and evaluate different means of DP with different privacy budgets, which shows how the trade-off varies between utility and privacy protection under different settings. Their study focuses on gradient perturbation mechanisms, \eg~DP-SGD, and uses Tensorflow Privacy to evaluate R\'{e}nyi DP among others~\cite{DBLP:conf/csfw/Mironov17}. It demonstrates that the privacy guarantees of DP for ML implementations may provide unacceptable balances of the trade-off between utility and privacy protection. They aim to find epsilon values that balance the trade-off for different DP approaches rather than for different DP tools. In another study, Tram\'{e}r \etal~\cite{DBLP:conf/iclr/TramerB21} provide an approach to improving performances of DP models using primarily Tensorflow and parts of the Opacus library. They point out that prior works have underestimated guarantees of utility and privacy protection as well as demonstrated that solid privacy may come at only a nominal cost of inaccuracy by tailoring the training to the data. However, their results, to the best of our knowledge, have not yet been generalized as a tool that can be easily leveraged by the developer community. Our study of the trade-off between utility and privacy-protection considers different DP-ML tools as well as facilitates the selection of DP-ML tools.

\subsection{Our contribution}

We study a critical aspect of data systems, which is the protection of privacy-sensitive information. Our study evaluates open-source differential privacy (DP) tools and services. The area of privacy protection has a well-recorded history of failed solutions. DP is a leading framework that offers qualitative guarantees for the protection of privacy-sensitive information. 
%
%
The implementation of tools for providing DP solutions has its own set of traps and pitfalls since it is a non-trivial effort to assure that all cryptographic and system aspects are well-addressed. 
A successful approach for addressing this challenge is to focus on open-source solutions because they can be scrutinized by a large community of developers. To the best of our knowledge, we are the first to offer a comprehensive study that compares the performance of these tools from the application utilization and system perspectives.

Through this work, we develop an evaluation framework to compare privacy-preserving tools to get a nuanced picture of the trade-offs in a data analysis where the tools are used. The framework is implemented on Docker~\cite{DBLP:journals/sigops/Boettiger15} that is compatible with the dominant operating systems of Windows, Linux, and iOS and offers thus the flexibility to programmers for software reuse purposes. The designed framework uses the proposed evaluation criteria to quantify an analysis' utility loss and system overhead compared to a non-private benchmark. Using the devised framework, we evaluate the most relevant open-source tools and compare their performance on DP data analysis. Through the evaluation results, we provide insights into the studied DP tools. Our study can facilitate the selection of DP tools by developers according to their needs and use-cases. 


\begin{itemize}
	\item For developers looking at accumulating general statistics about categorical or continuous datasets, Google DP seems promising with a margin of error from about 0.1\% to 2\% for simple queries (\texttt{SUM}, \texttt{COUNT}, \texttt{AVG}) with $\epsilon \gtrapprox 1.5$. Given the same set of queries and $\epsilon$ values, Smartnoise provides an error of about 0.5\% to 5\%. We also note that Smartnoise offers better accuracy for \texttt{HISTOGRAM} queries, with an error below ca. 10\% compared to Google DP's about 15\%, on \emph{Health Survey}.
	
	\item For developers looking at building DP machine learning models, both Opacus and Tensorflow Privacy (TFP) show promising results, obtaining data utility below around 6\% error for $\epsilon \geq 1.0$ on \emph{Parkinson} and \emph{Health Survey}. Nevertheless, TFP manages to obtain around two times better data utility than Opacus, given maximum data size and $\epsilon=3.0$. Diffprivlib on the other hand outperforms both tools on \emph{Health Survey}, given maximum data size and $\epsilon=3.0$. It should however be noted that Diffprivlib did not manage to generate any useful results on continuous data, and is also limited to linear regression and logistic regression models, while TFP and Opacus offer building custom neural networks, allowing developers to build complex models for a variety of problems.
	
\end{itemize}

For the sake of supporting the scientific process in the area and further development, we release our work as open-source software.\footnote{\url{https://github.com/anthager/dp-evaluation}} This enables the reuse of our work in further evaluation of the considered tools and beyond. We hope that this work can bring a landscape of the pros and cons of existing open-source privacy tools and an intuitive knowledge to practitioners on how to leverage them in their privacy protection service development, and ultimately narrow the gap between theoretical and applied research on DP.


\section{Evaluation Settings}
\label{sec:eval}
We describe our evaluation plan by explaining our 
%
%
criteria (Section~\ref{evaluation_criteria}) as well as propose our evaluation framework in Section~\ref{sec:frame} and present our experiment implementation in Section~\ref{sec:imp}.


\begin{table}[htp]
	\centering
	\renewcommand\arraystretch{1.15}
	\resizebox{0.75\textwidth}{!}{
		\begin{tabular}{|l|l|l|l|}
			\hline
			\multicolumn{2}{|l|}{\diagbox[width=11em,trim=l]} & Machine learning domain & Statistical query domain \\
			\hline
			\multicolumn{2}{|l|}{Environment} & \multicolumn{2}{l|}{\tabincell{l}{$\bullet$~Docker}} \\
			\hline
			\multicolumn{2}{|l|}{Task to evaluate} & \tabincell{l}{$\bullet$~Regression} & \tabincell{l}{$\bullet$~Sum\\$\bullet$~Count\\$\bullet$~Average\\$\bullet$~Histogram} \\
			\hline
			\multicolumn{2}{|l|}{Influencing factors} & \multicolumn{2}{l|}{\tabincell{l}{$\bullet$~Dataset size\\$\bullet$~Privacy budget,~$\epsilon$}} \\
			\hline
			\multirow{5}{*}{\tabincell{l}{Evaluation\\criterion}} & Utility & \tabincell{l}{Prediction accuracy\\reduction compared\\ with non-private\\ machine learning.} & \tabincell{l}{Decreased query\\ accuracy due to\\ privacy measures.} \\
			\cline{2-4}
			& Overhead & \multicolumn{2}{l|}{\tabincell{l}{Extra resource consumption induced by differential\\ privacy including:\\$\bullet$~Execution time\\$\bullet$~Memory consumption}} \\
			\hline
			\multicolumn{2}{|l|}{Database to evaluate on} & \multicolumn{2}{l|}{\tabincell{l}{$\bullet$~U.S. health reform monitoring survey data\\$\bullet$~UCI Parkinson data set}} \\
			\hline
		\end{tabular}
	}
	\caption{\label{table: evaluation strategy}Evaluation strategies in this work}
\end{table}

\subsection{Evaluation criterion}\label{evaluation_criteria}

Information on individual data might be disclosed from data analyses. However, tools that prevent this disclosure by leveraging differential privacy might ultimately cost analysis accuracy and increase the use of system resources~\cite{MicrosoftAzureWhitePapers,DBLP:journals/corr/abs-2011-05537}. Therefore, our focus is to study the difference between the DP and non-privacy protected (NP) results that these tools produce and how this difference varies between different tools.

We adopt data utility (also referred to as accuracy) and system overhead as evaluating metrics for the considered tools. \textit{Utility} refers to whether the data is still useful to conduct a specific functionality after the data is perturbed with DP measures. \Ie~how much an outcome deviates from the actual quantity it attempts to estimate. \Eg~what degree of accuracy reduction incurs when querying on the perturbed dataset compared with that on the original dataset. We define the system's \textit{overhead} as the additional time and memory it takes to complete a DP query or train a DP-ML model, versus the non-privacy protected (NP) query or ML model. Overhead is further divided into two metrics memory overhead and run-time overhead, which will be detailed below.

The metrics of \textit{utility} and \textit{overhead} illustrate what deviations can be expected from the DP results of the tools and allow for comparison amongst tools' performances. The criterion \emph{utility} is quantified by the deviation of the prediction error over several runs of the same experiment, using the \emph{Root Mean Square Percentage Error} (RMSPE). This is essentially the percentile difference between the DP and NP results shown in Equation~\ref{eq_rmpse}, where $N$ is the number of experiment runs, \emph{NP} is the NP benchmark result, and \emph{DP} is the DP result.

\begin{equation}\label{eq_rmpse}
\mbox{\large$\mbox{RMPSE} = \sqrt{\frac{\sum\limits_{n=1}^{N}\left(\frac{\mbox{\emph{NP - DP}}}{\mbox{\emph{NP}}}\right)^{2}}{N}} \cdot 100\ \%,$}
\end{equation}

\emph{Memory overhead} is measured by comparing the worst-case memory usage between DP and NP query/ML tasks to guarantee minimum system requirements for the tools. \Ie~the measurement shows the percentile difference between the worst-case memory usage of DP vs. NP query/ML results. Specifically, memory usage is recorded when the tools conduct DP and NP query/ML tasks. This criterion of memory overhead is considered in this evaluation since it can be noticed by users who care usability of the privacy tools. \Eg~lower memory usage ultimately improves speed due to less paging, fewer cache misses, and faster structure traversals, and it also improves stability by reducing virtual and physical out-of-memory aborts. Moreover, for specific tools that use an external database, including Smartnoise and Google DP, the memory usage of the database container is also recorded. This recording will show how the tools affect the memory usage during load on the database since we are running queries on a high frequency during this process.

\emph{Run-time overhead} is measured by comparing the time passed before and after entering the critical section, \ie~the part where the tool does an ML task or runs a query. To minimize external impact and obtain reliable and comparable results throughout the experiments, we aim to eliminate operations like initialization or saving results to the highest extent. Similar to the memory usage, we get the difference between the time passed before and after, comparing the DP and NP time usage. However, instead of comparing the maximum run-time, we average the results by applying RMSPE shown in Equation~\ref{eq_rmpse}.

\subsection{Evaluating framework}
\label{sec:frame}
\begin{figure*}[t!]
	\centering
	\makebox[\textwidth][c]{\includegraphics[width=0.9\textwidth]{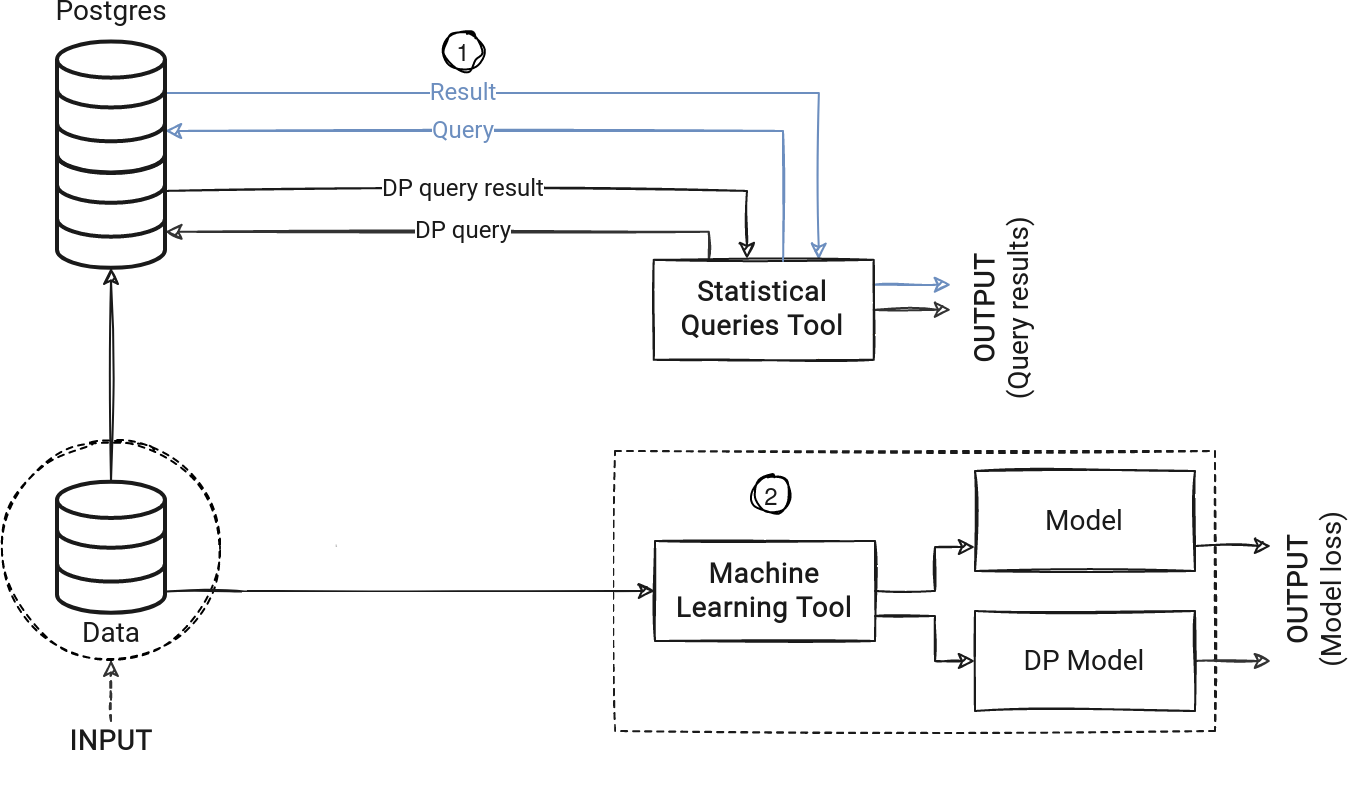}}
	\caption[A high-level overview of experiment flow]{A high-level overview of the experiment flow, showing {(1)} DP queries and non-privacy protected (NP) queries being conducted on Postgres and {(2)} ML tasks conducted for DP models and non-privacy protected models.}
	\label{fig_experiment_flow}
\end{figure*}

We construct our evaluating framework in pursuit of insights into the impact of DP tools' privacy measures on accuracy and system resource usage. To this end, we measure the difference between DP and NP results regarding our evaluation metrics: data utility and system overhead. In addition, we vary two parameters that affect these metrics, the dataset size (Table~\ref{table_dataset_sizes}) and privacy budget (Table~\ref{table_epsilons}), which is denoted by $\epsilon$, to further illuminate the privacy-utility trade-off induced by differential privacy measures.

We study the impact of $\epsilon$ and dataset size since they are the parameters that trade-off between privacy protection and data utility. We illuminate this trade-off to provide developers or DP service practitioners, \eg~healthcare institutions and companies, with practical results to make educated choices when applying these tools. The set of sizes for considered datasets where the evaluation experiments are conducted is listed below.

\begin{table}[!h]
	\centering
	\resizebox{0.75\textwidth}{!}{
		\begin{tabular}{c}
			\multicolumn{1}{c}{\bfseries Dataset sizes} \\
			\hline
			Health Survey size $\in$ \{1000, 2000, 3000, 4000, 5000, 6000, 7000, 8000, 9000, 9358\}\\
			Parkinson size $\in$ \{1000, 2000, 3000, 4000, 5000, 5499\}\\
			\hline
		\end{tabular}
	}
	\caption{List of dataset sizes for each dataset.}
	\label{table_dataset_sizes}
\end{table}

Our selection of $\epsilon$ values in the evaluation into takes consideration of recommendations both from research works and practical settings in the industry. While analytical research has evaluated DP algorithms using a privacy budget ranging from $0.01$ to $7$, practitioners prefer a narrow scale. \Eg~Apple Health in iOS \texttt{10.2} use $\epsilon=2$ for gathering what health data types are being edited by users.\footnote{\url{https://www.apple.com/privacy/docs/Differential\_Privacy\_Overview.pdf}} Microsoft, in collaboration with OpenDP, explains in their product \emph{Azure} that privacy budgets are typically set between $1$ and $3$ to limit the risk of re-identification.\footnote{\url{https://docs.microsoft.com/en-us/azure/machine-learning/concept-differential-privacy\#differential-privacy-metrics}} They state that $\epsilon$ values below $1$ provide full plausible deniability and that values above $1$ come with a higher risk of disclosing the actual data.

In order to cover a wide range of privacy-utility trade-off results, we use the practice of $\epsilon$ as guidelines for selecting a set of $\epsilon$ in our evaluation experiments. The set of $\epsilon$ values considered is listed in Table~\ref{table_epsilons}.

\begin{table}[!h]
	\centering
	\resizebox{0.65\textwidth}{!}{
		\begin{tabular}{c}
			\multicolumn{1}{c}{\bfseries $\epsilon$ values} \\
			\hline
			$\epsilon \in$ \{0.1, 0.25, 0.5, 0.75, 1.0, 1.25, 1.5, 1.75, 2.0, 2.25, 2.5, 2.75, 3.0\}\\
			\hline
		\end{tabular}
	}
	\caption{The privacy budget, $\epsilon$, values}
	\label{table_epsilons}
\end{table}

Since different DP tools function and conduct computation with different techniques, we group the tools by the service that they offer into two domains, \emph{statistical queries} and \emph{machine learning}. This categorization allows for a reasonable comparison of tool performance within each group of tools.

For the evaluation of statistical query tools, we select a set of queries to conduct on each column in the dataset as listed in Table~\ref{table_queries}. Note that the \texttt{HISTOGRAM} queries are only conducted on columns composed of categorical values since the statistics of the categorical values can be sorted into buckets. For machine learning tools, we carry out linear regression tasks since regression is the only service that all the considered ML tools have in common.

\begin{table}[!h]
	\centering
	\begin{tabular}{c}
		\multicolumn{1}{c}{\bfseries Queries} \\
		\hline
		Queries $qs \in$ \{\texttt{SUM, AVG, COUNT, HISTOGRAM}\}\\
		\hline
	\end{tabular}
	\caption{List of queries.}
	\label{table_queries}
\end{table}

To clarify how we conduct our experiments for the different tools, we present a high-level diagram shown in Figure~\ref{fig_experiment_flow} illustrating how data flows through the different tools and generates results.

\subsection{Experiment implementation}
\label{sec:imp}
\begin{figure}[!ht]
	\centering
	\makebox[\textwidth][c]{\includegraphics[width=0.65\textwidth]{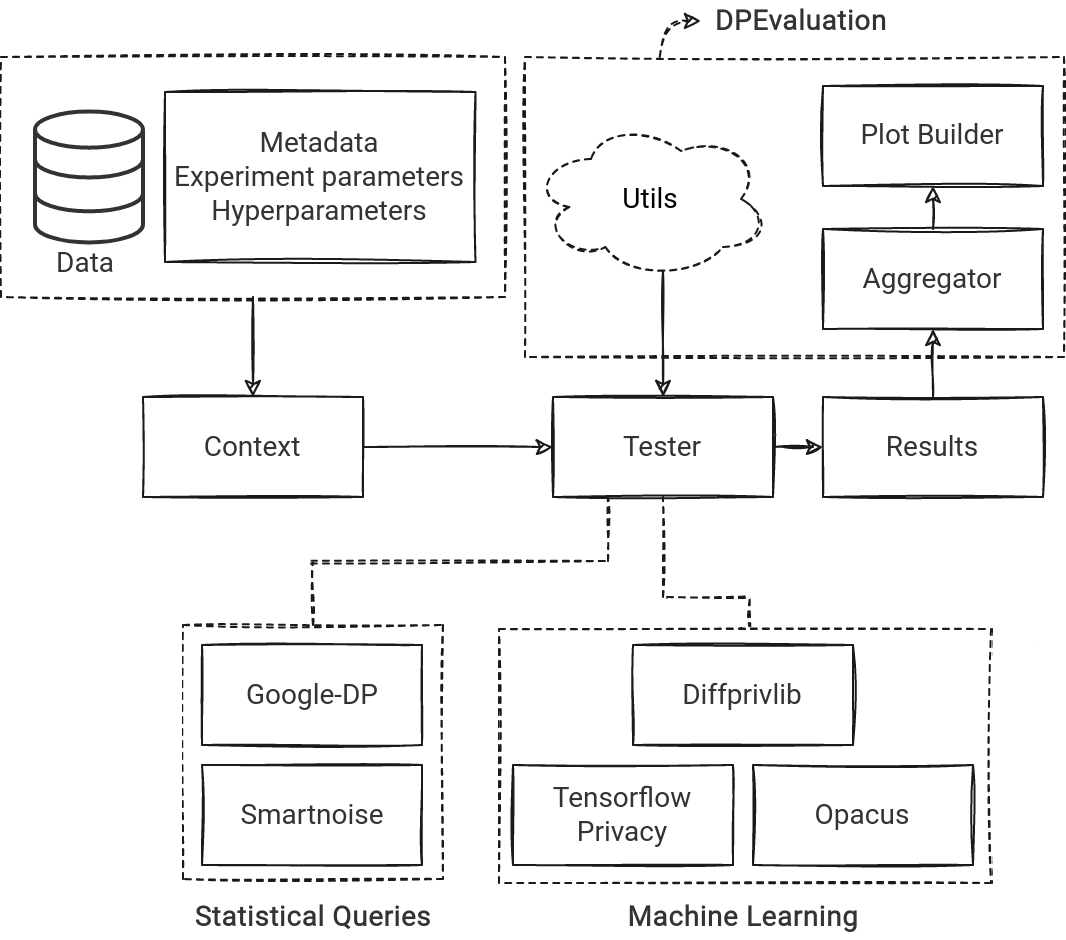}}
	\caption[High-level overview of evaluation framework]{High-level overview of experiment implementation, showing how data flows in the evaluation framework.}
	\label{fig_system_architecture}
\end{figure}

We aim to enable the reuse of our framework to the full extent and allow users to develop and test our code in any environment without installing and handling dependencies locally. Therefore, we build a collective framework for all the tools, focusing on usability and portability.

To make sure that our code works across different environments, we develop the evaluation framework on Docker~\cite{DBLP:journals/sigops/Boettiger15} that packages each tool \texttt{package} and its dependencies in a virtual container. Docker also eases the memory usage measurements by exposing a RESTful API running on the host system through a UNIX socket, from which metrics such as memory usage can be fetched.

Each tool evaluated is packaged in its own docker image together with all the packages the tool depends on, along with the framework code. This implementation empowers evaluations using the framework to run on any Linux, Windows, or macOS computer. The complete evaluation, therefore, only requires Docker version \texttt{20.10.6} or higher.

To visualize how the framework is constructed, we present a diagram in Figure~\ref{fig_system_architecture}. It shows the dataflow in the framework as follows: (1) Data and Metadata (with the experiment- and hyper parameters) are loaded in \texttt{Context} class, which is (2) initialized in \texttt{Tester} script for the respective tool. (3) \texttt{Tester} makes use of various utils from a collective package \texttt{DPEvaluation}. (4) The results are saved after running \texttt{Tester}, whose outcomes are collected by (5) \texttt{Plot Builder} with \texttt{Aggregator}, both of which are parts of the collective utils package.

\section{Evaluation Results}
\label{sec:res}
This section presents the results of our evaluation, where we show to what level prevailing functionality is affected by the use of differential privacy (DP). The results of statistical queries appear in Section~\ref{sq:results} and machine learning results appear in Section~\ref{ml:results}. We summarize the results in Section~\ref{sec:summary}, where we search for emerging patterns that are related to the tools' performance.  

%
%
%


\subsection{Statistical tools assessment}\label{sq:results}

This section evaluates two tools with statistical query services, namely Google Differential Privacy (Google DP) and OpenDP Smartnoise, using the two considered databases as shown in Table~\ref{table_dataset_sizes}. The evaluation varies the privacy budget $\epsilon$ and data size during the experiments and looks into how the query results of \texttt{SUM}, \texttt{AVERAGE}, \texttt{COUNT}, and \texttt{HISTOGRAM} change when the DP mechanism is integrated. In the evaluation, each query runs $20$ times, of which two extreme results are removed, and the remaining 18 are averaged for analysis.

Generally, the results reflect the trend that utility increases given larger $\epsilon$ values and data sizes for both tools, yet no obverse connections among memory overhead, $\epsilon$ values, and data sizes can be observed. There also exists an opposite impact of data size on the run-time overhead for the two tools that, while larger data size brings an increase in run time for Smartnoise, it acts conversely for Google DP, though significant irregularities exist. In comparing the two query tools, we observe that Google DP offers better query accuracy than Smartnoise in all query types except \texttt{HISTOGRAM} and that Google DP induces significantly less run-time than Smartnoise when comparing their DP queries against the benchmark. The overall results indicate an advantage of Google DP on Smartnoise under limited conditions. We detail the quantified evaluation results as follows.

\subsubsection{Data utility}\label{sq:exp1}

The evaluation in this part studies the differential privacy (DP) query tools' impact on utility by comparing how DP query results differ from non-privacy-protected ones using different settings.
We anticipate that a higher privacy budget, $\epsilon$, and a larger data size will provide better utility since less noise is injected under such conditions. However, as detailed below, though the experimental results match our anticipation, we notice local irregularities, \eg~in the case of \texttt{HISTOGRAM} queries.

\begin{figure}[!h]
	\centering
	\subfloat[]{\label{fig:exp1:gdp:sum:H}\includegraphics[width=0.25\textwidth]{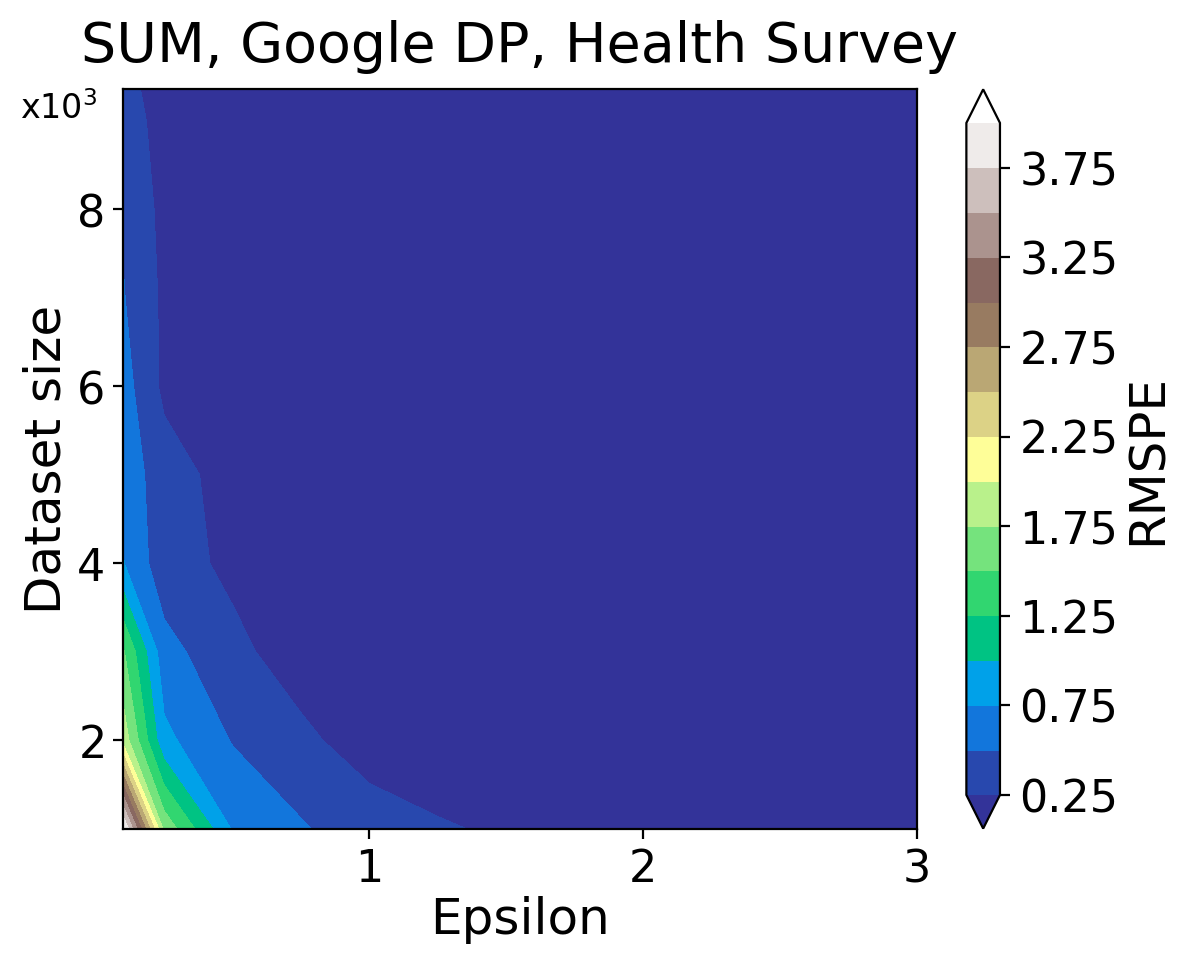}}
	\subfloat[]{\label{fig:exp1:gdp:count:H}\includegraphics[width=0.25\textwidth]{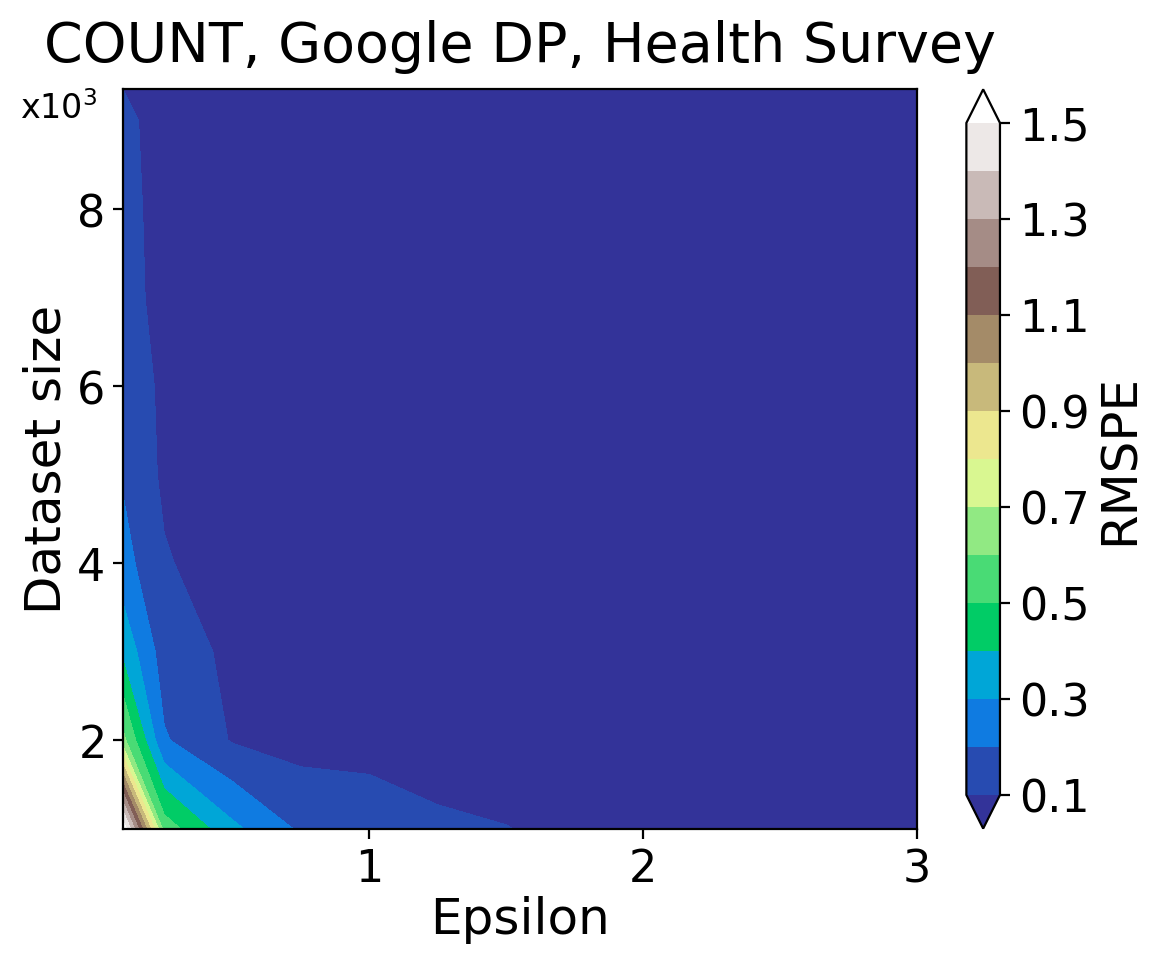}}
	\subfloat[]{\label{fig:exp1:gdp:avg:H}\includegraphics[width=0.25\textwidth]{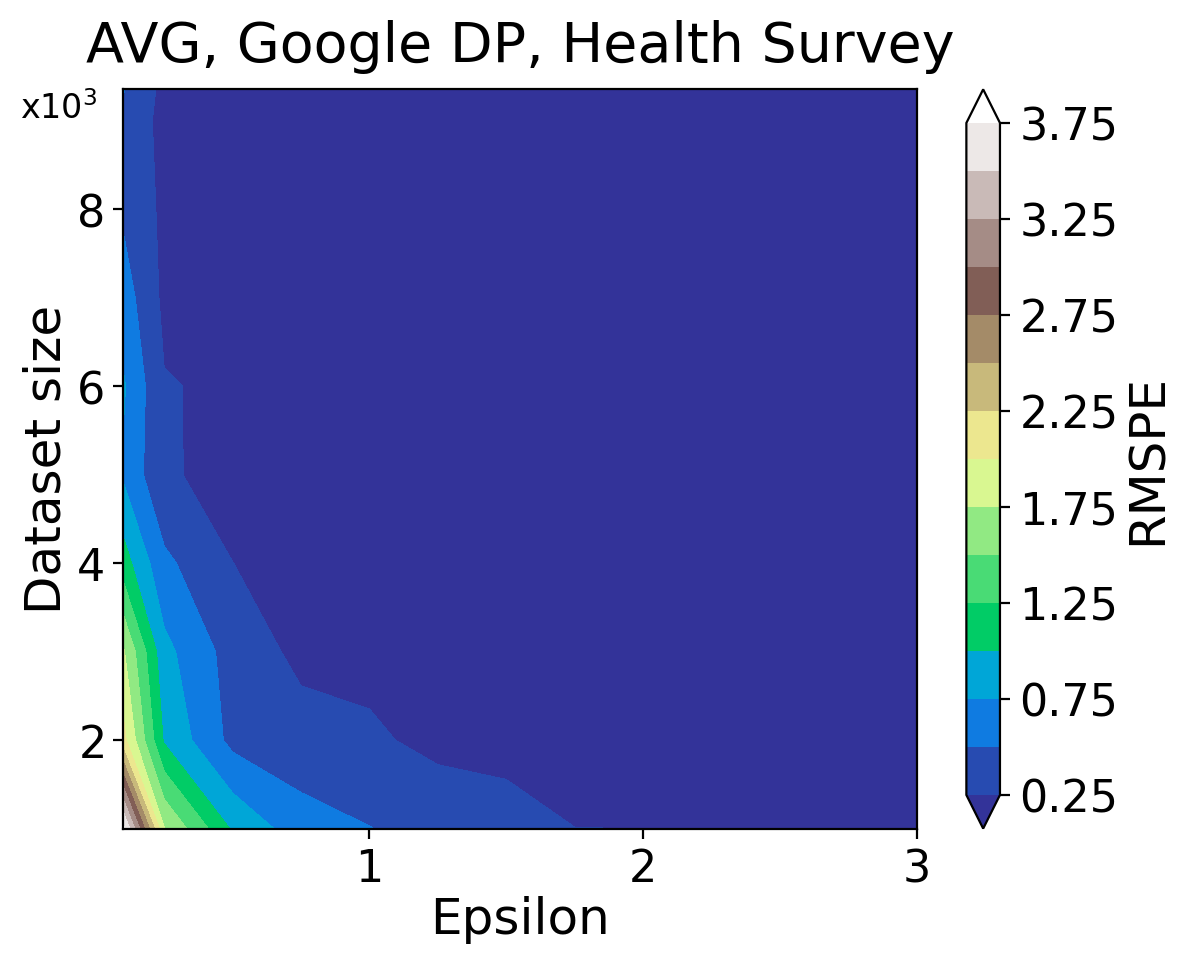}}
	\subfloat[]{\label{fig:exp1:gdp:hist:H}\includegraphics[width=0.25\textwidth]{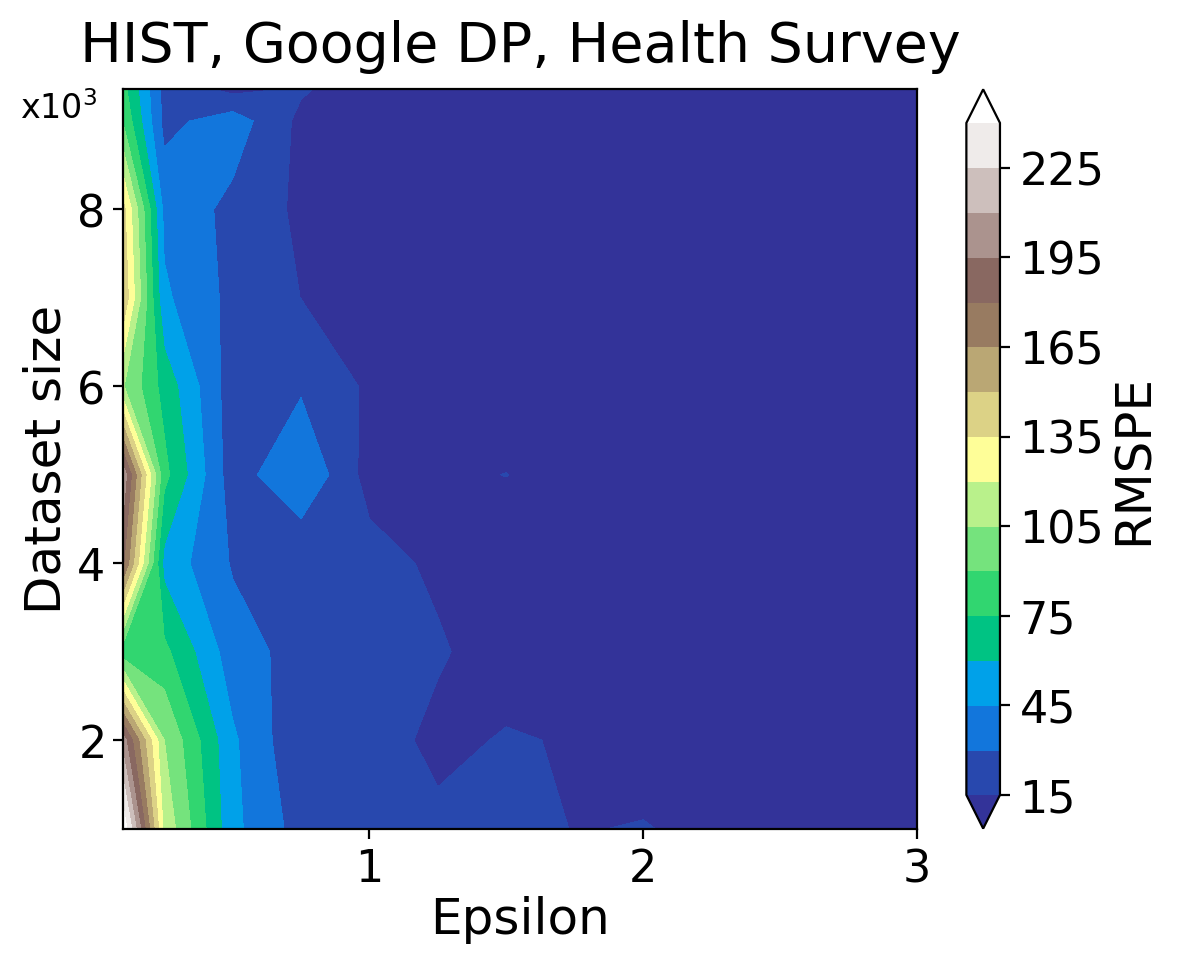}}
	\\
	\subfloat[]{\label{fig:exp1:smart:sum:H}\includegraphics[width=0.25\textwidth]{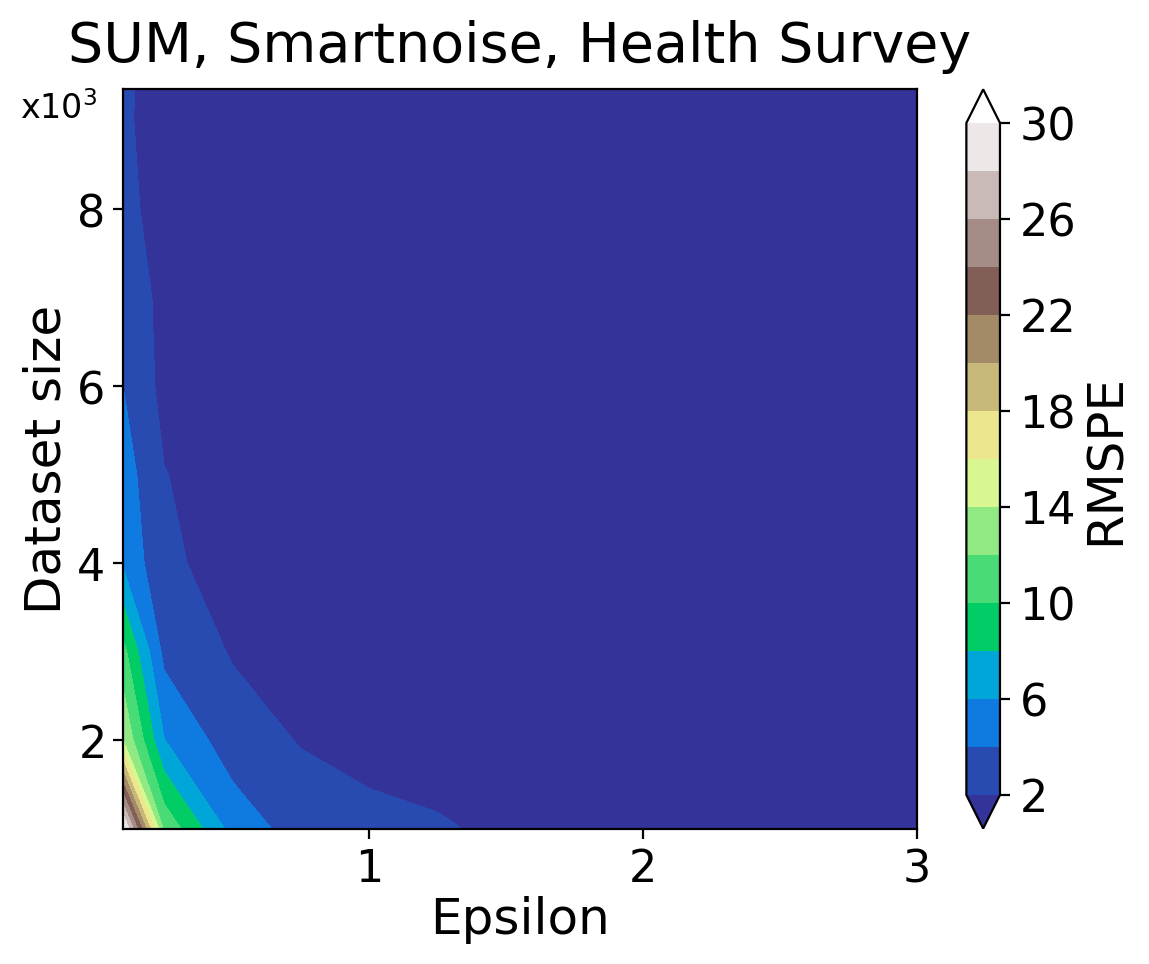}}
	\subfloat[]{\label{fig:exp1:smart:count:H}\includegraphics[width=0.25\textwidth]{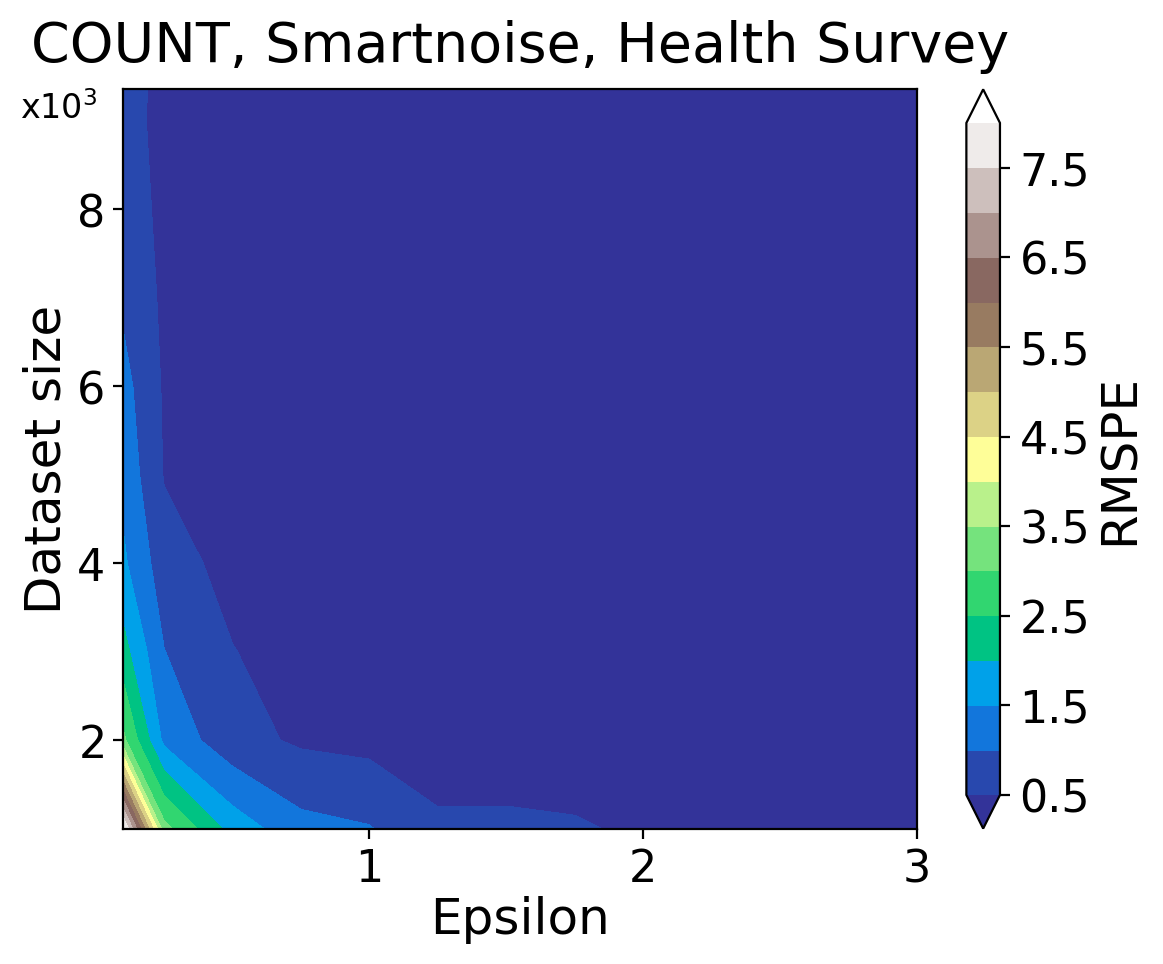}}
	\subfloat[]{\label{fig:exp1:smart:avg:H}\includegraphics[width=0.25\textwidth]{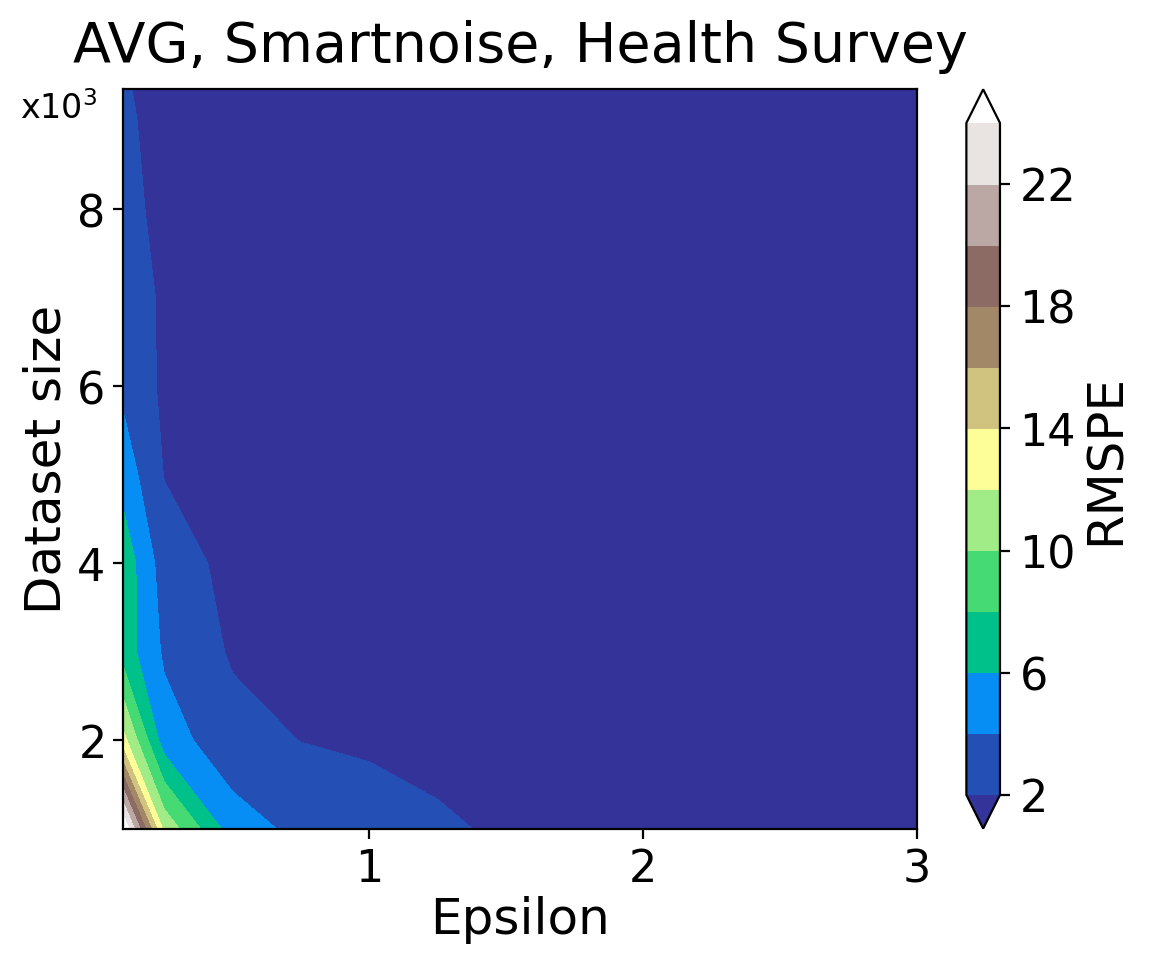}}
	\subfloat[]{\label{fig:exp1:smart:hist:H}\includegraphics[width=0.25\textwidth]{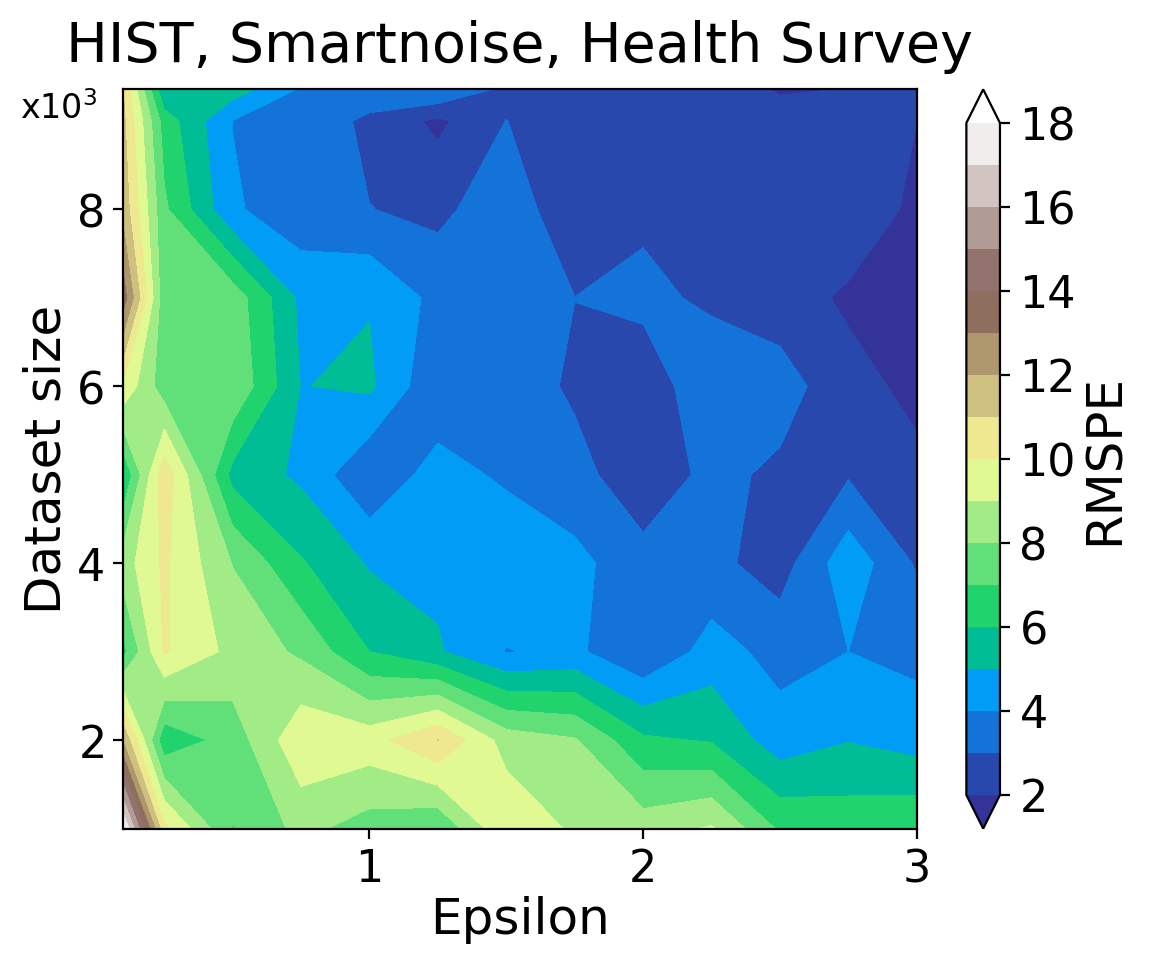}}
	\\
	\subfloat[]{\label{fig:exp1:gdp:sum:P}\includegraphics[width=0.25\textwidth]{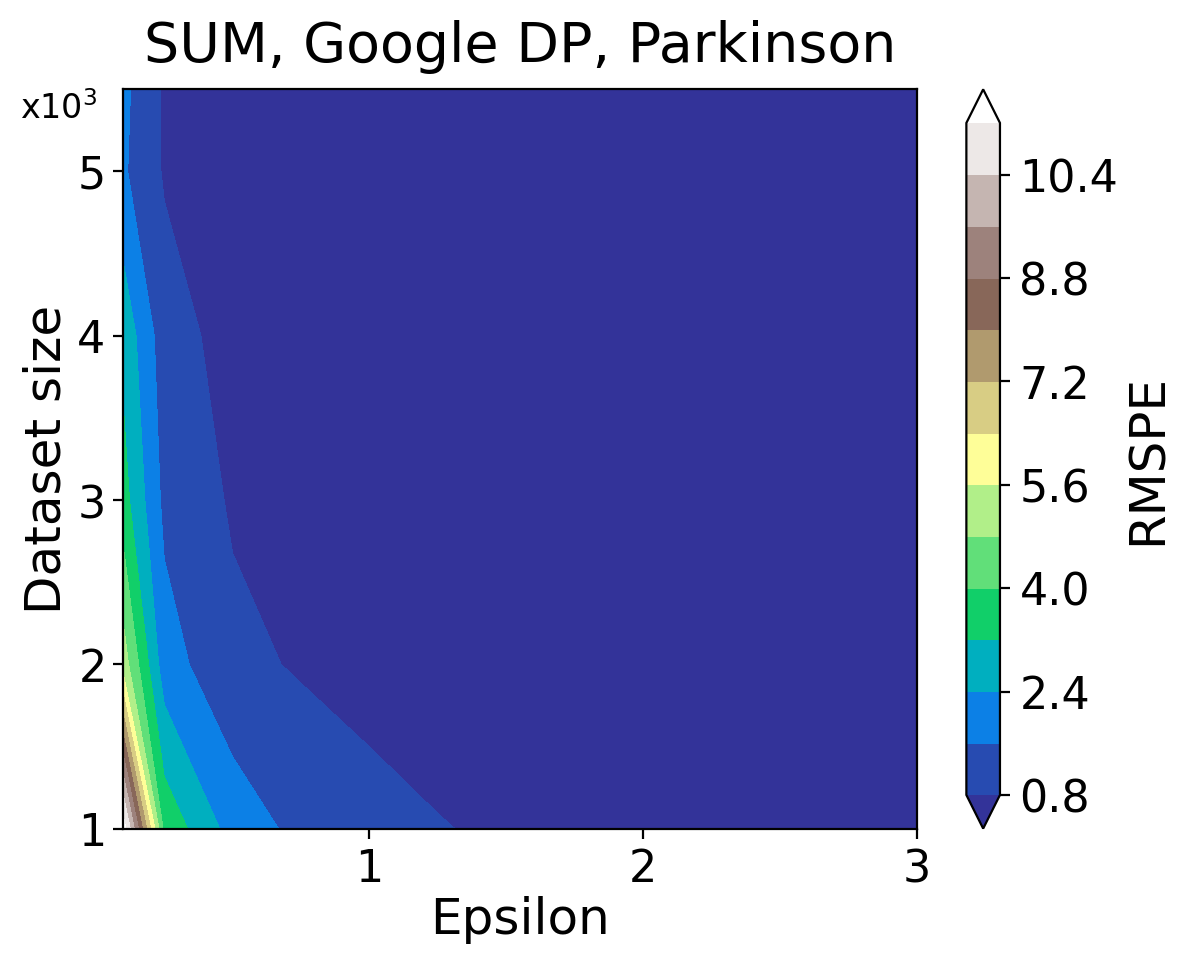}}
	\subfloat[]{\label{fig:exp1:gdp:count:P}\includegraphics[width=0.25\textwidth]{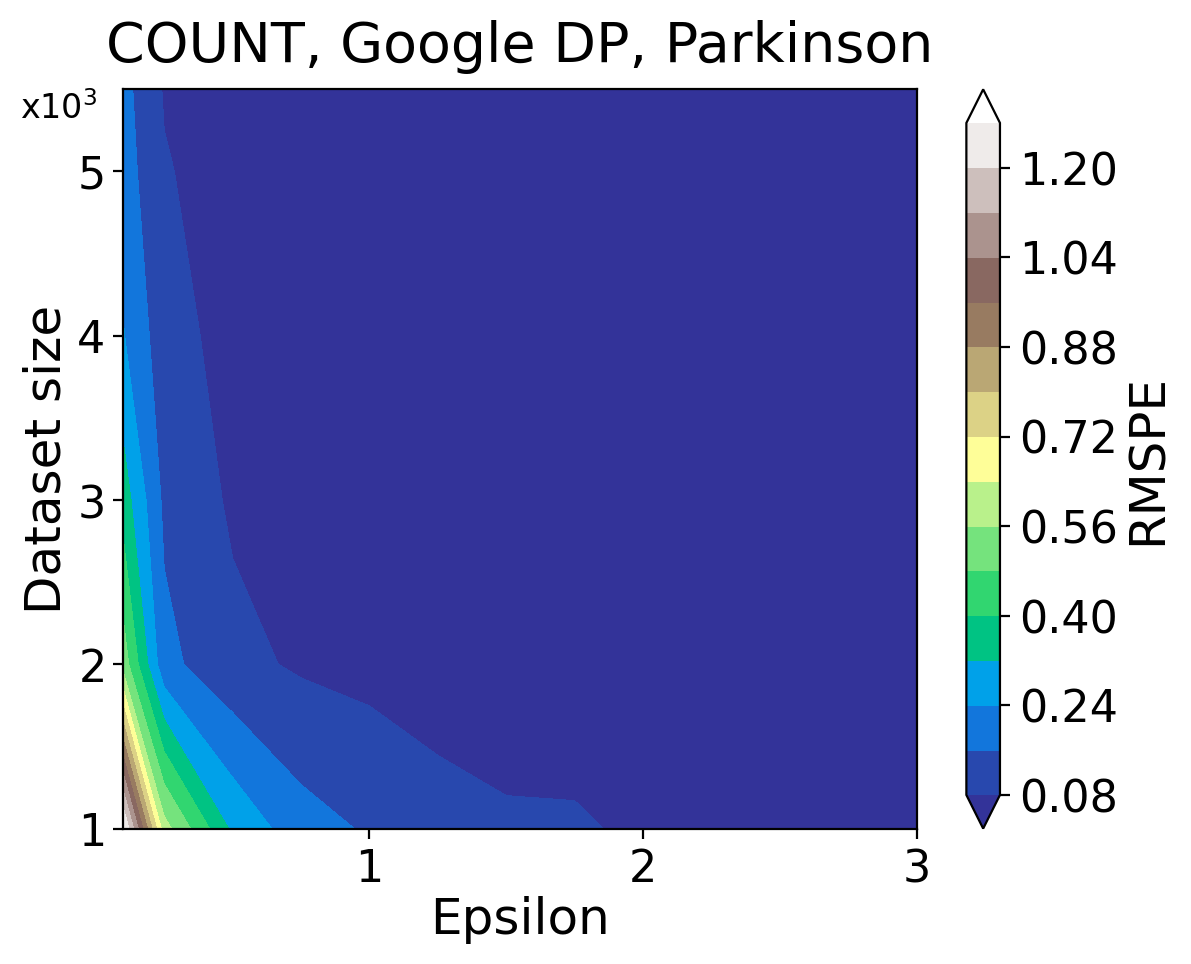}}
	\subfloat[]{\label{fig:exp1:gdp:avg:P}\includegraphics[width=0.25\textwidth]{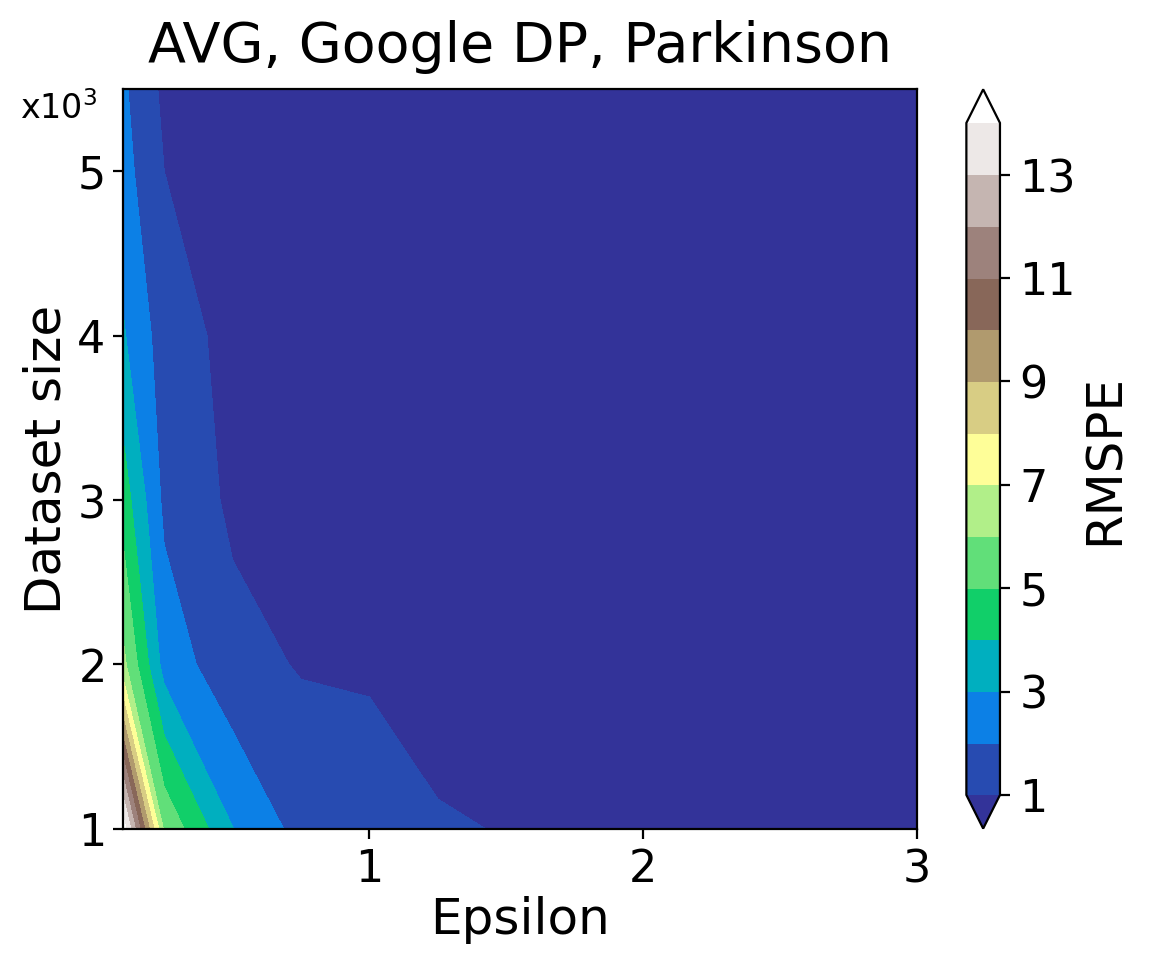}}
	\subfloat[]{\label{fig:exp1:gdp:hist:P}\includegraphics[width=0.25\textwidth]{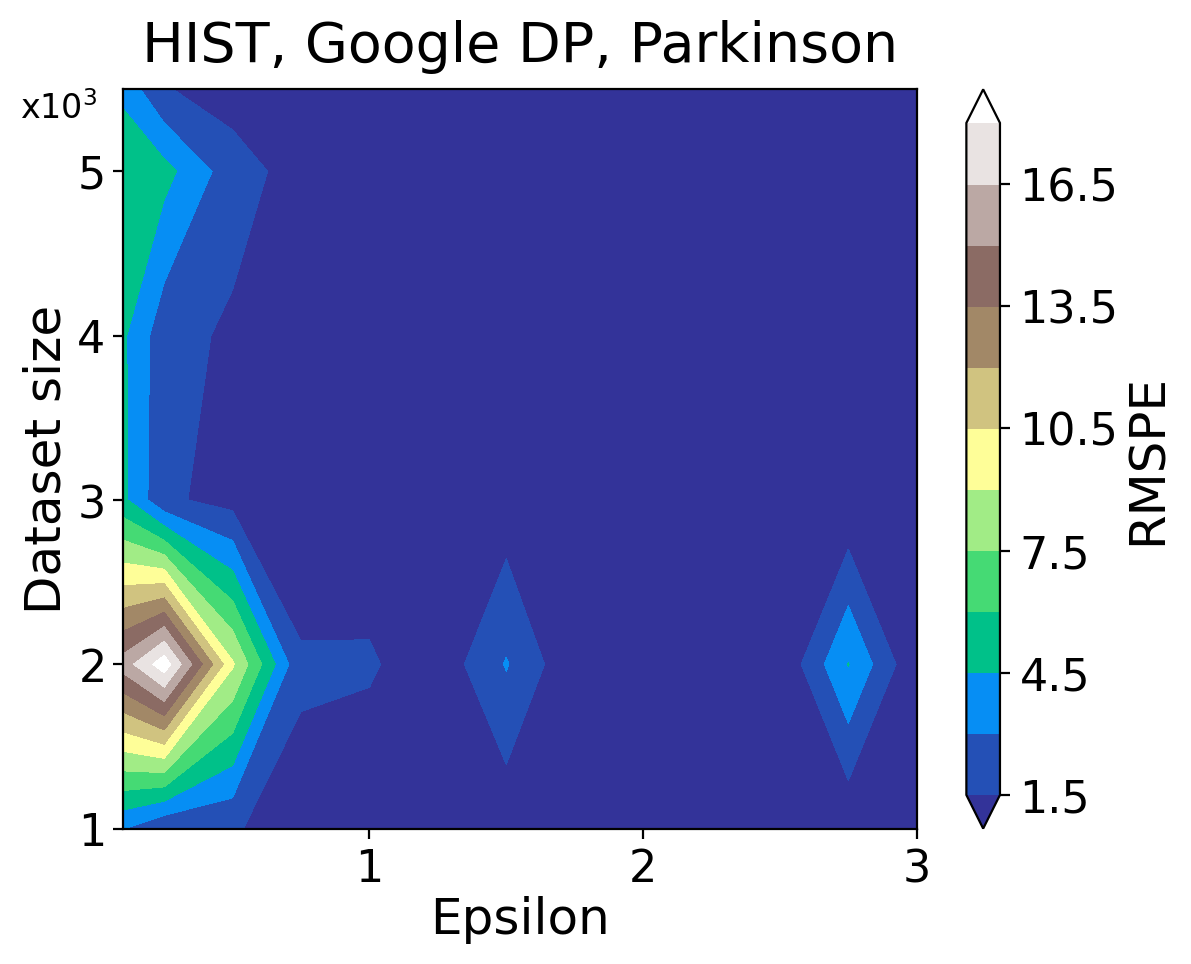}}
	\\
	\subfloat[]{\label{fig:exp1:smart:sum:P}\includegraphics[width=0.25\textwidth]{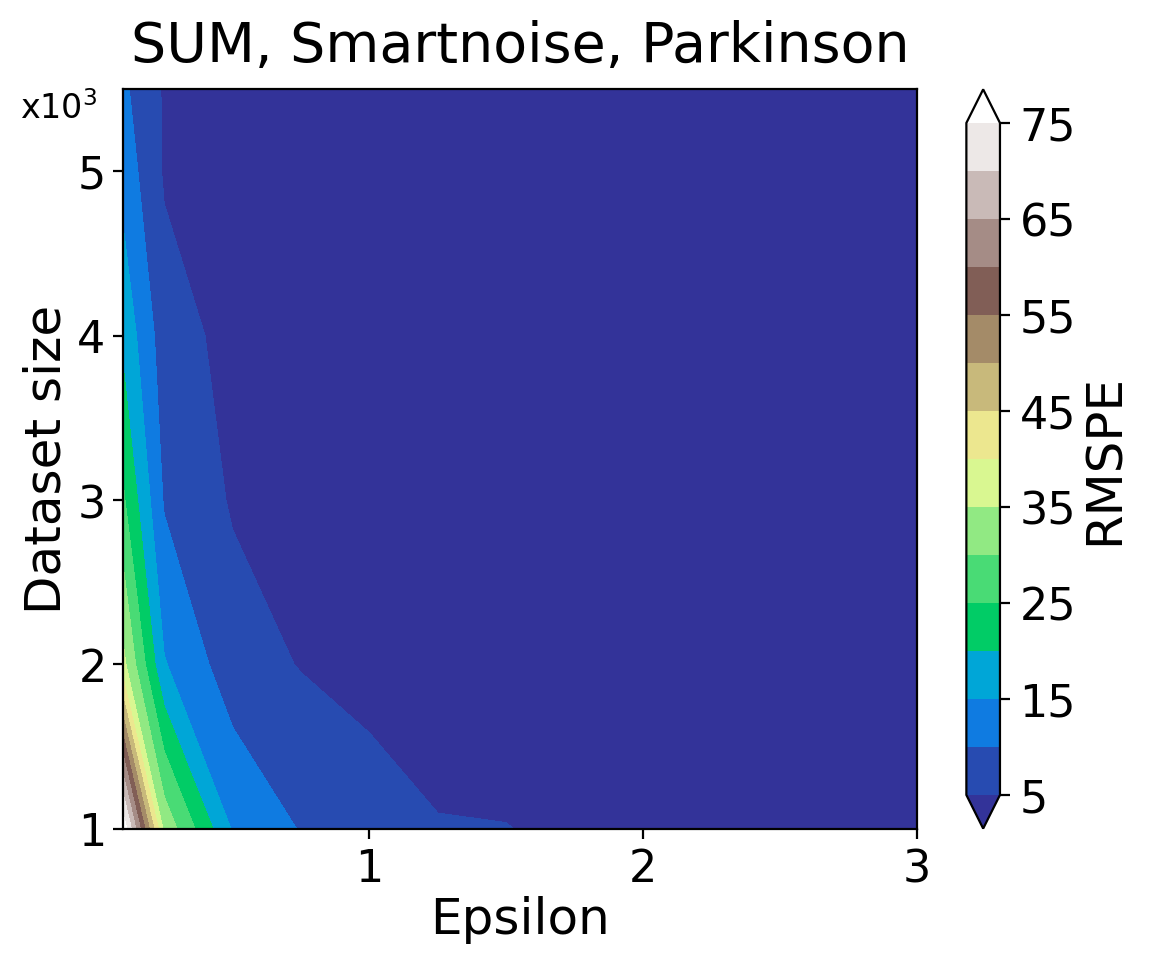}}
	\subfloat[]{\label{fig:exp1:smart:count:P}\includegraphics[width=0.25\textwidth]{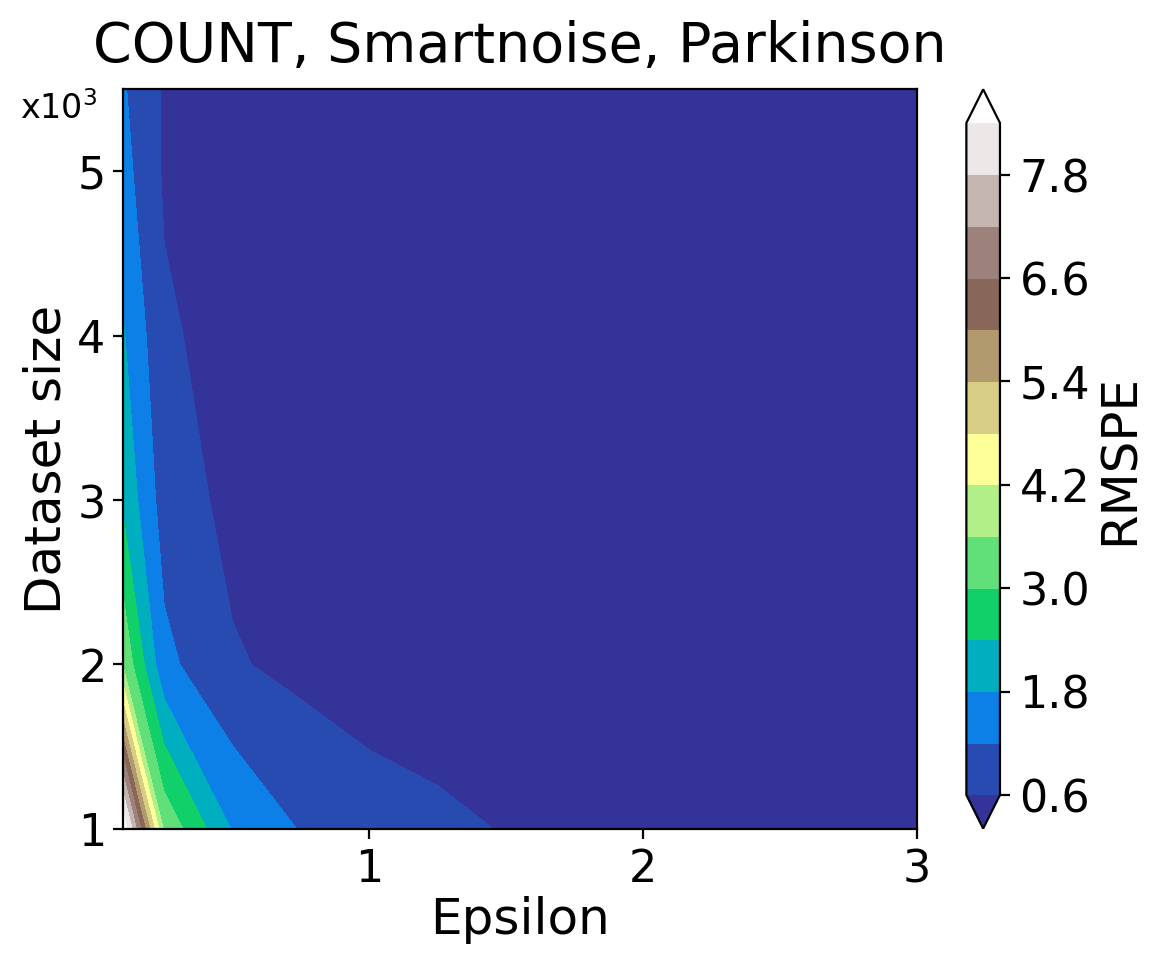}}
	\subfloat[]{\label{fig:exp1:smart:avg:P}\includegraphics[width=0.25\textwidth]{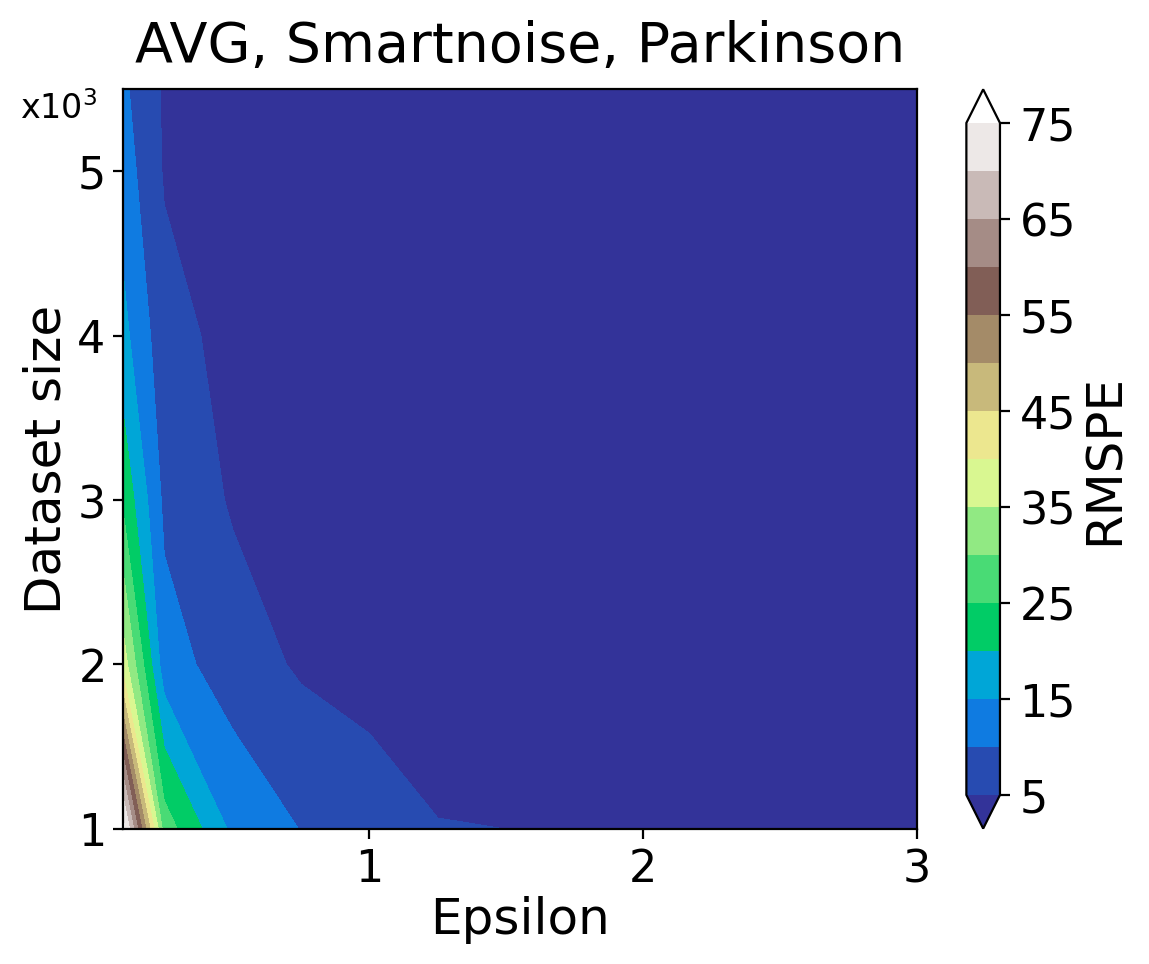}}
	\subfloat[]{\label{fig:exp1:smart:hist:P}\includegraphics[width=0.25\textwidth]{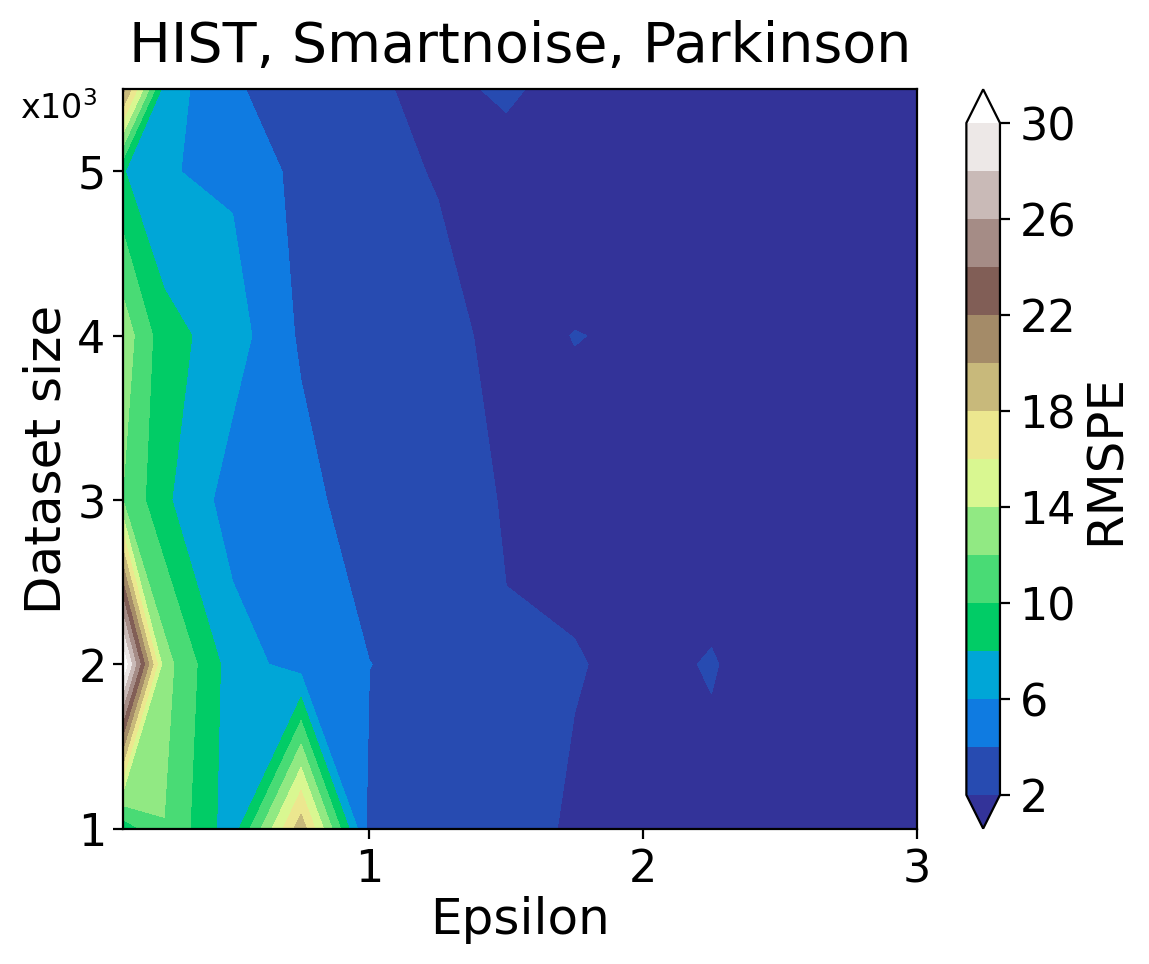}}
	
	\caption[Results of Experiment 1. Data utility for statistical query tools]{Contour plots for the evaluation of statistical query tools on utility when DP is integrated, for different data sizes (Table \ref{table_dataset_sizes}), $\epsilon$ values (Table \ref{table_epsilons}), and queries (Table \ref{table_queries}). RMSPE is defined in Section~\ref{evaluation_criteria}.}
	
	\label{fig:exp1:contour}
\end{figure}

Figure~\ref{fig:exp1:contour} shows contour plots for each tool's performance regarding different $\epsilon$ values and data sizes. Darker shades of blue in the plots indicate a lower RMSPE, corresponding to higher utility. Conversely, lighter shades indicate a larger RMSPE, corresponding to lower utility. Note that the plots have different RMSPE scales, implying that a shade in one plot (which indicates an RMSPE value) does not necessarily correspond to the same shade in another plot.

The contour plots demonstrate that, for simple queries of \texttt{COUNT}, \texttt{SUM}, and \texttt{AVG}, Google DP bears RMSPE between 0.1\% and 20\%, indicating 0.1\%-20\% worse than the benchmark, while Smartnoise performs between 0.2\% and 350\% over the considered parameter ranges of $\epsilon$ and data size. However, Smartnoise performs better than Google DP on the \texttt{HISTOGRAM} queries with RMSPE between 0.5\% and 60\%, compared to Google DP's between 0.2\% and 250\%. Such results imply Google DP's advantage over simple query types while converse in the \texttt{HISTOGRAM} query. We also observe that \texttt{HISTOGRAM} queries generally have a more significant impact on data utility than other types of queries for Google DP and Smartnoise. A possible reason might be that \texttt{HISTOGRAM} queries expose more information about the dataset properties; as a result, more noise is injected into the \texttt{HISTOGRAM} results to guarantee the privacy of individuals, which in turn reduces data utility.

Moreover, the results show that \texttt{HISTOGRAM} queries obtain better results on \emph{Parkinson} than on \emph{Health Survey} for both tools, which is expected since the categorical columns in \emph{Parkinson} have fewer bins than \emph{Health Survey} data, thus exposing less information about \emph{Parkinson} data properties. Consequently, less noise is injected into the results on \emph{Parkinson} data in pursuing individuals' privacy and higher query accuracy is obtained.

\begin{figure}[!h]
	\centering
	\subfloat[]{\label{fig:exp1:gdp:H:7000}\includegraphics[width=0.25\textwidth]{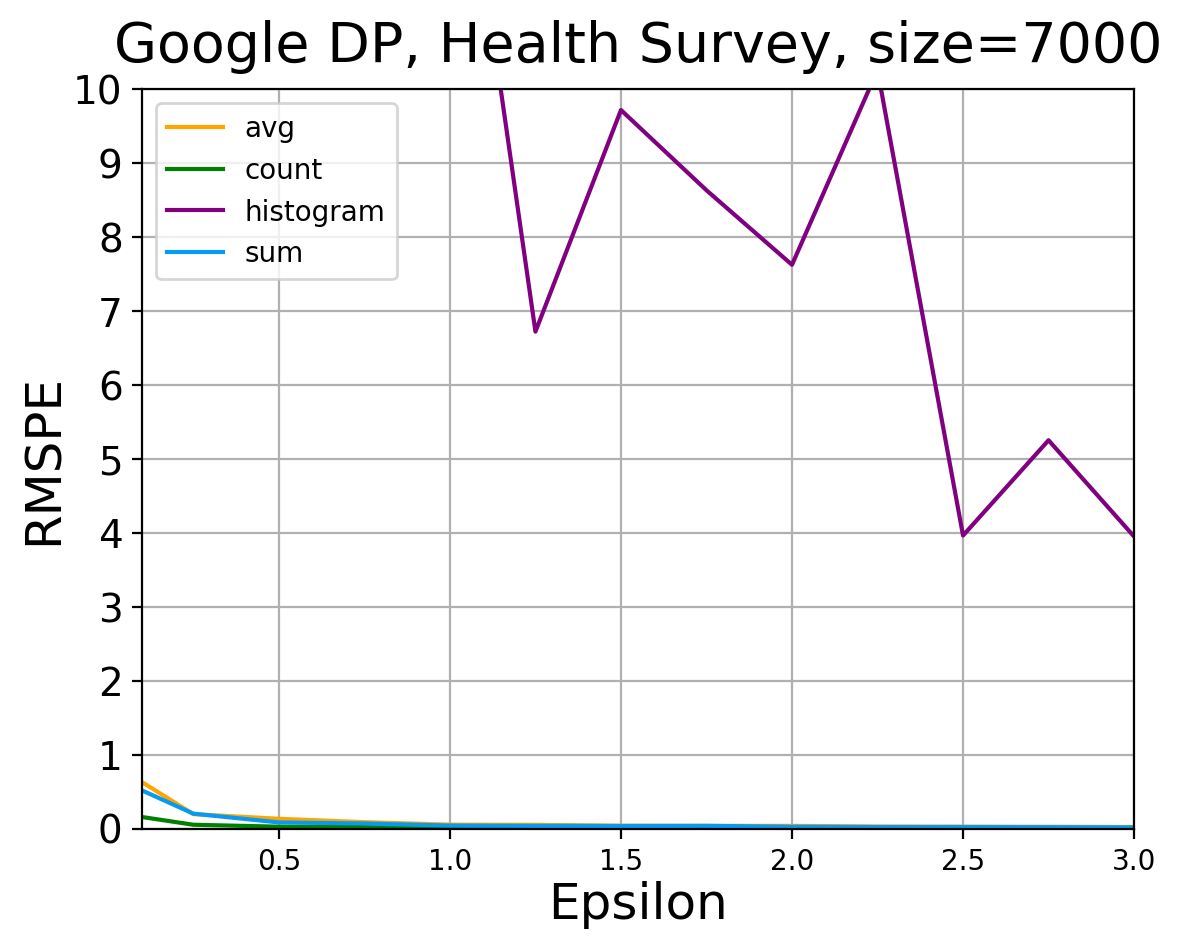}}
	\subfloat[]{\label{fig:exp1:gdp:H:8000}\includegraphics[width=0.25\textwidth]{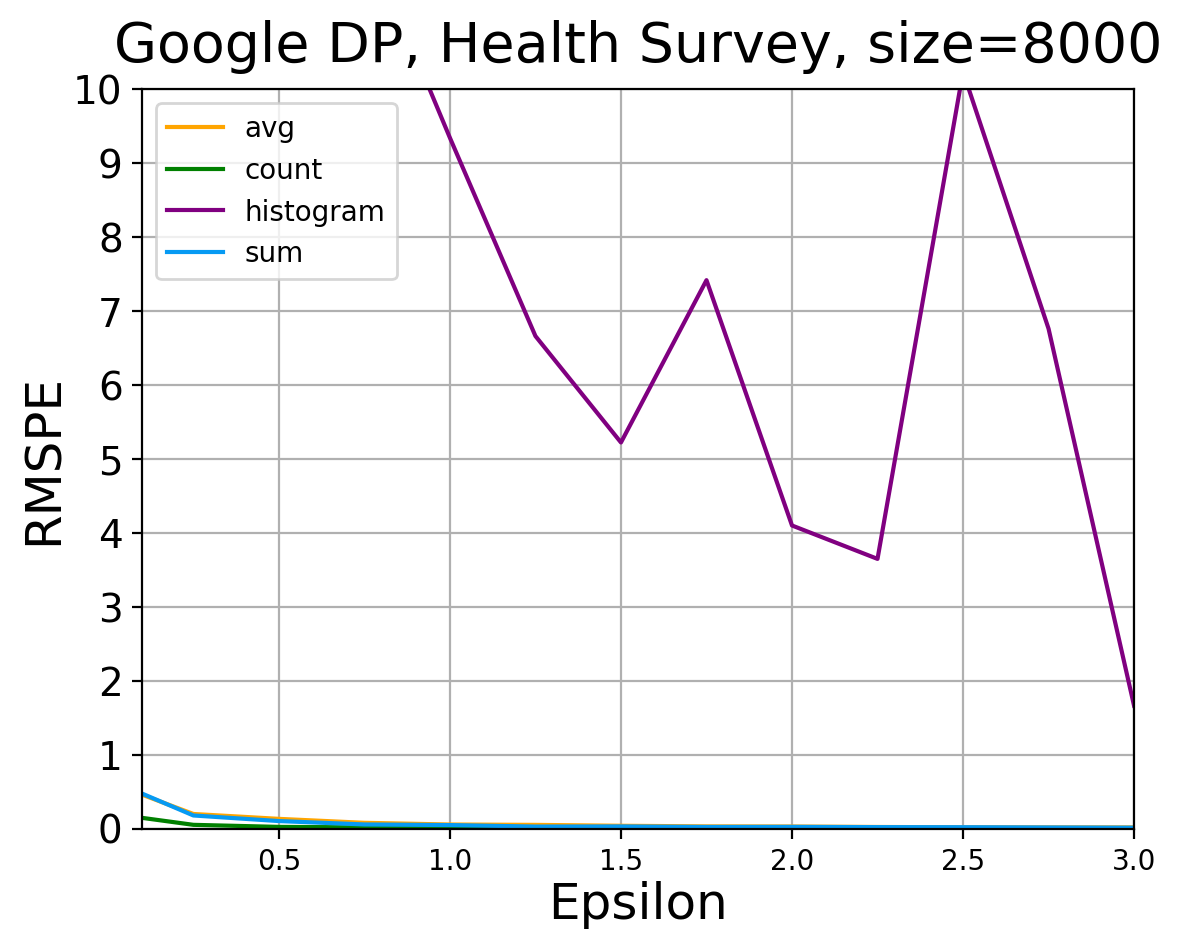}}
	\subfloat[]{\label{fig:exp1:gdp:H:9000}\includegraphics[width=0.25\textwidth]{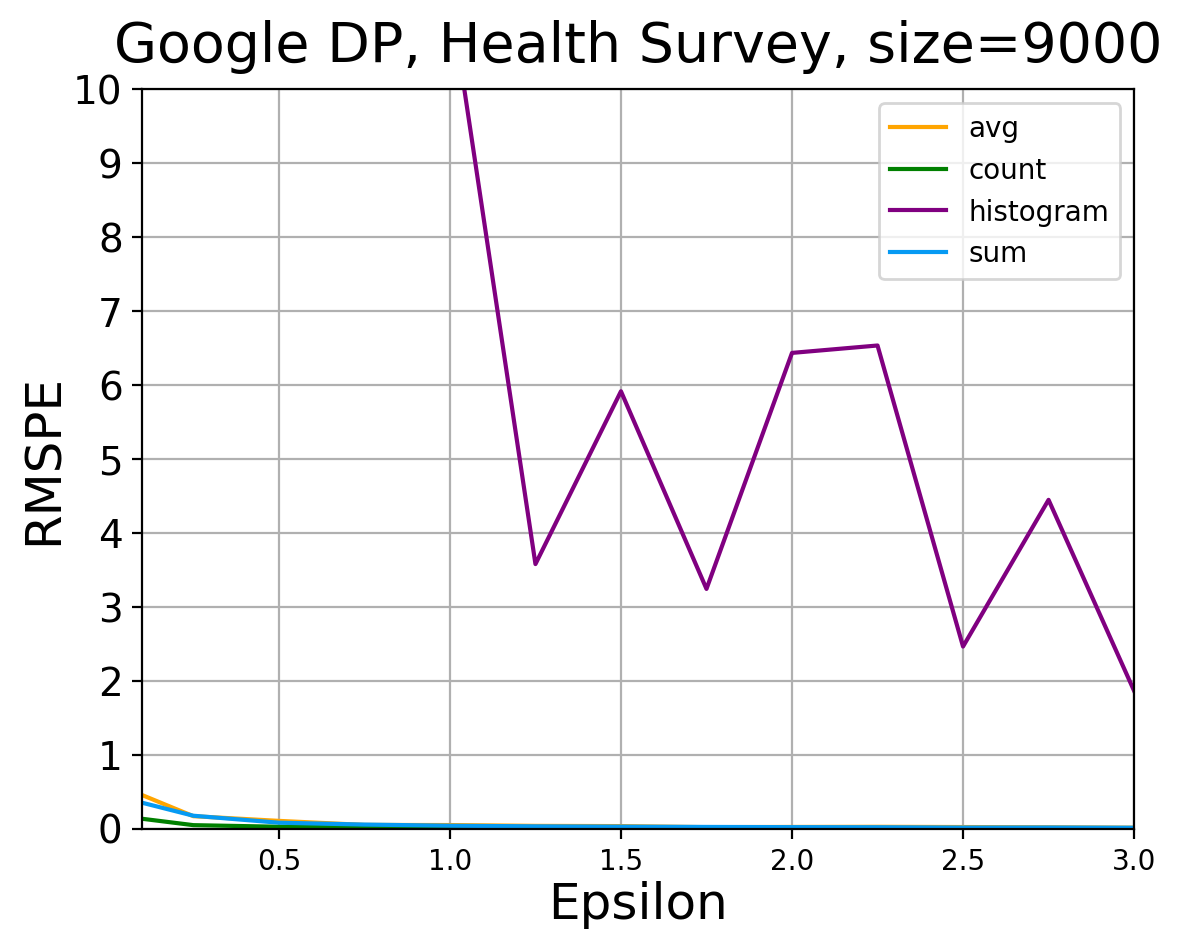}}
	\subfloat[]{\label{fig:exp1:gdp:H:9358}\includegraphics[width=0.25\textwidth]{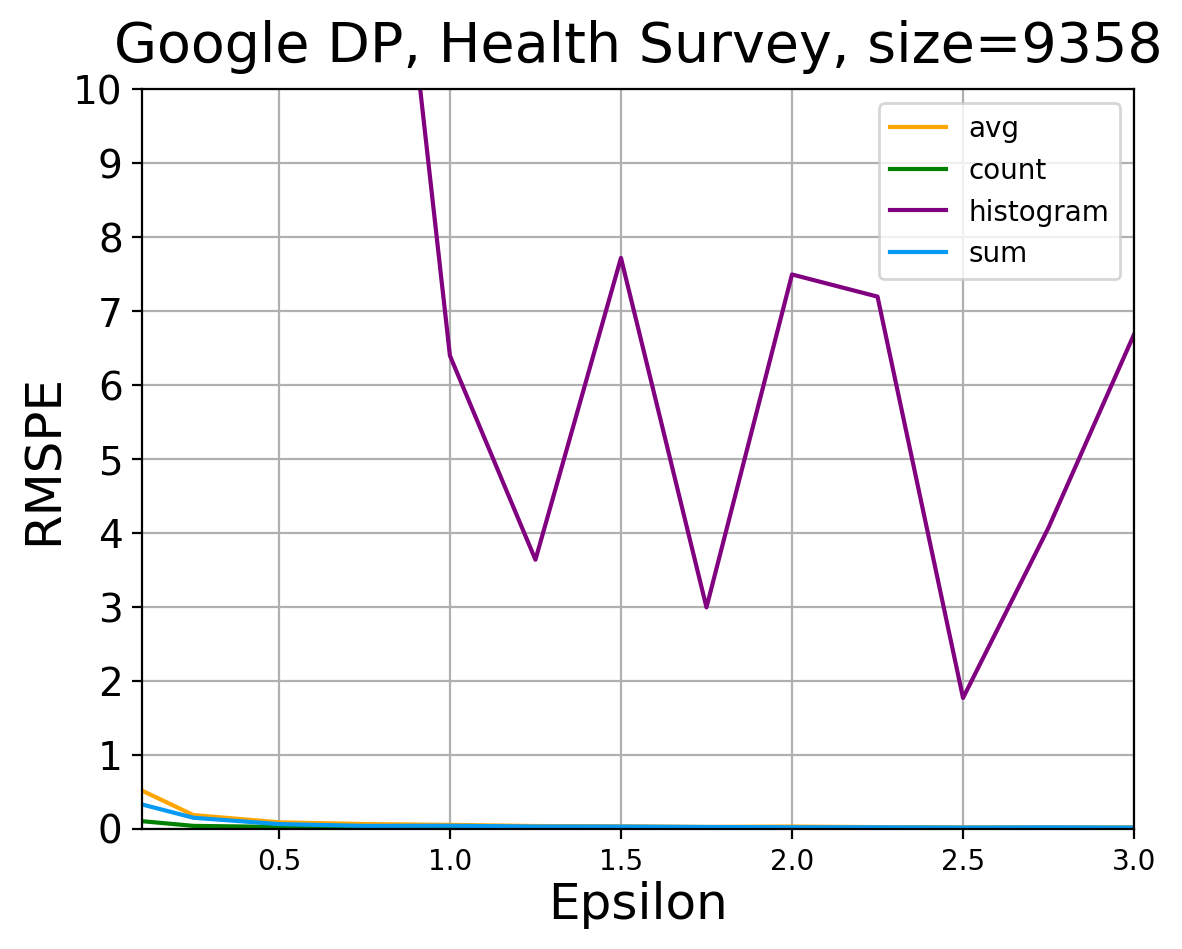}}
	\\
	\subfloat[]{\label{fig:exp1:smart:H:7000}\includegraphics[width=0.25\textwidth]{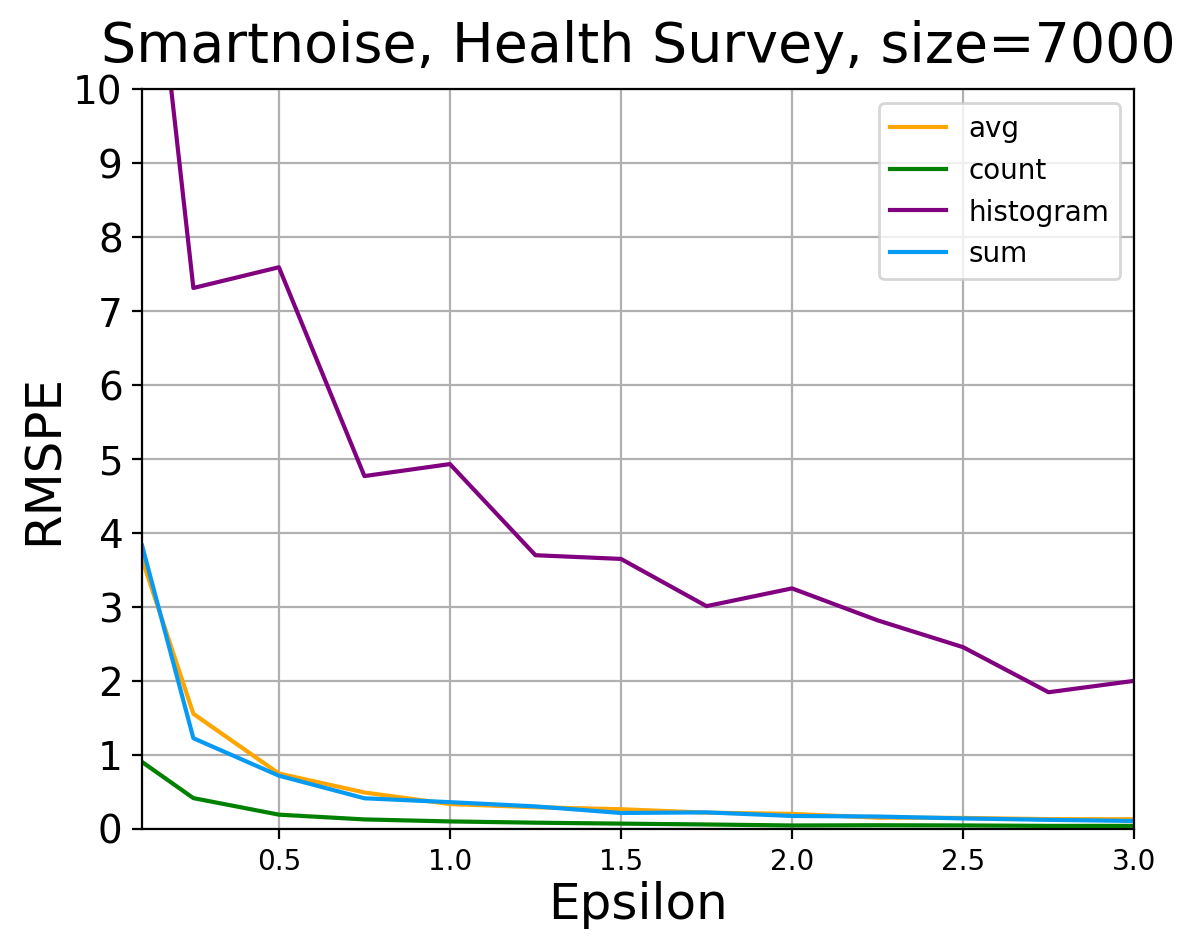}}
	\subfloat[]{\label{fig:exp1:smart:H:8000}\includegraphics[width=0.25\textwidth]{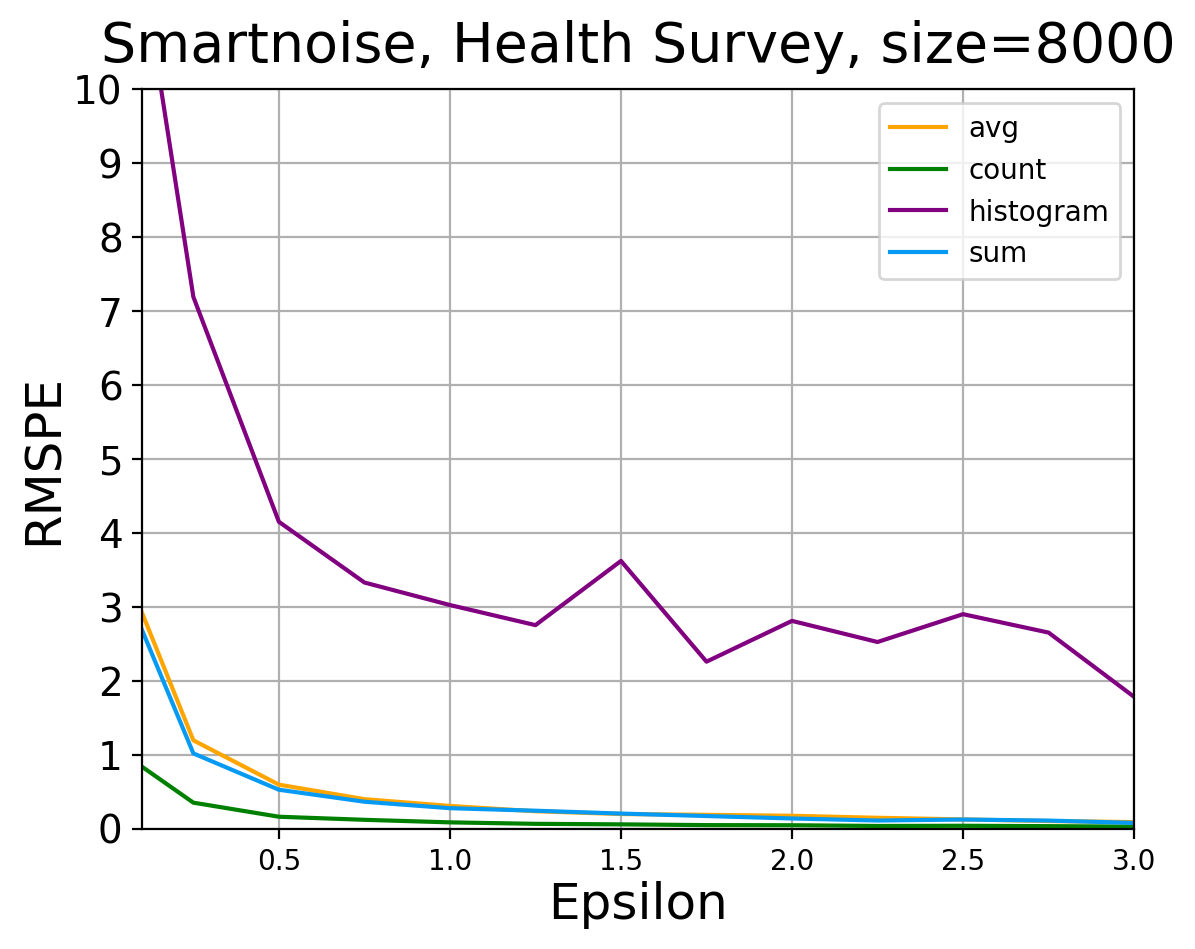}}
	\subfloat[]{\label{fig:exp1:smart:H:9000}\includegraphics[width=0.25\textwidth]{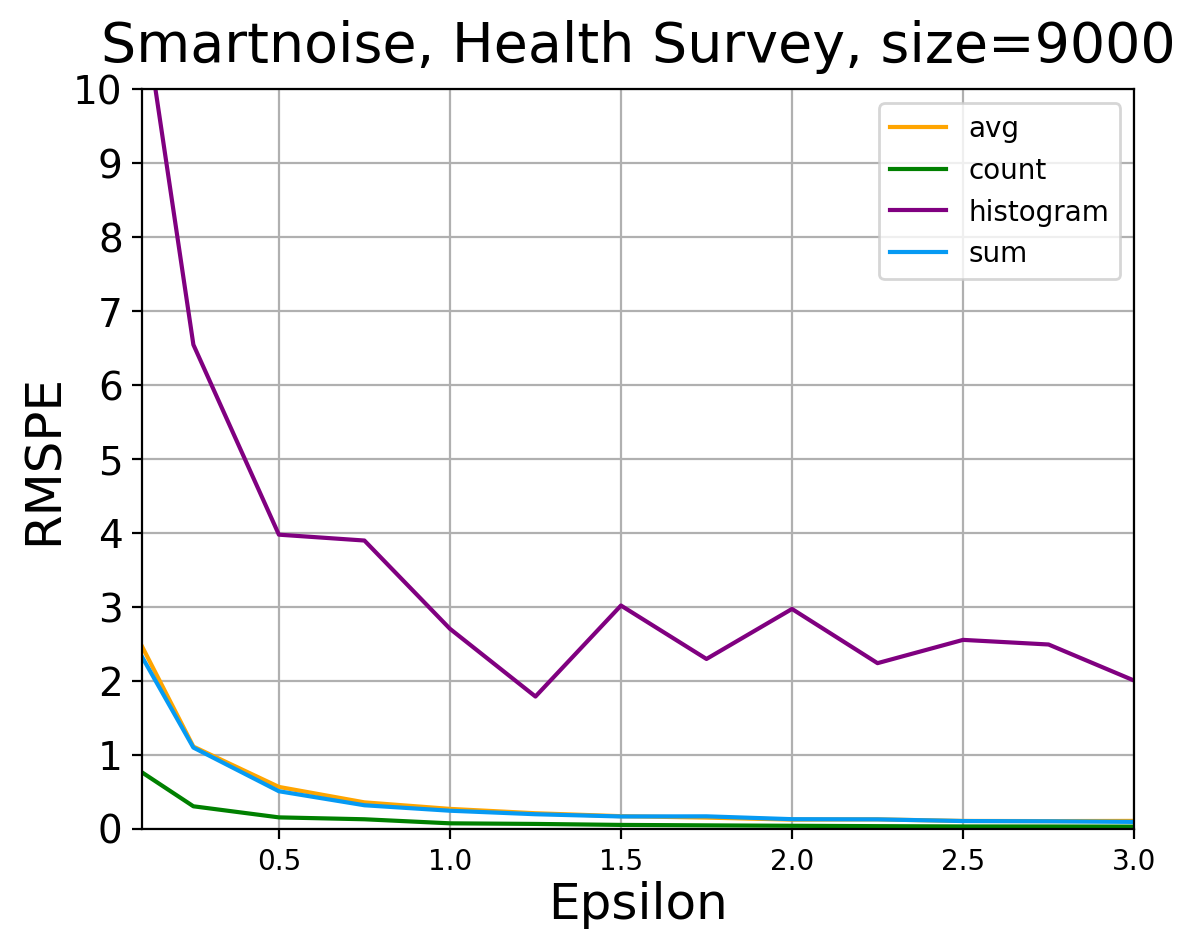}}
	\subfloat[]{\label{fig:exp1:smart:H:9358}\includegraphics[width=0.25\textwidth]{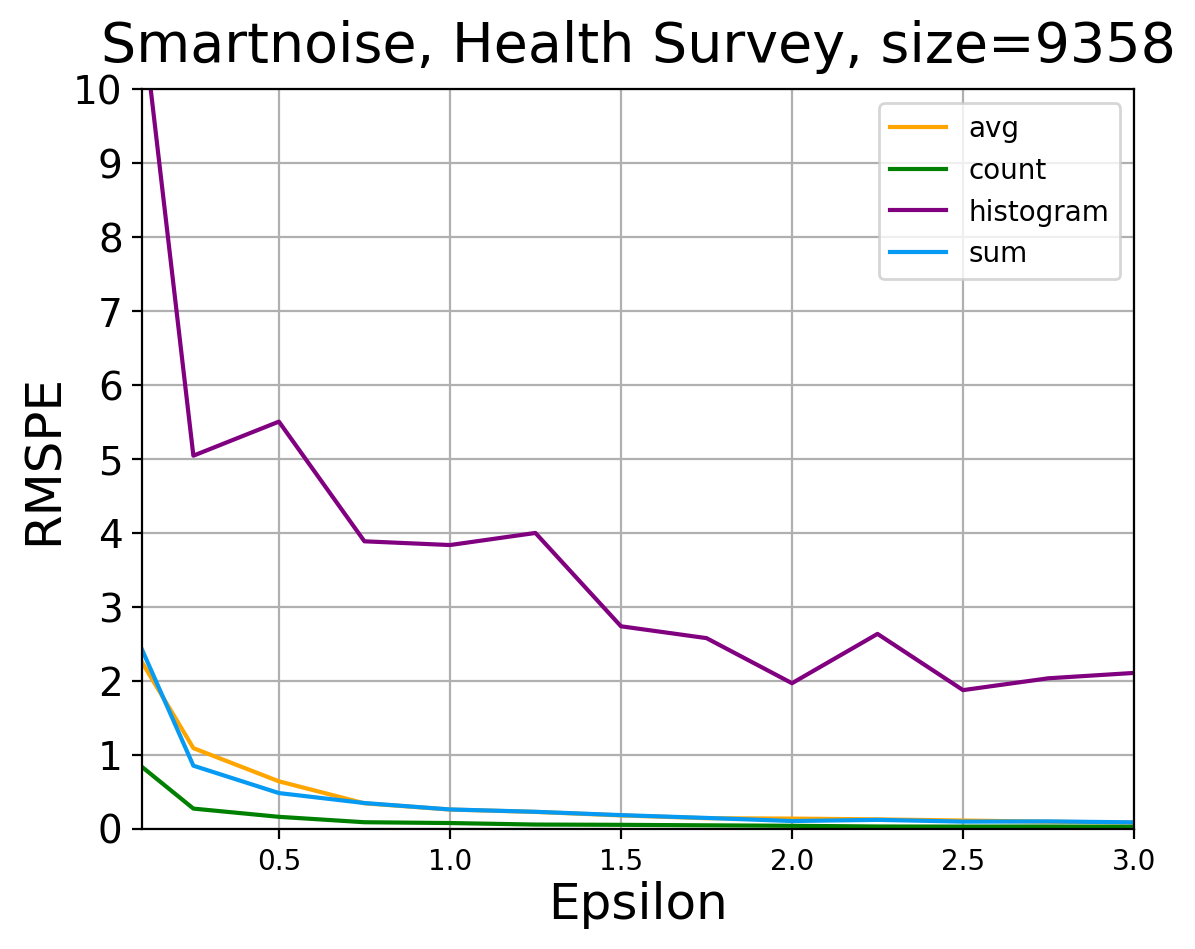}}
	\\
	\subfloat[]{\label{fig:exp1:gdp:P:3000}\includegraphics[width=0.25\textwidth]{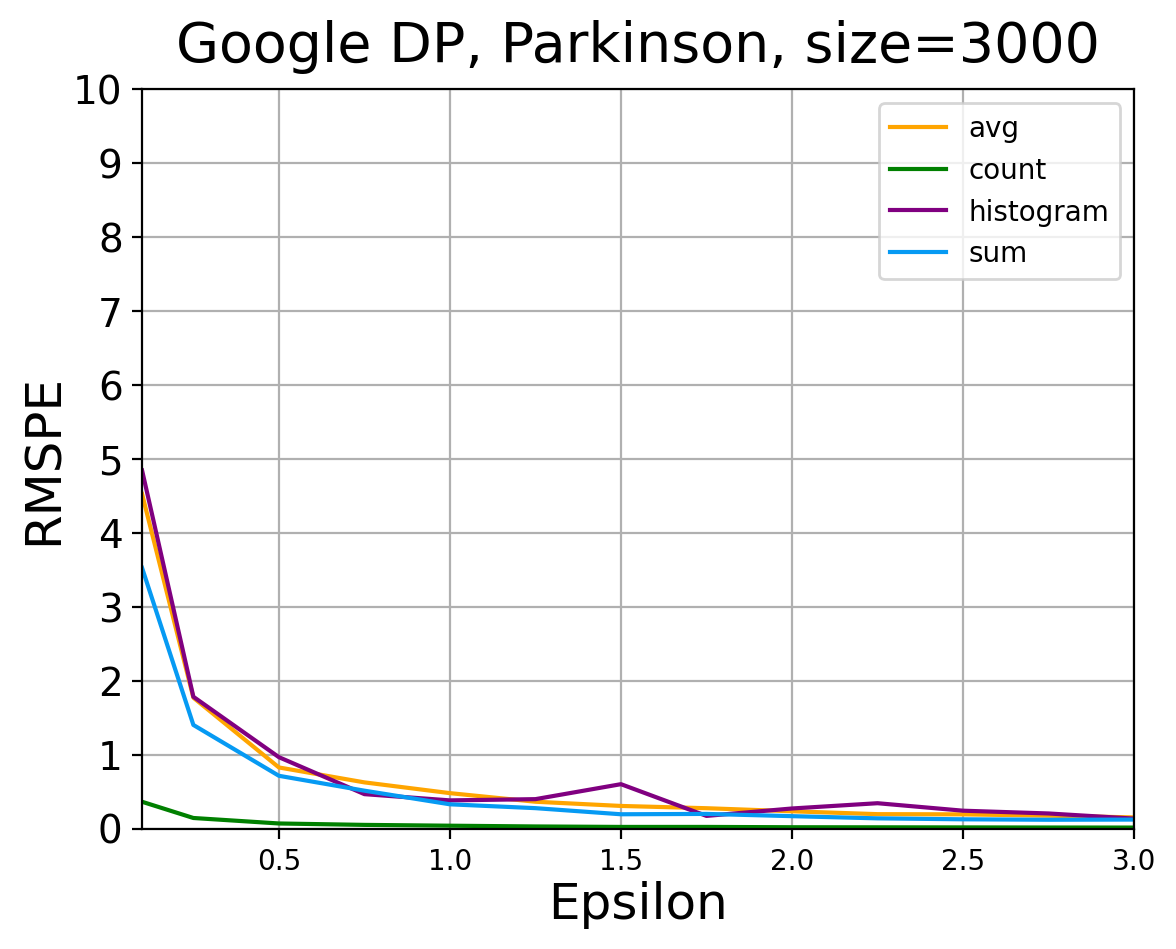}}
	\subfloat[]{\label{fig:exp1:gdp:P:4000}\includegraphics[width=0.25\textwidth]{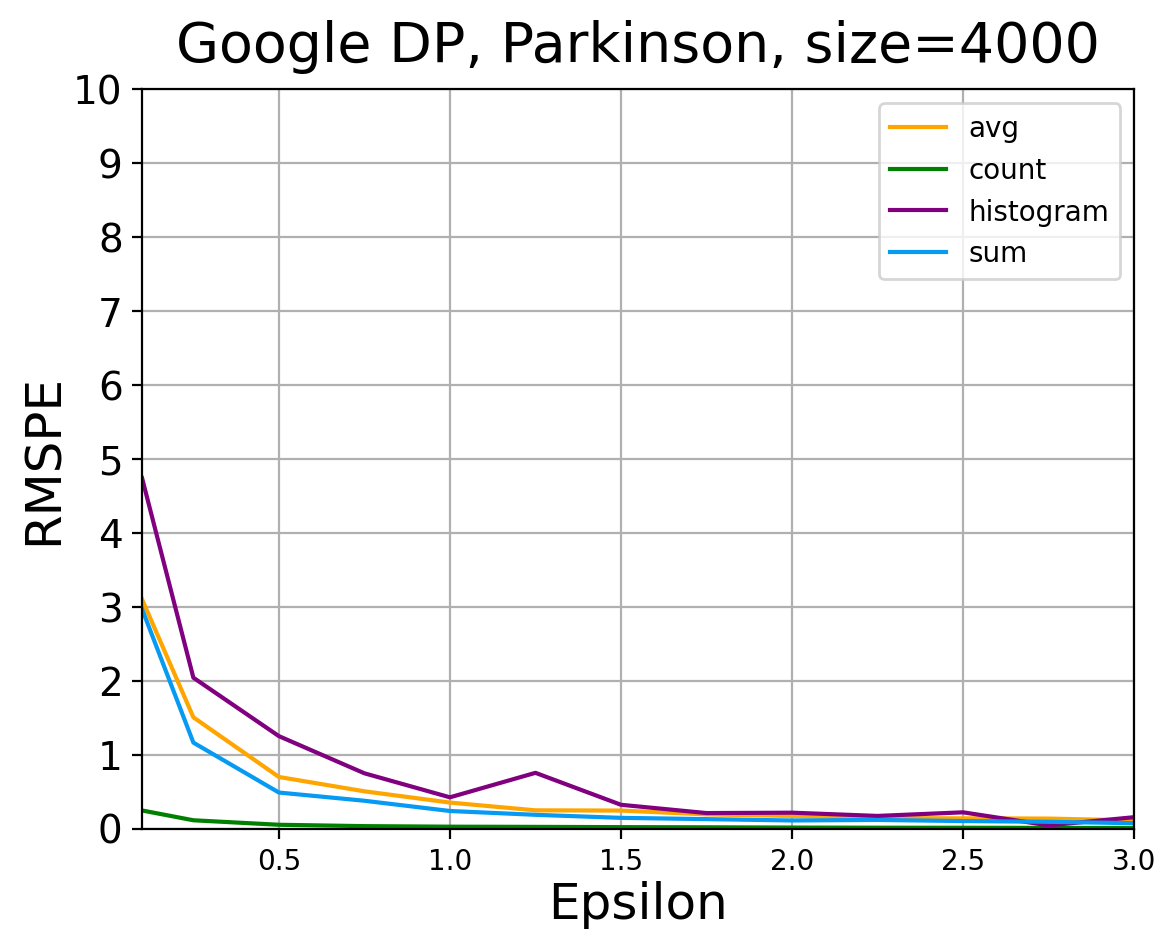}}
	\subfloat[]{\label{fig:exp1:gdp:P:5000}\includegraphics[width=0.25\textwidth]{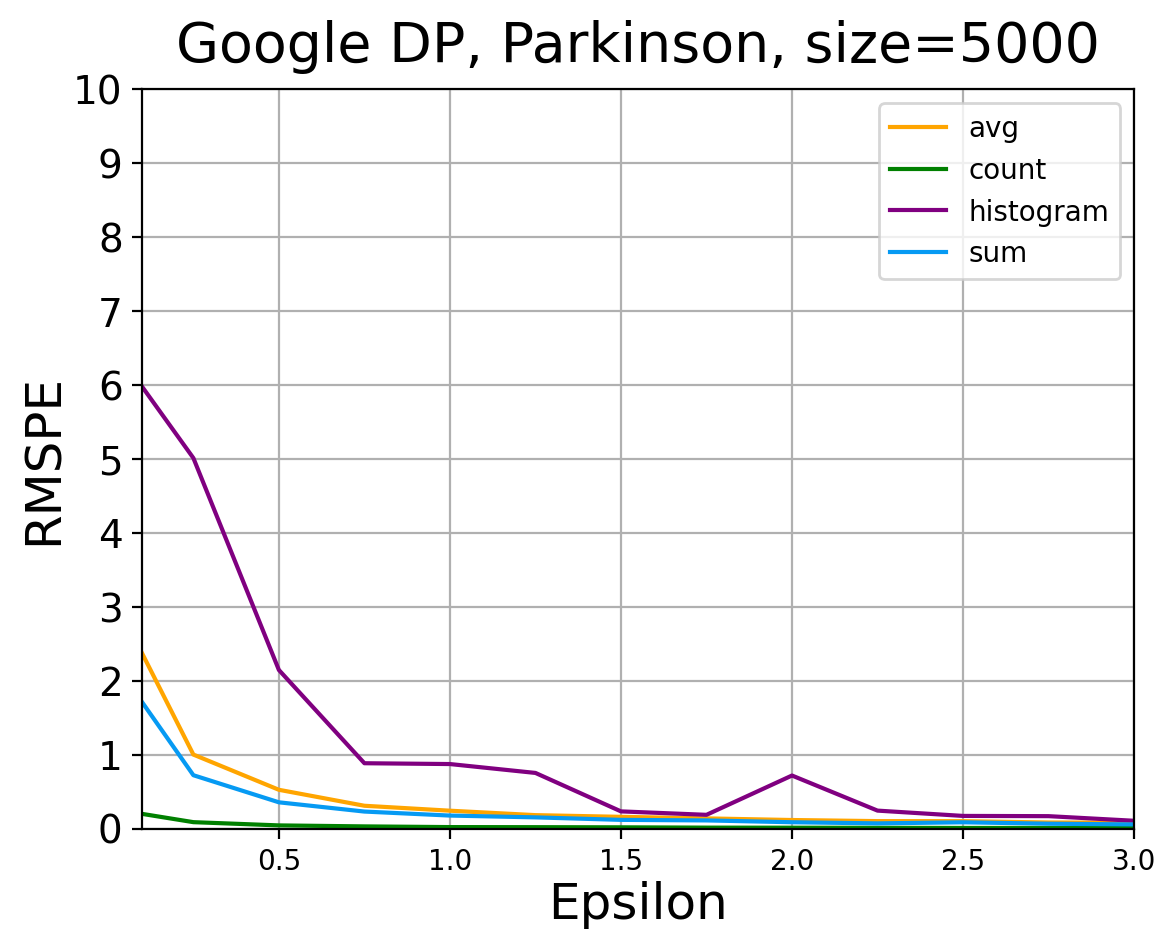}}
	\subfloat[]{\label{fig:exp1:gdp:P:5499}\includegraphics[width=0.25\textwidth]{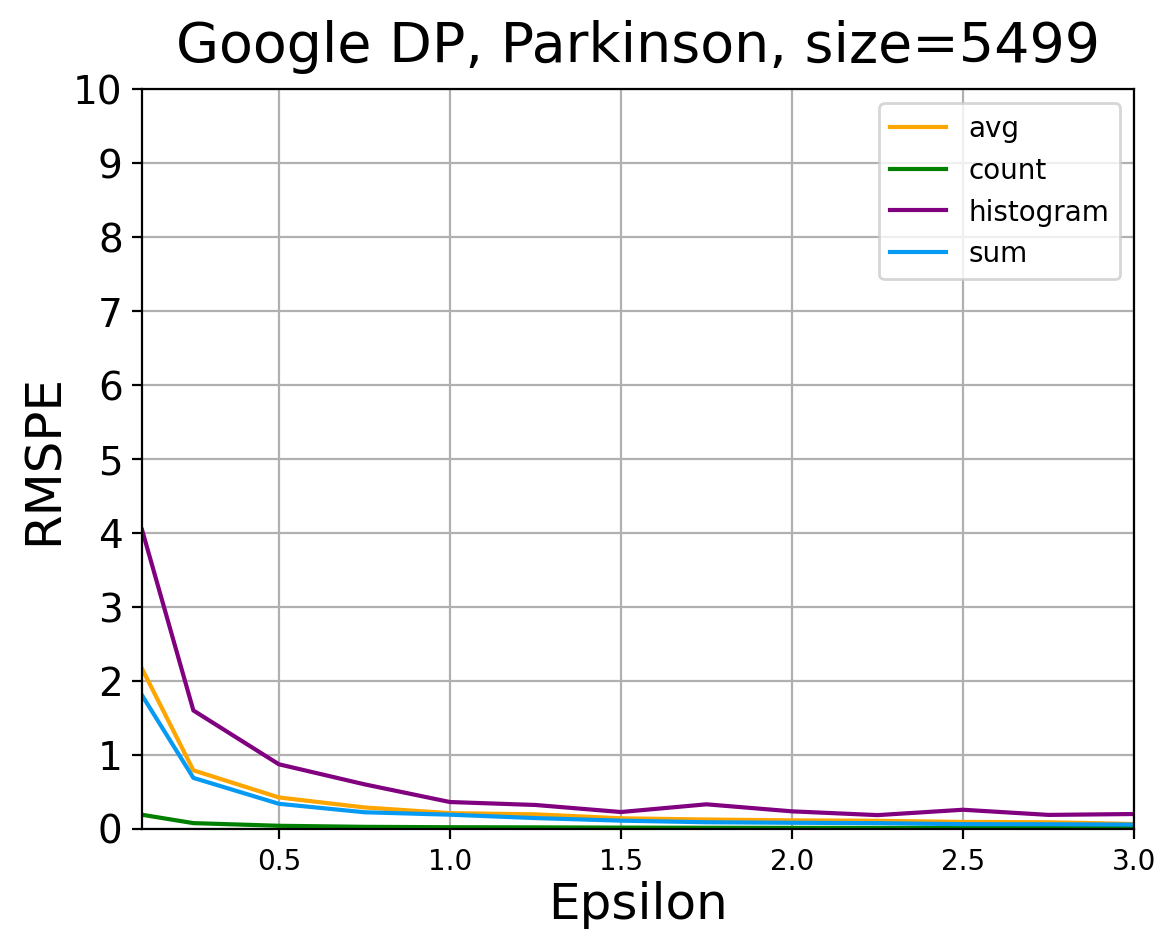}}
	\\
	\subfloat[]{\label{fig:exp1:smart:P:3000}\includegraphics[width=0.25\textwidth]{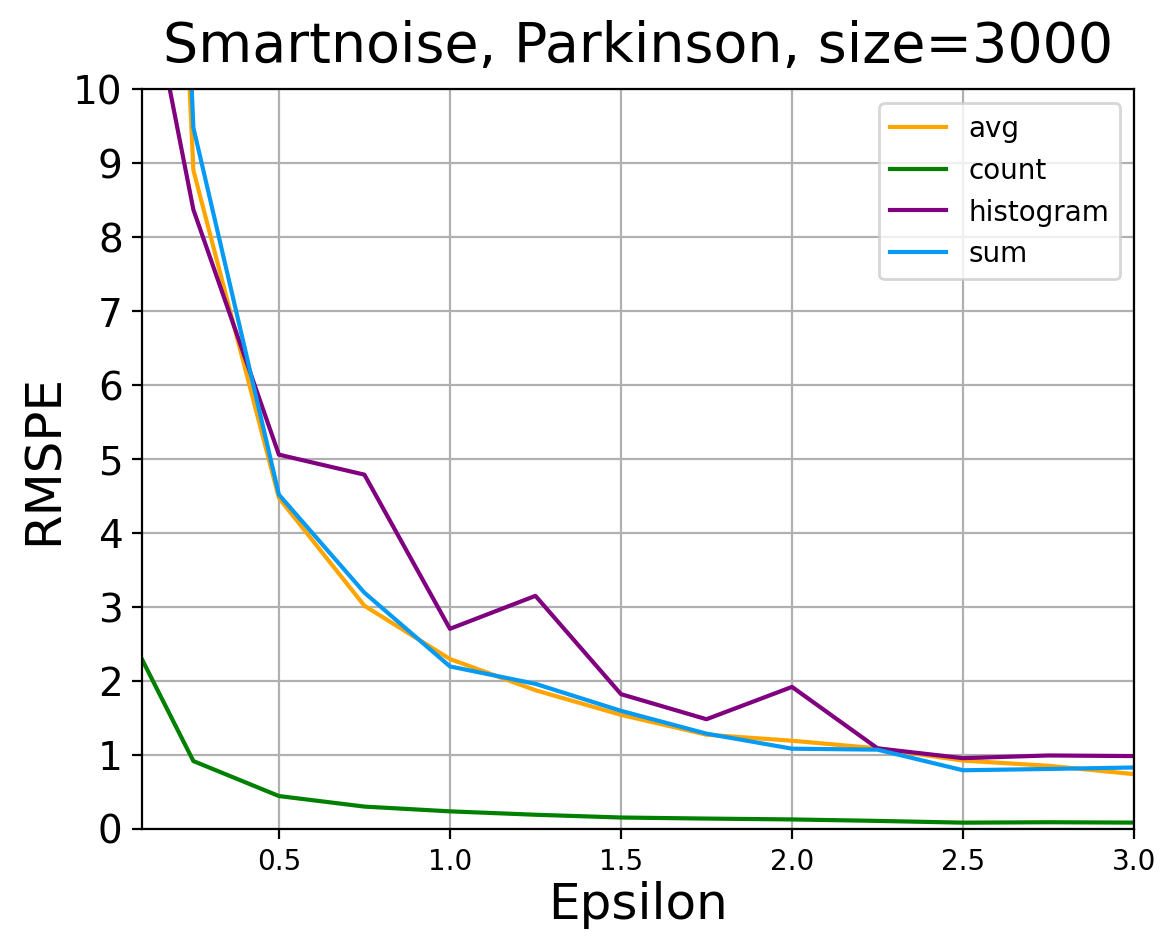}}
	\subfloat[]{\label{fig:exp1:smart:P:4000}\includegraphics[width=0.25\textwidth]{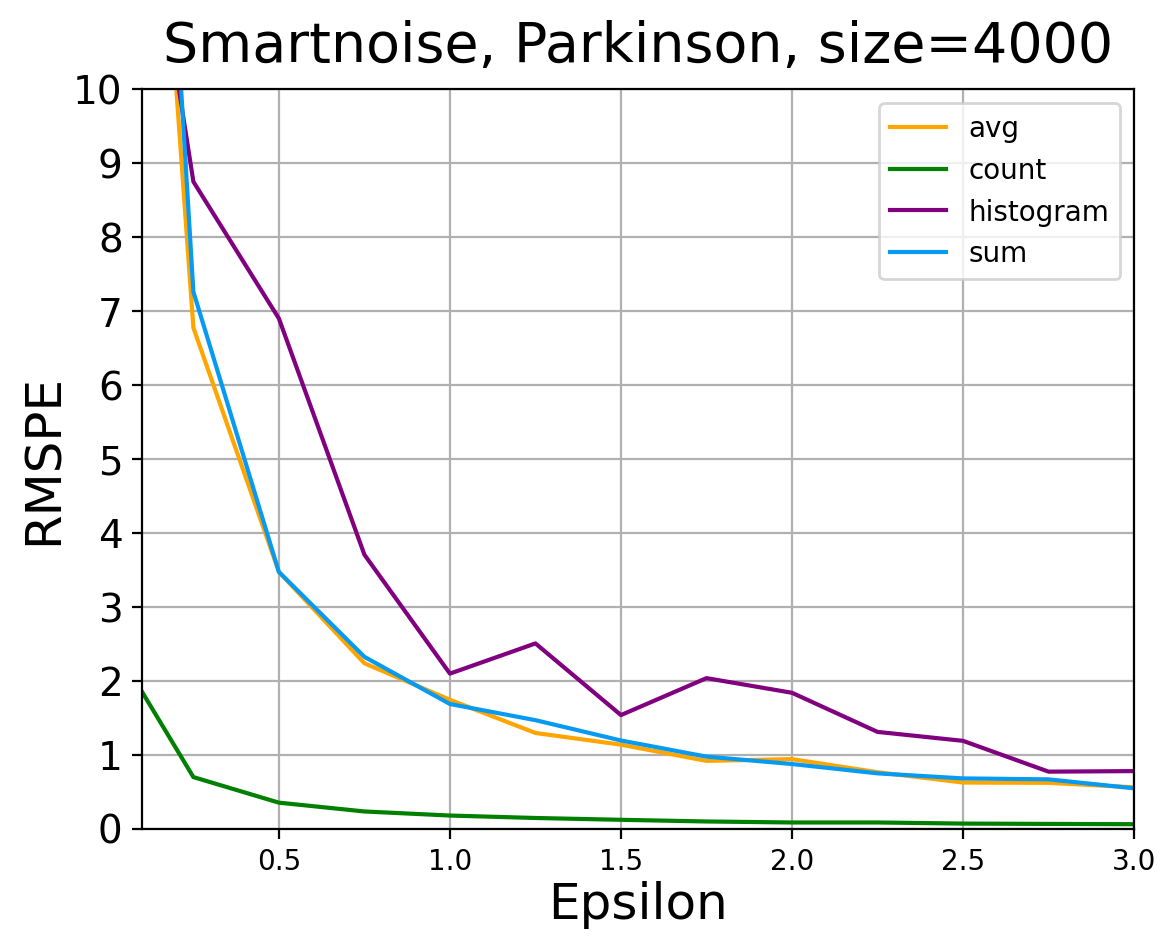}}
	\subfloat[]{\label{fig:exp1:smart:P:5000}\includegraphics[width=0.25\textwidth]{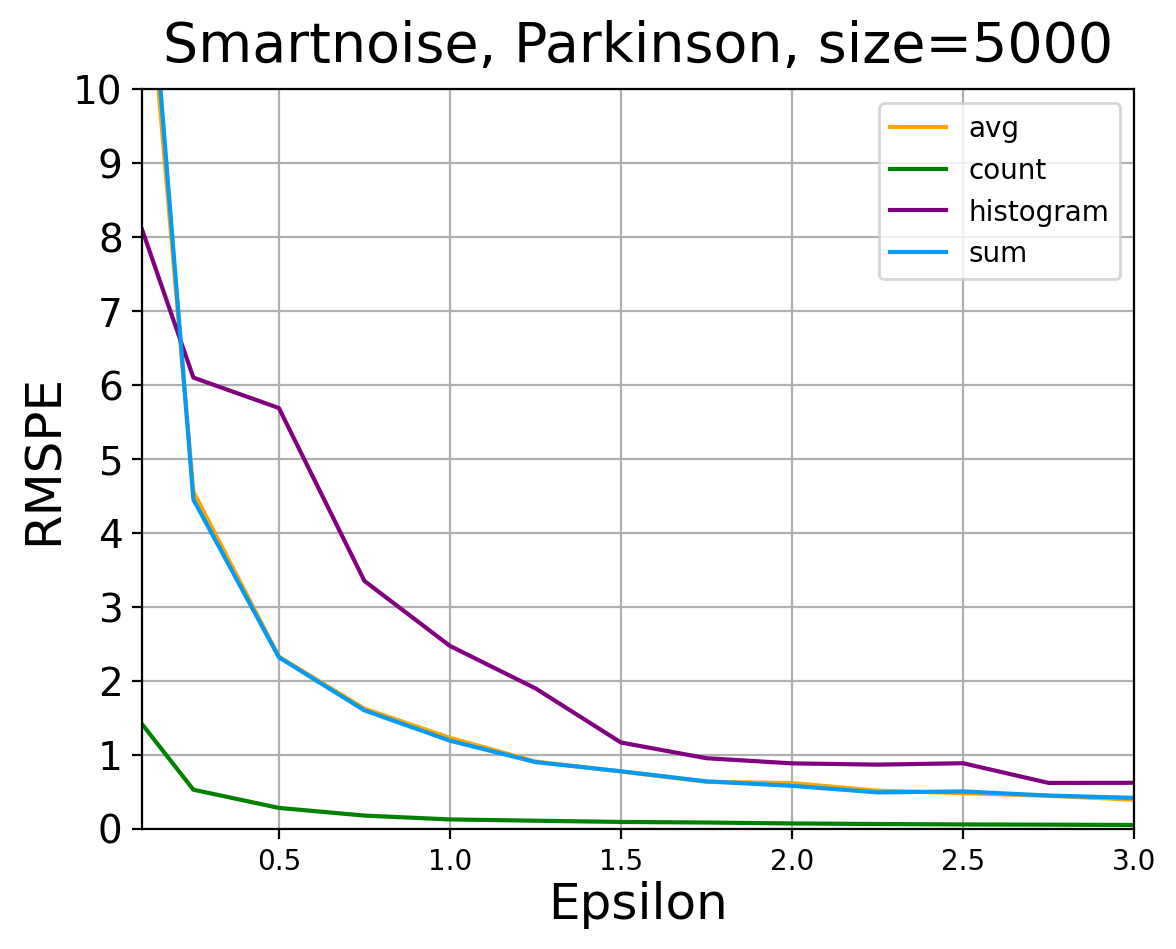}}
	\subfloat[]{\label{fig:exp1:smart:P:5499}\includegraphics[width=0.25\textwidth]{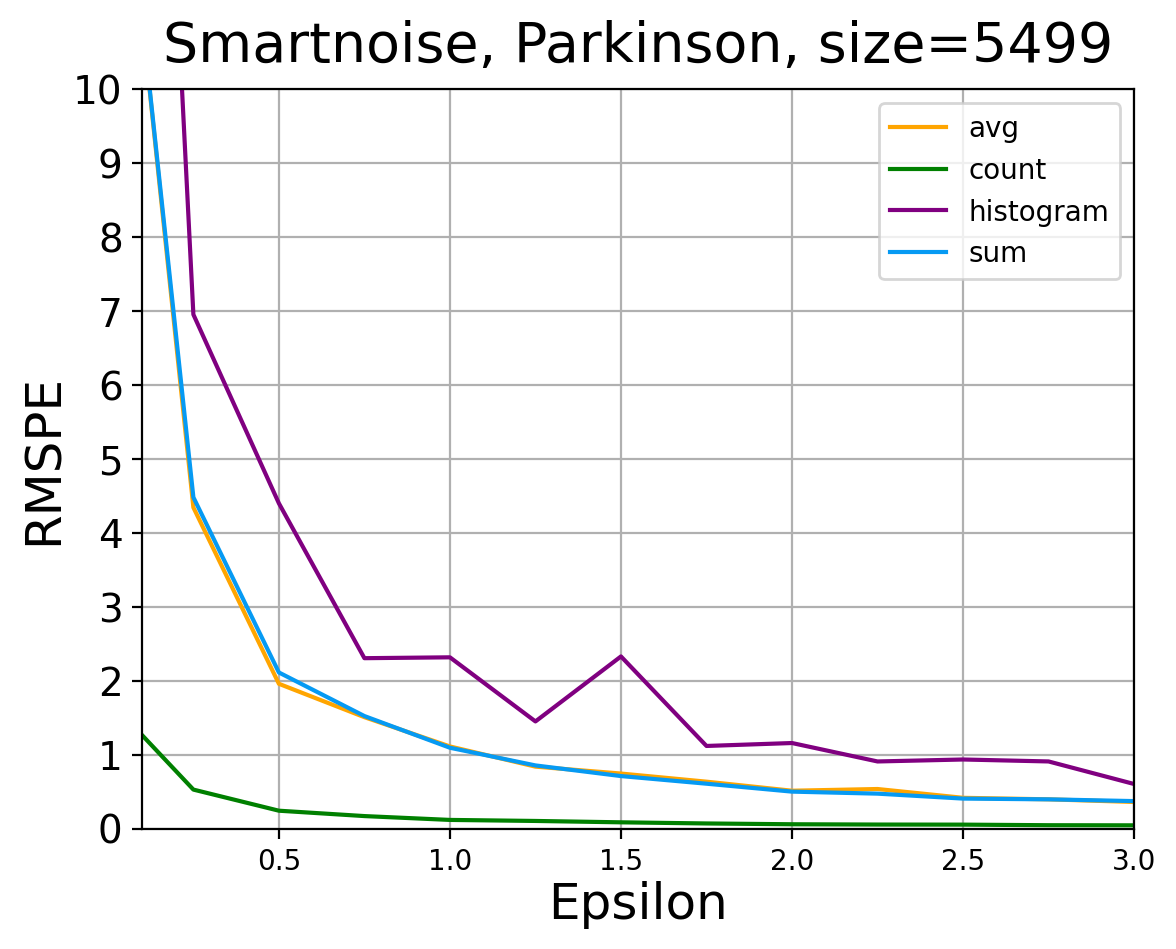}}
	
	\caption[Results of Experiment 1. Data utility for statistical query tools]{The evaluation results of statistical query tools on utility for different data sizes (Table \ref{table_dataset_sizes}), $\epsilon$ values (Table \ref{table_epsilons}), and queries (Table \ref{table_queries}). RMSPE is defined in Section~\ref{evaluation_criteria}.}
	
	\label{fig:exp1:line}
\end{figure}

Figure~\ref{fig:exp1:line} shows more details on the results that higher $\epsilon$ values decrease the RMSPE~(see the definition in Section~\ref{evaluation_criteria}), implying higher utility of DP queries is obtained where higher $\epsilon$ values improve accuracy for all queries of \texttt{SUM}, \texttt{AVERAGE}, \texttt{COUNT}, and \texttt{HISTOGRAM} for both of our two considered datasets. This observation is quite explicit, especially for results on \emph{Parkinson}. As anticipated, the relationship that data utility grows with the increase in data size can also be observed. Generally, the results on \emph{Parkinson} are more consistent with our anticipation, while those for \texttt{HISTOGRAM} queries on \emph{Health Survey} data show marginal levels of fluctuation.

\subsubsection{Run-time overhead}\label{sq:exp2}

In this section, we illustrate how the run-time differs between conducting differential private (DP) queries and non-private ones by testing the tools of Google Differential Privacy and Smartnoise. Note that we vary the settings of $\epsilon$ and data size to gain evaluating results under different conditions.

For this evaluation, we anticipate that the run-time overhead might increase when DP is integrated since DP requires additional computations to conduct the query. Intuitively, the run-time for DP queries is also expected to grow with an increase in data size since more information will be processed. The results, as detailed below, demonstrate that Google DP poses less run-time than Smartnoise, while how the two tools are impacted by DP differs. \Ie~Smartnoise experiences an increase when data size rises, while Google DP reacts in the opposite. Beyond that, we observe apparent fluctuations in the results and no clear relationship between $\epsilon$ and run-time.

\begin{figure}[!h]
	\centering
	\subfloat[]{\label{fig:exp2:gdp:sum:H}\includegraphics[width=0.25\textwidth]{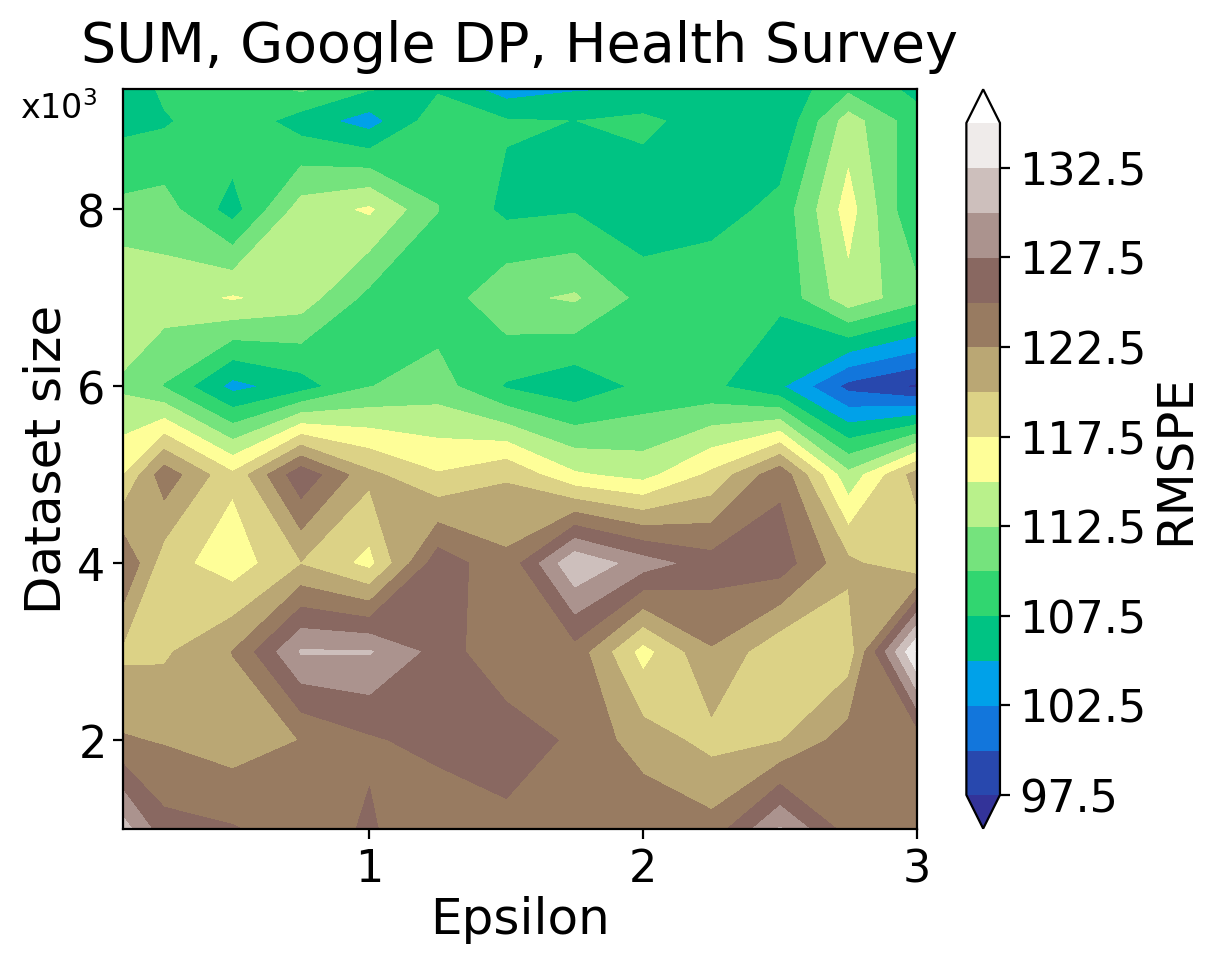}}
	\subfloat[]{\label{fig:exp2:gdp:count:H}\includegraphics[width=0.25\textwidth]{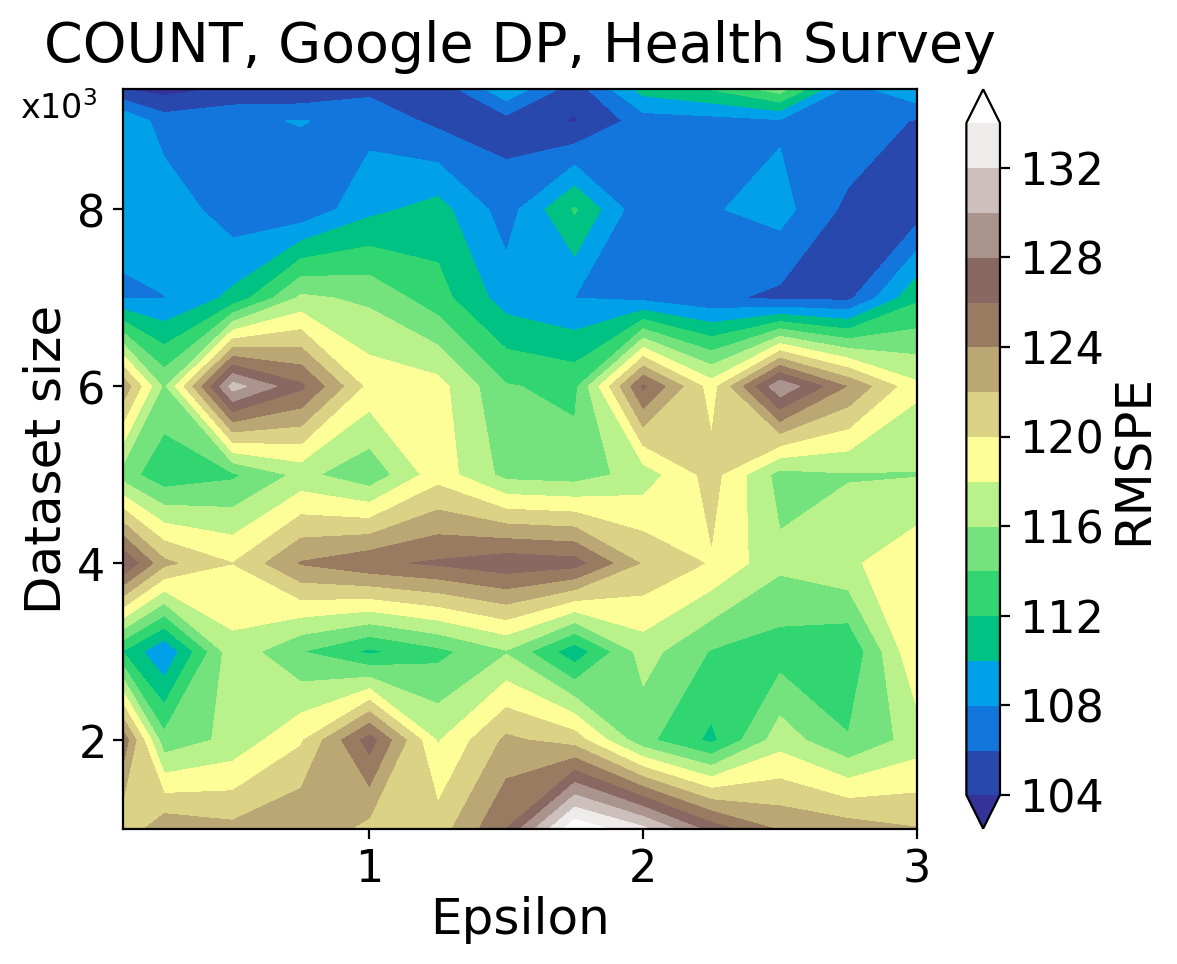}}
	\subfloat[]{\label{fig:exp2:gdp:avg:H}\includegraphics[width=0.25\textwidth]{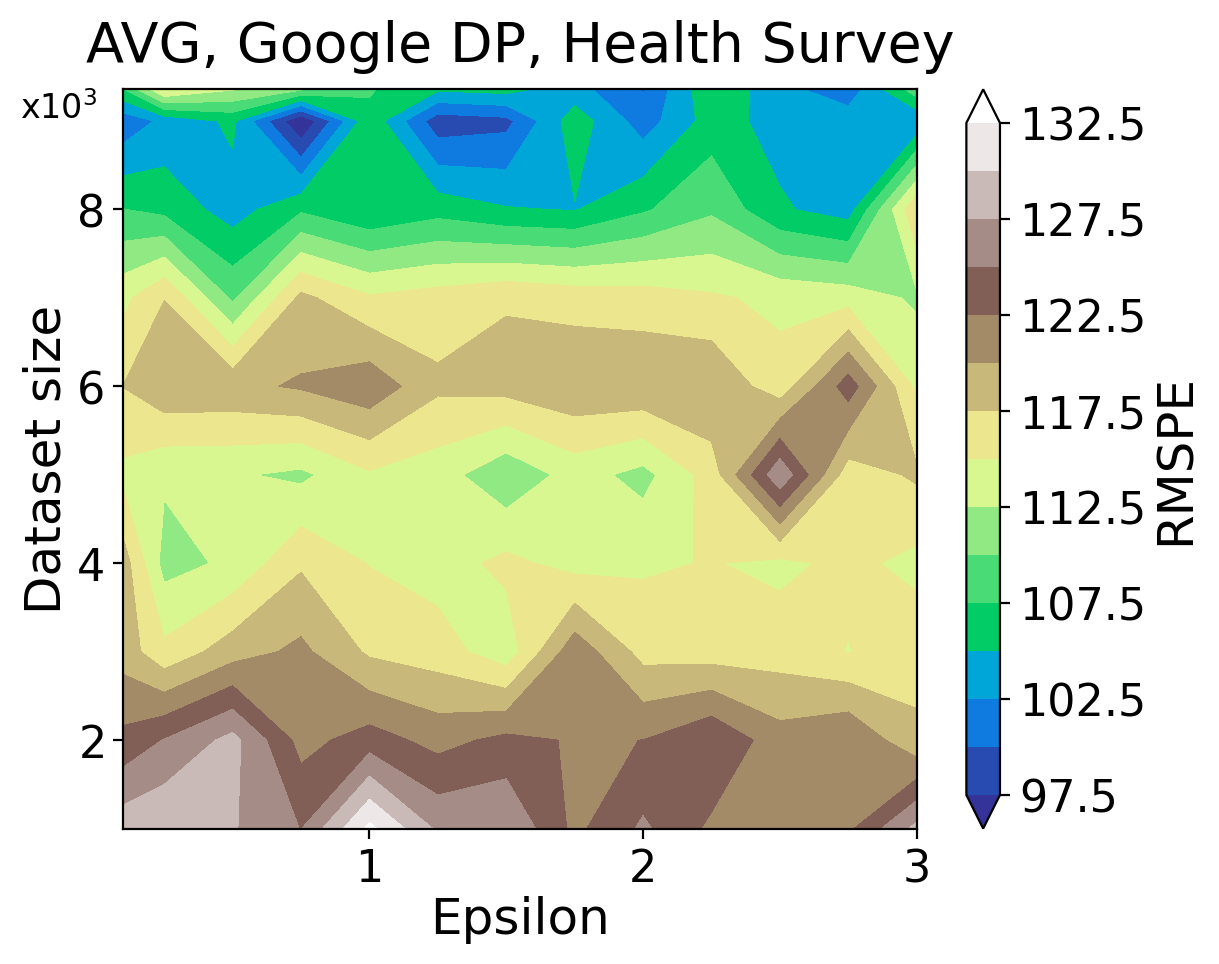}}
	\subfloat[]{\label{fig:exp2:gdp:hist:H}\includegraphics[width=0.25\textwidth]{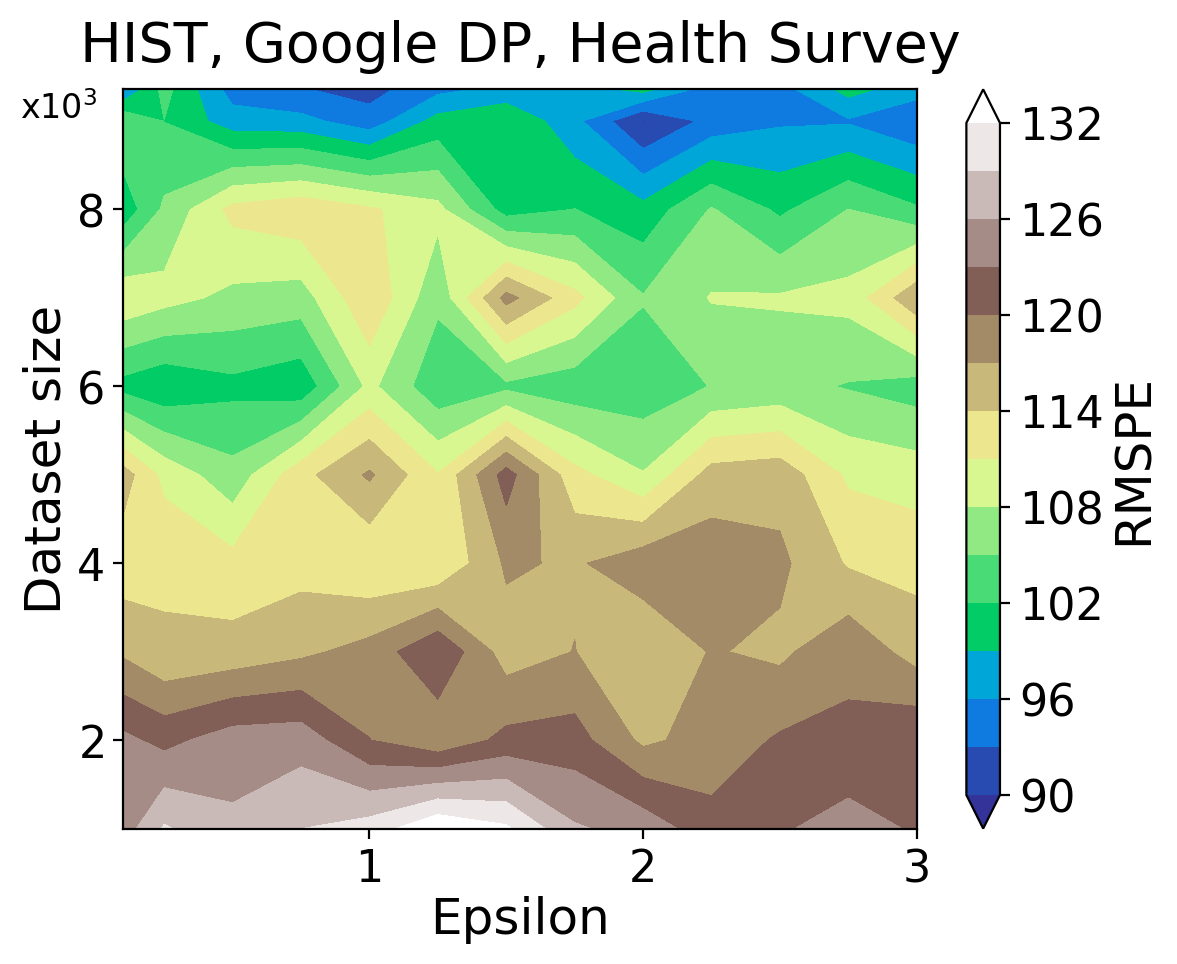}}
	\\
	\subfloat[]{\label{fig:exp2:smart:sum:H}\includegraphics[width=0.25\textwidth]{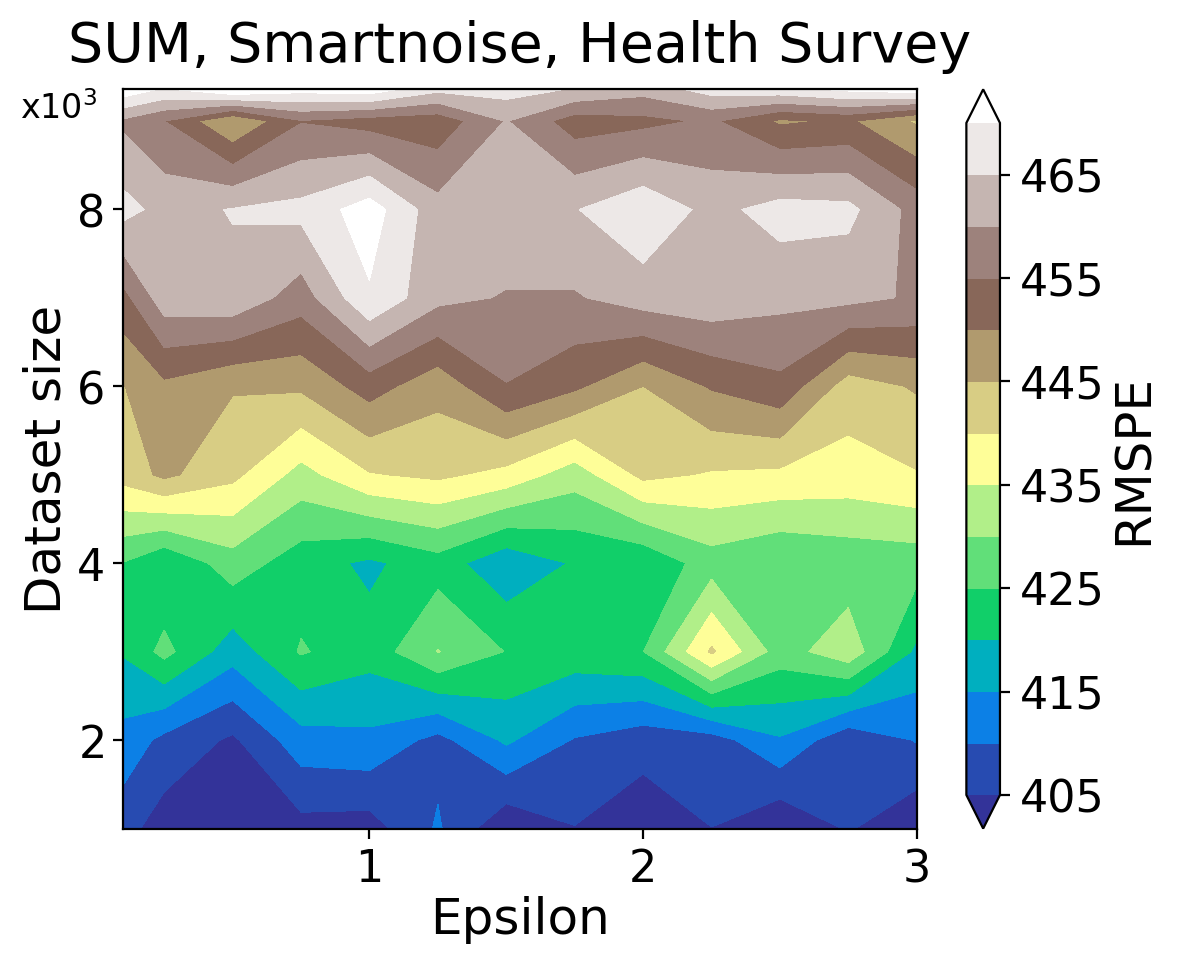}}
	\subfloat[]{\label{fig:exp2:smart:count:H}\includegraphics[width=0.25\textwidth]{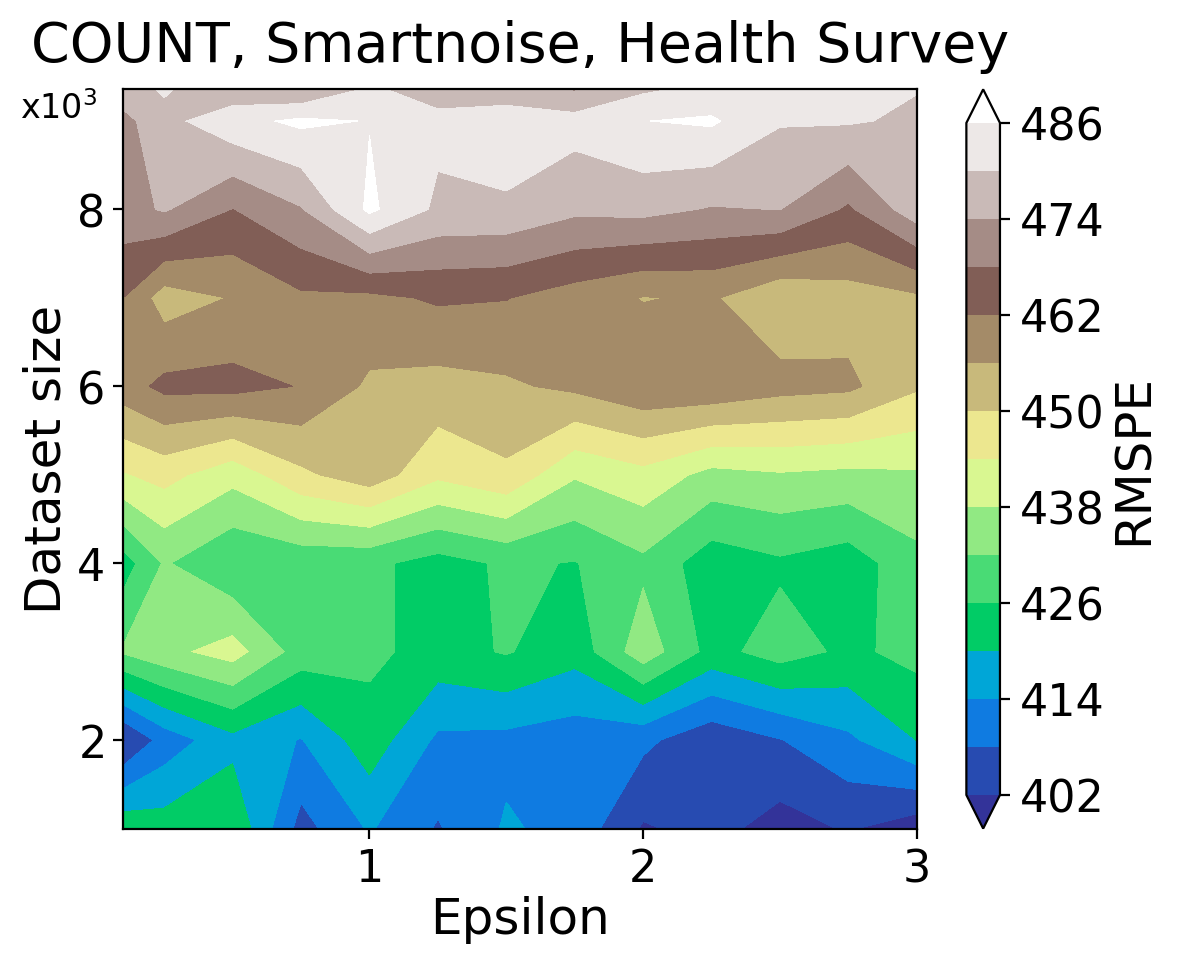}}
	\subfloat[]{\label{fig:exp2:smart:avg:H}\includegraphics[width=0.25\textwidth]{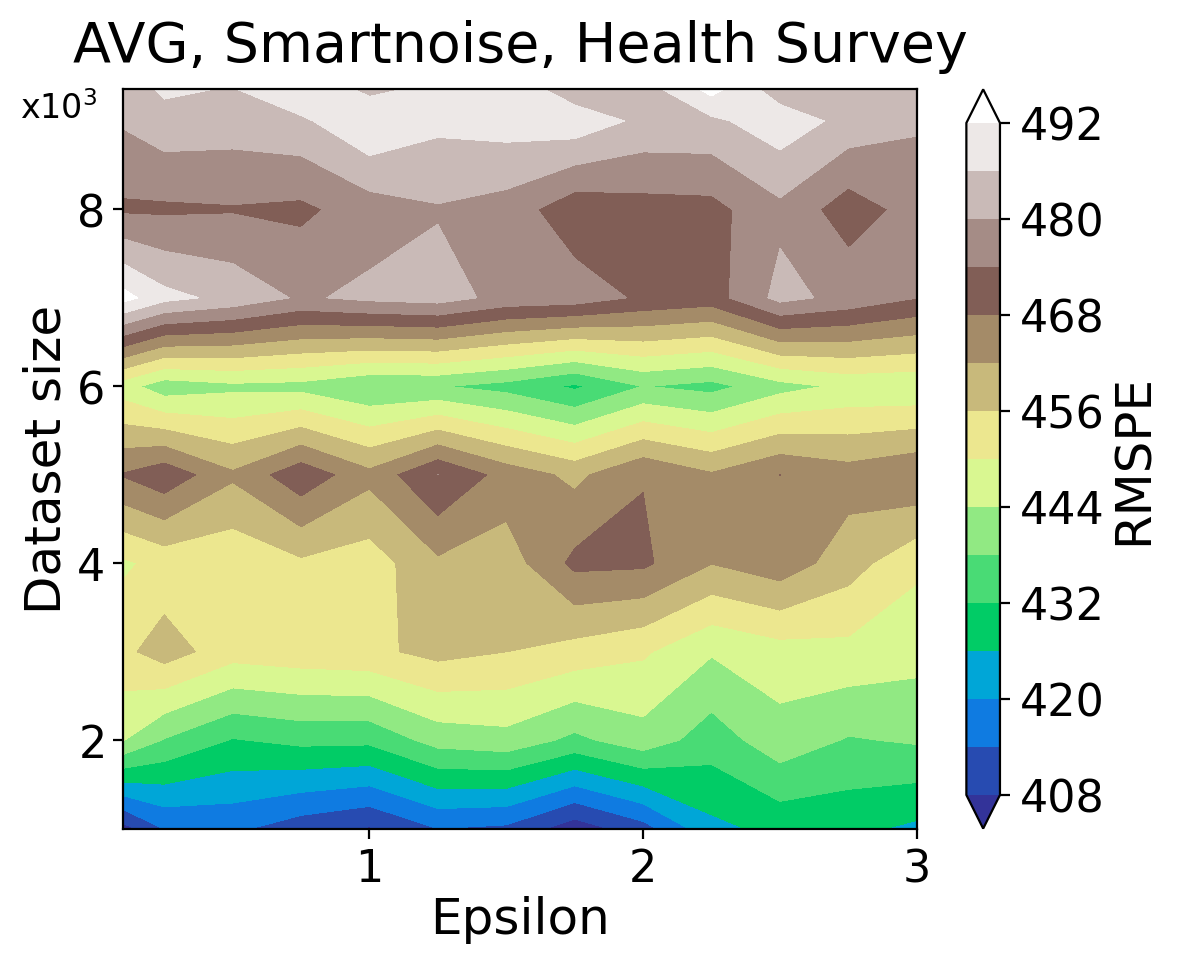}}
	\subfloat[]{\label{fig:exp2:smart:hist:H}\includegraphics[width=0.25\textwidth]{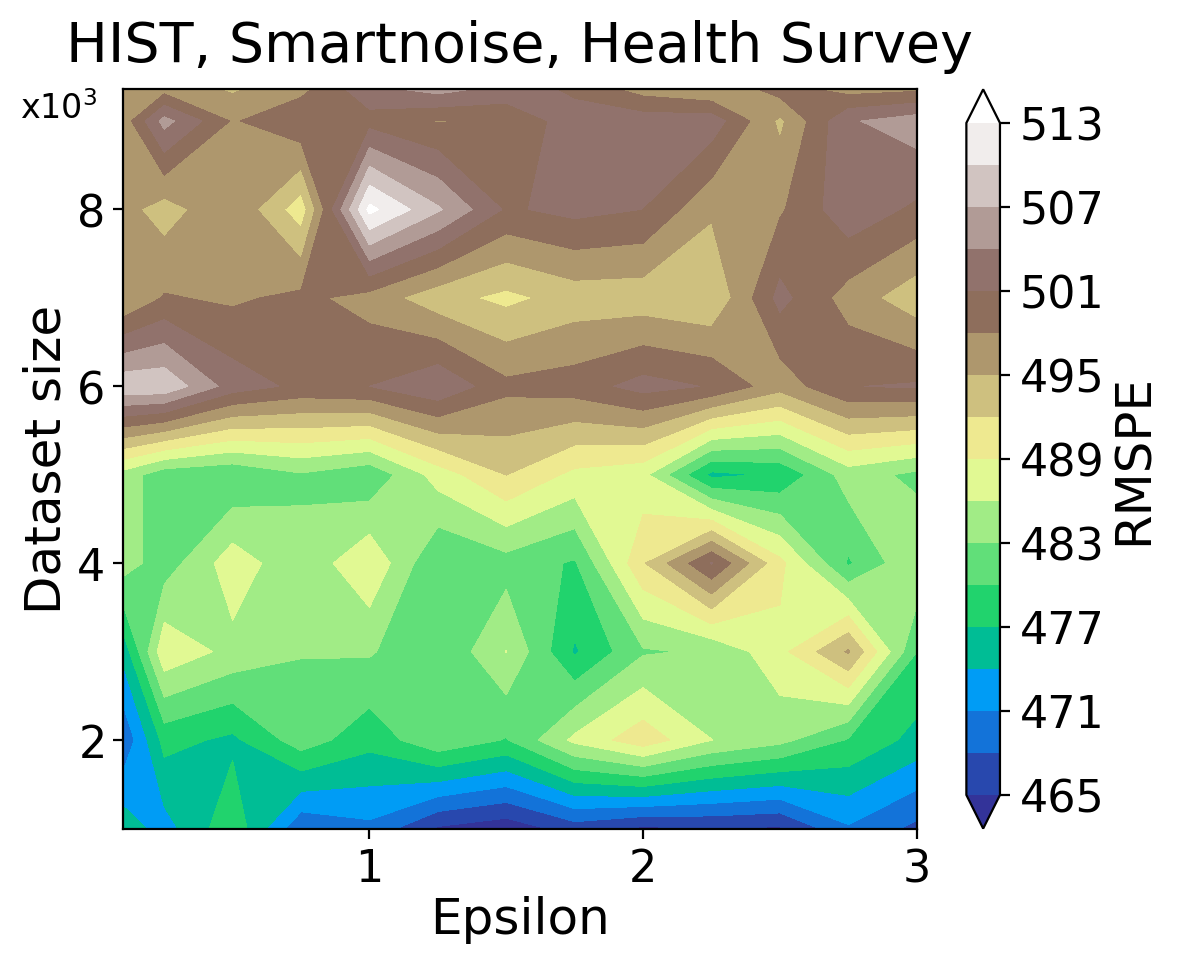}}
	\\
	\subfloat[]{\label{fig:exp2:gdp:sum:P}\includegraphics[width=0.25\textwidth]{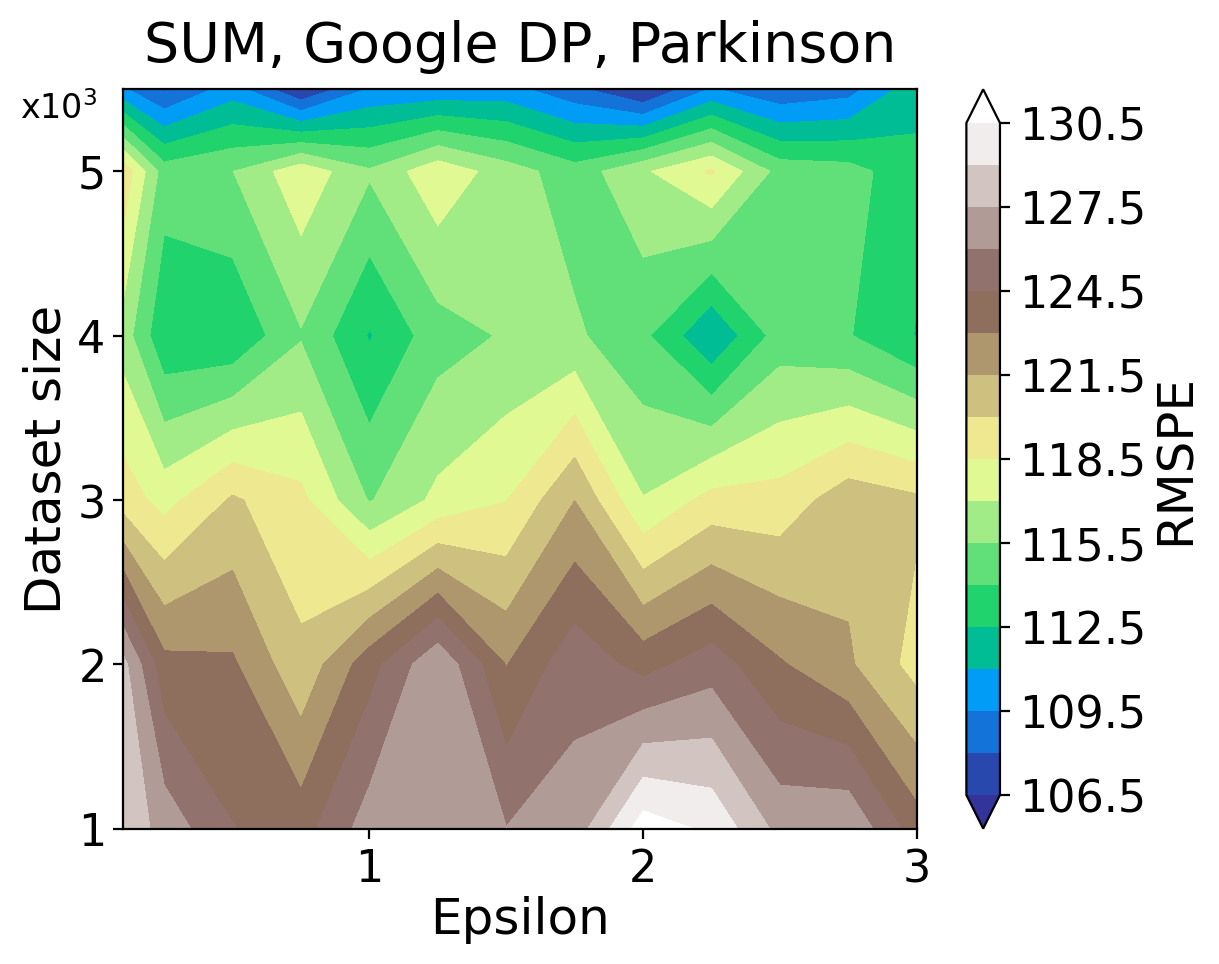}}
	\subfloat[]{\label{fig:exp2:gdp:count:P}\includegraphics[width=0.25\textwidth]{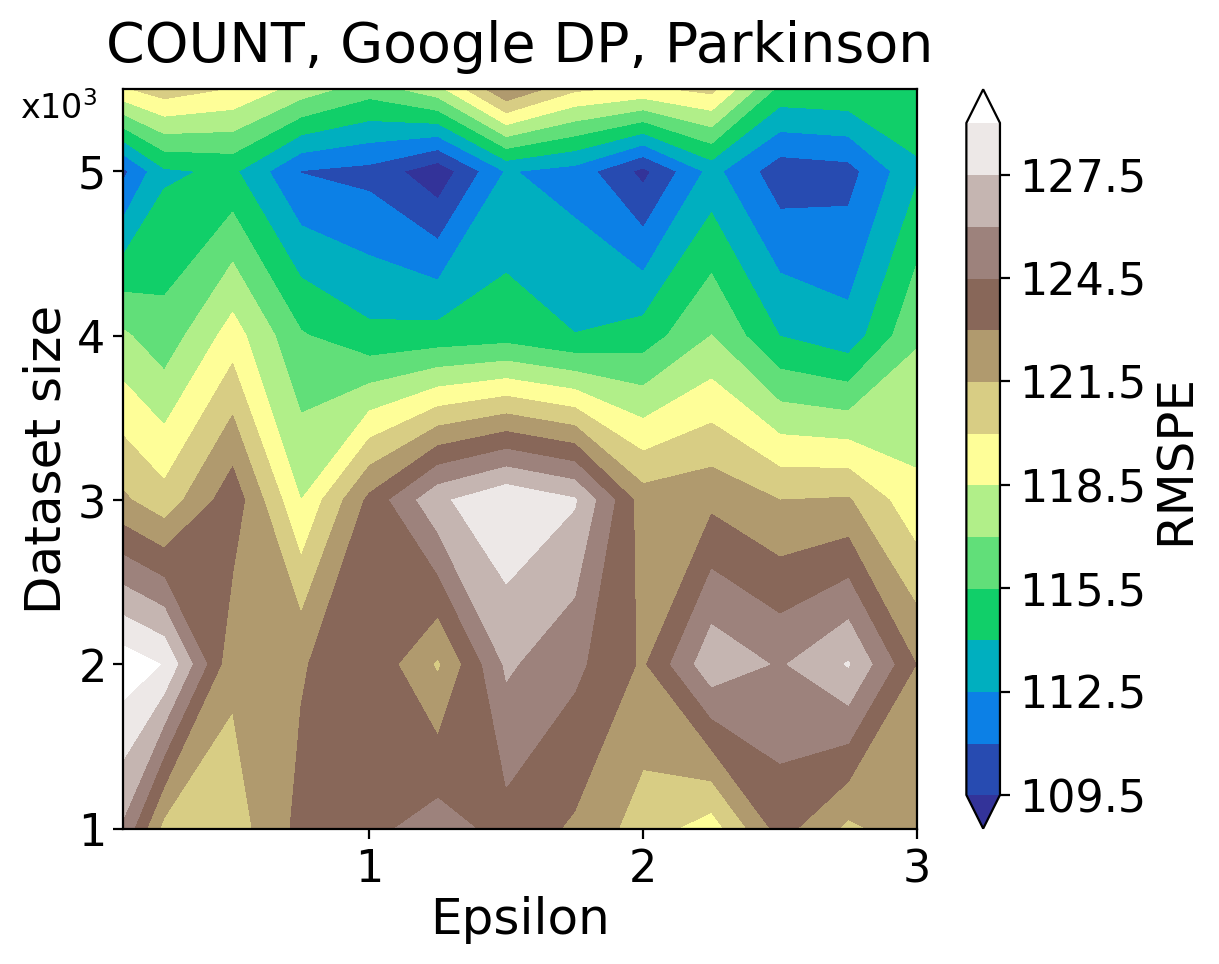}}
	\subfloat[]{\label{fig:exp2:gdp:avg:P}\includegraphics[width=0.25\textwidth]{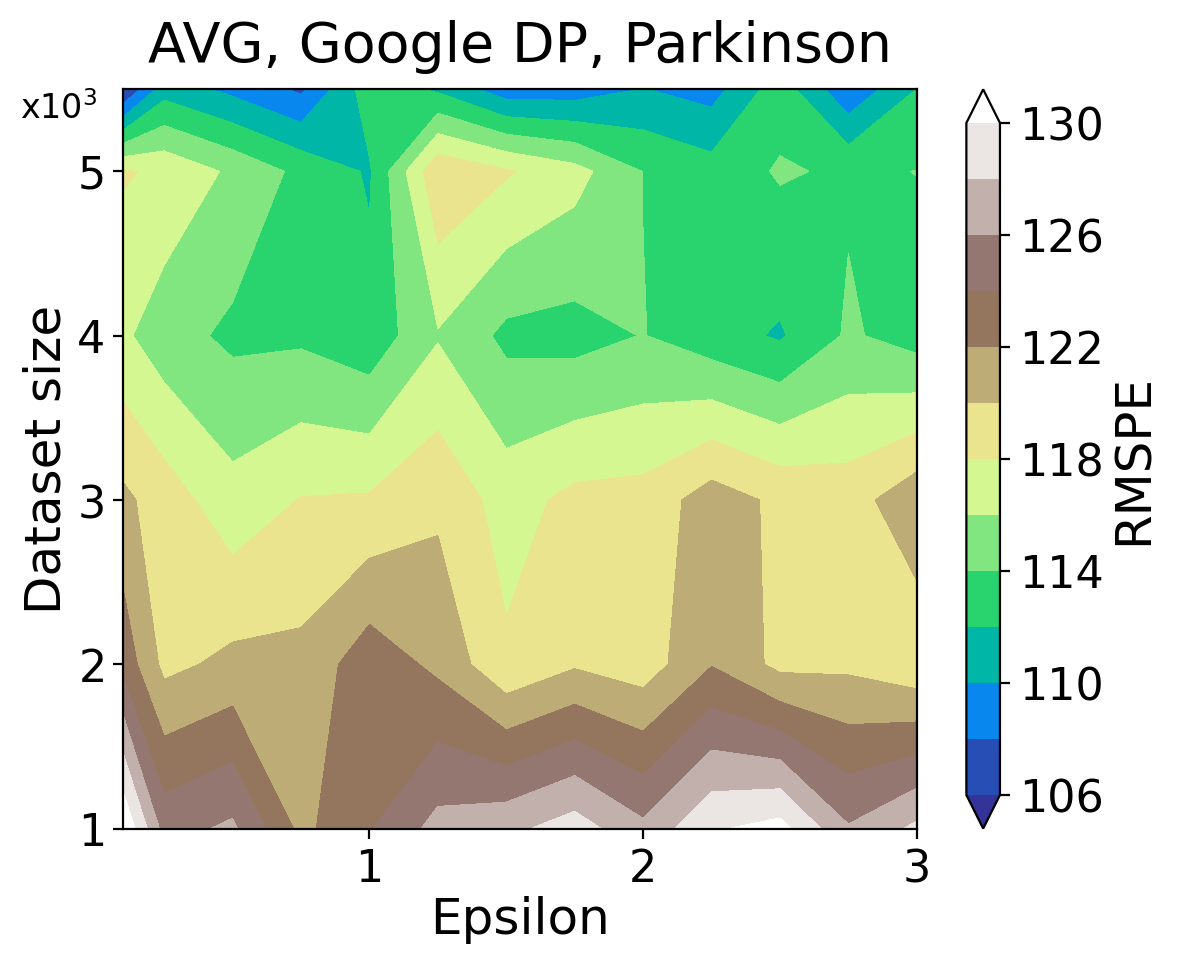}}
	\subfloat[]{\label{fig:exp2:gdp:hist:P}\includegraphics[width=0.25\textwidth]{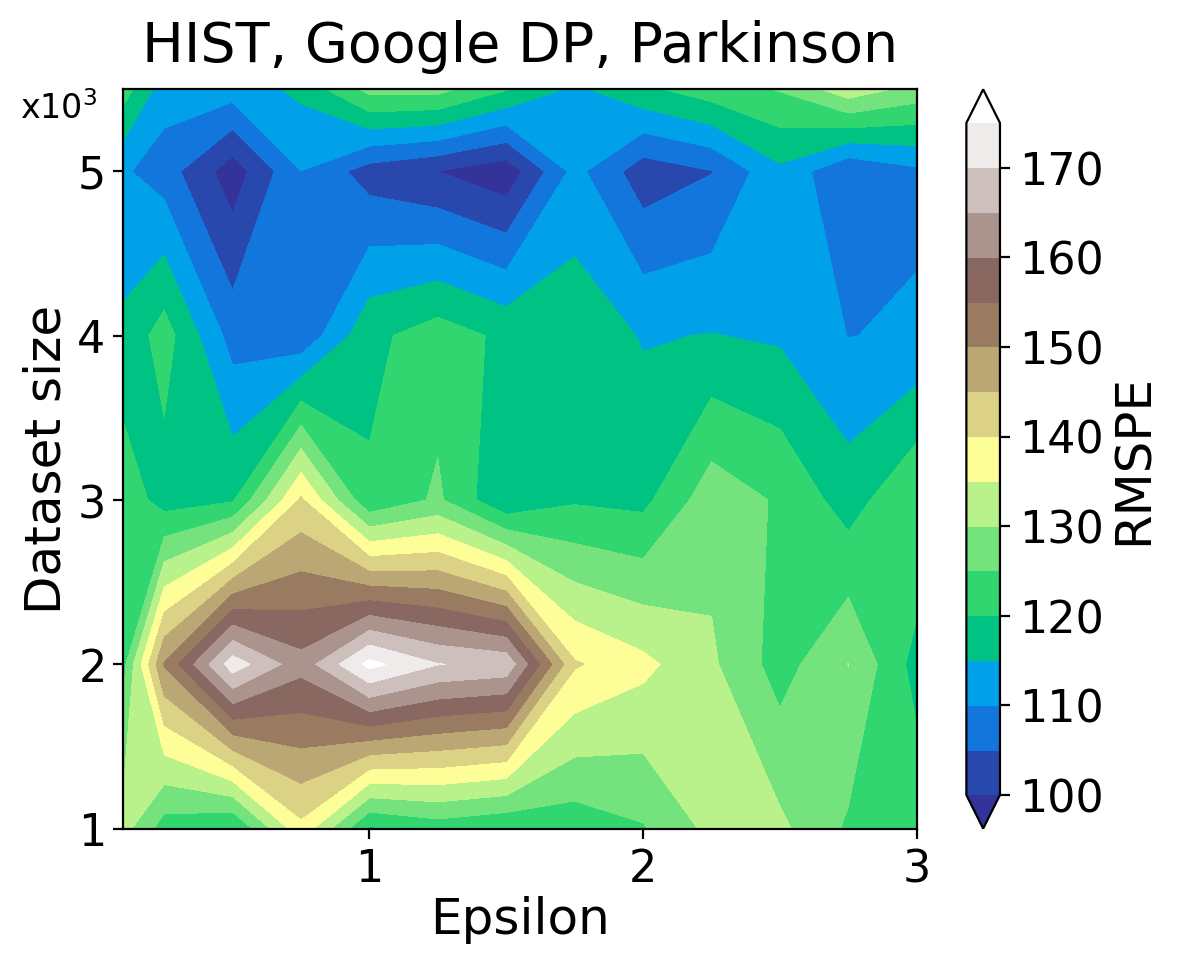}}
	\\
	\subfloat[]{\label{fig:exp2:smart:sum:P}\includegraphics[width=0.25\textwidth]{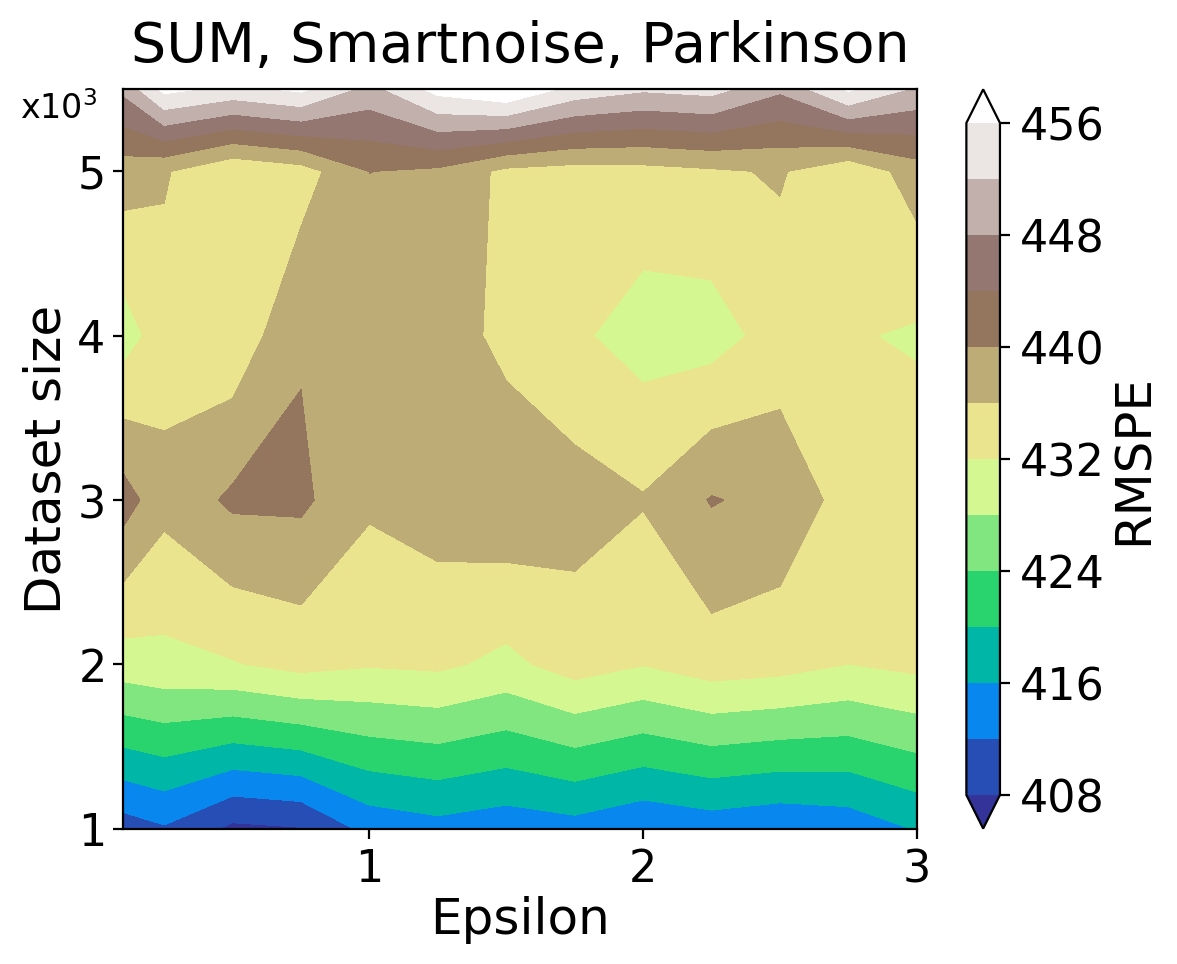}}
	\subfloat[]{\label{fig:exp2:smart:count:P}\includegraphics[width=0.25\textwidth]{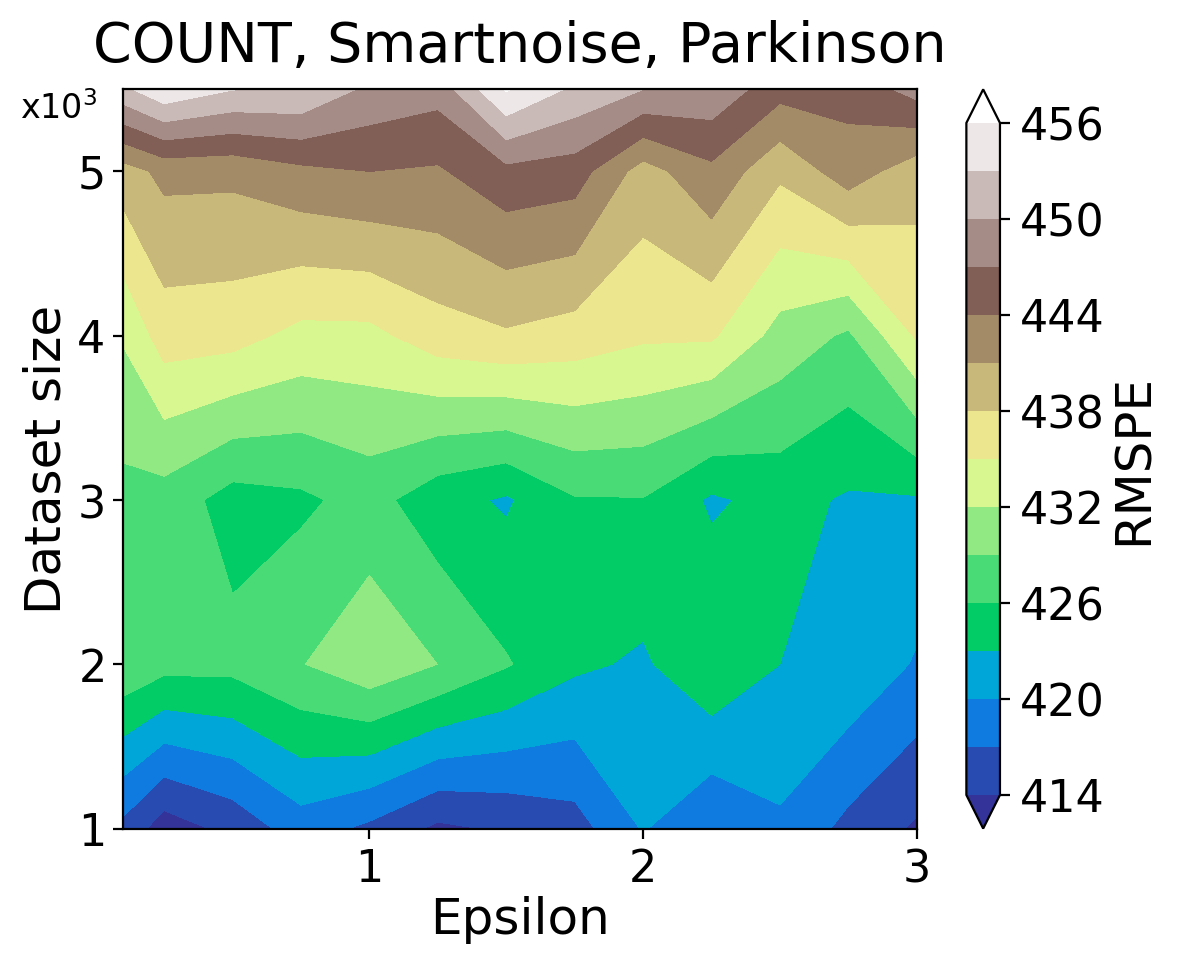}}
	\subfloat[]{\label{fig:exp2:smart:avg:P}\includegraphics[width=0.25\textwidth]{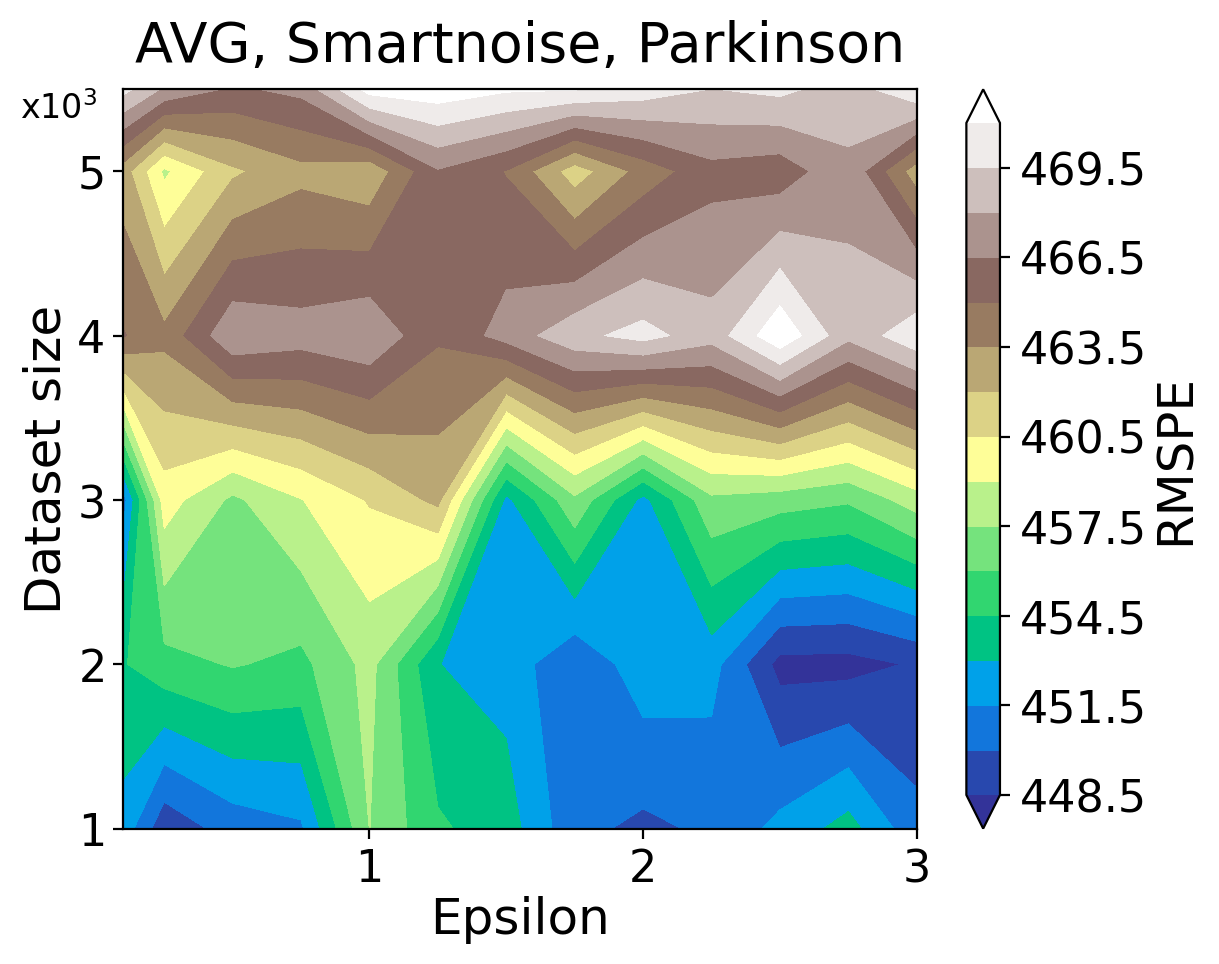}}
	\subfloat[]{\label{fig:exp2:smart:hist:P}\includegraphics[width=0.25\textwidth]{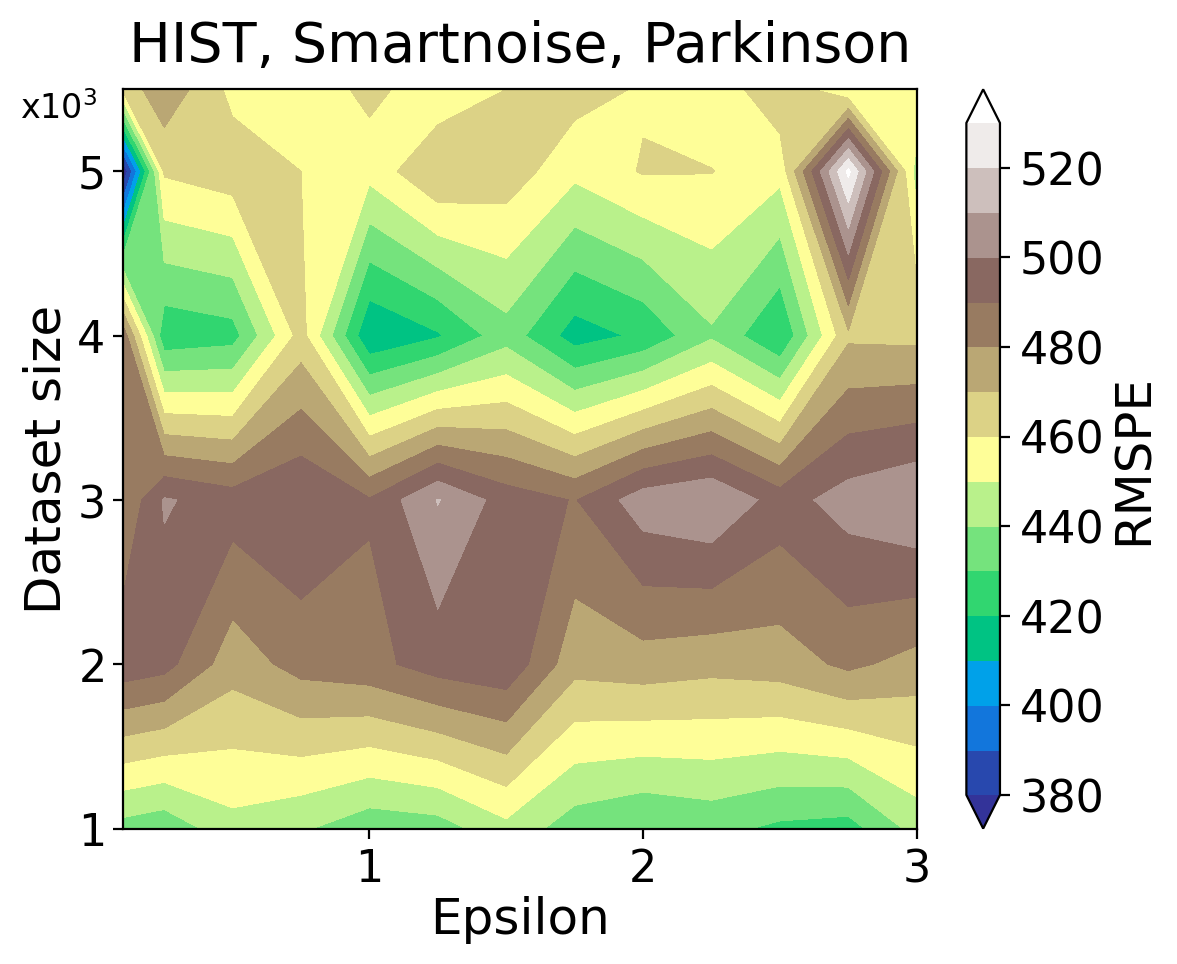}}
	
	\caption[Results of Experiment 2. Run-time overhead for statistical query tools]{Contour plots for the evaluation of statistical query tools on run-time overhead, for different data sizes (Table \ref{table_dataset_sizes}), $\epsilon$ values (Table \ref{table_epsilons}), and queries (Table \ref{table_queries}). RMSPE is defined in Section~\ref{evaluation_criteria}.}
	
	\label{fig:exp2:contour}
\end{figure}

In Figure~\ref{fig:exp2:contour}, we provide contour plots for each tool's run-time overhead regarding $\epsilon$ and data size. The plots show that large data sizes generally increase the run-time for Smartnoise, \ie~between 400\% to 500\% RMSPE for the smallest data sizes and about 450\% to 550\% for the largest. In contrast, Google DP performs does not follow our anticipation with RMPSE between 110\% to 150\% for the smallest data size, and 100\% to 130\% for the largest. Overall, Google DP outperforms Smartnoise, which runs around 400\% to 500\% slower when conducting DP queries, compared to Google DP which runs around 100\% slower. A possible reason is that Google DP performs more efficient DP calculations, using a plugin inside the database compiled to native code (Section~\ref{architecture_google-dp}). Therefore, the run-time might improve, since no additional layer operates between the database and the analyst that conducts the queries. In comparison, Smartnoise implements pre-processing of queries before communicating with the database (Section~\ref{architecutre_smartnoise}), which might negatively impact the run-time.

\begin{figure}[!h]
	\centering
	\subfloat[]{\label{fig:exp2:gdp:H:1000}\includegraphics[width=0.25\textwidth]{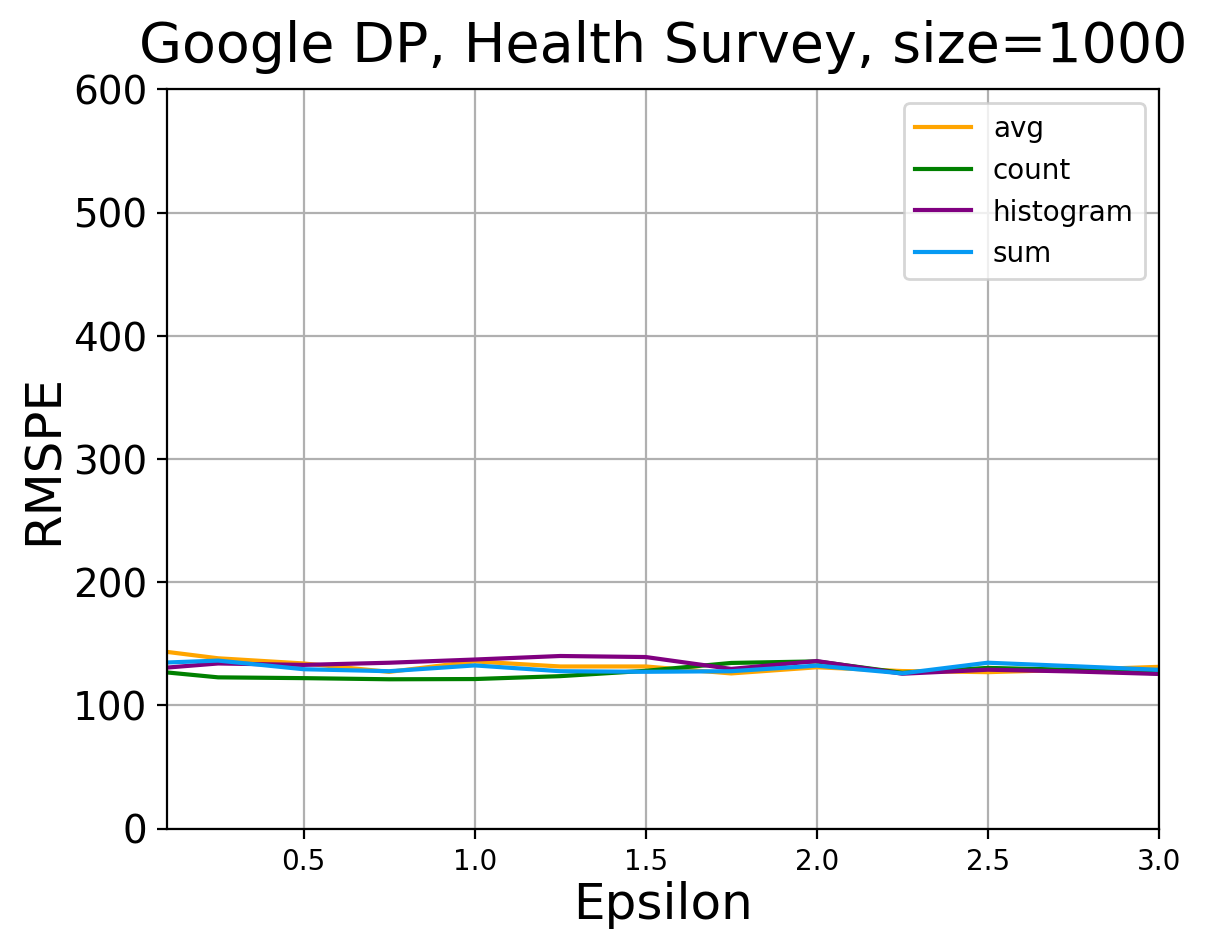}}
	\subfloat[]{\label{fig:exp2:gdp:H:2000}\includegraphics[width=0.25\textwidth]{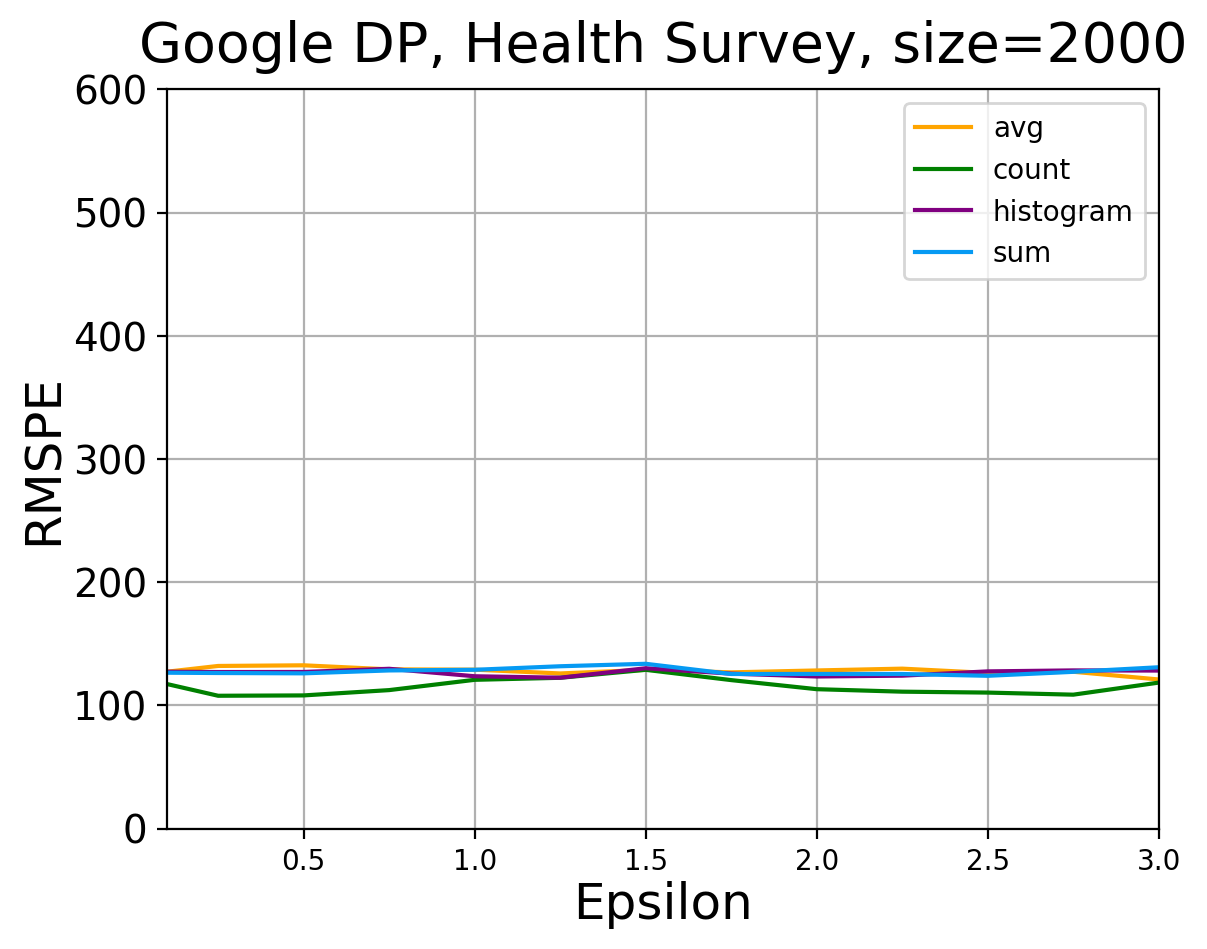}}
	\subfloat[]{\label{fig:exp2:gdp:H:9000}\includegraphics[width=0.25\textwidth]{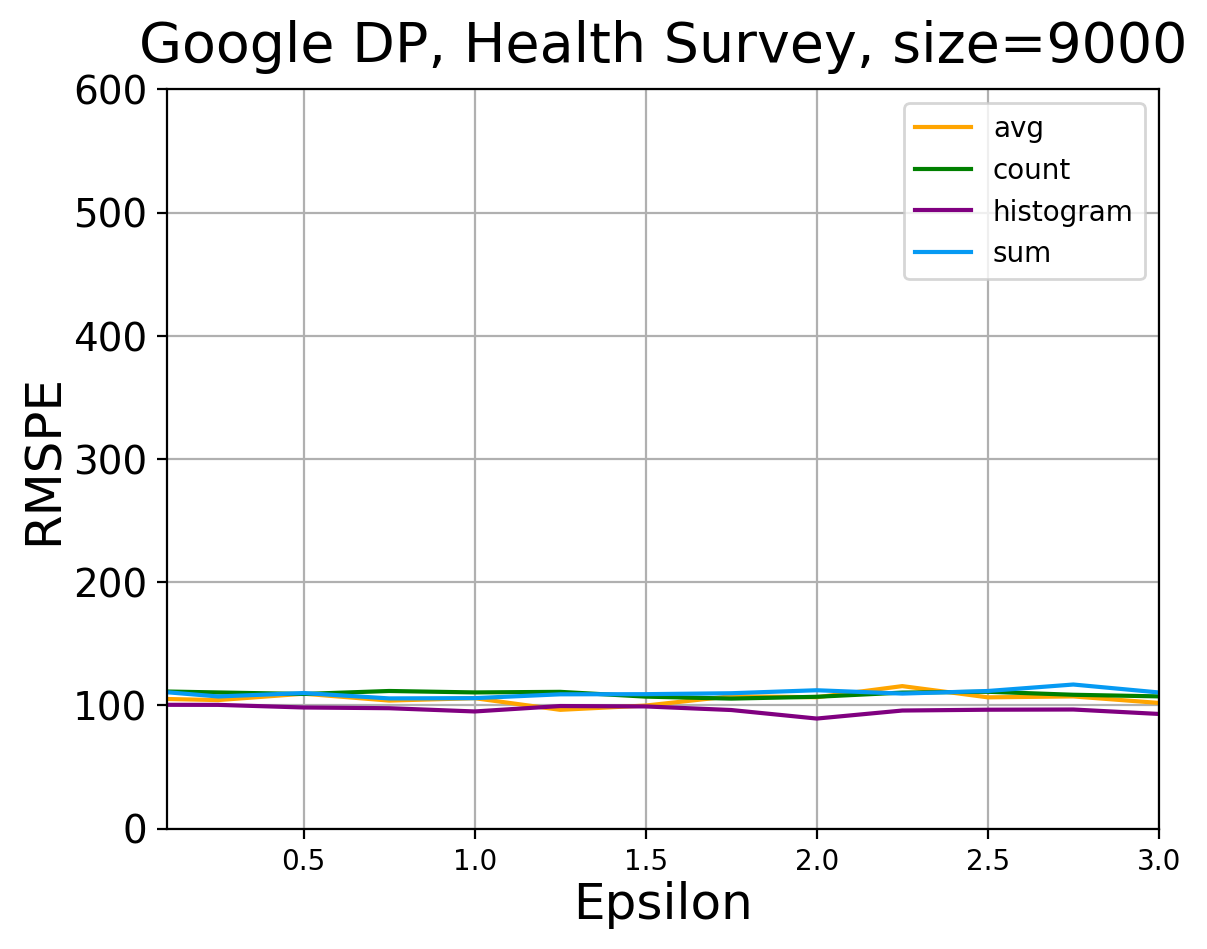}}
	\subfloat[]{\label{fig:exp2:gdp:H:9358}\includegraphics[width=0.25\textwidth]{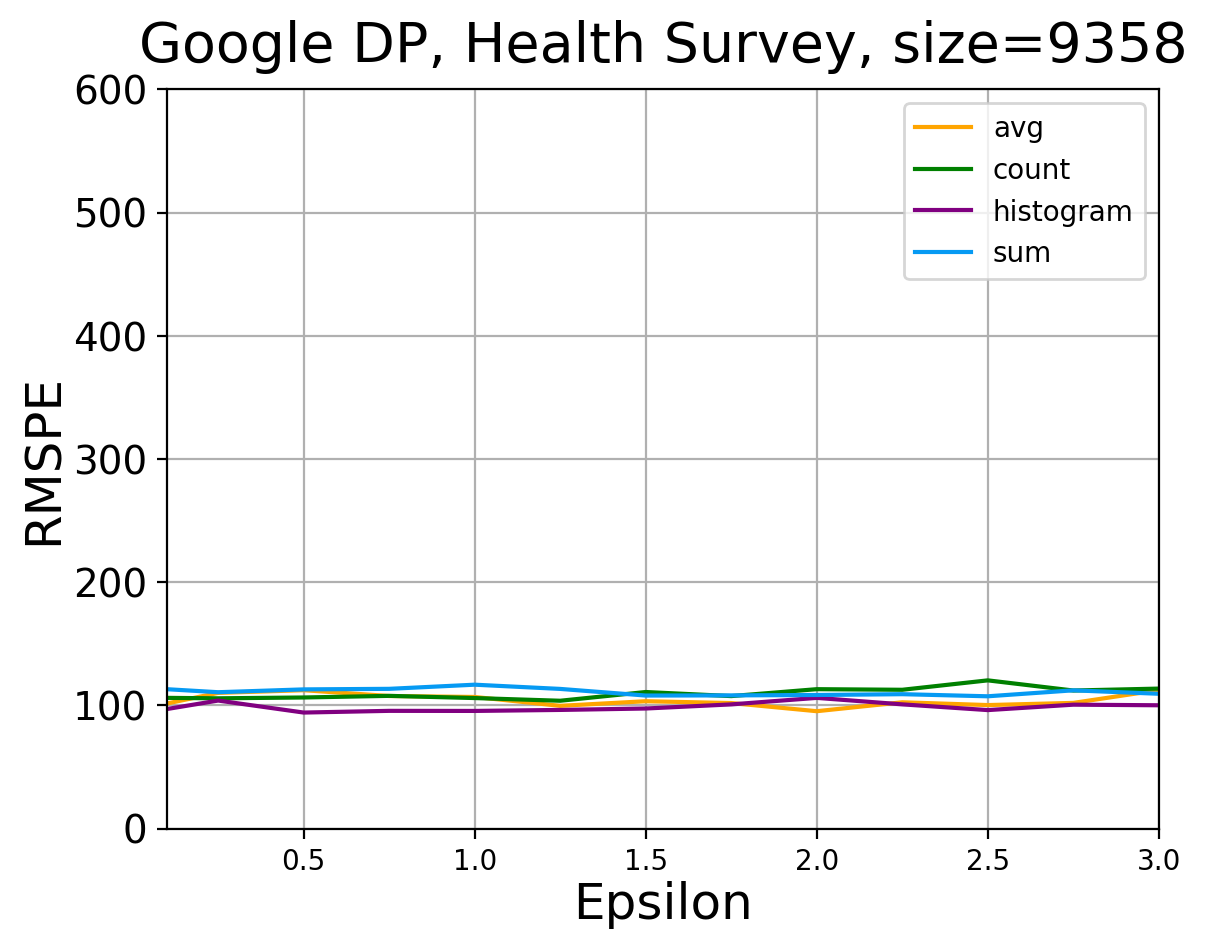}}
	\\
	\subfloat[]{\label{fig:exp2:smart:H:1000}\includegraphics[width=0.25\textwidth]{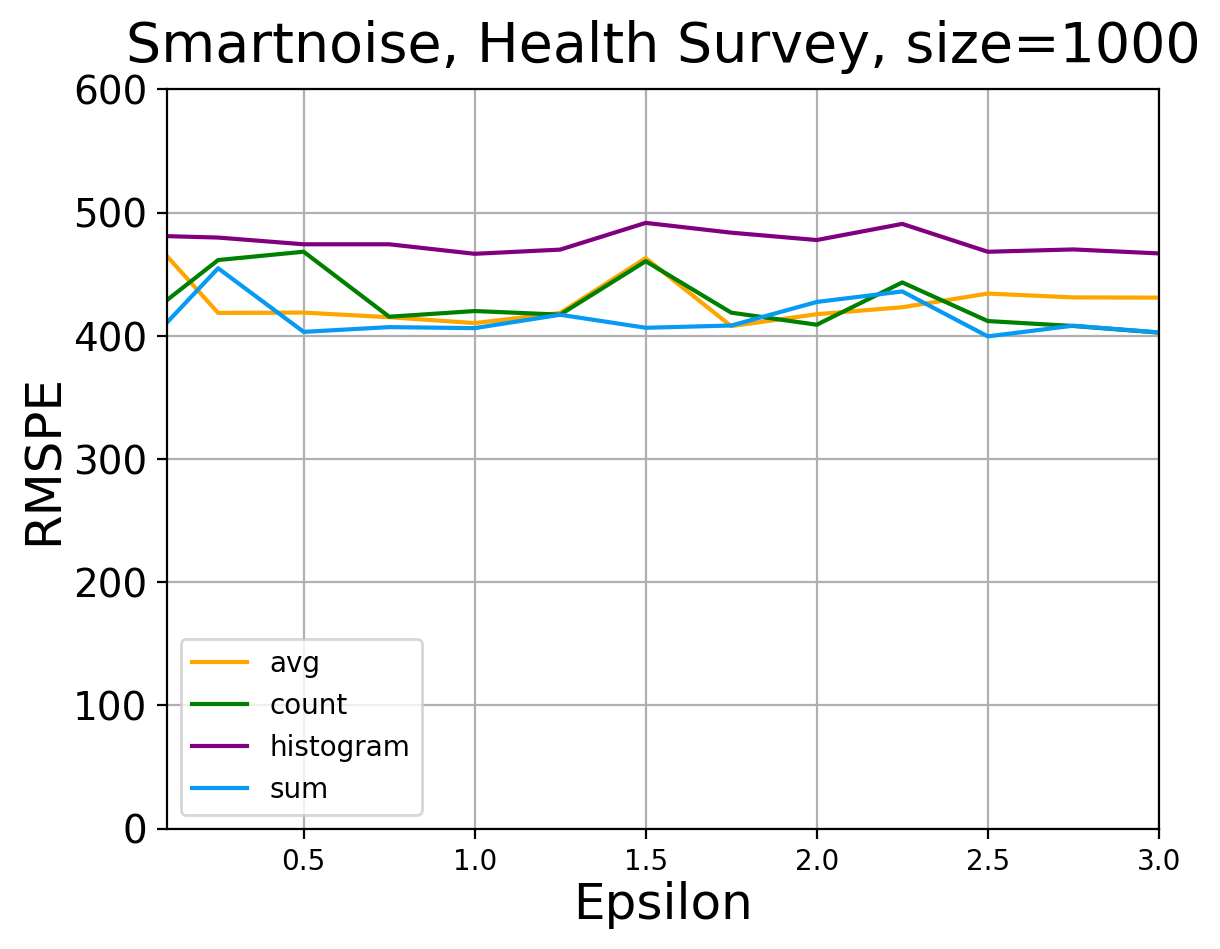}}
	\subfloat[]{\label{fig:exp2:smart:H:2000}\includegraphics[width=0.25\textwidth]{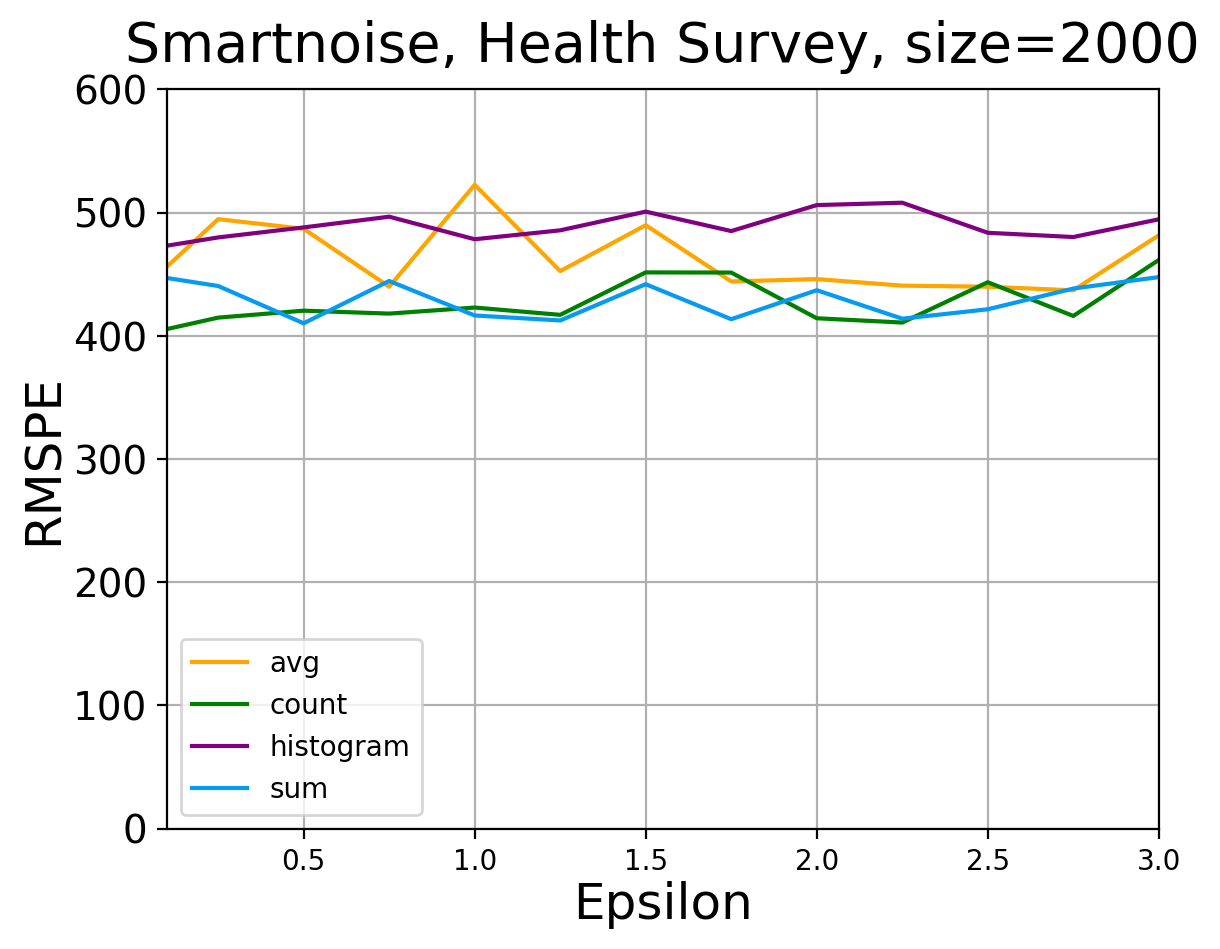}}
	\subfloat[]{\label{fig:exp2:smart:H:9000}\includegraphics[width=0.25\textwidth]{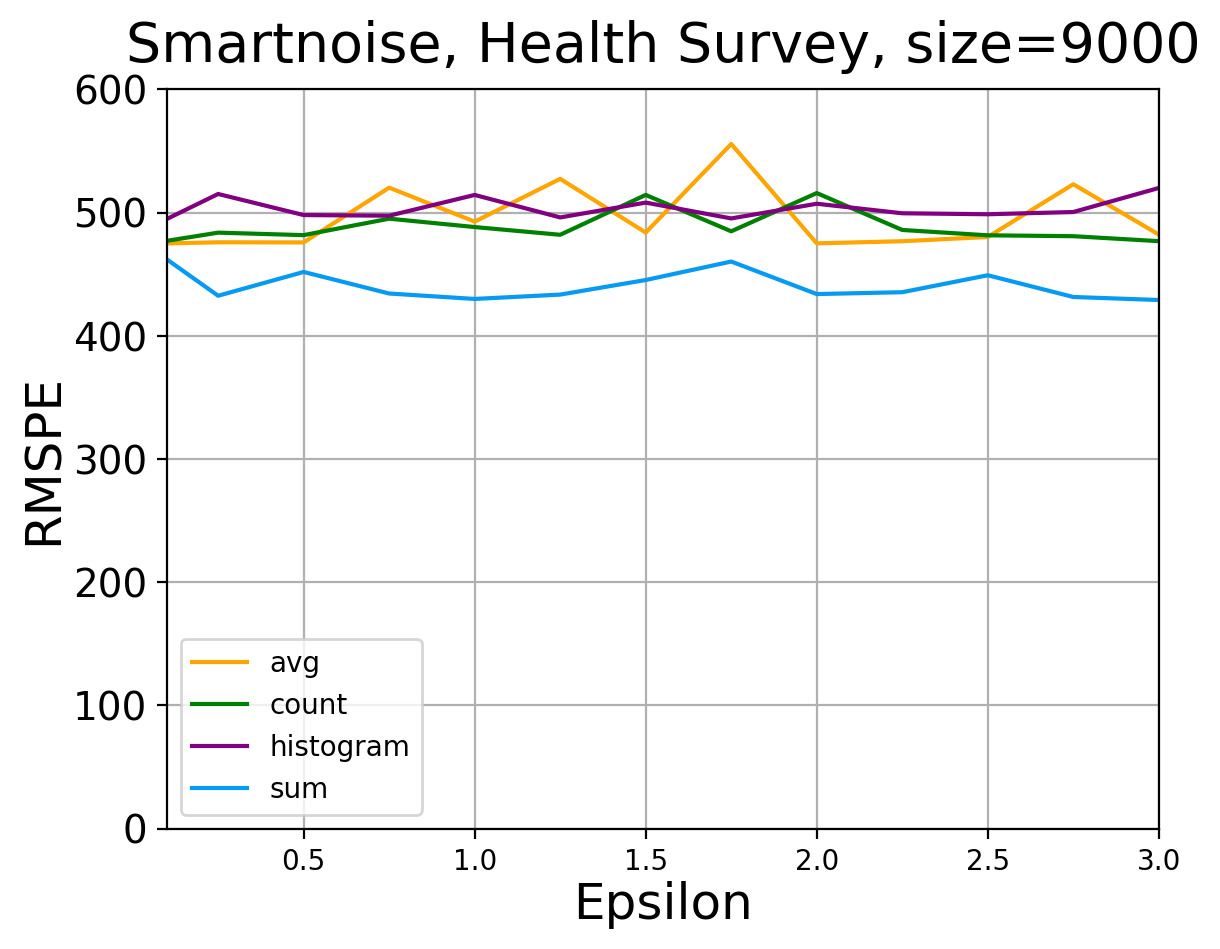}}
	\subfloat[]{\label{fig:exp2:smart:H:9358}\includegraphics[width=0.25\textwidth]{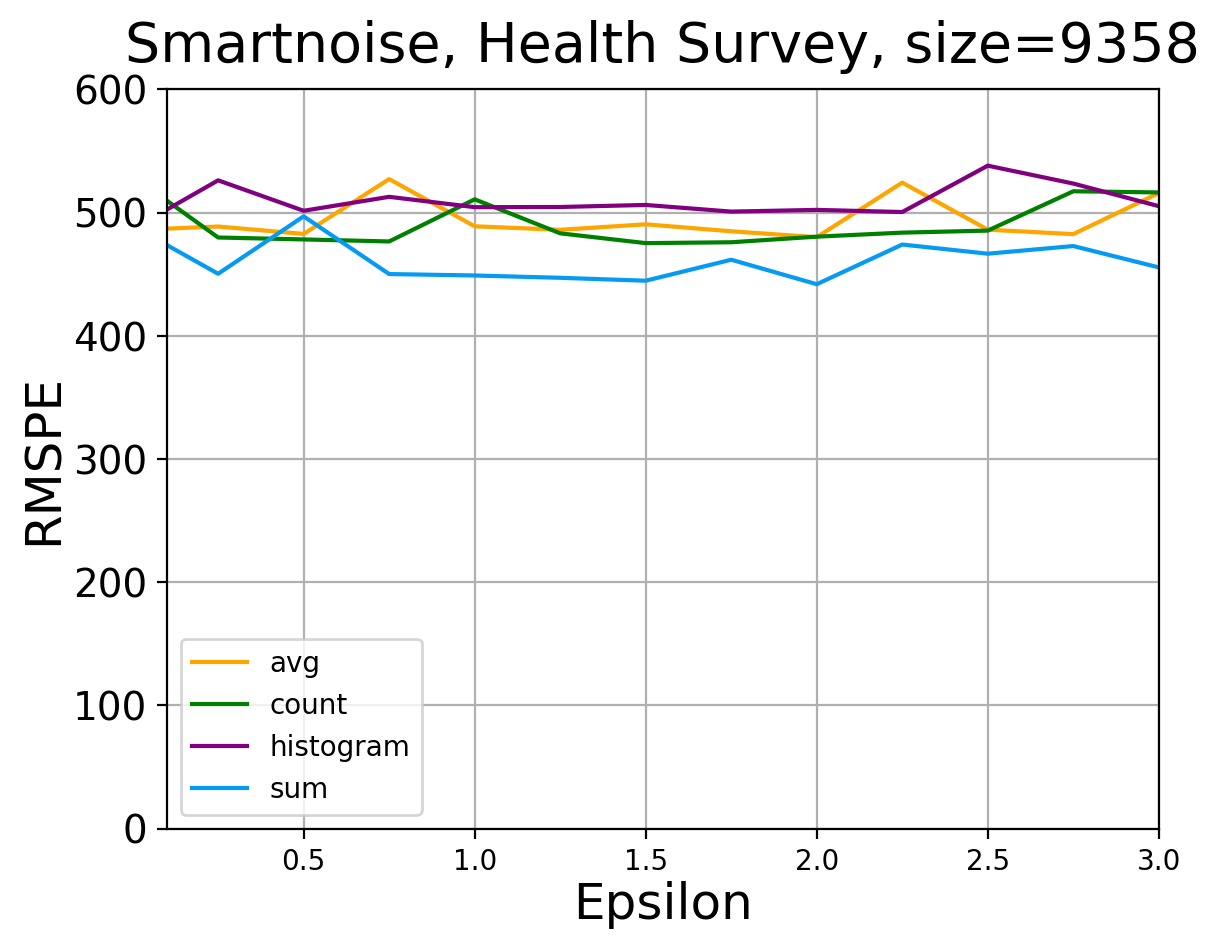}}
	\\
	\subfloat[]{\label{fig:exp2:gdp:P:1000}\includegraphics[width=0.25\textwidth]{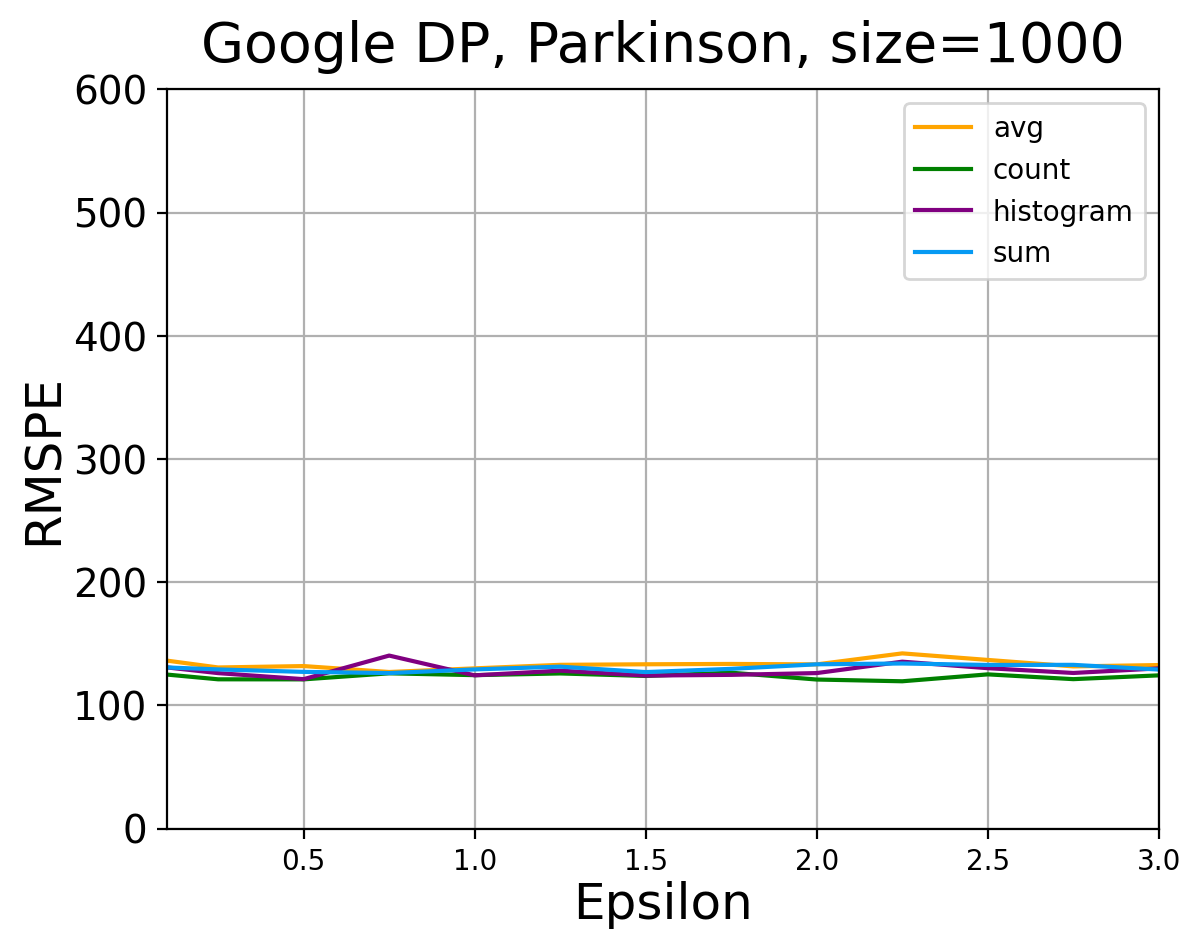}}
	\subfloat[]{\label{fig:exp2:gdp:P:2000}\includegraphics[width=0.25\textwidth]{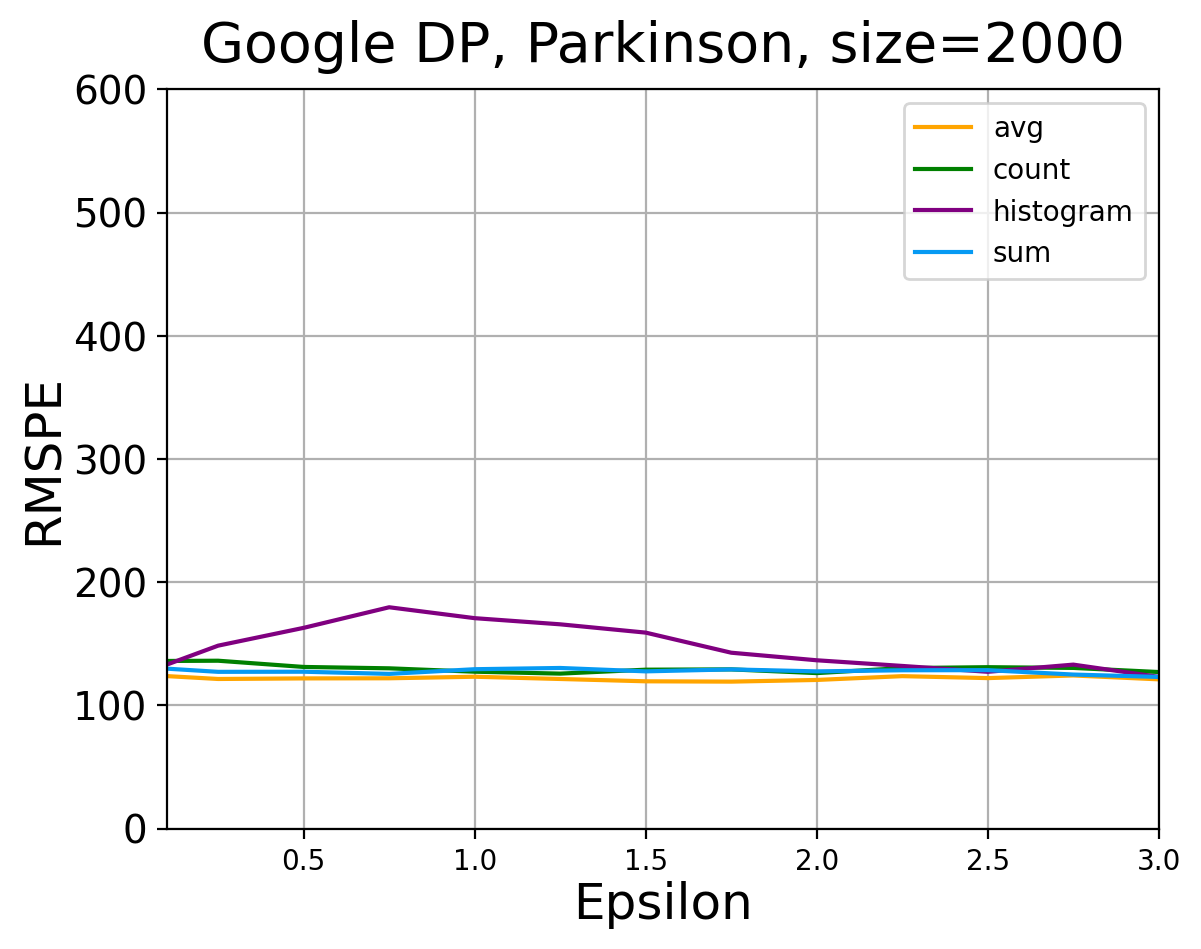}}
	\subfloat[]{\label{fig:exp2:gdp:P:5000}\includegraphics[width=0.25\textwidth]{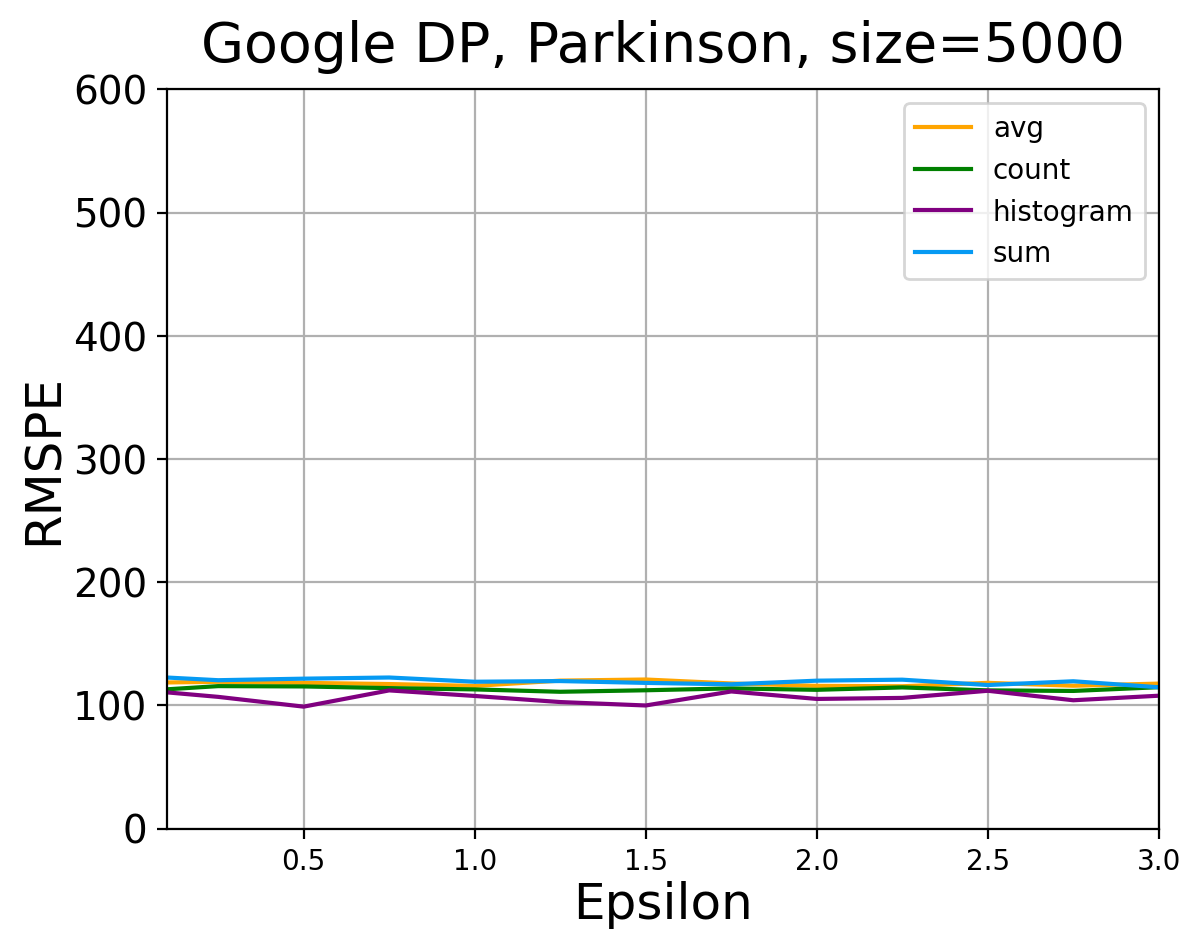}}
	\subfloat[]{\label{fig:exp2:gdp:P:5499}\includegraphics[width=0.25\textwidth]{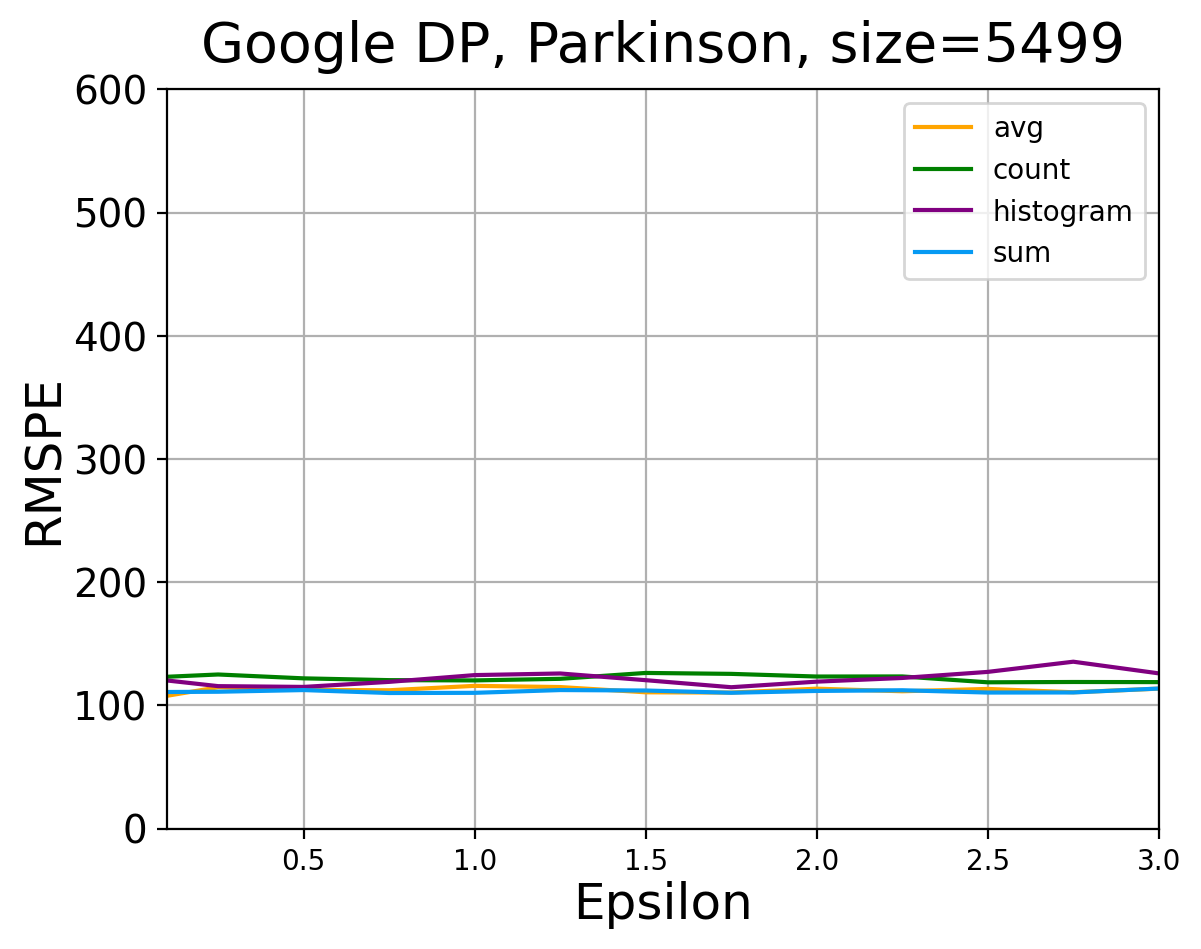}}
	\\
	\subfloat[]{\label{fig:exp2:smart:P:1000}\includegraphics[width=0.25\textwidth]{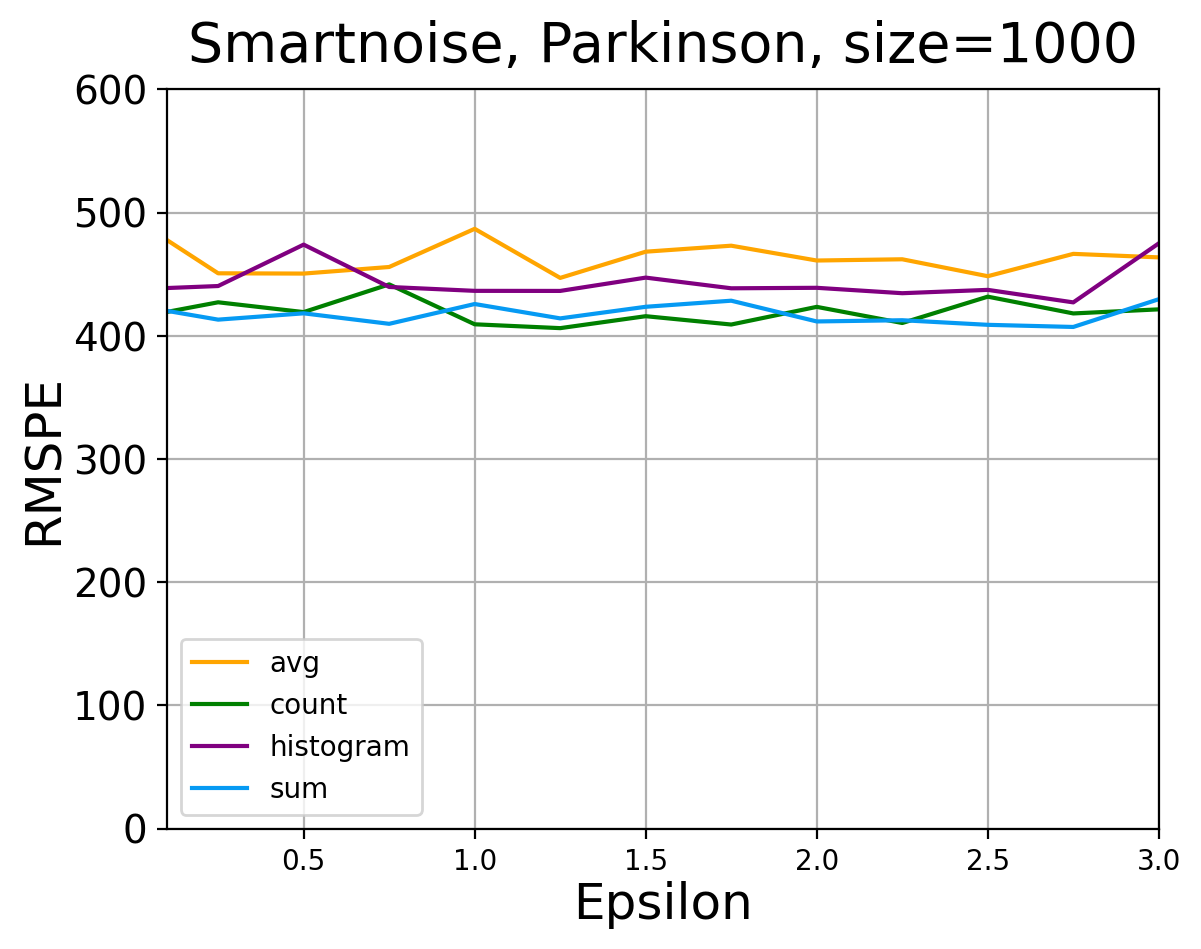}}
	\subfloat[]{\label{fig:exp2:smart:P:2000}\includegraphics[width=0.25\textwidth]{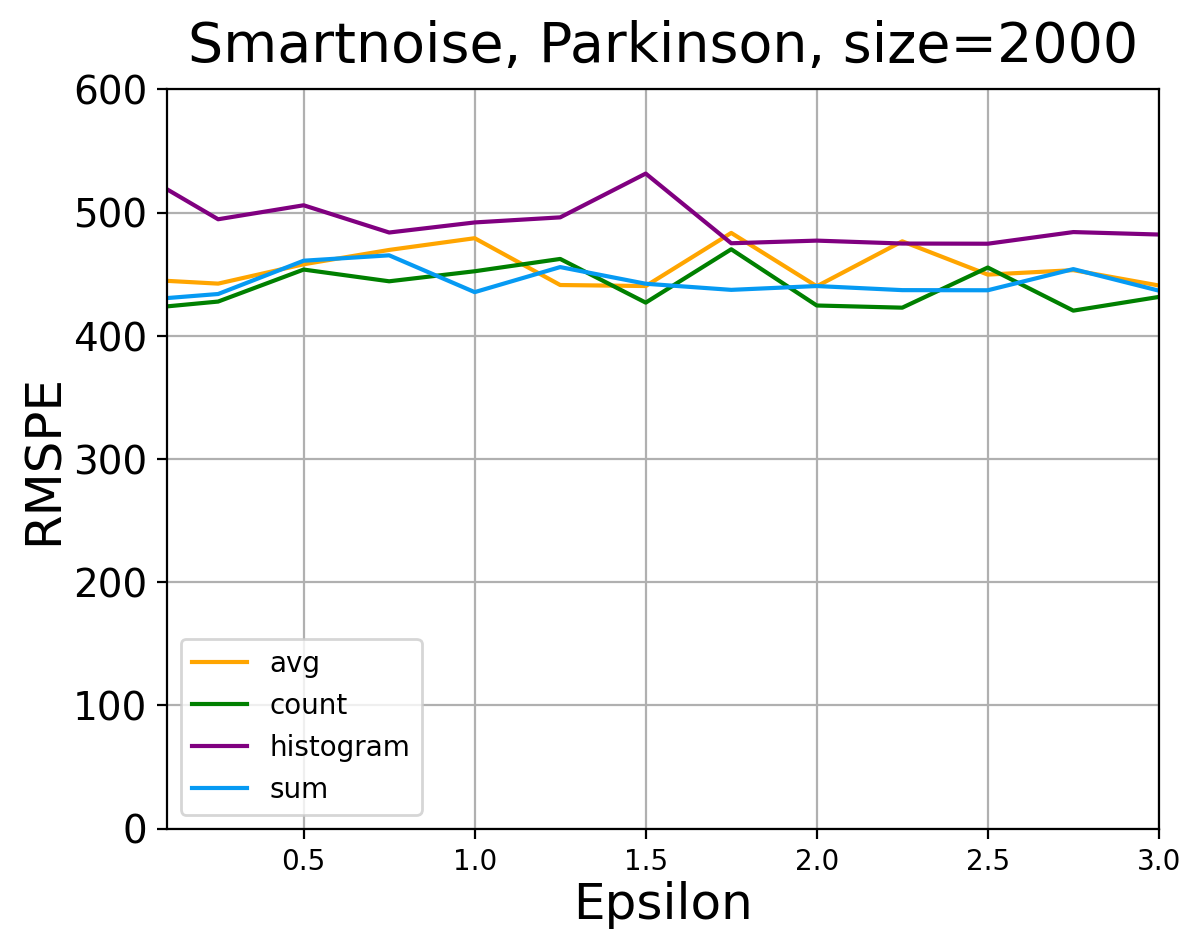}}
	\subfloat[]{\label{fig:exp2:smart:P:5000}\includegraphics[width=0.25\textwidth]{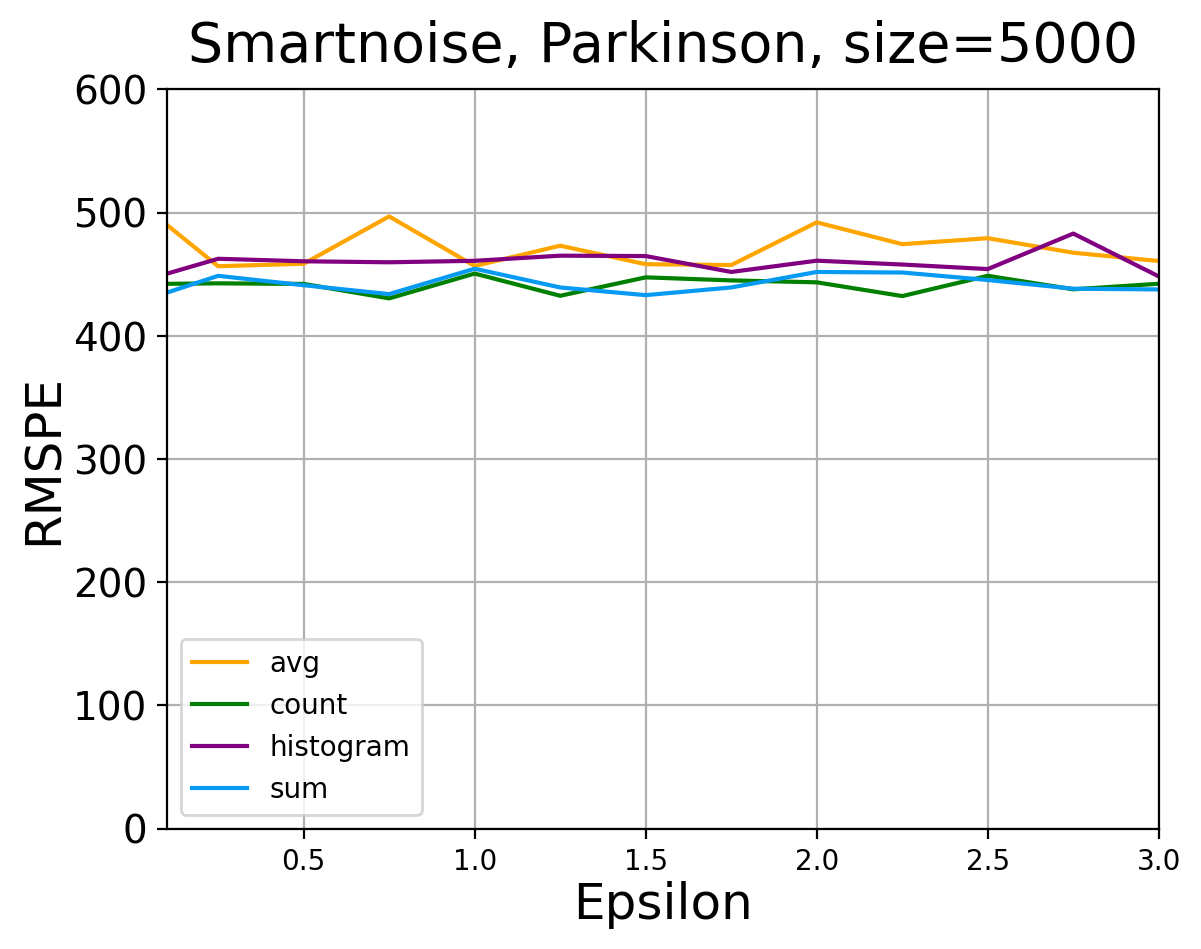}}
	\subfloat[]{\label{fig:exp2:smart:P:5499}\includegraphics[width=0.25\textwidth]{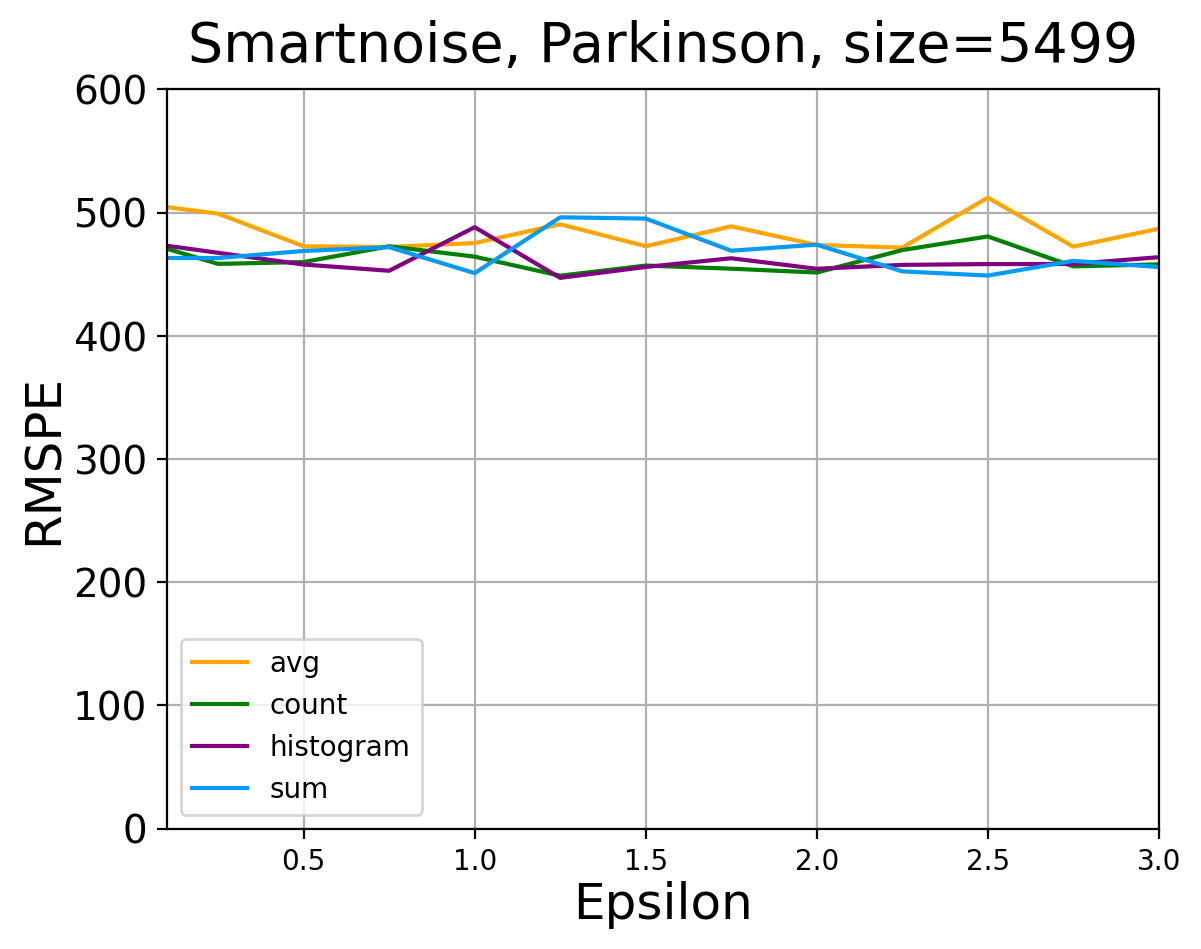}}
	
	\caption[Results of Experiment 2. Run-time overhead for statistical query tools]{The evaluation results of statistical query tools on run-time overhead when DP is integrated, for different data sizes (Table \ref{table_dataset_sizes}), $\epsilon$ values (Table \ref{table_epsilons}), and queries (Table \ref{table_queries}). RMSPE is defined in Section~\ref{evaluation_criteria}.}
	
	\label{fig:exp2:line}
\end{figure}

Figure~\ref{fig:exp2:line} shows how $\epsilon$ impacts query accuracy for both Google DP and Smartnoise for both data sets. It reveals that higher $\epsilon$ values do not necessarily decrease or increase the RMSPE, indicating that $\epsilon$ values do not generally impact the run-time of DP queries. Though slight local fluctuations exist, this observation is quite explicit and holds for both tools and datasets, especially obvious for Google DP on \emph{Health Survey}.

\subsubsection{Memory overhead}\label{sq:exp3}

This section investigates how DP impacts memory usage when running statistical queries on Google DP and Smartnoise. We use various $\epsilon$ and data sizes to see how the results differ under different settings. To gain a nuanced result, we measure the memory usage in both the container of the Postgres database where noise is added and processed data are stored and the container that processes the issued private or non-private queries to the database.

In this evaluation, we anticipate the memory overhead for running queries to increase when DP is integrated, and that the memory overhead grows as the data size rises since more data are involved in the calculation procedures. As elaborated below, the results show that DP generally poses an additional memory usage of less than 3\% for the operations in the Postgres database, while the processing of private queries poses 5\%-40\% extra memory consumption. However, the results disclose no clear relationship between $\epsilon$, data size, and memory overhead, and fluctuations of different levels exist throughout the results. In comparison, there is no apparent advantage of one tool over another regarding memory overhead, yet Google DP slightly outperforms Smartnoise on the \emph{Health Survey} data composed of categorical variables.

\begin{figure}[!ht]
	\centering
	\subfloat[]{\label{fig:exp3:gdp:H}\includegraphics[width=0.25\textwidth]{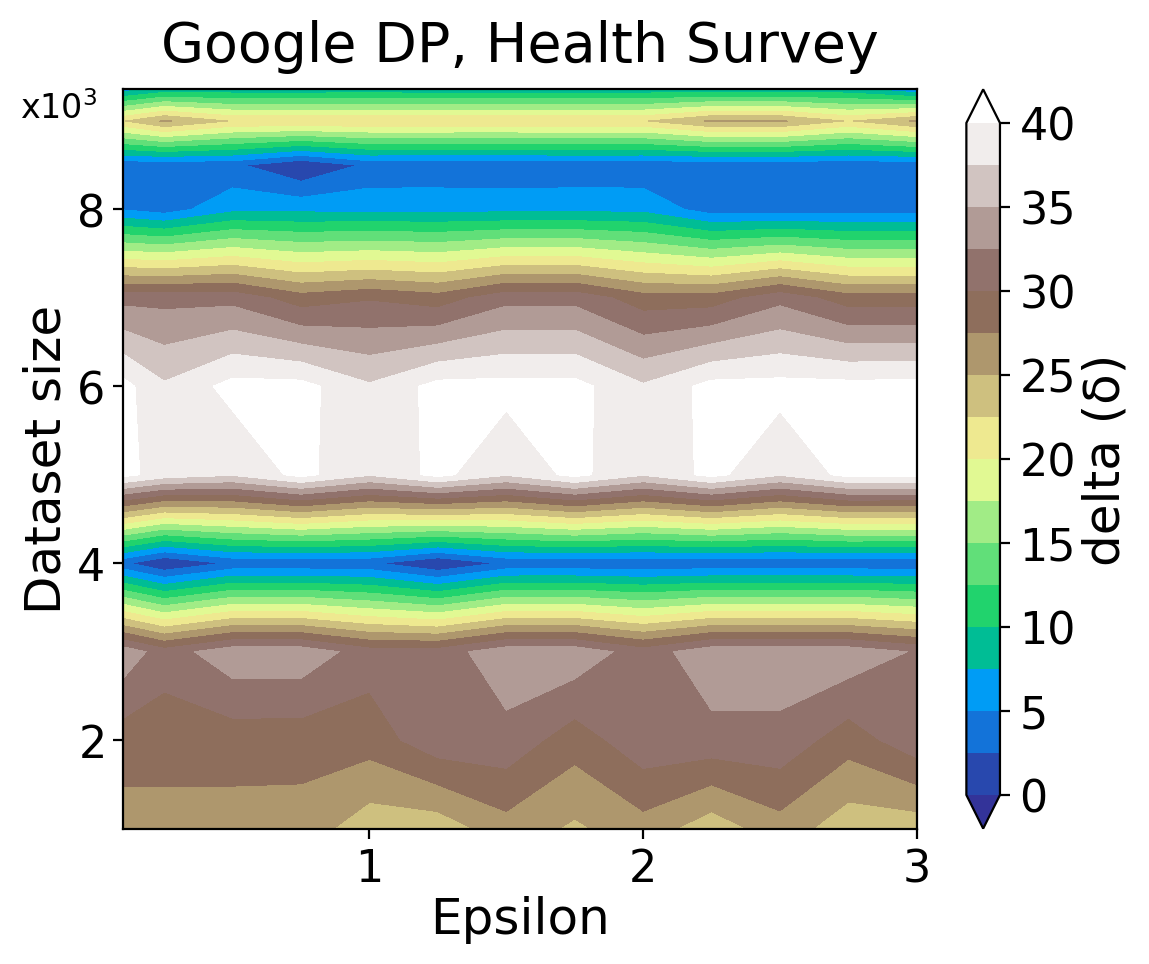}}
	\subfloat[]{\label{fig:exp3:gdp:post:H}\includegraphics[width=0.25\textwidth]{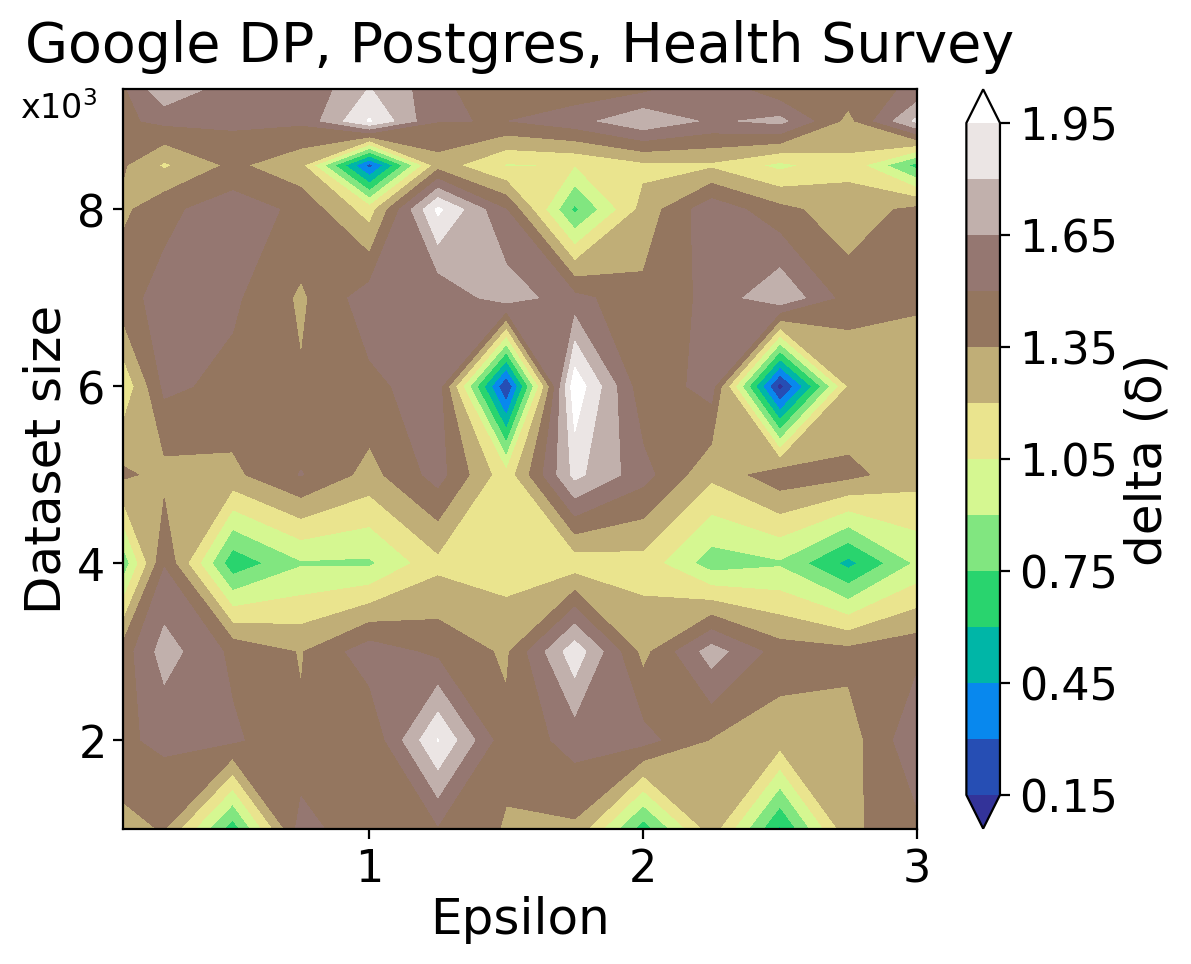}}
	\subfloat[]{\label{fig:exp3:smart:H}\includegraphics[width=0.25\textwidth]{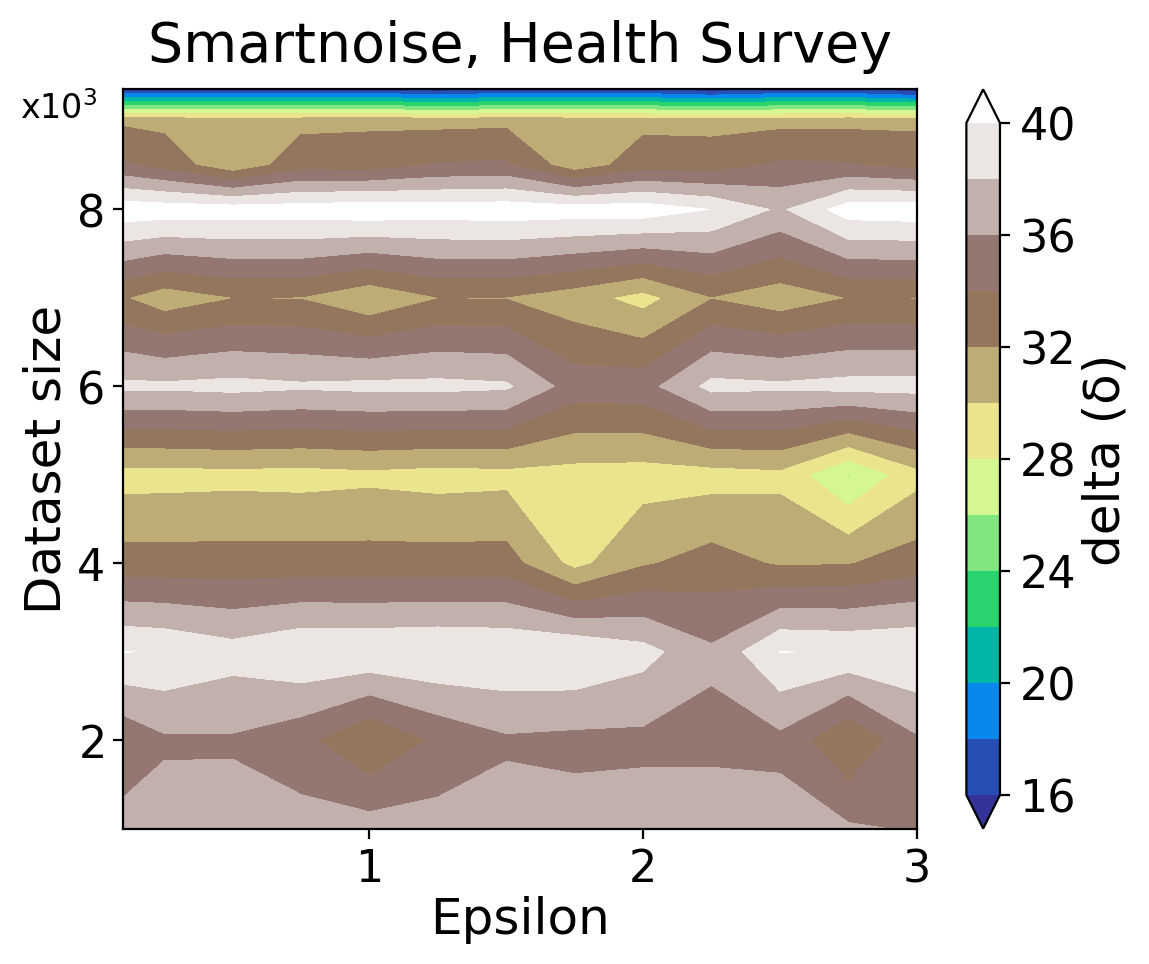}}
	\subfloat[]{\label{fig:exp3:smart:post:H}\includegraphics[width=0.25\textwidth]{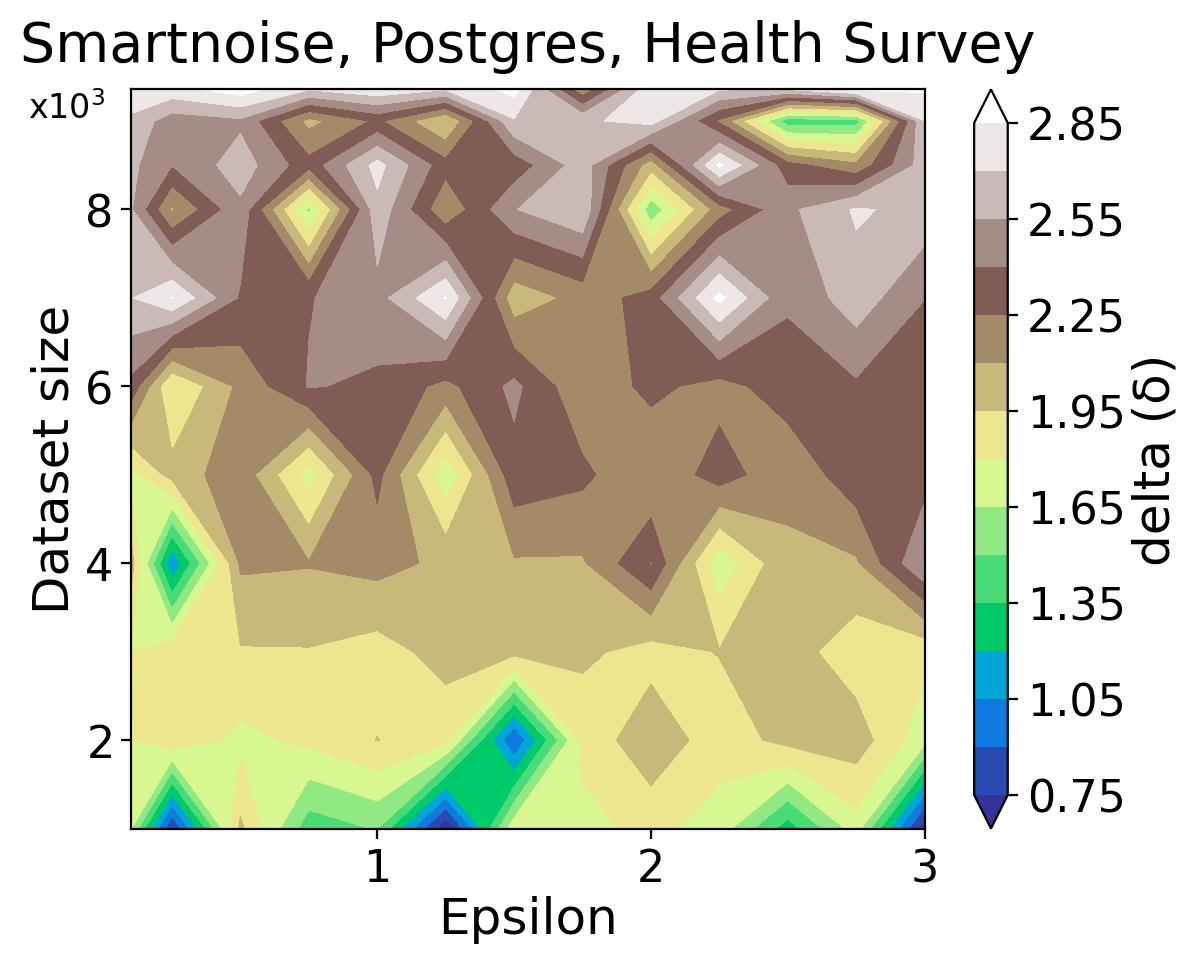}}
	\\
	\subfloat[]{\label{fig:exp3:gdp:P}\includegraphics[width=0.25\textwidth]{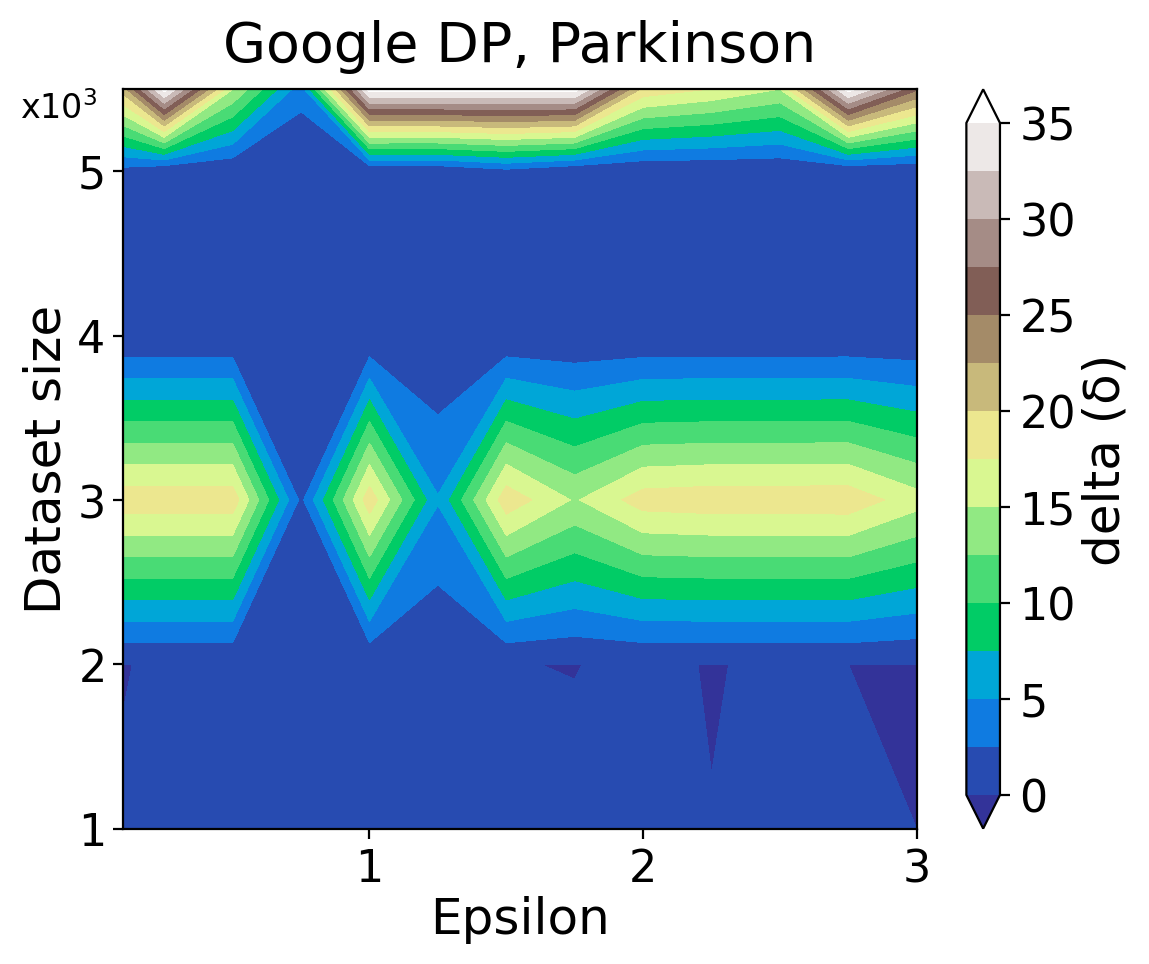}}
	\subfloat[]{\label{fig:exp3:gdp:post:P}\includegraphics[width=0.25\textwidth]{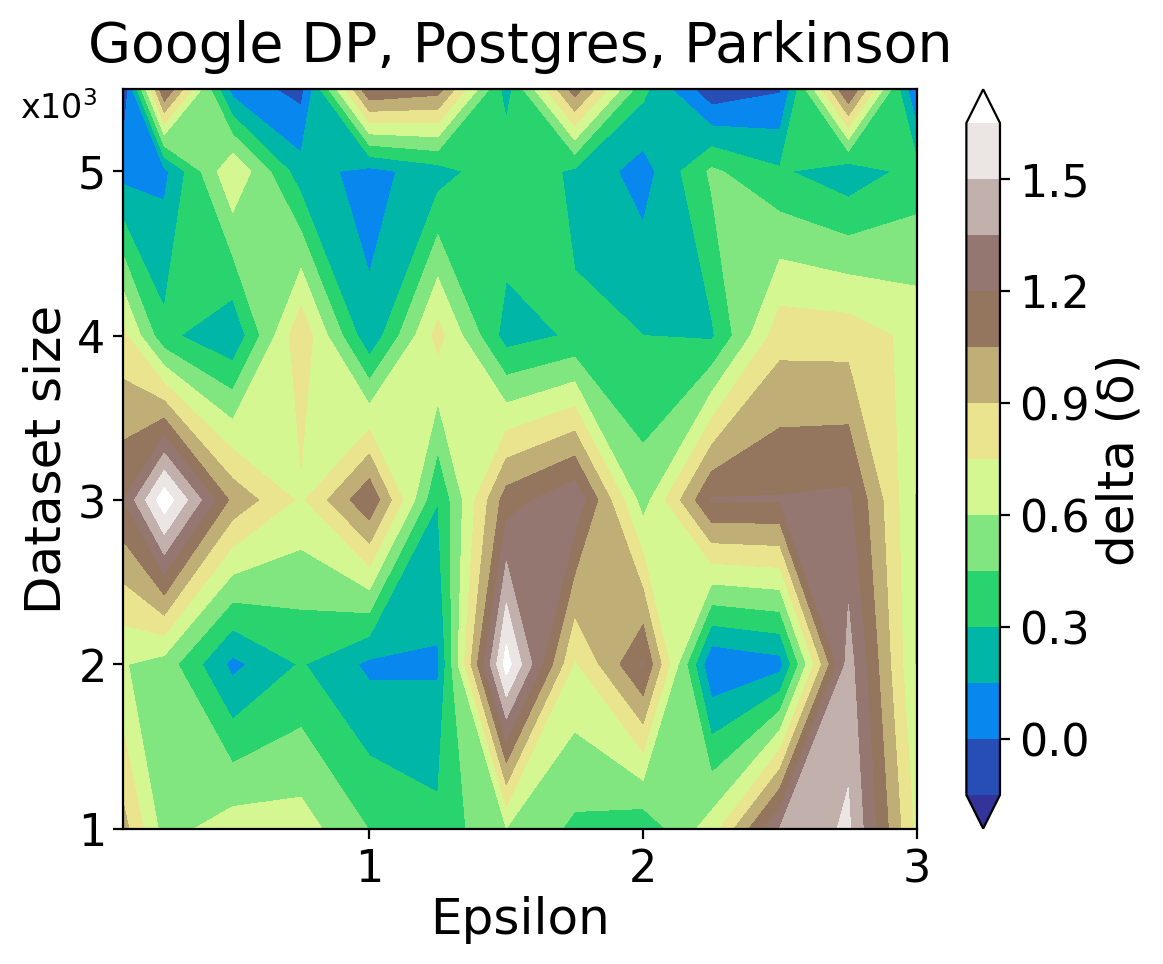}}
	\subfloat[]{\label{fig:exp3:smart:P}\includegraphics[width=0.25\textwidth]{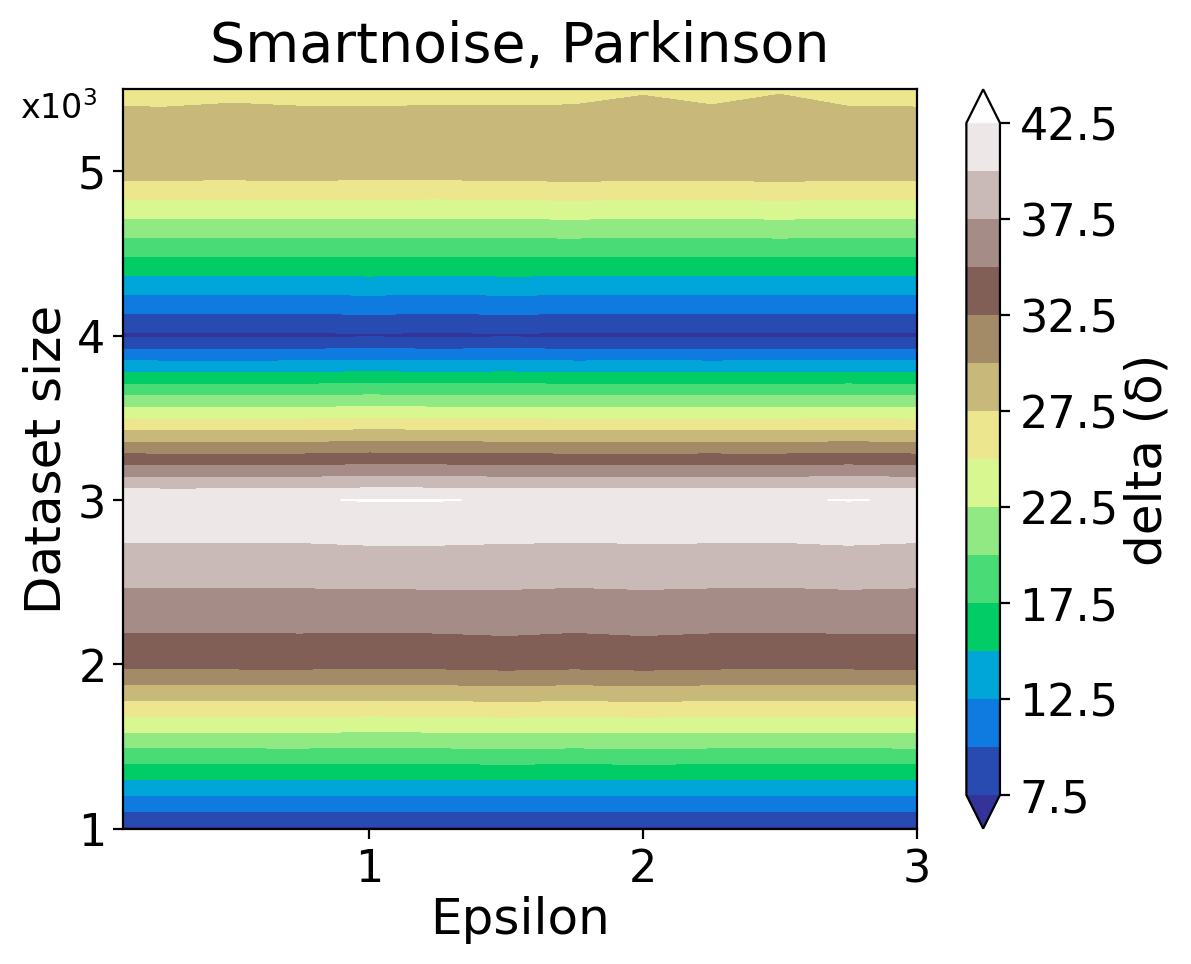}}
	\subfloat[]{\label{fig:exp3:smart:post:P}\includegraphics[width=0.25\textwidth]{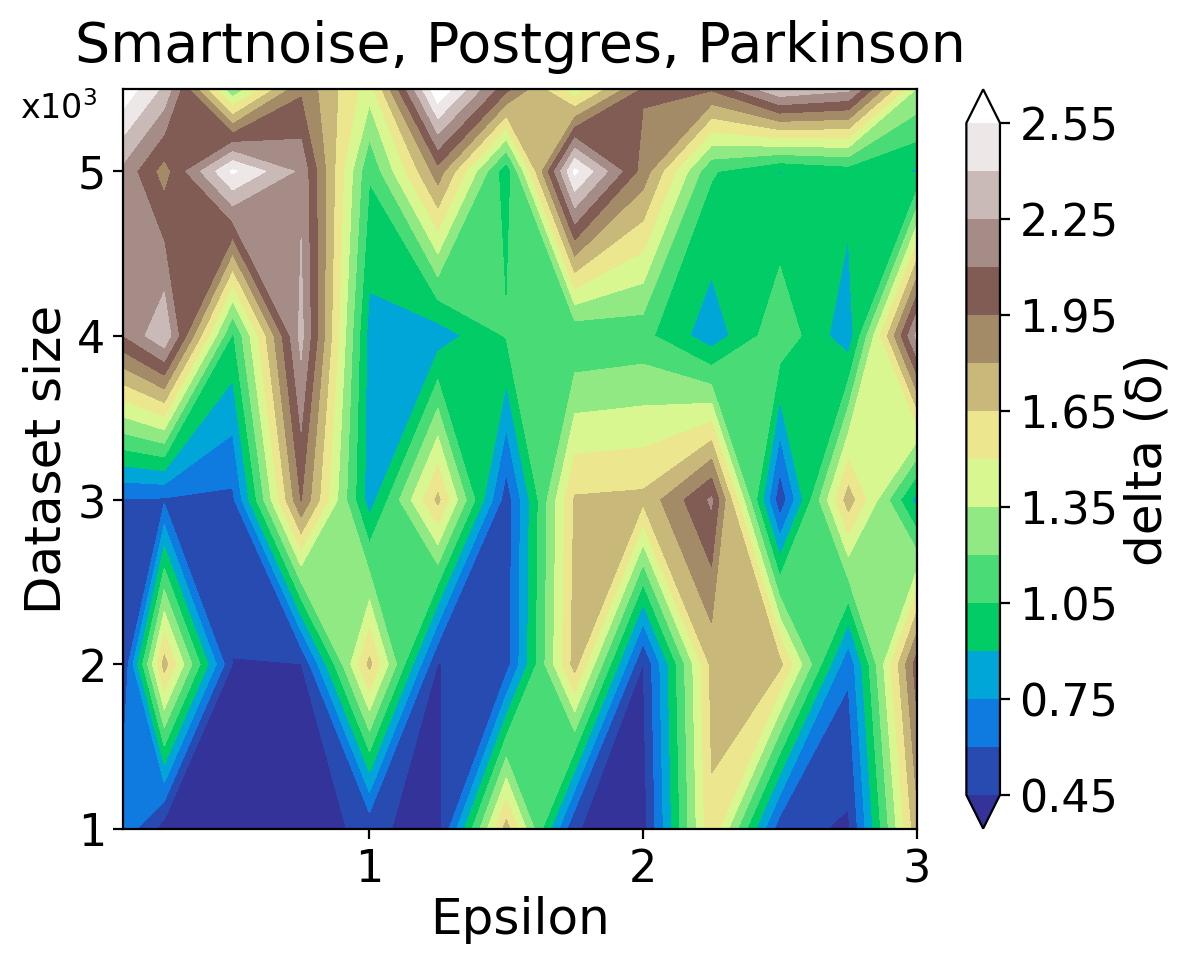}}
	
	\caption[Results of Experiment 3. Memory overhead for statistical query tools]{Contour plots for the evaluation of statistical query tools on memory overhead for different data sizes (Table \ref{table_dataset_sizes}), $\epsilon$ values (Table \ref{table_epsilons}), and queries (Table \ref{table_queries}). $\delta$ is defined in Section~\ref{evaluation_criteria}.}
	
	\label{fig:exp3:contour}
\end{figure}

Figure~\ref{fig:exp3:contour} shows contour plots for each tool's performance on memory overhead regarding different $\epsilon$ values and data sizes. While the plots reveal no explicit patterns, the results on \texttt{HISTOGRAM} manifest a slight trend for the Postgres database operation in Smartnoise that memory overhead grows with an increase in data size (figures~\ref{fig:exp3:smart:post:H} and~\ref{fig:exp3:smart:post:P}), which holds for both data sets. However, we also observe a significant impact of data size on the query processing than that of $\epsilon$ (figures~\ref{fig:exp3:gdp:H},~\ref{fig:exp3:smart:H},~\ref{fig:exp3:gdp:P}, and~\ref{fig:exp3:smart:P}), where the value of $\epsilon$ does not necessarily affect memory overhead, and the data size irregularly influences the memory overhead.

\begin{figure}[!ht]
	\centering
	\subfloat[]{\label{fig:exp3:gdp:post:H:7000}\includegraphics[width=0.25\textwidth]{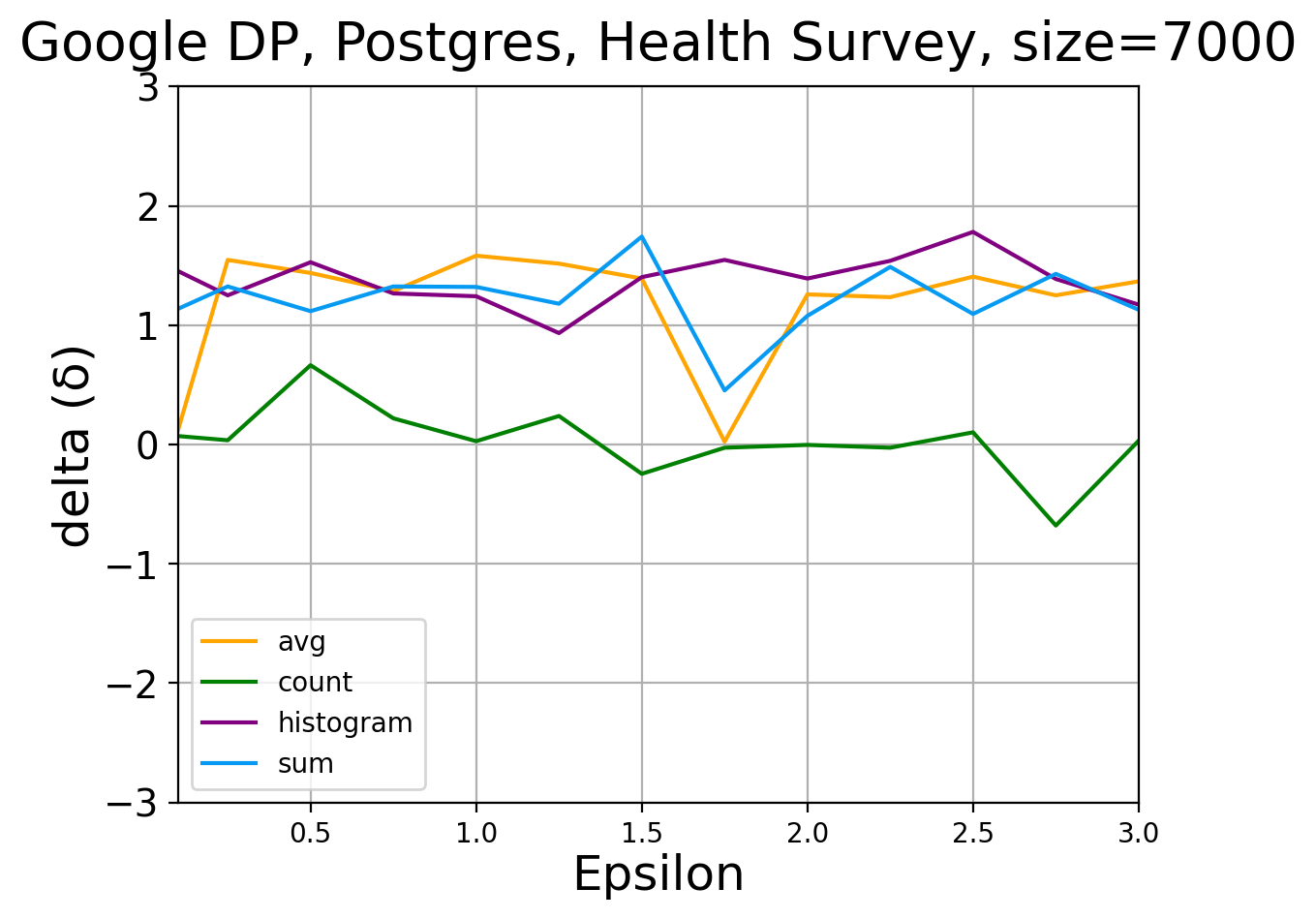}}
	\subfloat[]{\label{fig:exp3:gdp:post:H:8000}\includegraphics[width=0.25\textwidth]{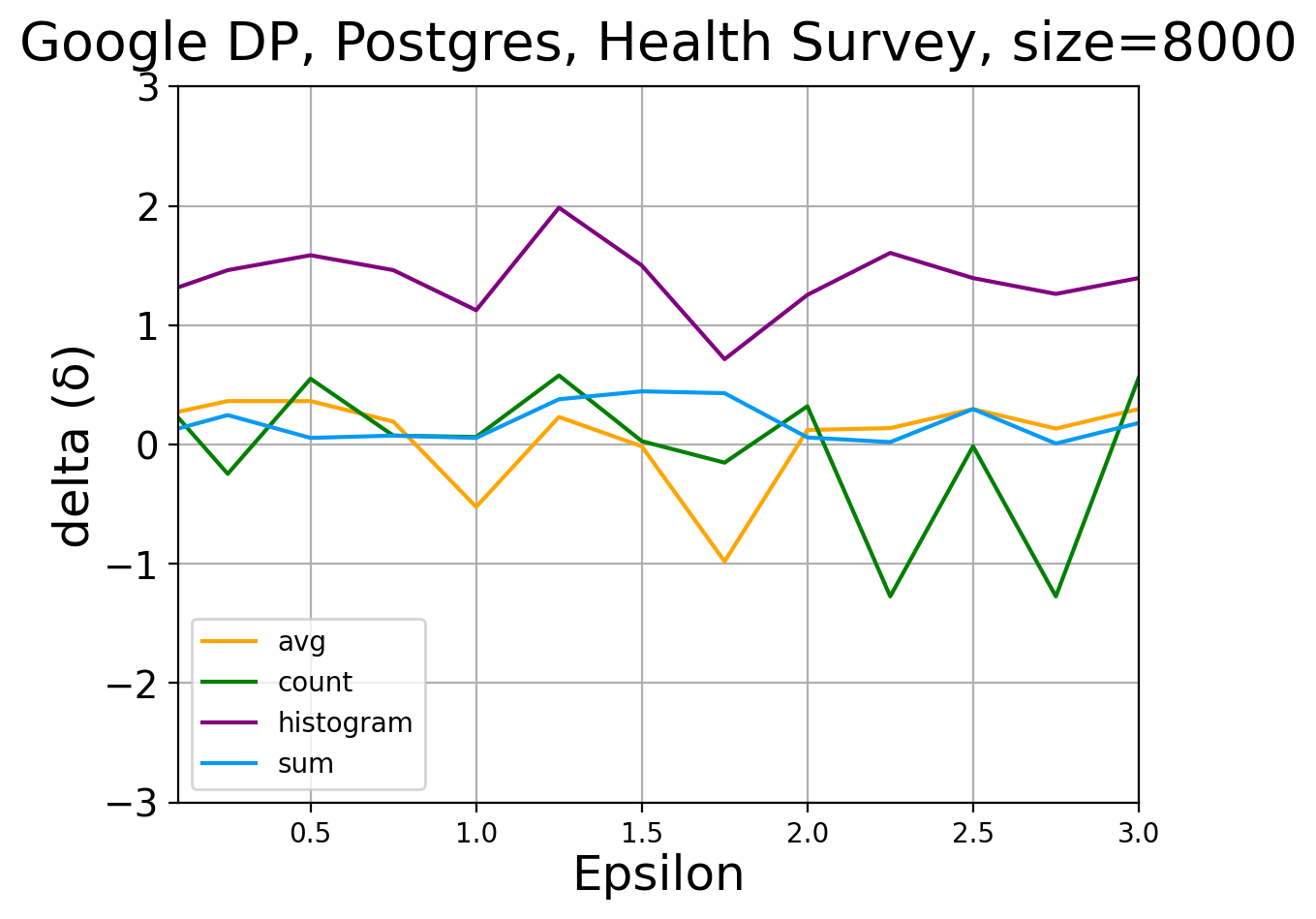}}
	\subfloat[]{\label{fig:exp3:gdp:post:H:9000}\includegraphics[width=0.25\textwidth]{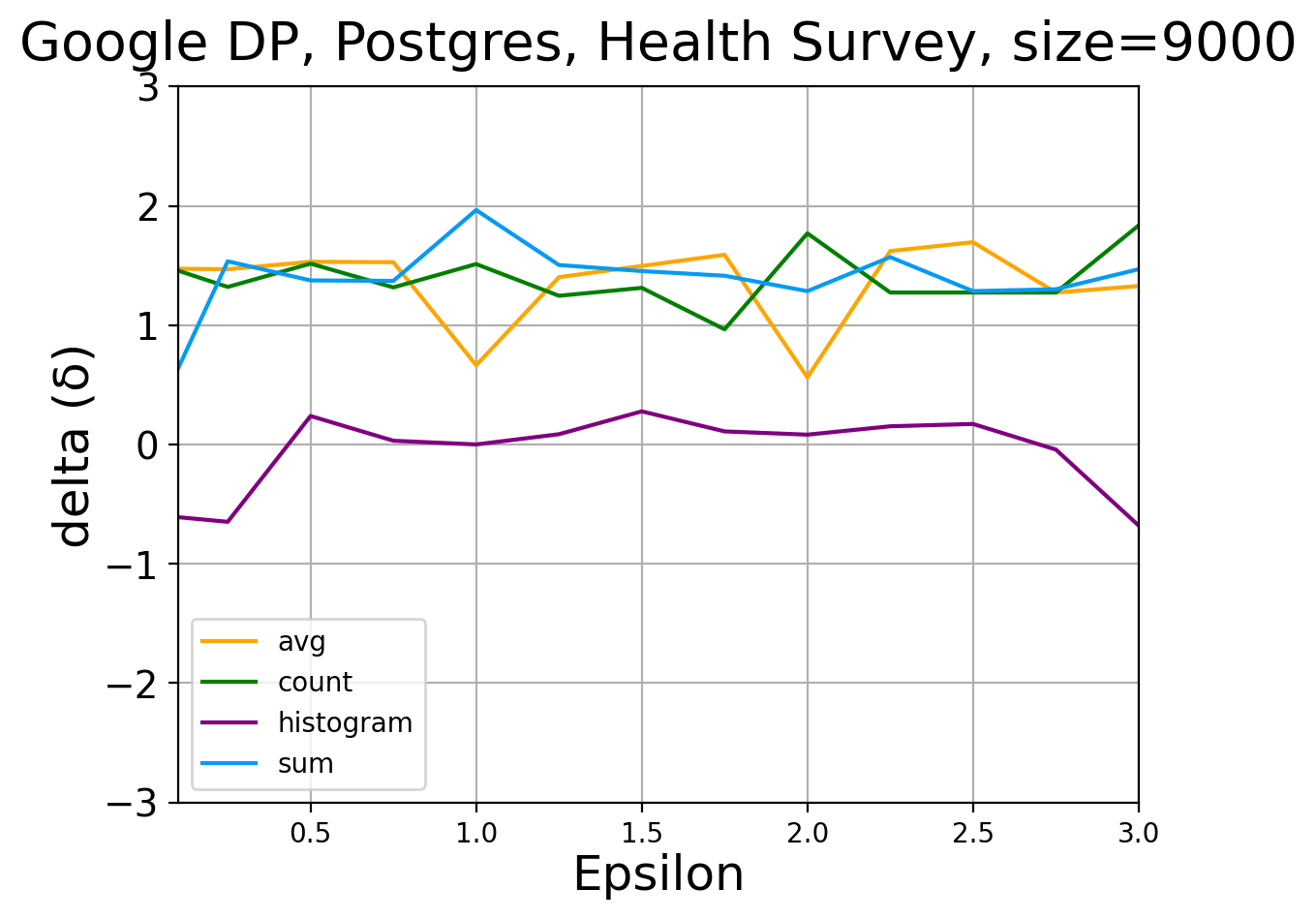}}
	\subfloat[]{\label{fig:exp3:gdp:post:H:9358}\includegraphics[width=0.25\textwidth]{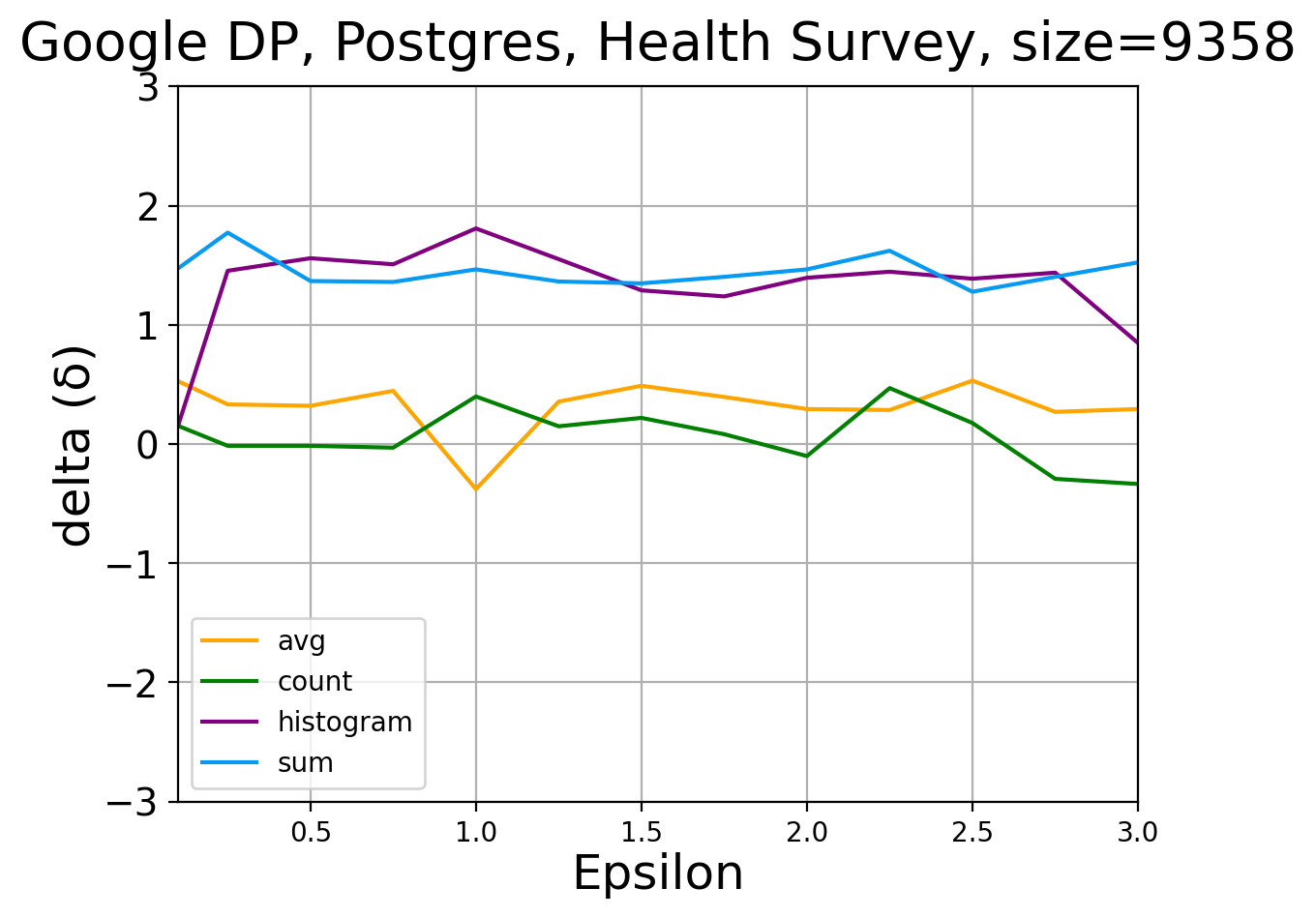}}
	\\
	\subfloat[]{\label{fig:exp3:smart:post:H:7000}\includegraphics[width=0.25\textwidth]{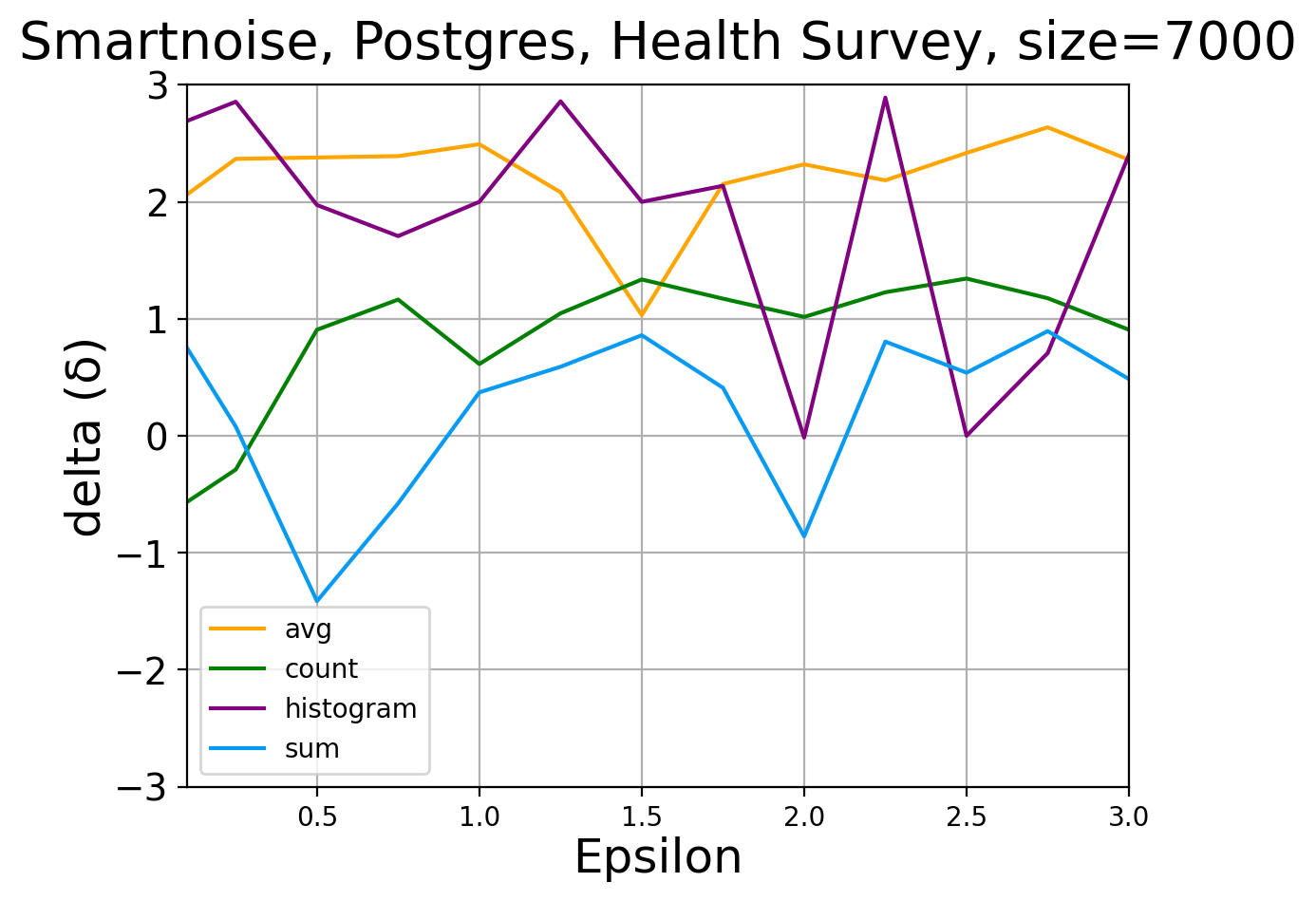}}
	\subfloat[]{\label{fig:exp3:smart:post:H:8000}\includegraphics[width=0.25\textwidth]{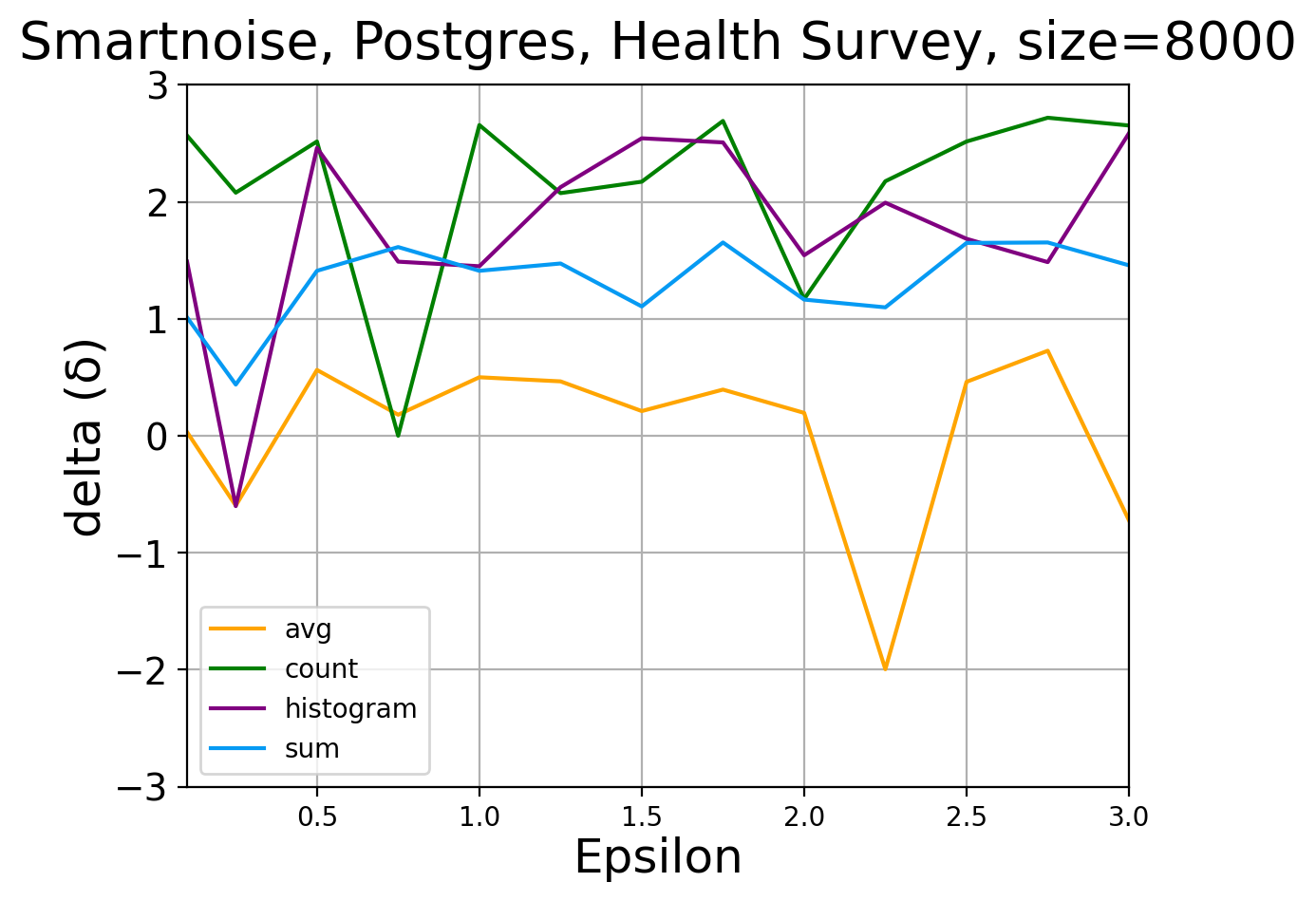}}
	\subfloat[]{\label{fig:exp3:smart:post:H:9000}\includegraphics[width=0.25\textwidth]{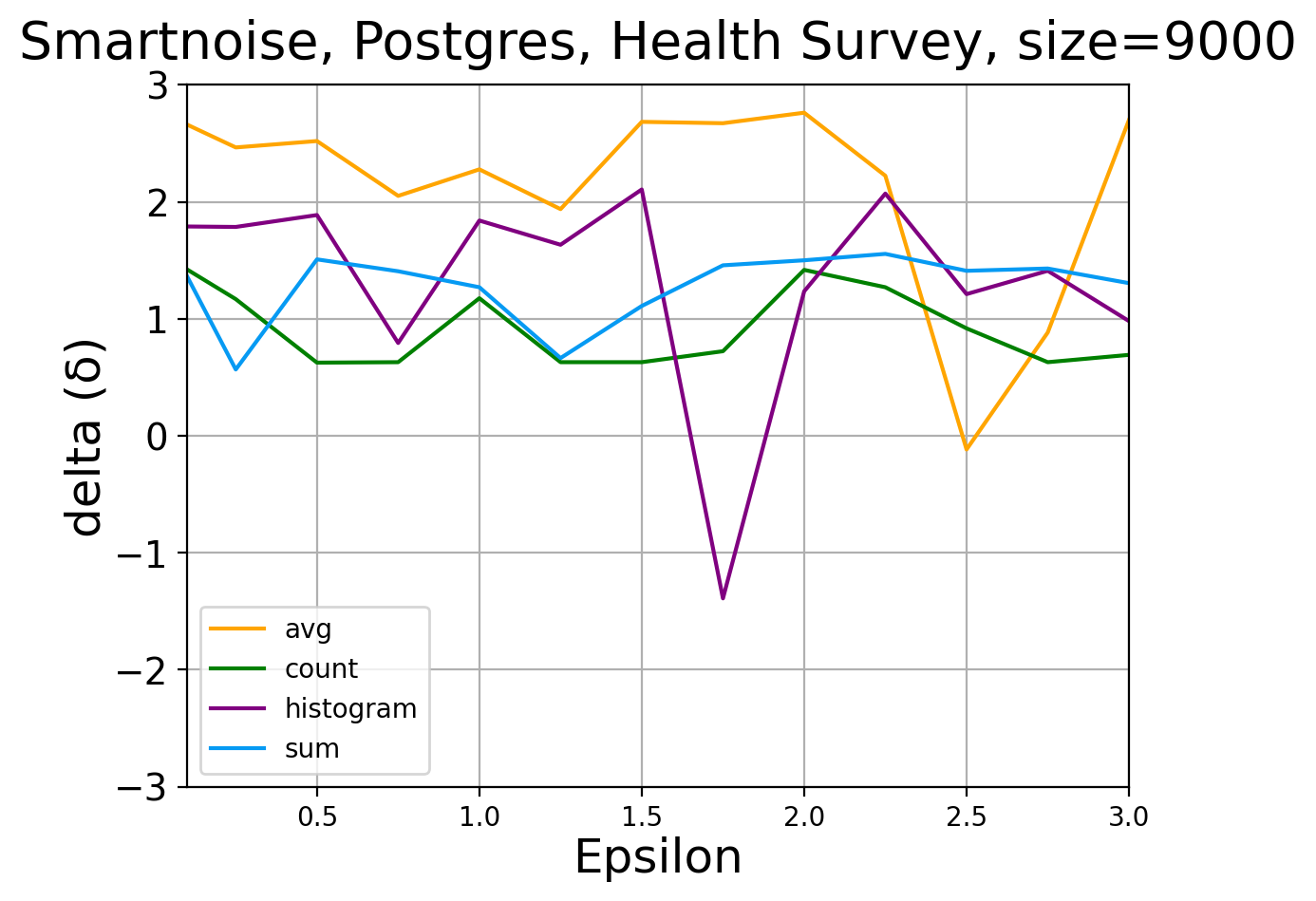}}
	\subfloat[]{\label{fig:exp3:smart:post:H:9358}\includegraphics[width=0.25\textwidth]{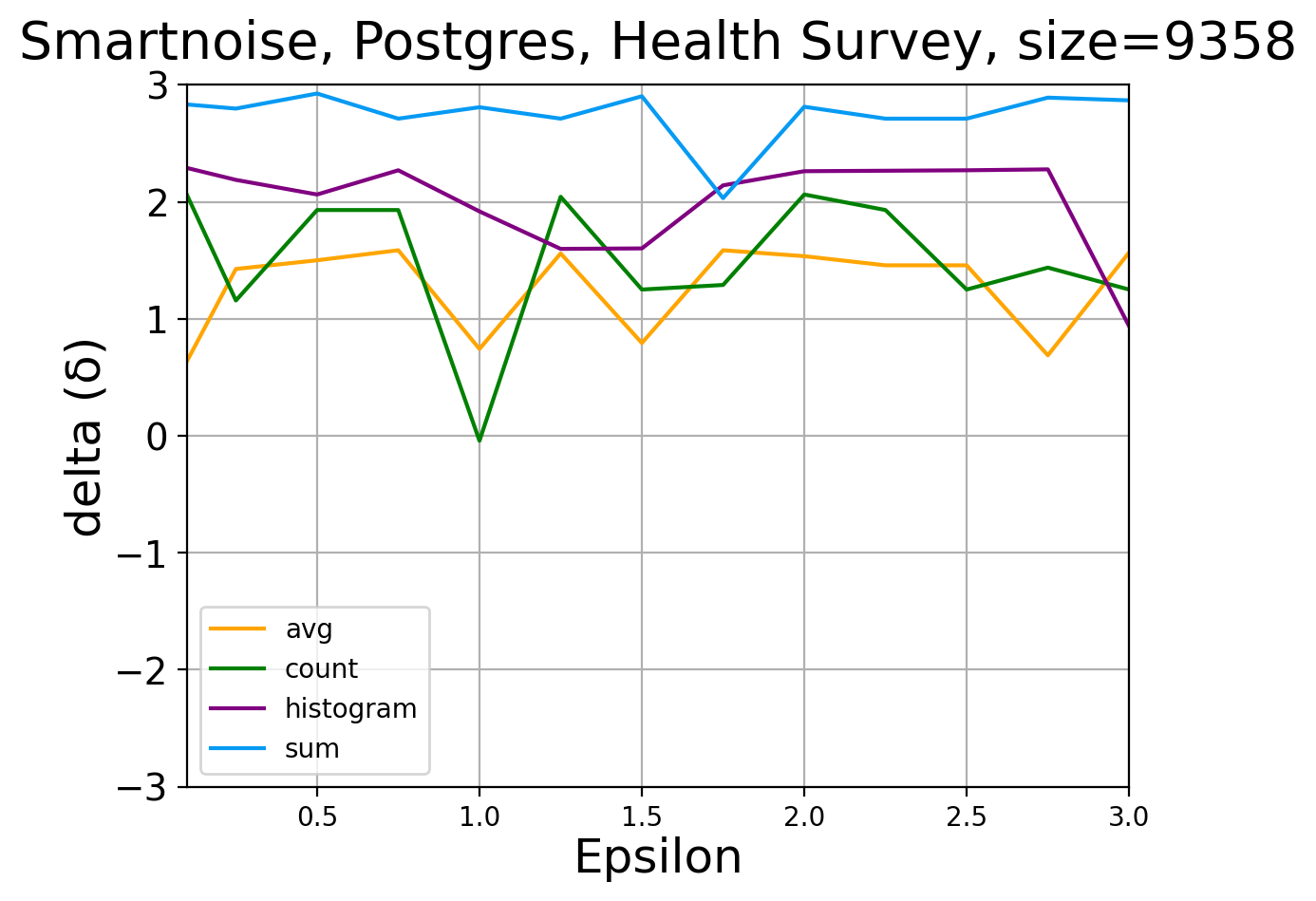}}
	\\
	\subfloat[]{\label{fig:exp3:gdp:post:P:3000}\includegraphics[width=0.25\textwidth]{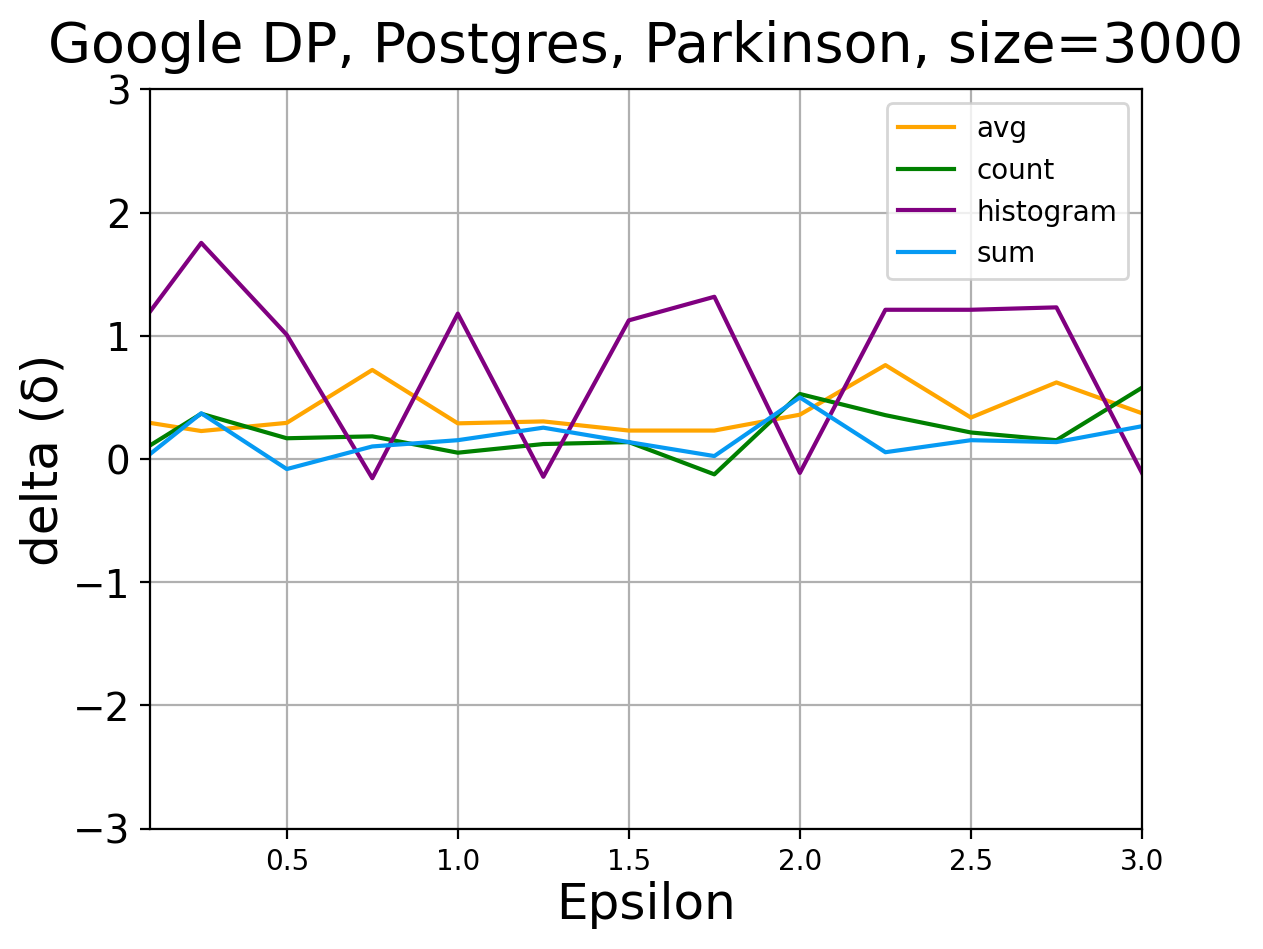}}
	\subfloat[]{\label{fig:exp3:gdp:post:P:4000}\includegraphics[width=0.25\textwidth]{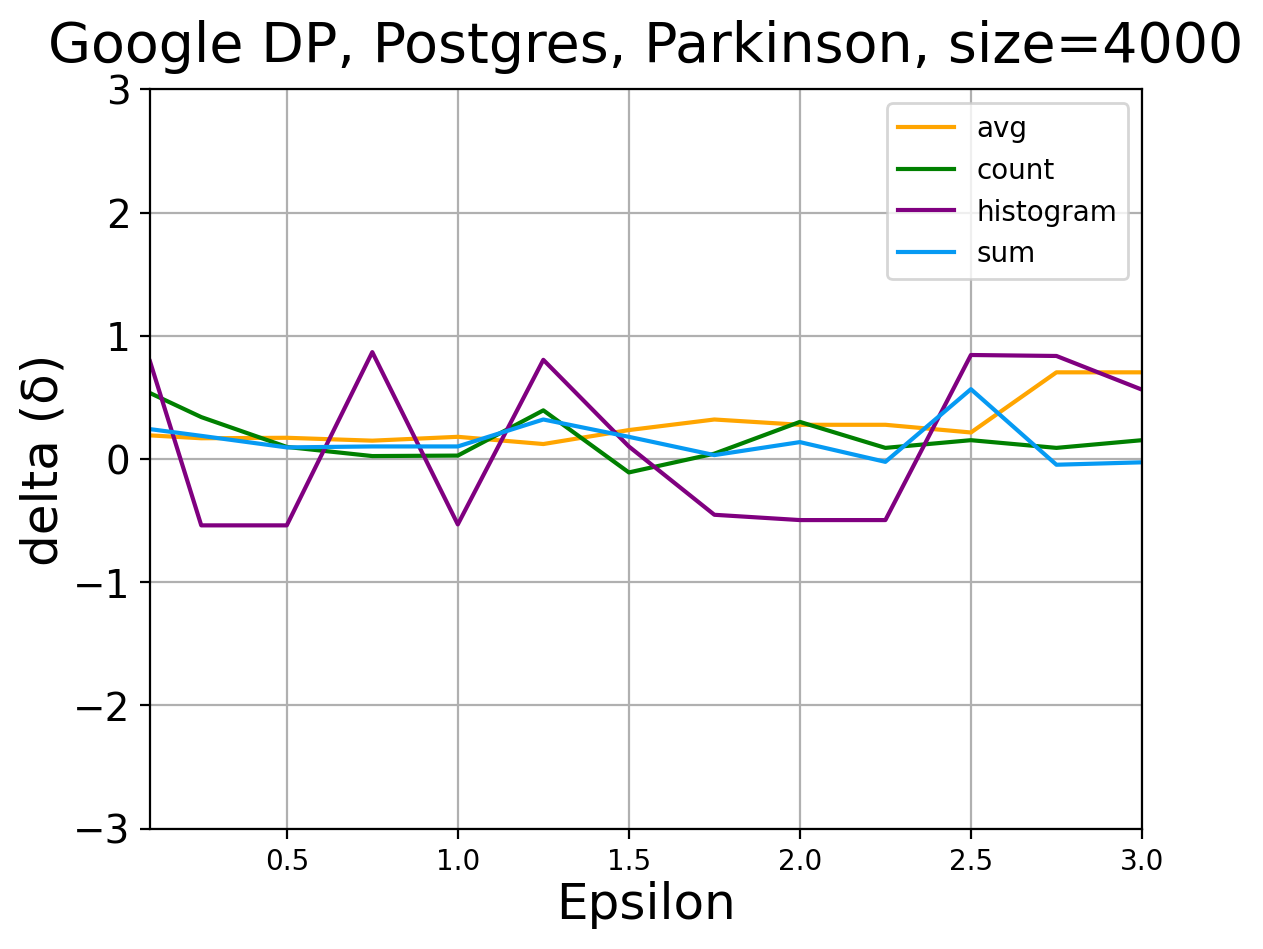}}
	\subfloat[]{\label{fig:exp3:gdp:post:P:5000}\includegraphics[width=0.25\textwidth]{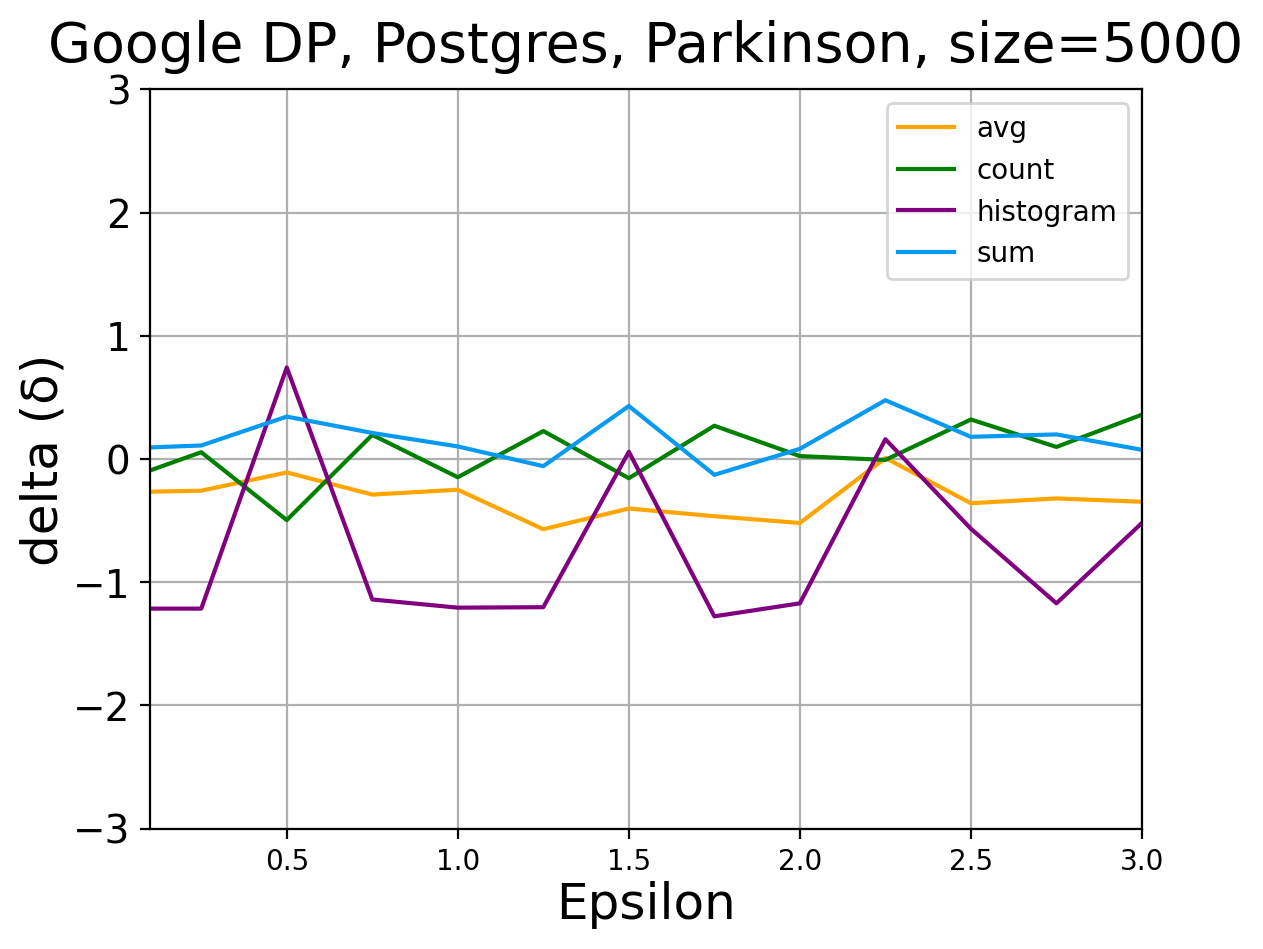}}
	\subfloat[]{\label{fig:exp3:gdp:post:P:5499}\includegraphics[width=0.25\textwidth]{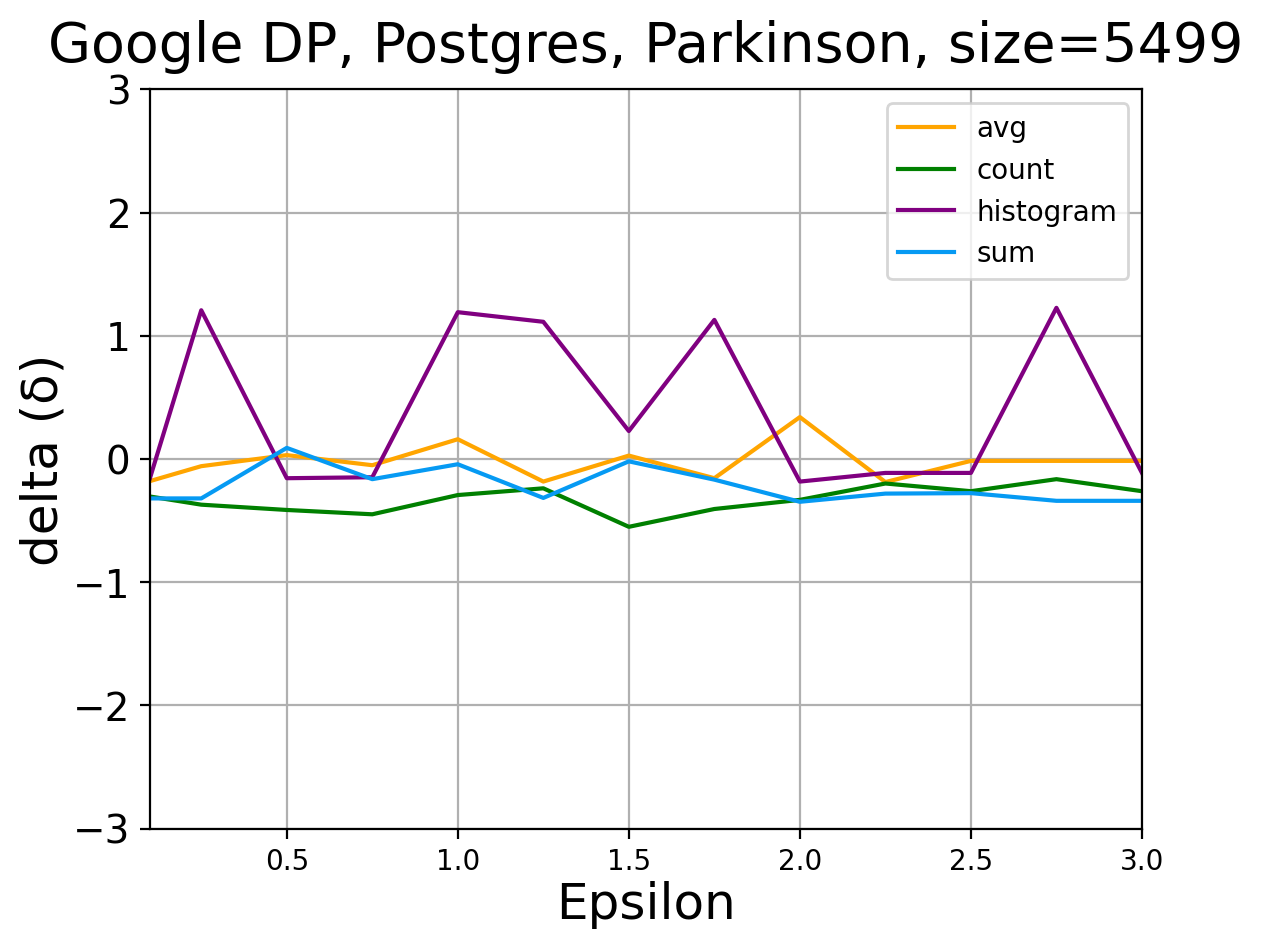}}
	\\
	\subfloat[]{\label{fig:exp3:smart:post:P:3000}\includegraphics[width=0.25\textwidth]{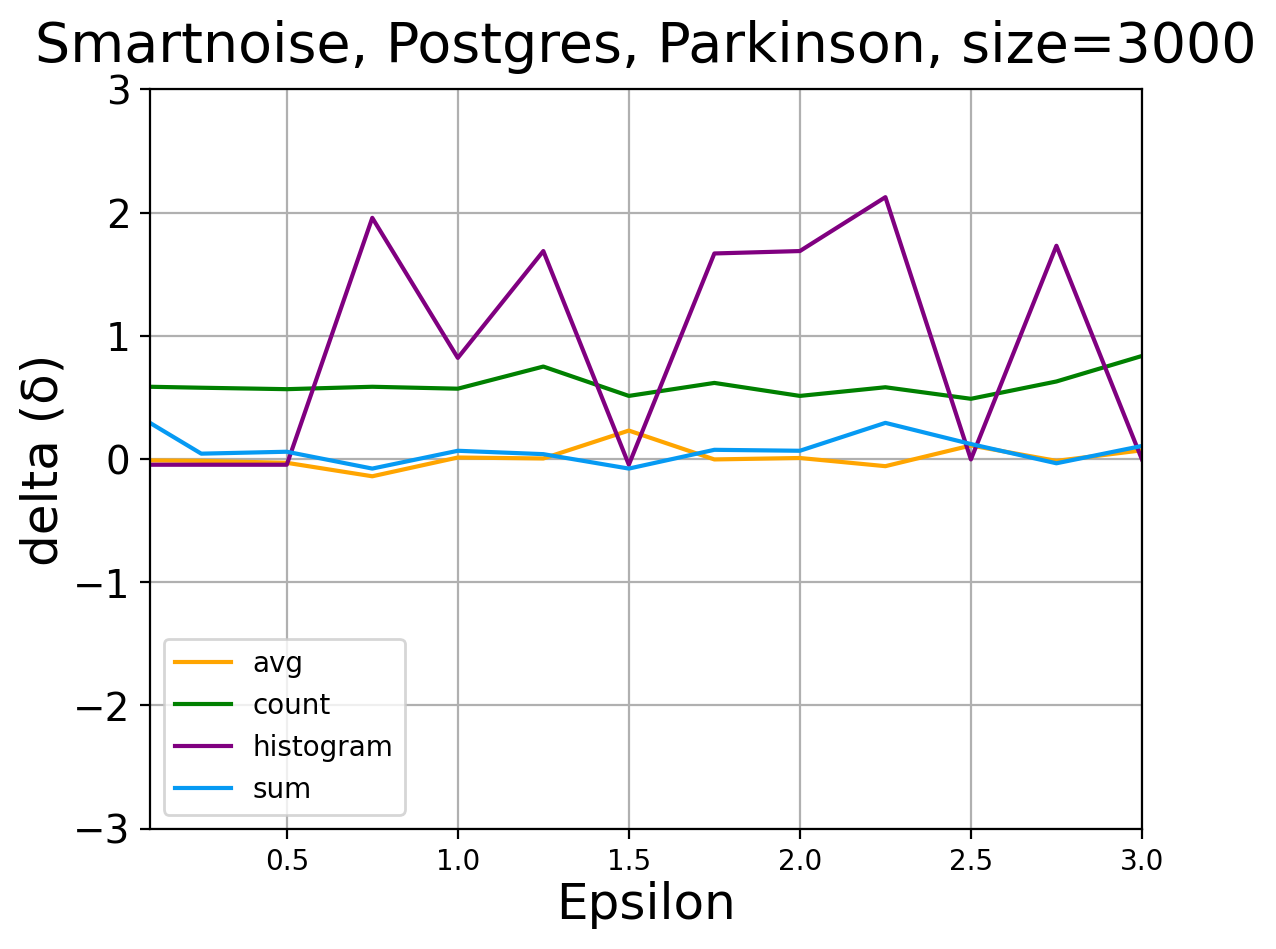}}
	\subfloat[]{\label{fig:exp3:smart:post:P:4000}\includegraphics[width=0.25\textwidth]{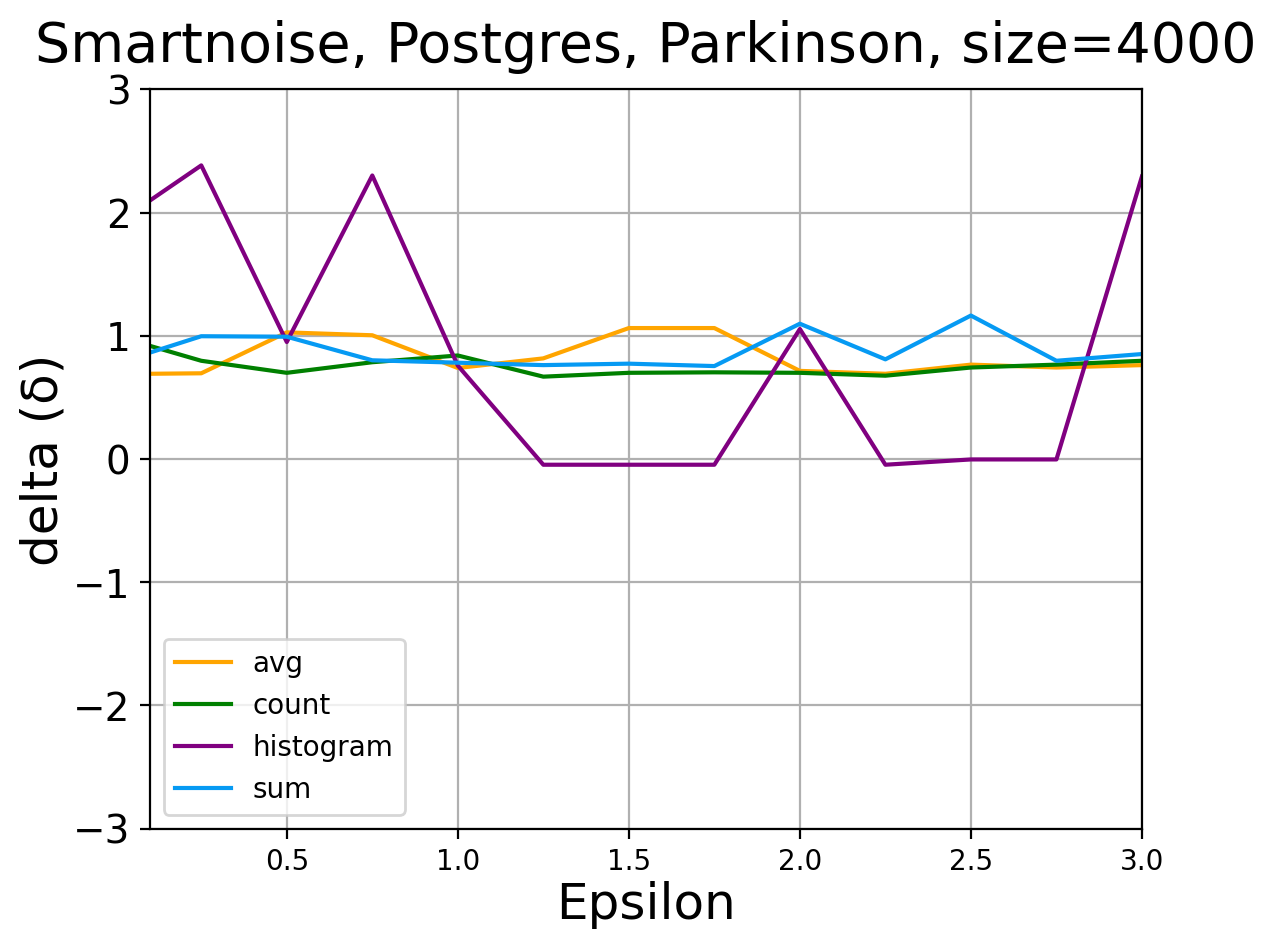}}
	\subfloat[]{\label{fig:exp3:smart:post:P:5000}\includegraphics[width=0.25\textwidth]{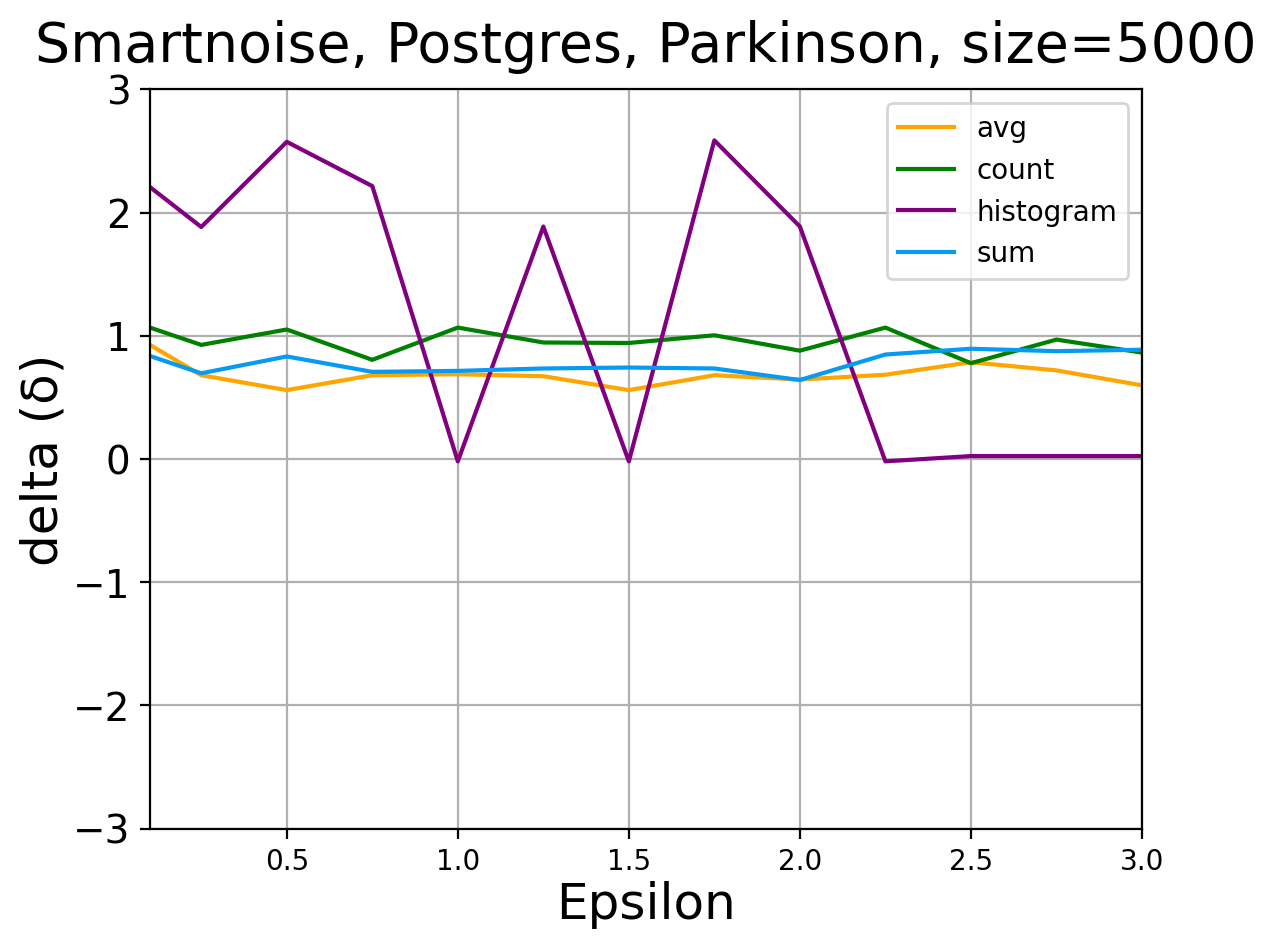}}
	\subfloat[]{\label{fig:exp3:smart:post:P:5499}\includegraphics[width=0.25\textwidth]{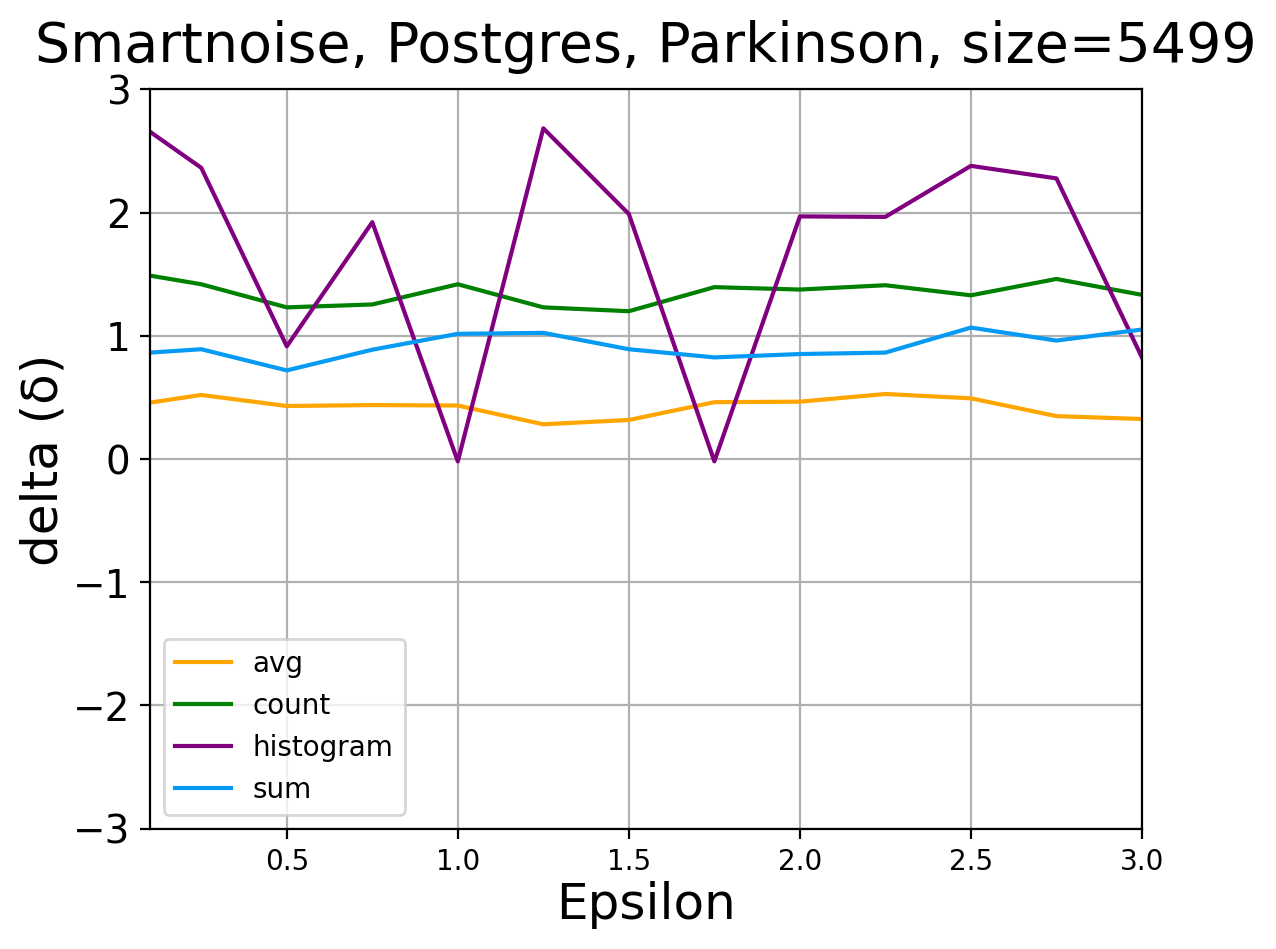}}
	
	\caption[Results of Experiment 3. Memory overhead for statistical query tools]{The evaluation results of statistical query tools on memory overhead when DP is integrated, for different data sizes (Table \ref{table_dataset_sizes}), $\epsilon$ values (Table \ref{table_epsilons}), and queries (Table \ref{table_queries}). Specifically, the results on memory overhead in the database (Postgres) that the queries are conducted on. $\delta$ is defined in Section~\ref{evaluation_criteria}.}
	
	\label{fig:exp3:post:line}
\end{figure}

Figure~\ref{fig:exp3:post:line} details the impact of DP on the Postgres database, from which we cannot summarize any clear correlations between $\epsilon$, data size, and memory overhead. In general, the induced memory usage on the operation of the Postgres database is notably low ($\lessapprox 3\%$) for both tools and data sets, which indicates a marginal impact. In comparison, though irregularities exist, Google DP shows lower peaks (less than 2) in the delta axis than Smartnoise (between 2 and 3) in the \emph{Health Survey} experiments, implying an advantage of Google DP on categorical data set over Smartnoise. In contrast, this advantage is not evident in the \emph{Parkinson} evaluation.

\begin{figure}[!ht]
	\centering
	\subfloat[]{\label{fig:exp3:gdp:H:7000}\includegraphics[width=0.25\textwidth]{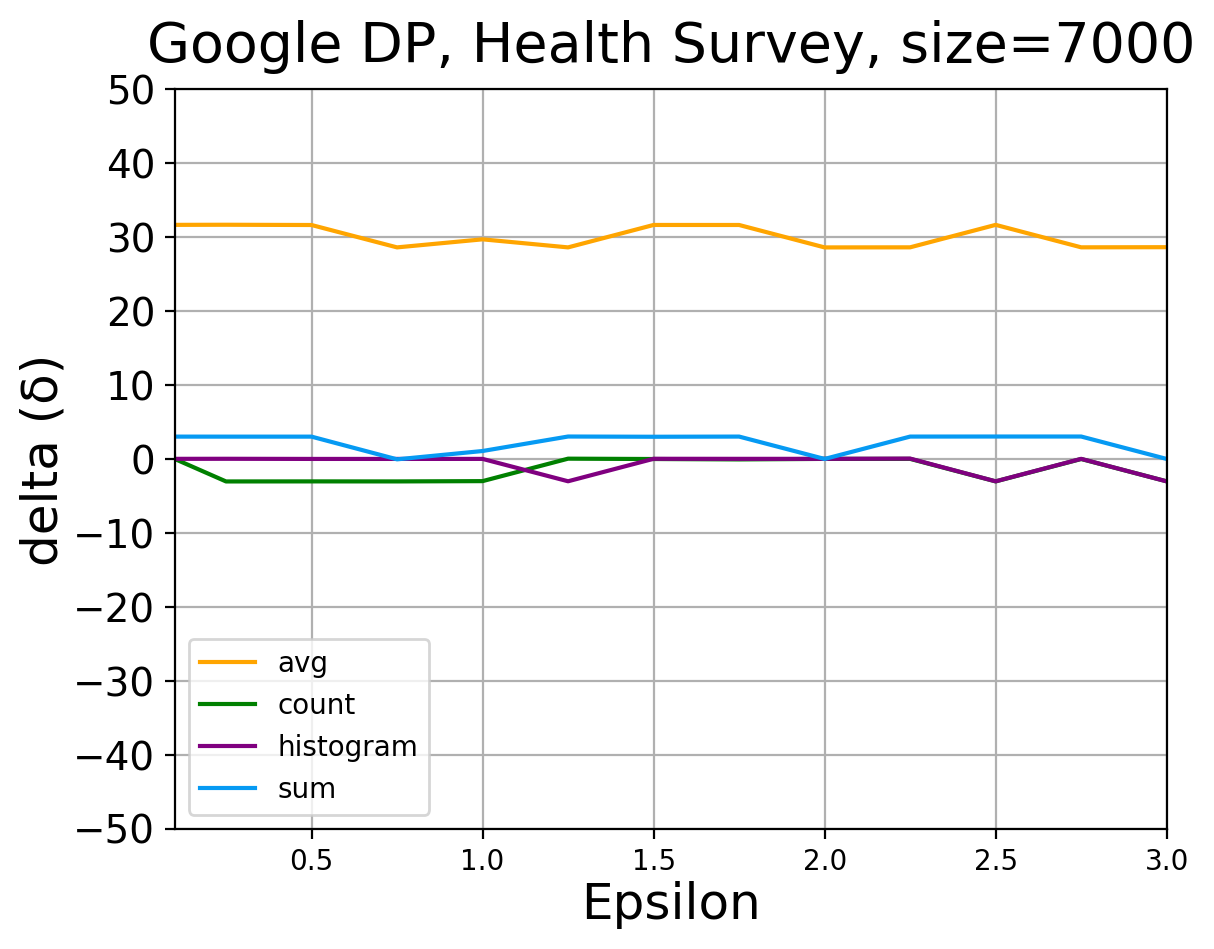}}
	\subfloat[]{\label{fig:exp3:gdp:H:8000}\includegraphics[width=0.25\textwidth]{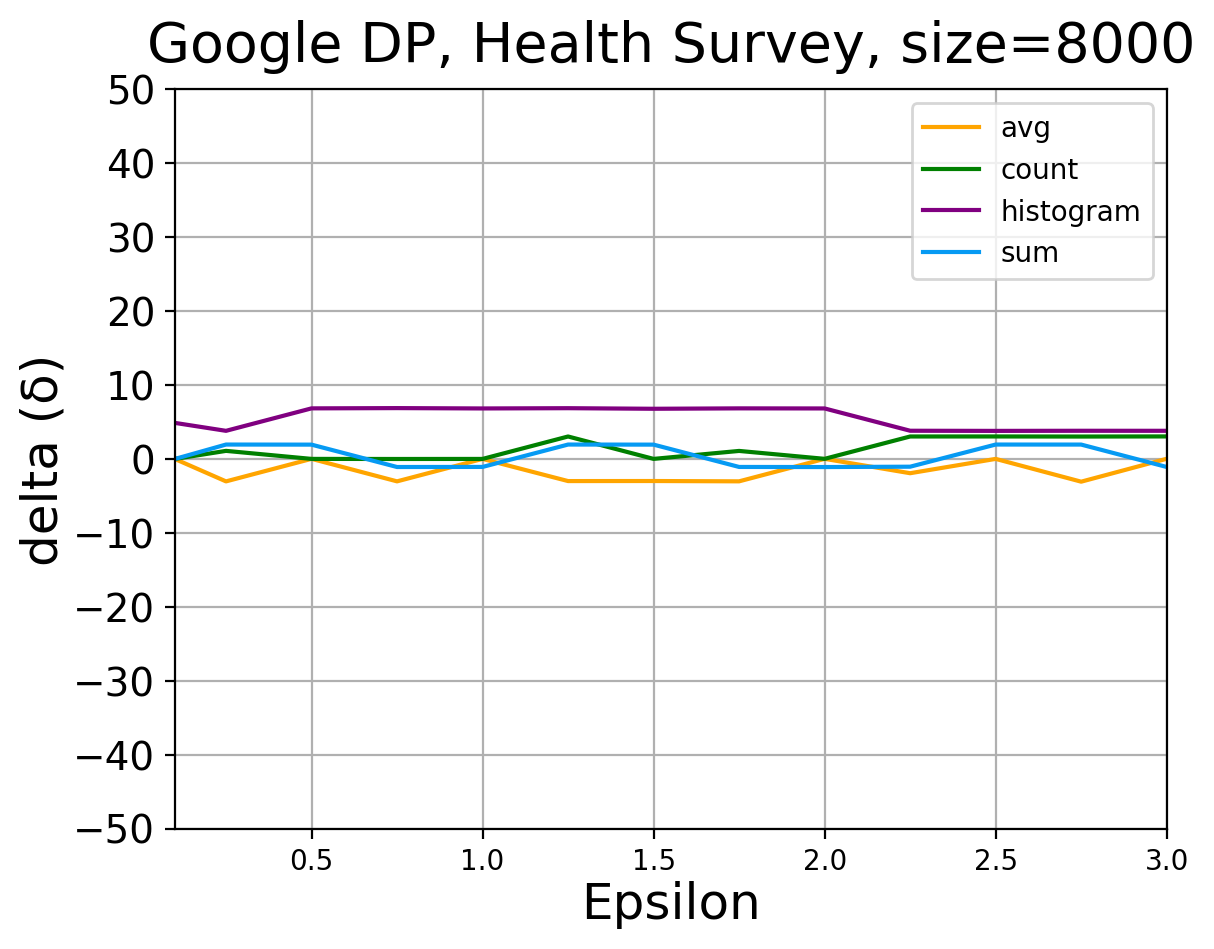}}
	\subfloat[]{\label{fig:exp3:gdp:H:9000}\includegraphics[width=0.25\textwidth]{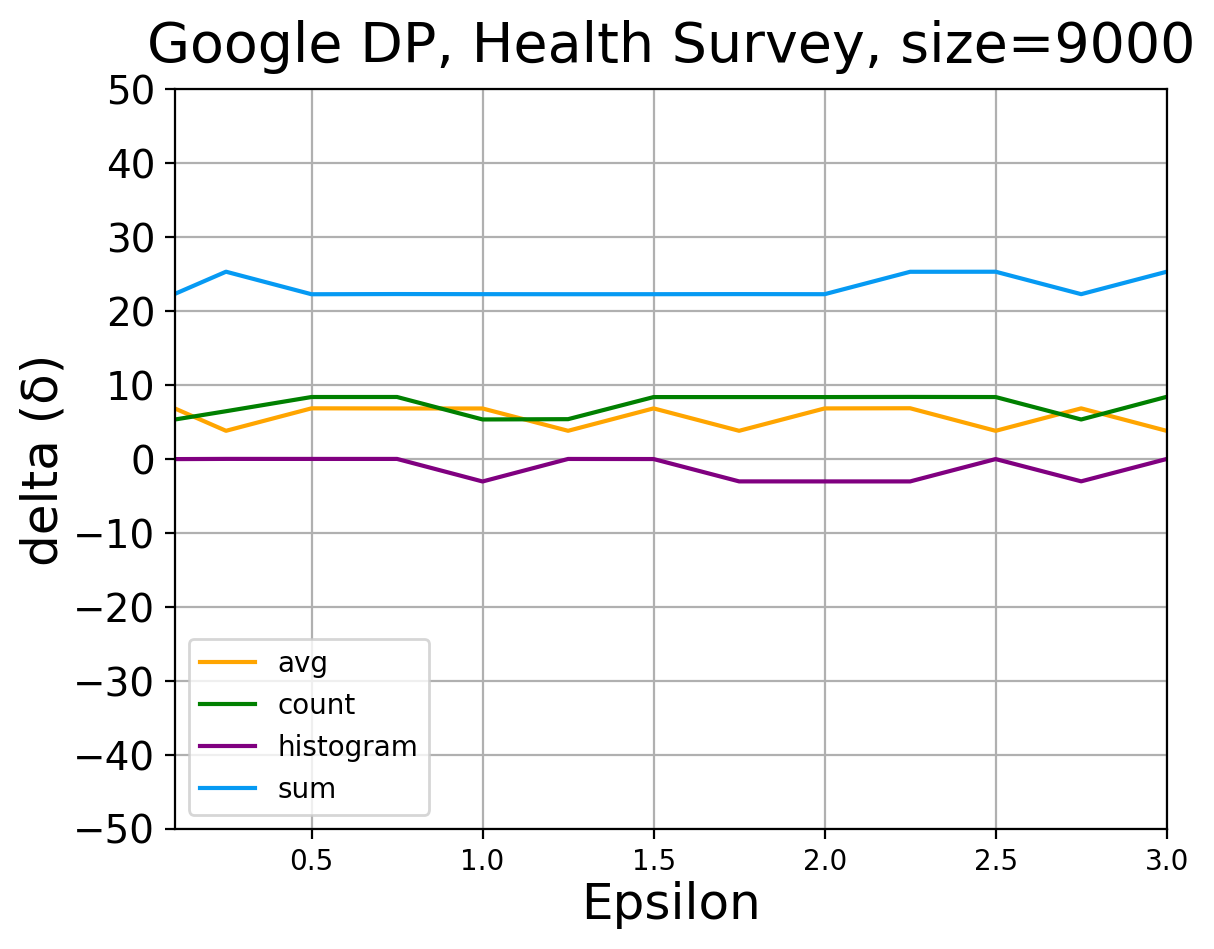}}
	\subfloat[]{\label{fig:exp3:gdp:H:9358}\includegraphics[width=0.25\textwidth]{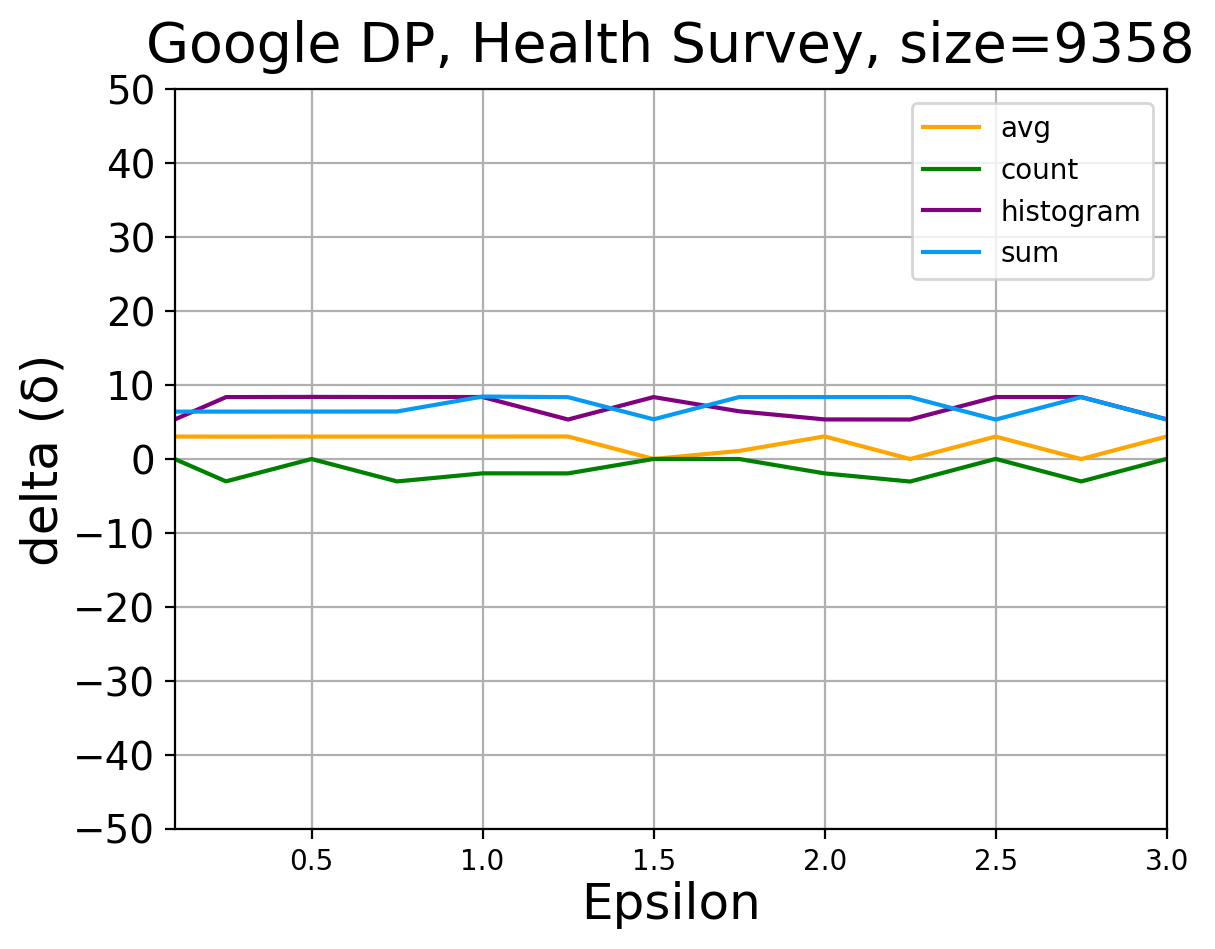}}
	\\
	\subfloat[]{\label{fig:exp3:smart:H:7000}\includegraphics[width=0.25\textwidth]{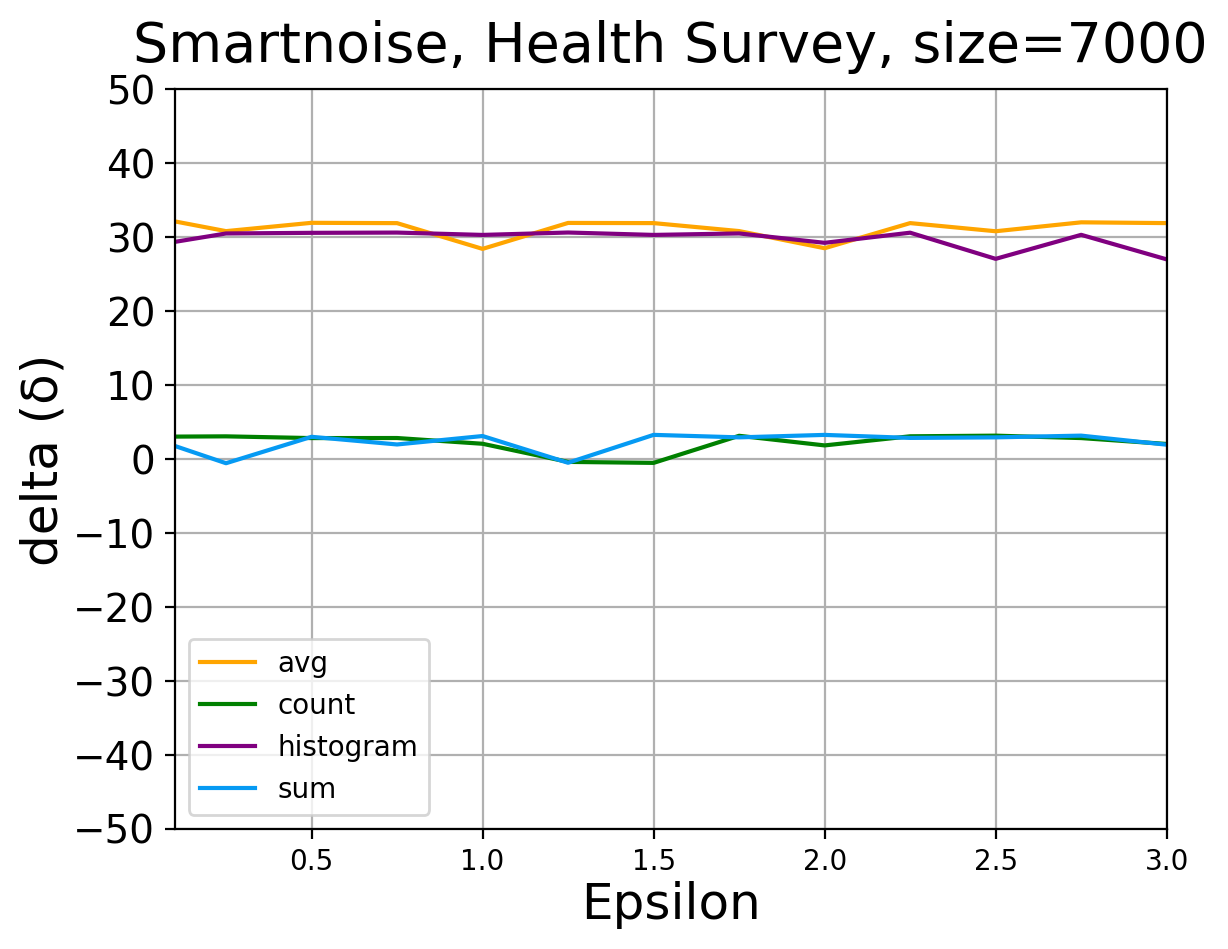}}
	\subfloat[]{\label{fig:exp3:smart:H:8000}\includegraphics[width=0.25\textwidth]{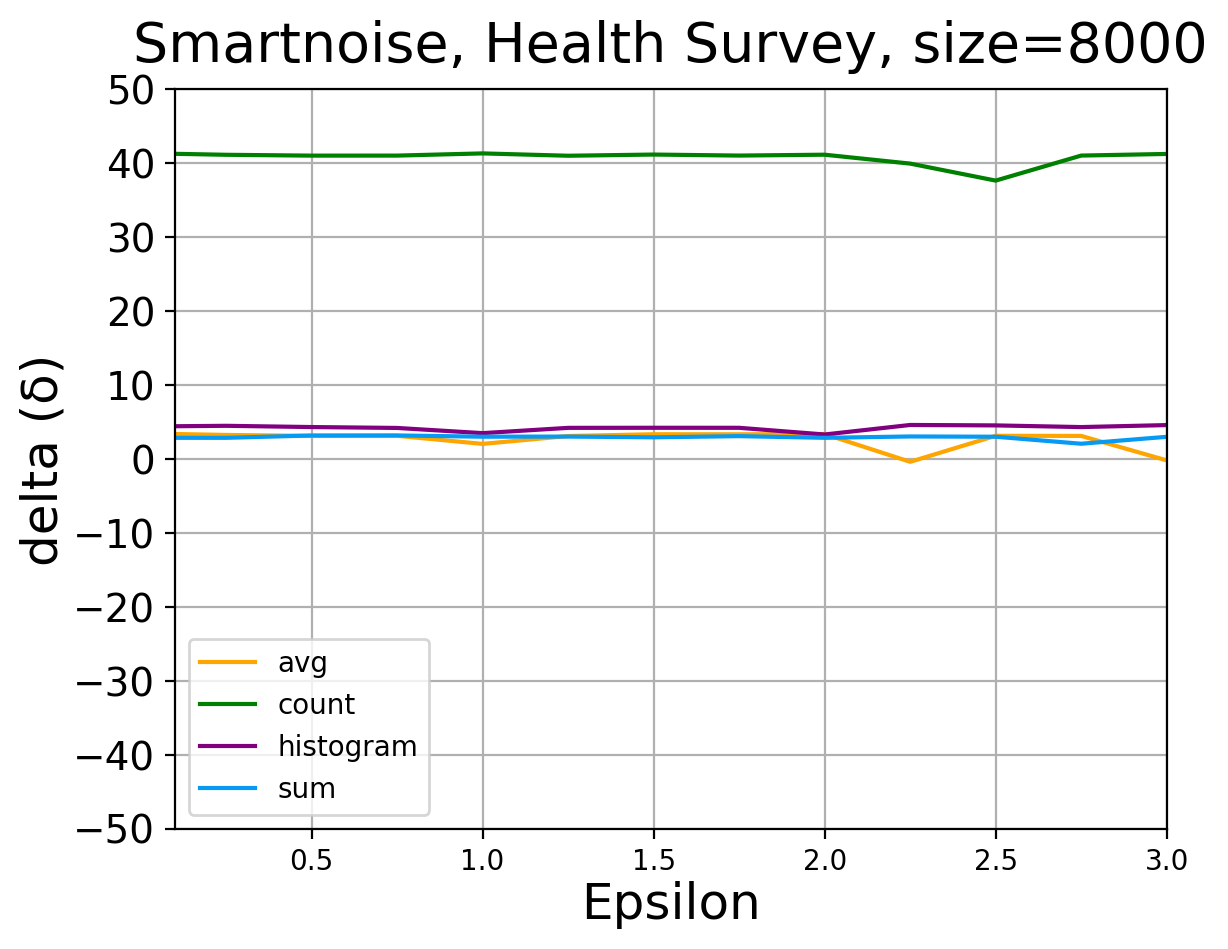}}
	\subfloat[]{\label{fig:exp3:smart:H:9000}\includegraphics[width=0.25\textwidth]{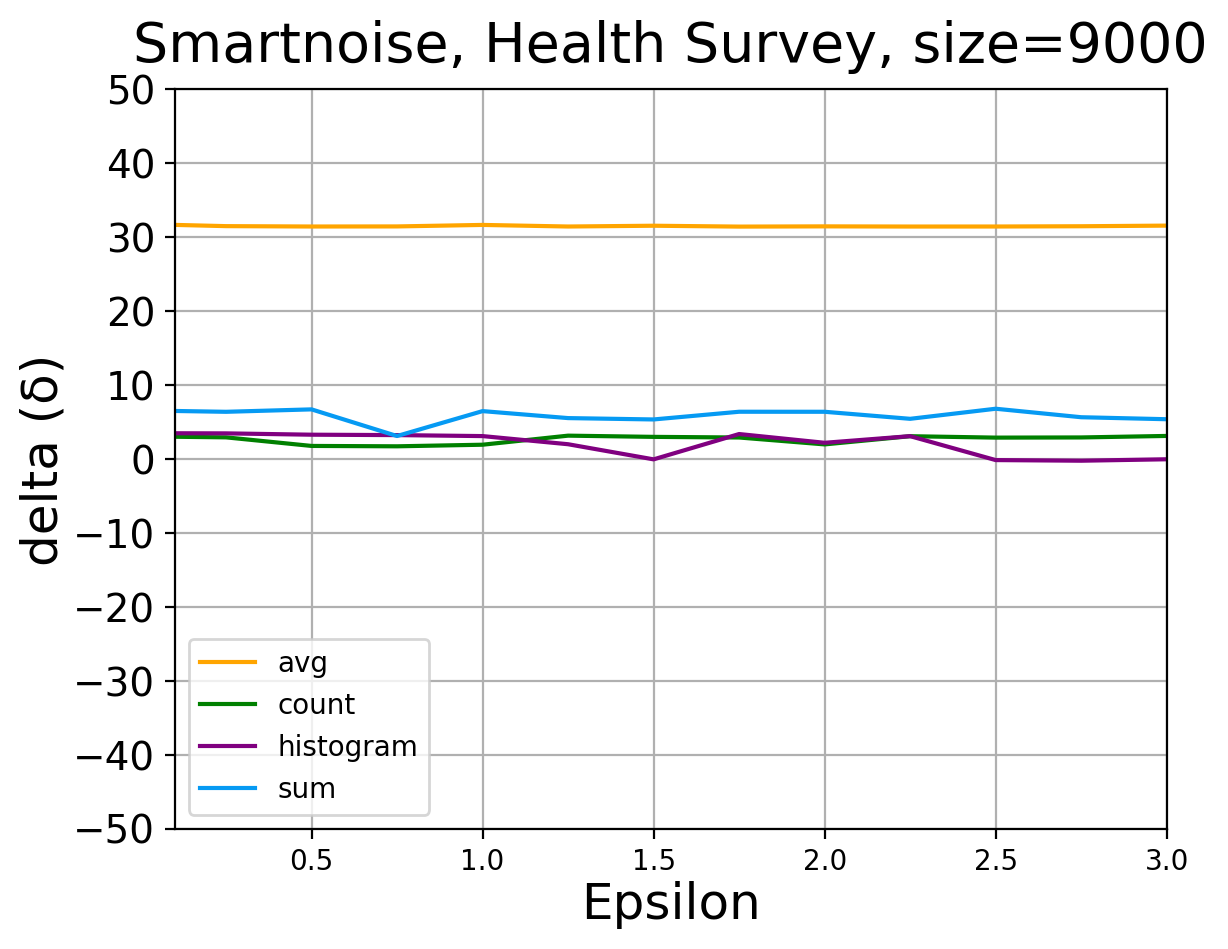}}
	\subfloat[]{\label{fig:exp3:smart:H:9358}\includegraphics[width=0.25\textwidth]{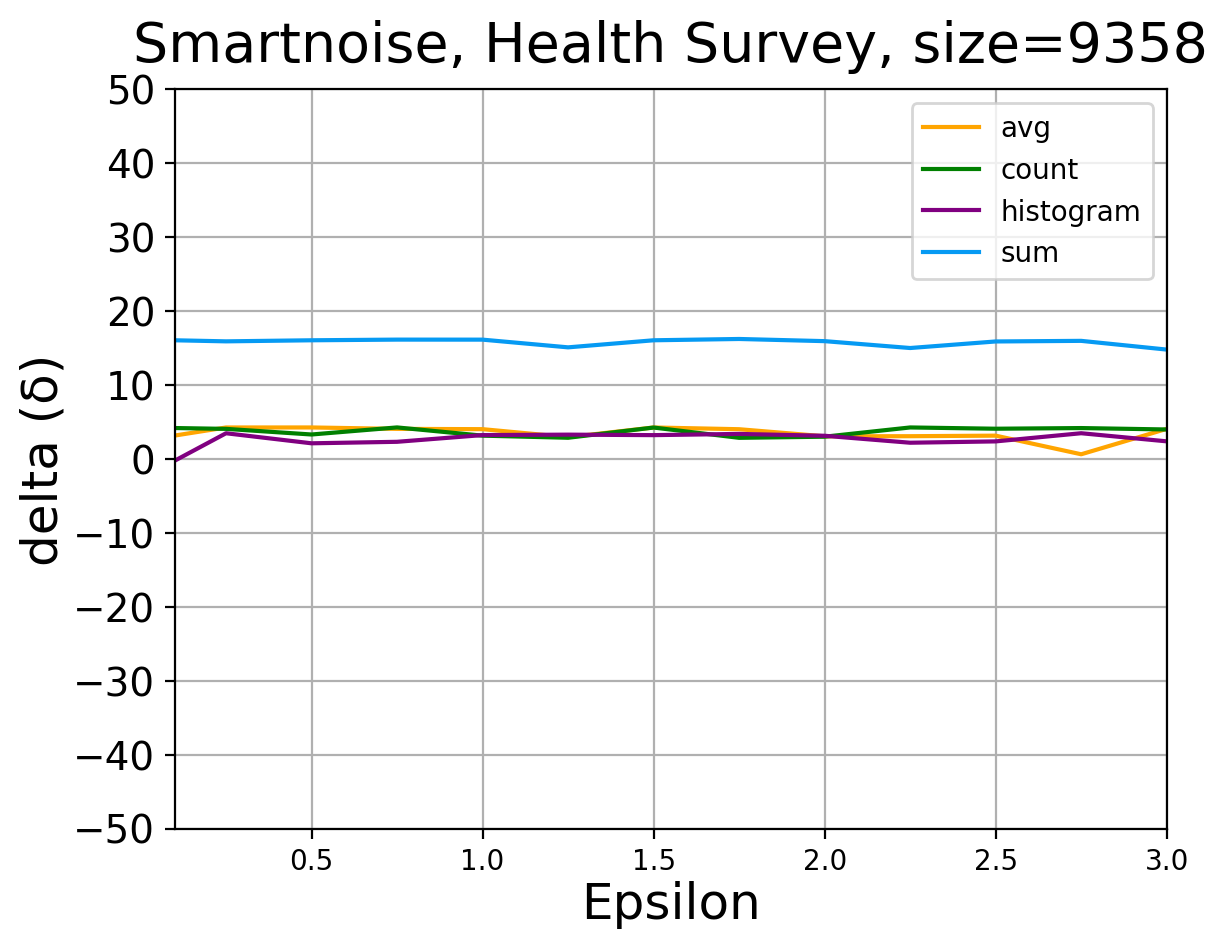}}
	\\
	\subfloat[]{\label{fig:exp3:gdp:P:3000}\includegraphics[width=0.25\textwidth]{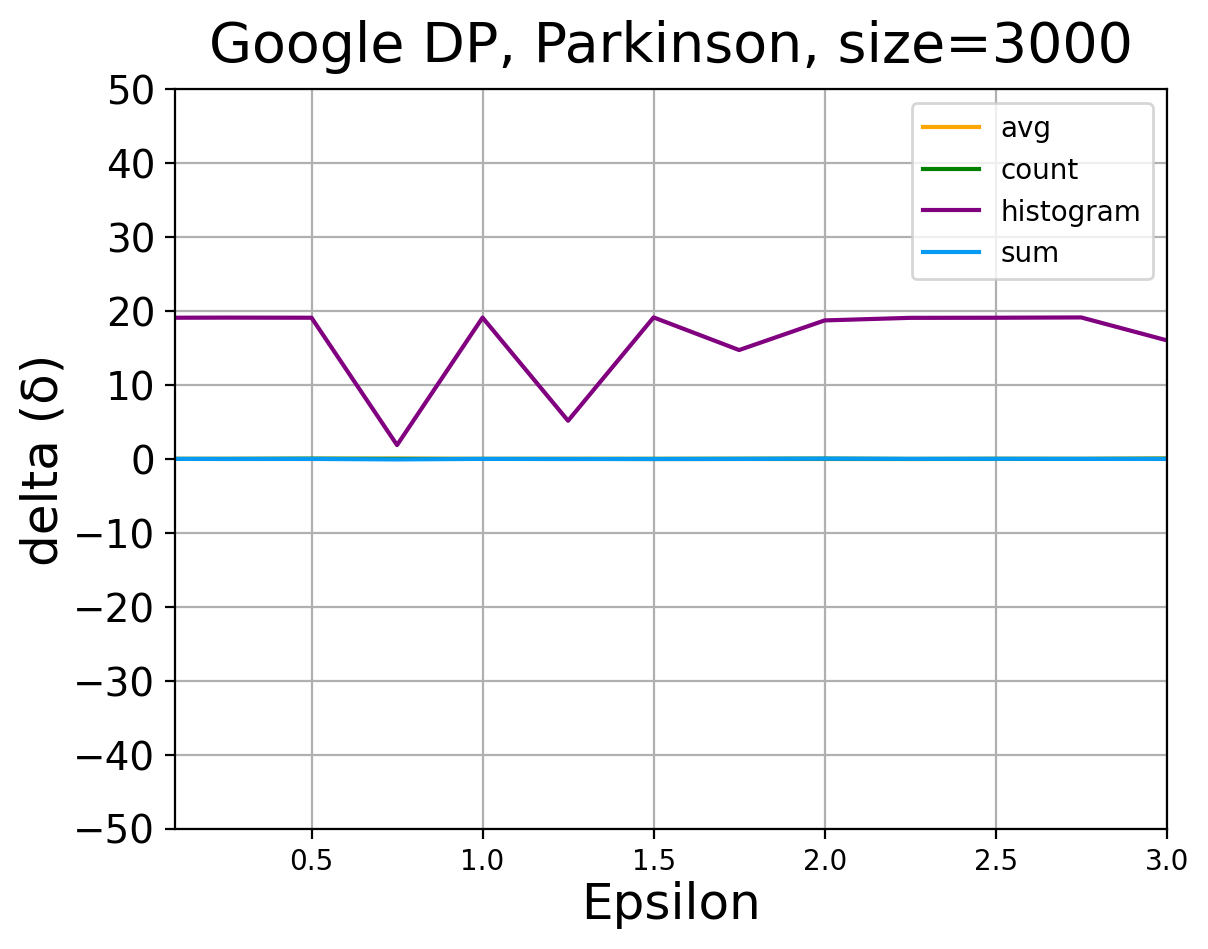}}
	\subfloat[]{\label{fig:exp3:gdp:P:4000}\includegraphics[width=0.25\textwidth]{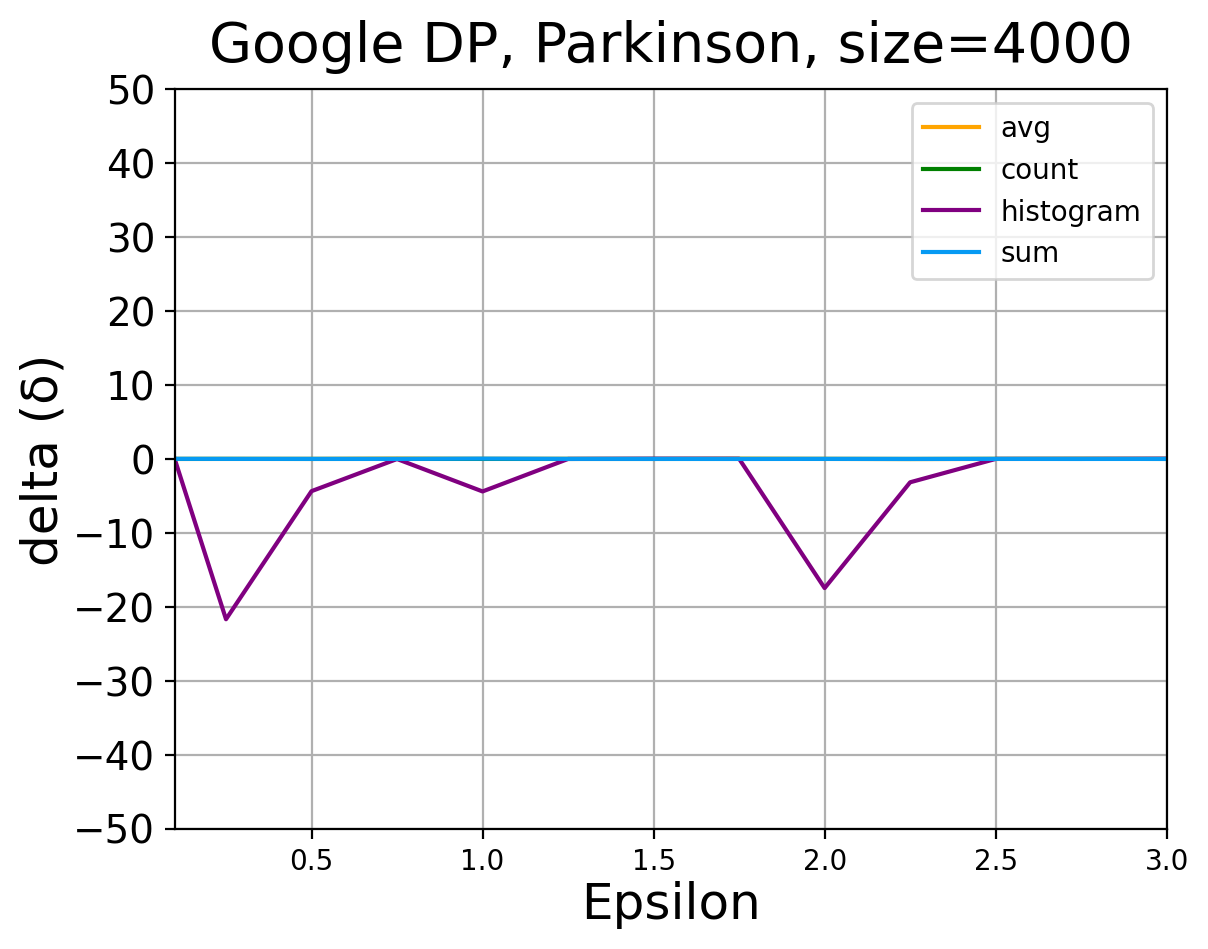}}
	\subfloat[]{\label{fig:exp3:gdp:P:5000}\includegraphics[width=0.25\textwidth]{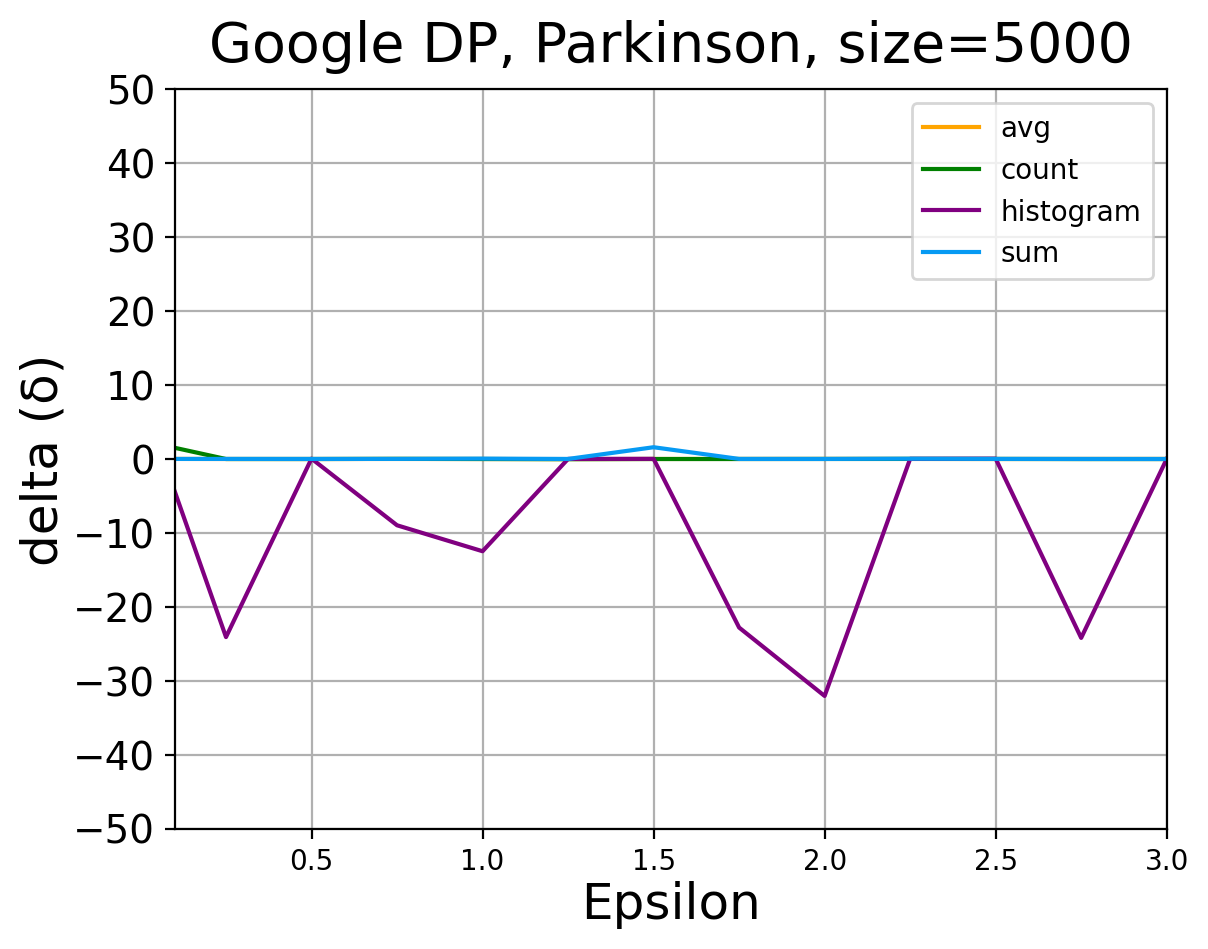}}
	\subfloat[]{\label{fig:exp3:gdp:P:5499}\includegraphics[width=0.25\textwidth]{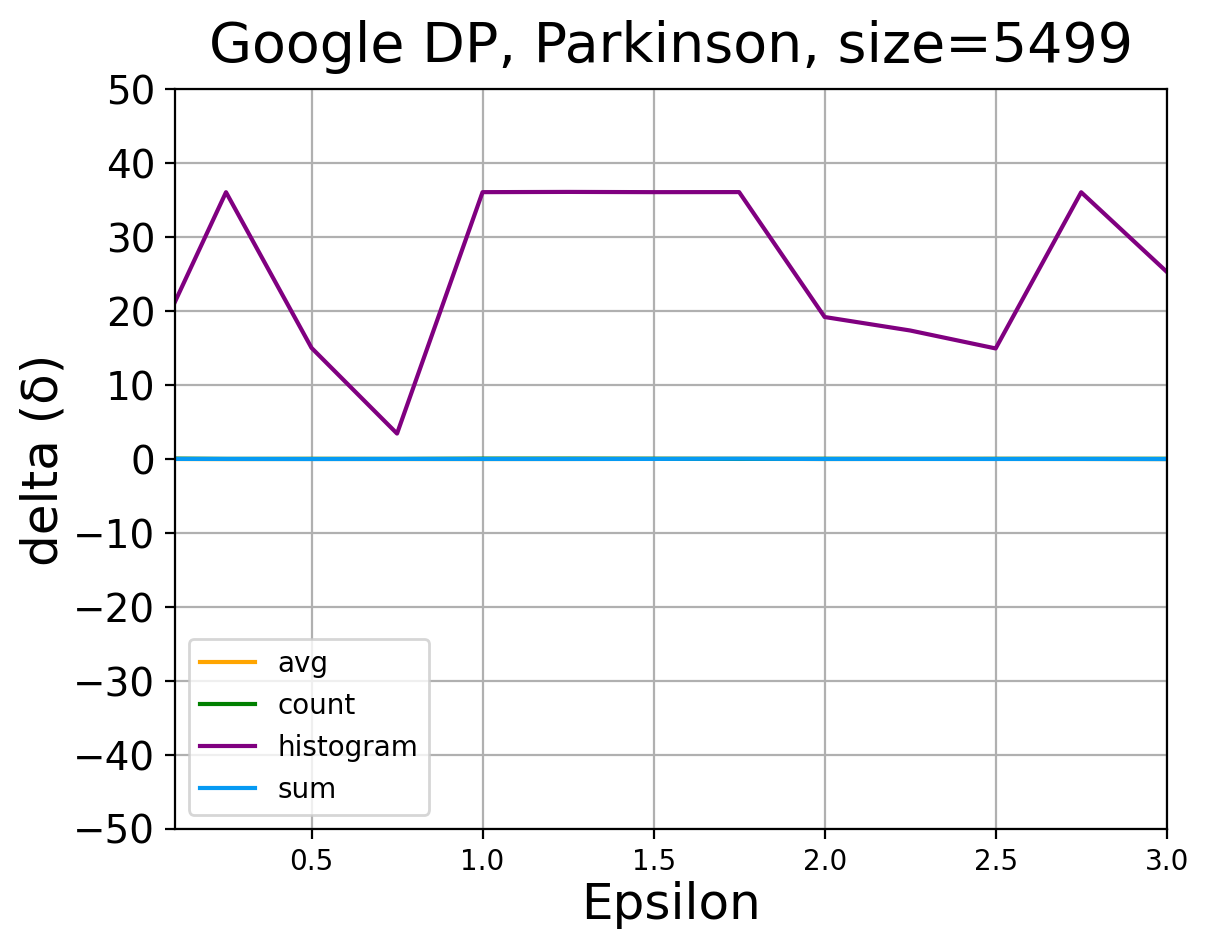}}
	\\
	\subfloat[]{\label{fig:exp3:smart:P:3000}\includegraphics[width=0.25\textwidth]{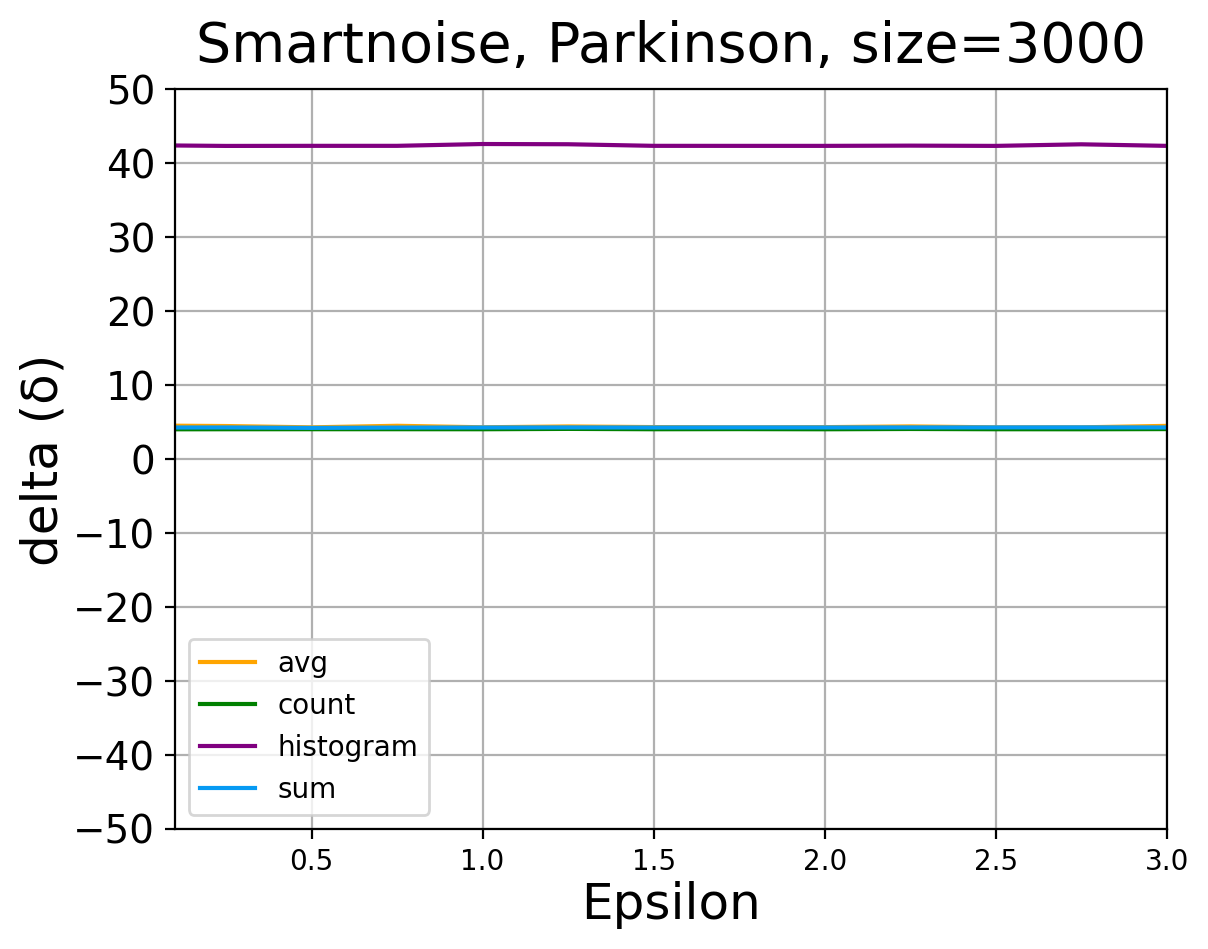}}
	\subfloat[]{\label{fig:exp3:smart:P:4000}\includegraphics[width=0.25\textwidth]{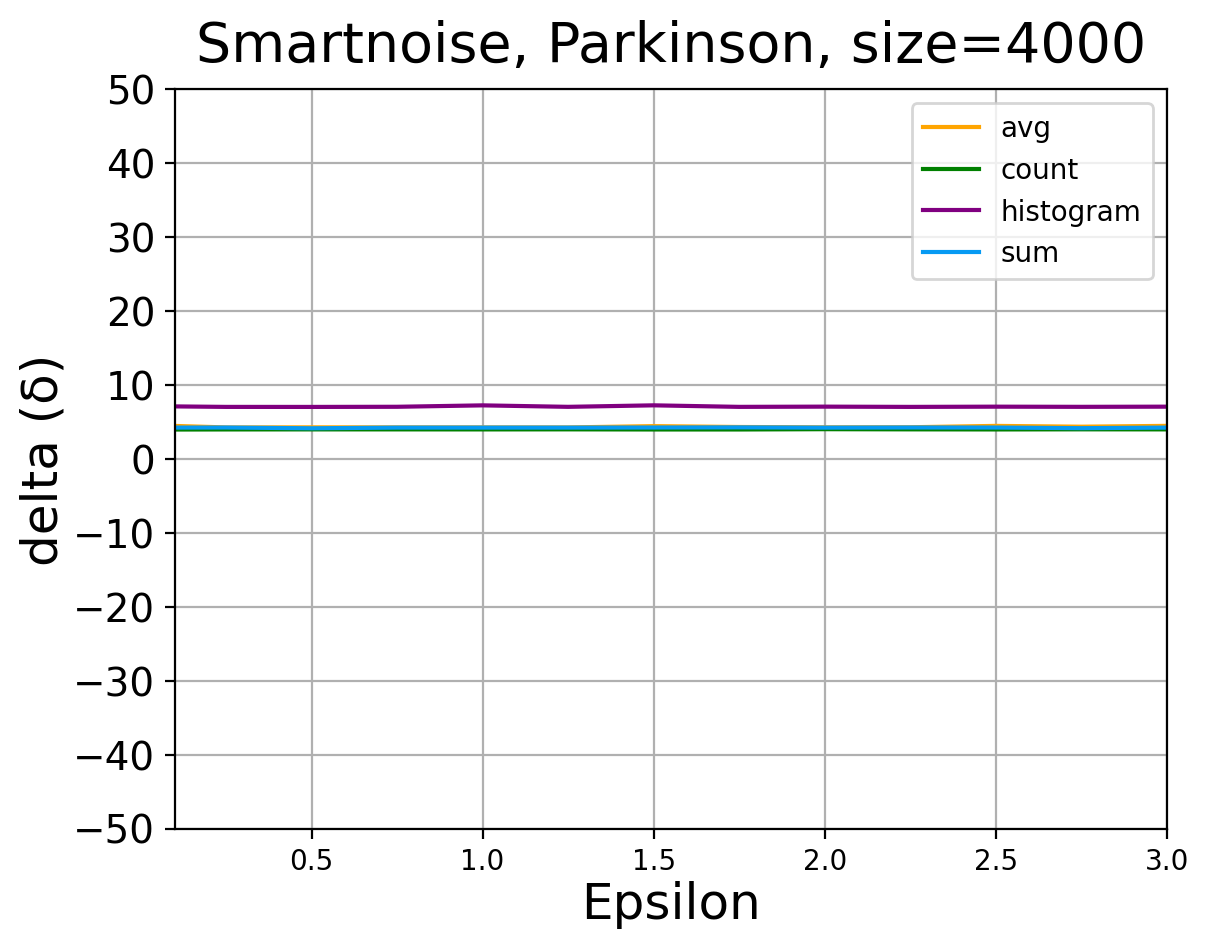}}
	\subfloat[]{\label{fig:exp3:smart:P:5000}\includegraphics[width=0.25\textwidth]{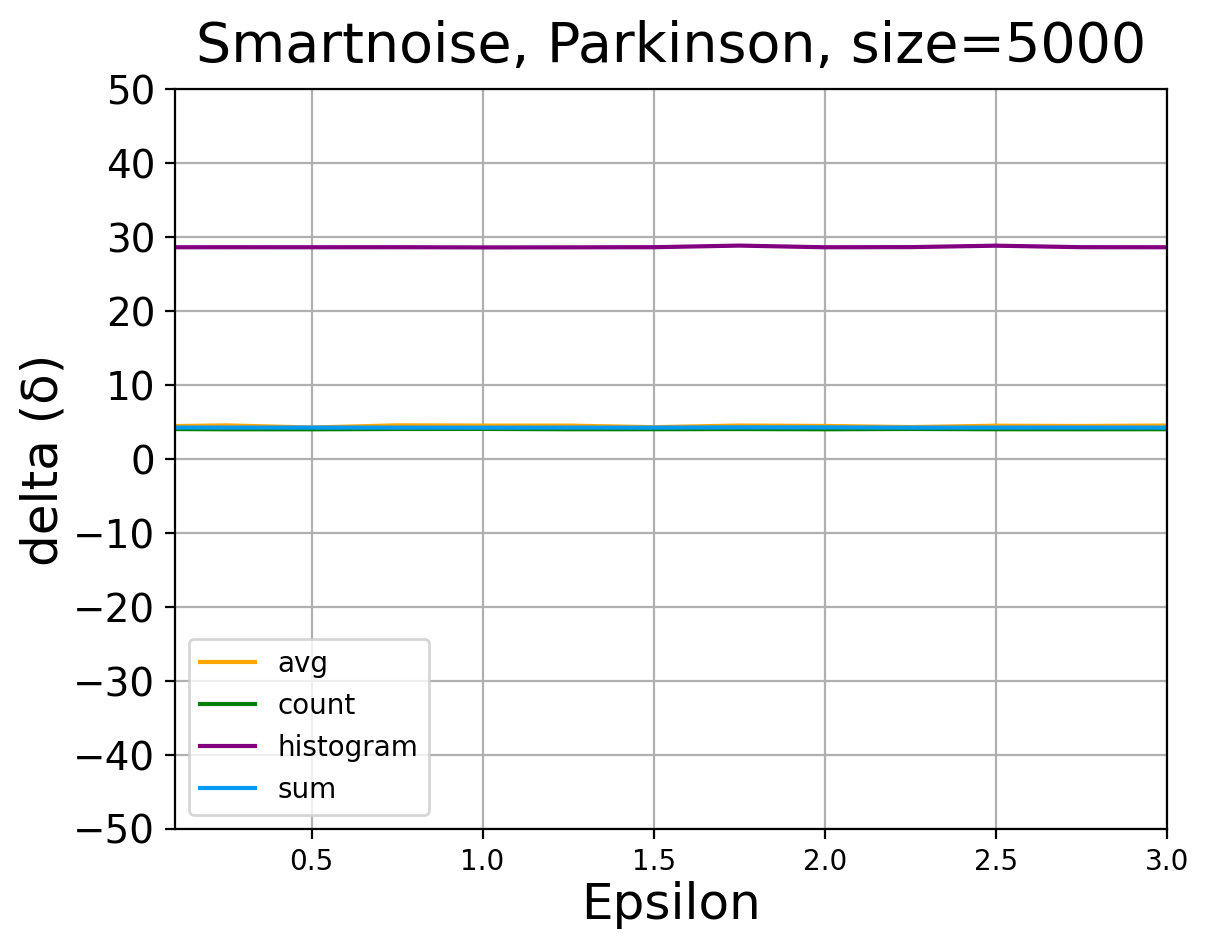}}
	\subfloat[]{\label{fig:exp3:smart:P:5499}\includegraphics[width=0.25\textwidth]{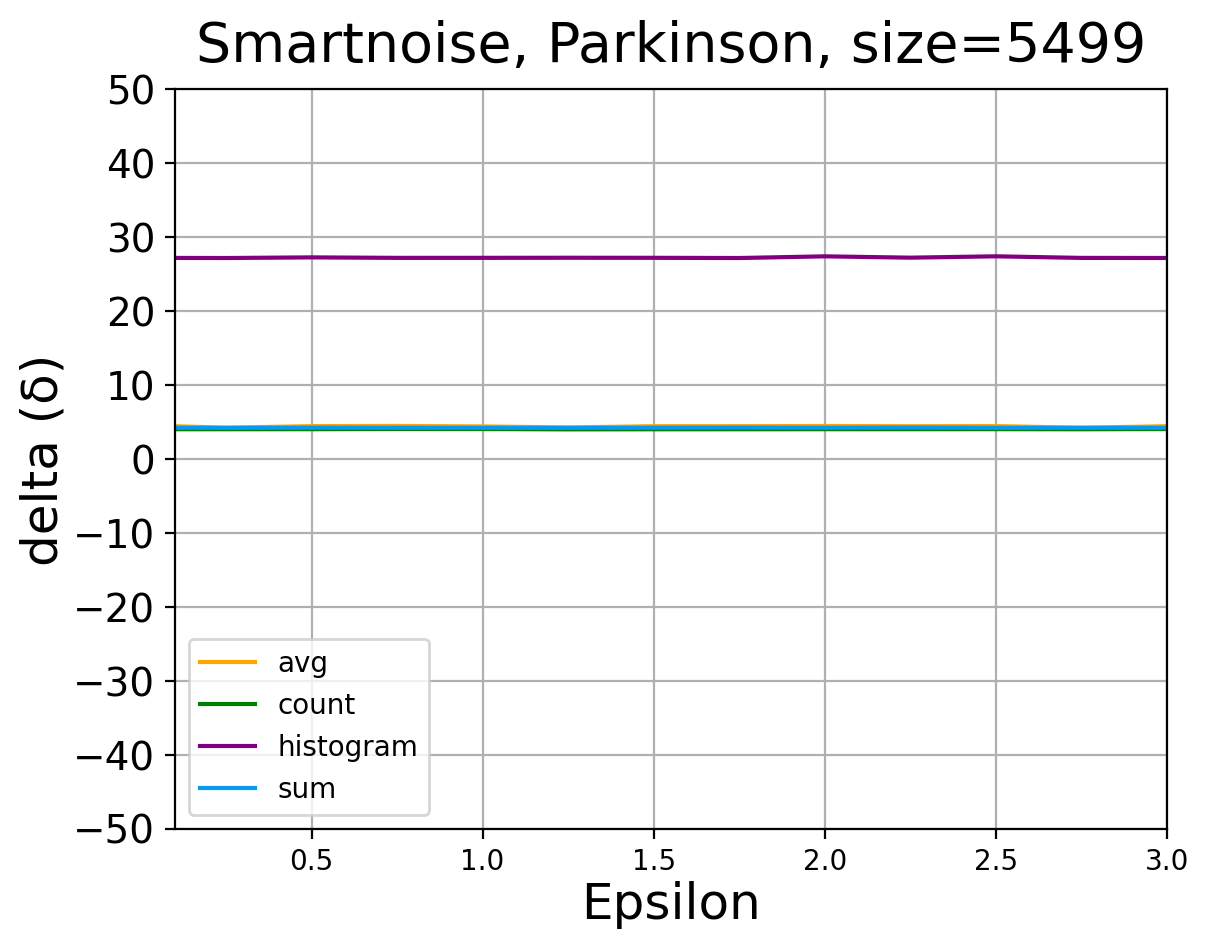}}
	
	\caption[Results of Experiment 3. Memory overhead for statistical query tools]{The evaluation results of statistical query tools on memory overhead when DP is integrated, for different data sizes (Table \ref{table_dataset_sizes}), $\epsilon$ values (Table \ref{table_epsilons}), and queries (Table \ref{table_queries}). $\delta$ is defined in Section~\ref{evaluation_criteria}.}
	
	\label{fig:exp3:line}
\end{figure}

Figure~\ref{fig:exp3:line} describes how DP impacts the processing of queries issued to the Postgres database regarding memory overhead. We cannot observe any general relationship between $\epsilon$, data size, and memory overhead through this figure. However, the results show an overhead of 5\%-40\%, indicating more memory consumption of querying procedures than database operations with less than 3\% memory overhead (see Figure~\ref{fig:exp3:post:line}). The results also suffer from fluctuations, especially for the \texttt{HISTOGRAM} query on \emph{Parkinson} by Google DP, whereas Smartnoise performs more stably though no better in that case.

\subsection{Machine learning tools assessment}\label{ml:results}

We evaluate three machine learning tools with the provision of DP service, \ie~Tensorflow Privacy, Opacus, and Diffprivlib, and investigate how their private models differ from non-private ones learned from two data sets as shown in Table~\ref{table: evaluation strategy}. This evaluation considers the regression model for all the tools, which we instantiate as a linear regression model since regression is the only functionality the considered tools hold in common. Furthermore, we vary the privacy budget $\epsilon$ and data size in the evaluation to see how the results differ regarding utility, run-time overhead, and memory overhead defined in Section~\ref{evaluation_criteria}. Note that each model training runs ten times, of which two extrema results are removed, and the remaining eight are averaged for analysis.

The results generally manifest the trend that the integration of DP in model training induces model accuracy reduction, and this reduction lowers with an increase in either $\epsilon$ or data size, which holds for all tools on both \emph{Parkinson} and \emph{Health Survey} data except Diffprivlib on \emph{Parkinson} data, where no useful result is obtained. We also observe that Tensorflow Privacy poses less memory overhead for the larger data sizes of the considered datasets, while no clear relationship exists between $\epsilon$, data size, and run-time overhead. In comparison, Opacus induces less model accuracy reduction than Tensorflow Privacy and Diffprivlib, given $\epsilon\leq0.5$ for both data sets, while Tensorflow Privacy outperforms Opacus and Diffprivlib on the continuous data set of \emph{Parkinson} within a wide range of privacy budget ($0.5\leq\epsilon\leq3.0$). The results also indicate that Tensorflow Privacy poses less run-time, and Opacus adds less memory usage when DP is integrated into model training. The quantitative evaluating results are detailed in the following.

\subsubsection{Data utility}\label{ml:exp4}

Evaluation in this section investigates how a differentially private machine-learning model differs from a benchmark regarding model accuracy to show the trade-off between privacy and utility when a machine learning task is combined with differential privacy (DP) under different experimental settings.

In this evaluation, we expect that higher $\epsilon$ values and larger data sizes will provide better utility for the considered data sets since such conditions cause less noise added during DP model training. As anticipated, the results show that the trained machine learning model brings better accuracy as $\epsilon$ and data size increase. However, along with this expected trend, there also exists local irregularities, and Diffprivlib incurs severe accuracy reduction on the \texttt{Parkinson} data making it far from useful in that case. We further detail the evaluation results below.

\begin{figure}[!h]
	\centering
	\subfloat[]{\label{fig:exp4:tf:H}\includegraphics[width=0.25\textwidth]{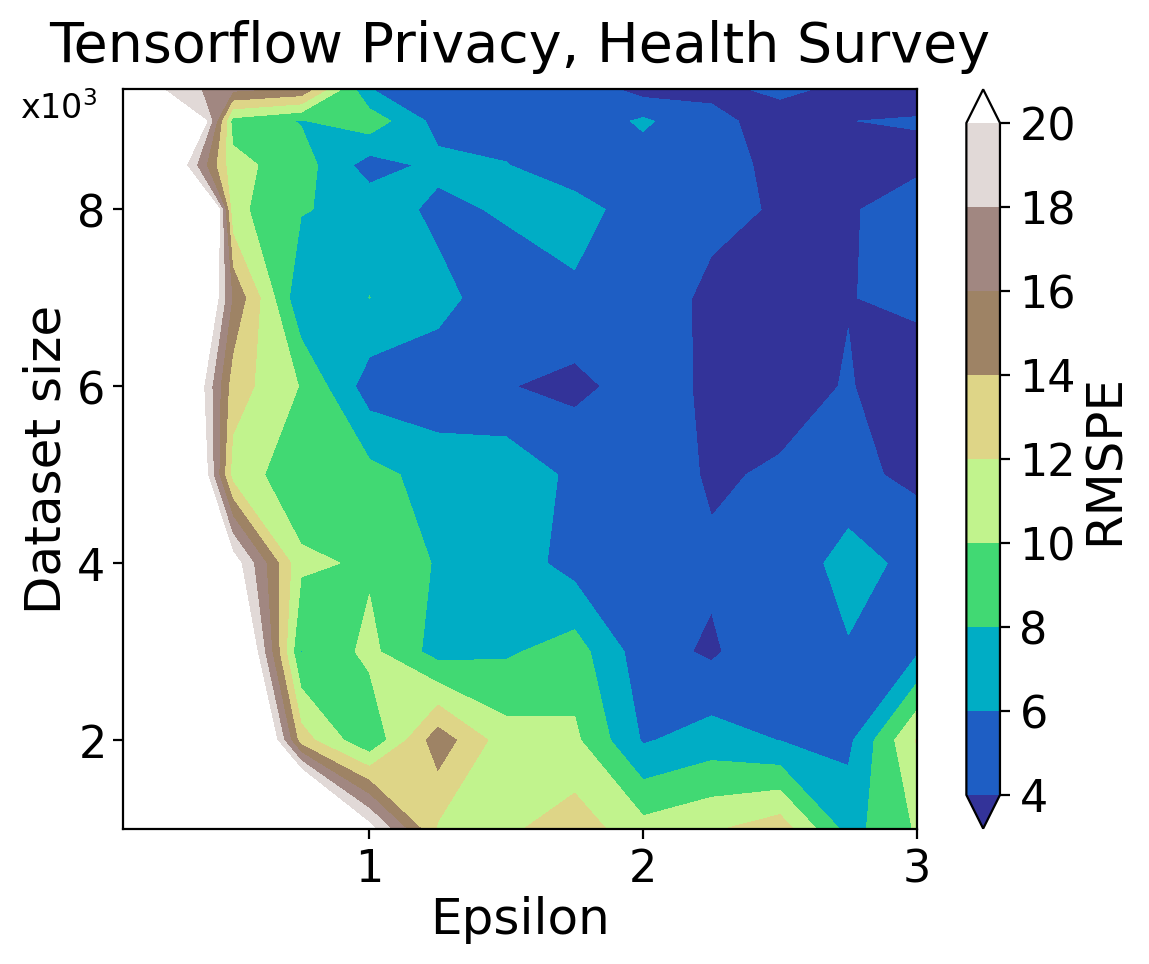}}
	\subfloat[]{\label{fig:exp4:opa:H}\includegraphics[width=0.25\textwidth]{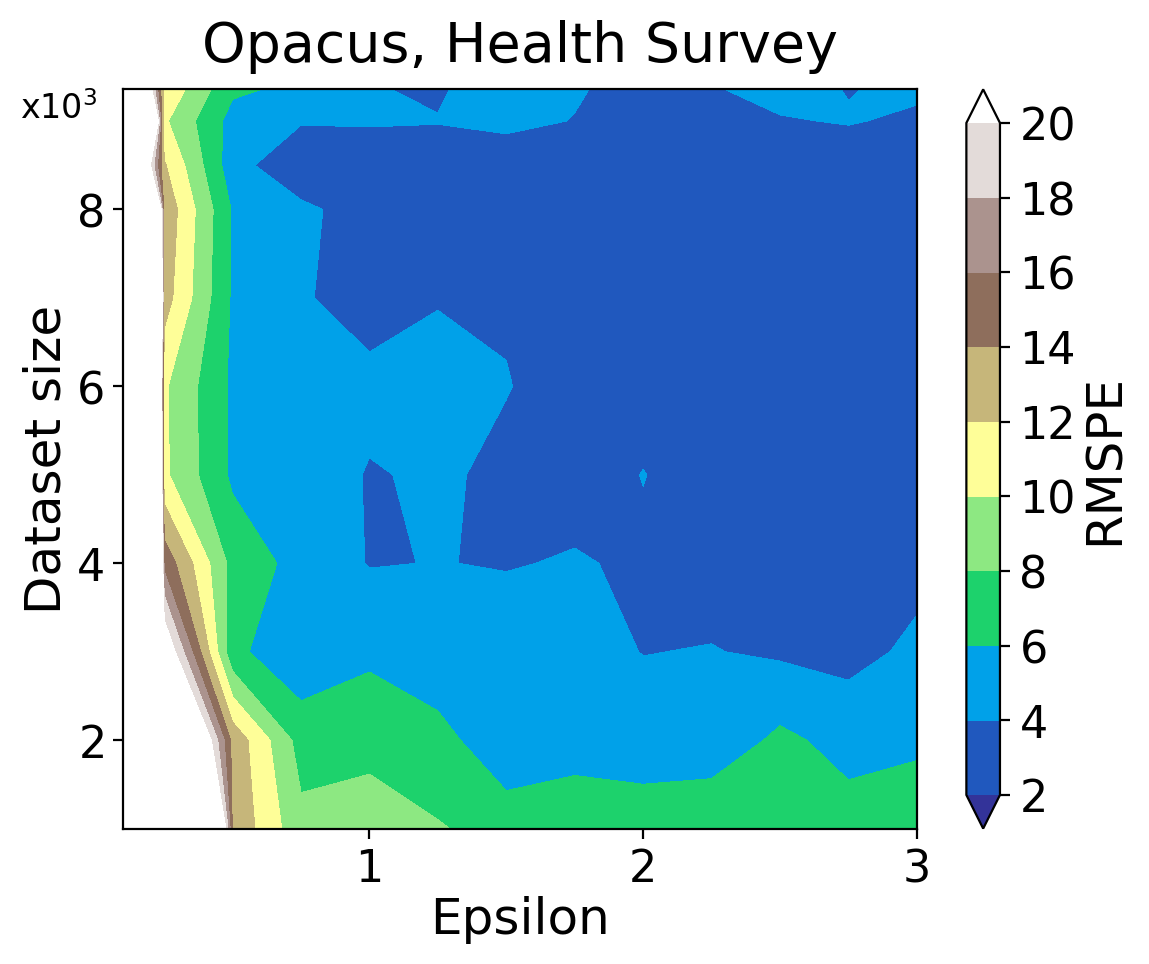}}
	\subfloat[]{\label{fig:exp4:diff:H}\includegraphics[width=0.25\textwidth]{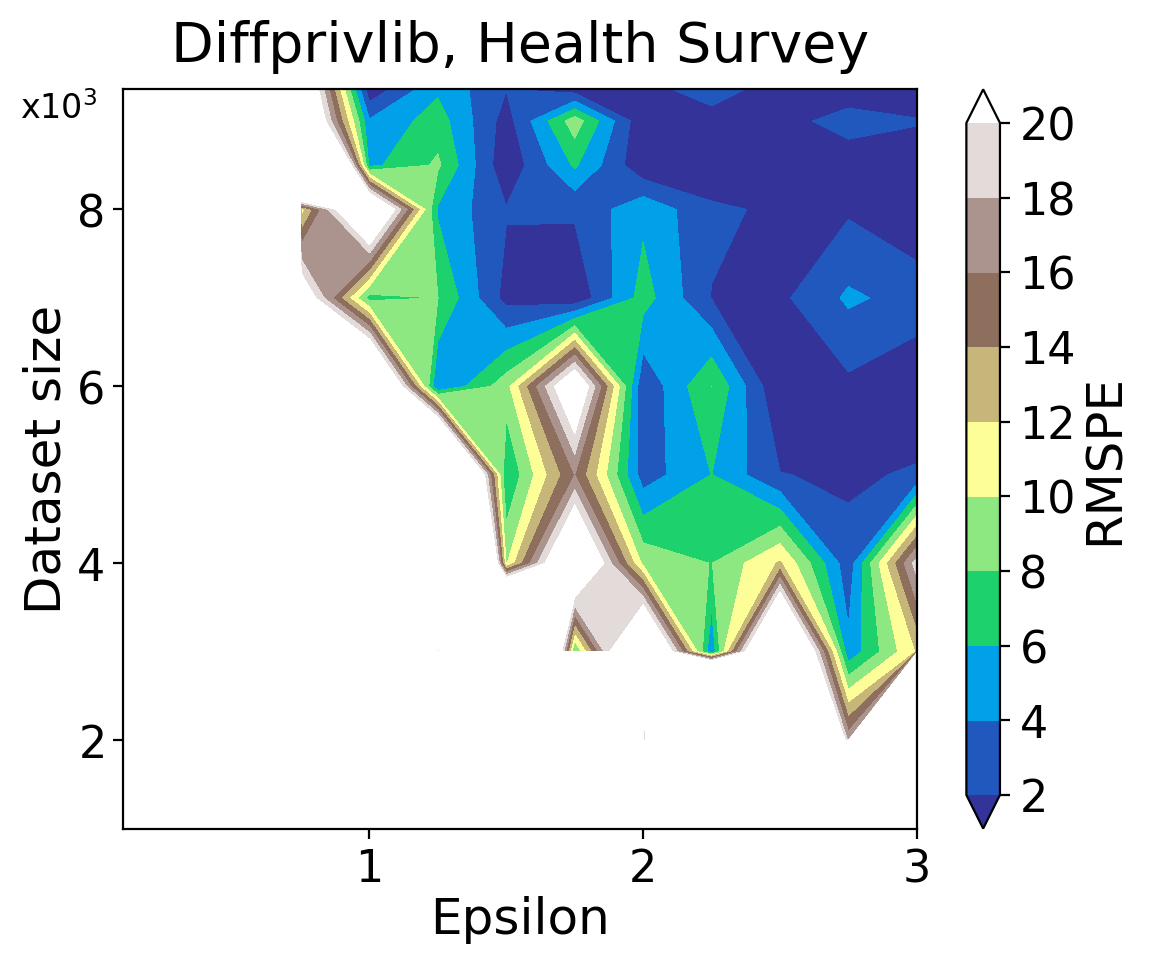}}
	\\
	\subfloat[]{\label{fig:exp4:tf:P}\includegraphics[width=0.25\textwidth]{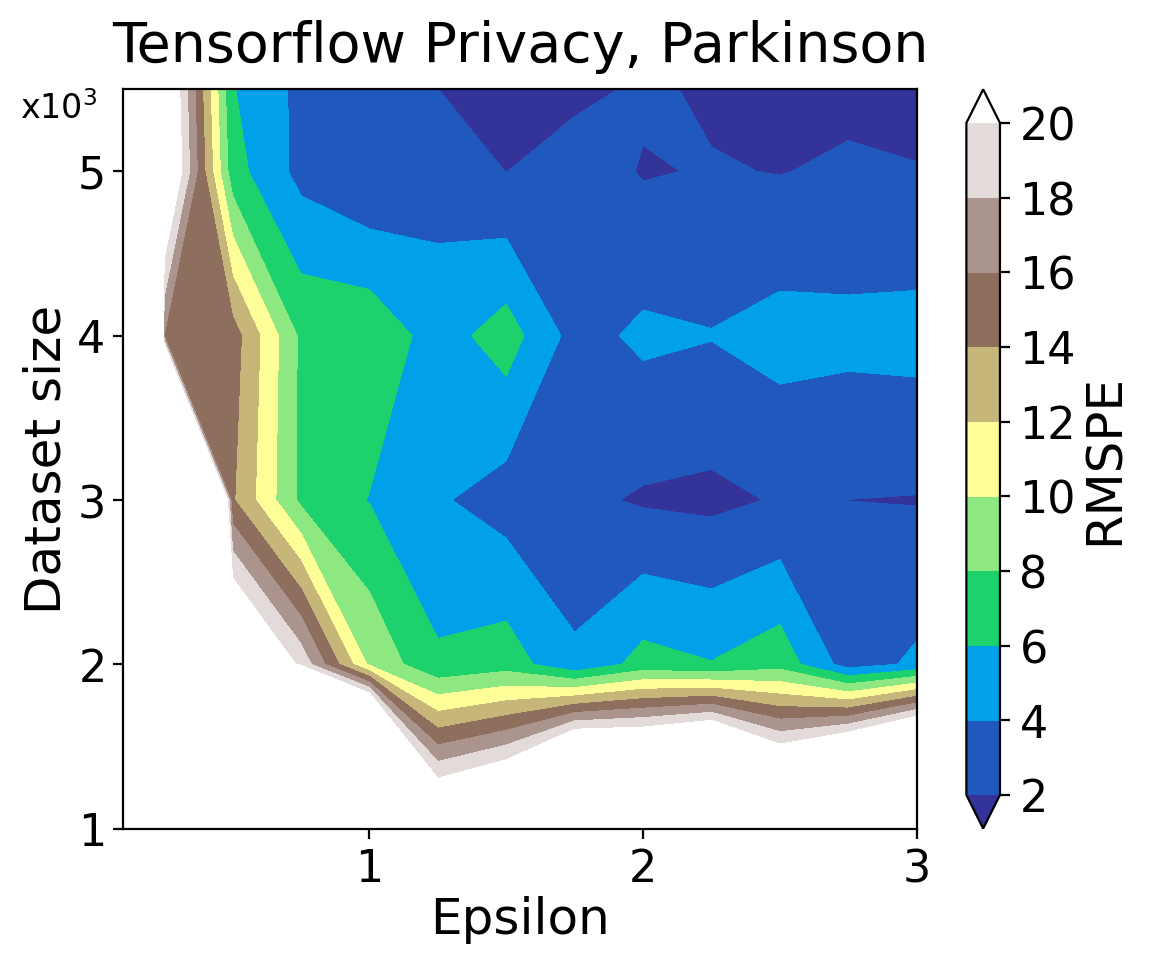}}
	\subfloat[]{\label{fig:exp4:opa:P}\includegraphics[width=0.25\textwidth]{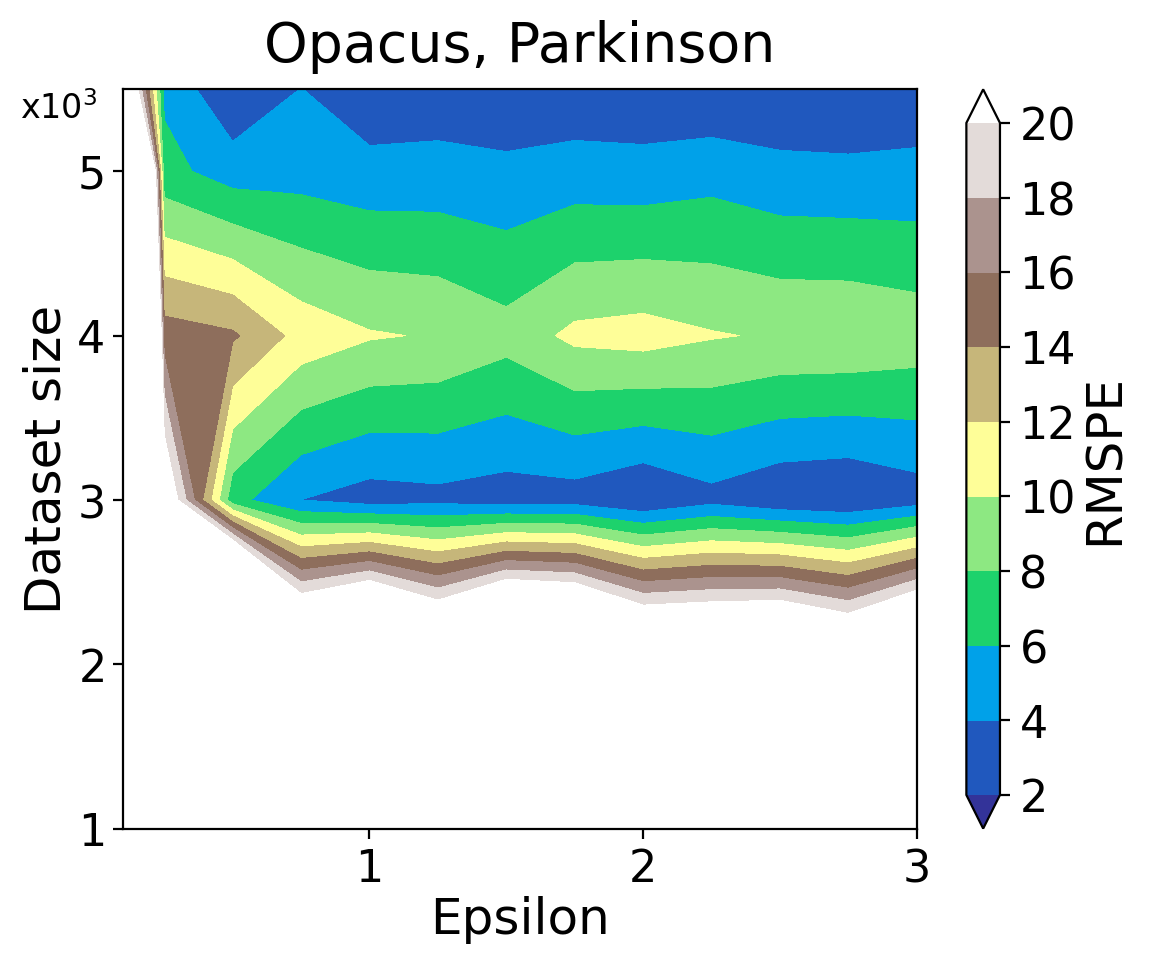}}
	\subfloat[]{\label{fig:exp4:diff:P}\includegraphics[width=0.25\textwidth]{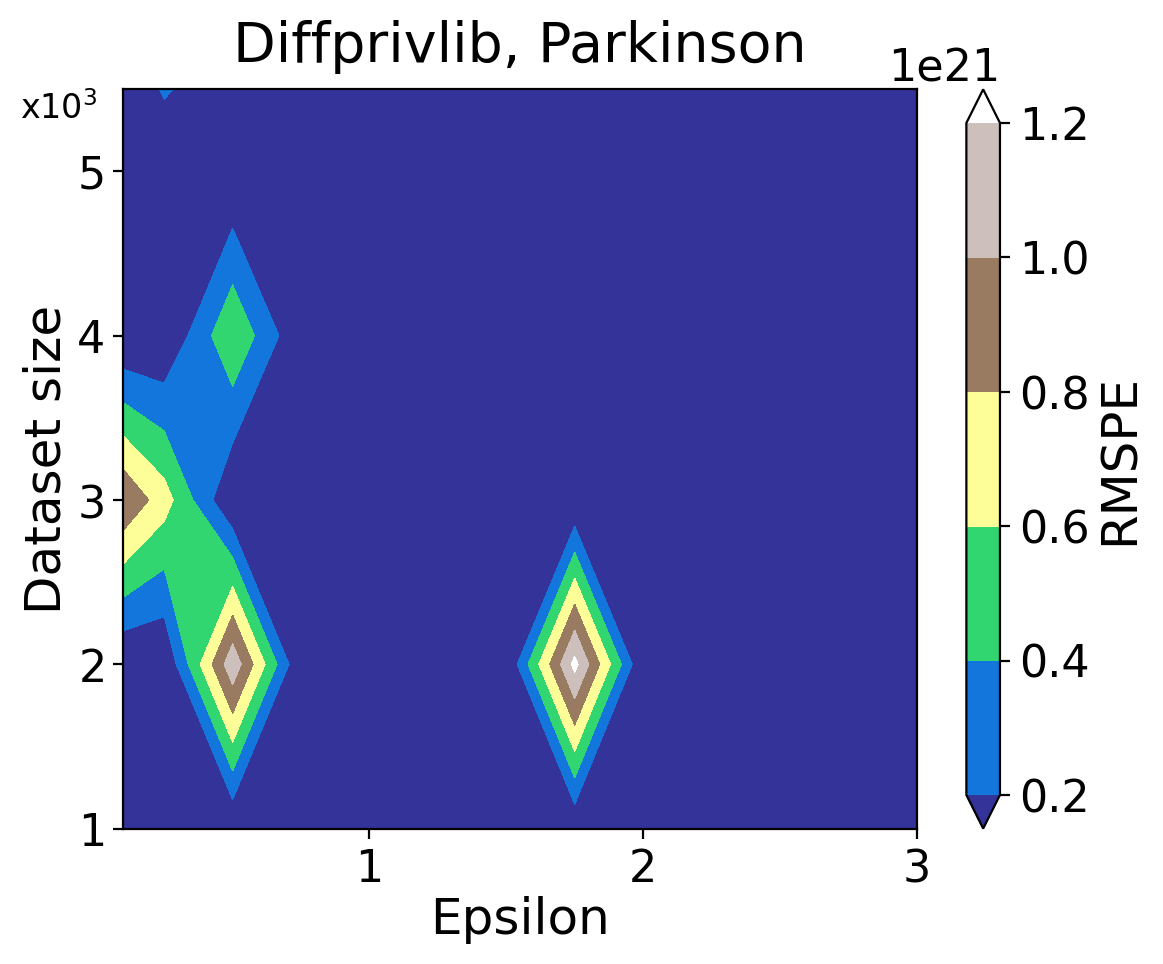}}
	\caption[Results of Experiment 4]{Contour plots for the evaluation of DP ML tools on data utility for different data sizes (Table \ref{table_dataset_sizes}) and $\epsilon$ values (Table \ref{table_epsilons}). RMSPE is defined in Section~\ref{evaluation_criteria}.}
	\label{fig:exp4:line}
\end{figure}

The contour plots in Figure~\ref{fig:exp4:line} describe how the learned model's utility measured by RMSPE varies regarding $\epsilon$ and data size. In general, the plots corroborate that utility grows with larger $\epsilon$ and data size, though there are irregularities in the \emph{Parkinson} results by Opacus, where the data size 4000 presents a slightly increased RMSPE than the lower data size. Also, $\epsilon$ does not necessarily affect Opacus' modeling utility on \emph{Parkinson} when $\epsilon\geq1.0$. We also observe that in the \emph{Health survey} experiments, Opacus exhibits marginal less RMSPE (generally $\leq6$) than Tensorflow Privacy (generally $\leq10$), both of which perform much better than Diffprivlib ($\geq10$ under most of the settings); However, Tensorflow Privacy provides obvious less RMSPE than Opacus and Diffprivlib on \emph{Parkinson} data, as shown in figures~\ref{fig:exp4:tf:P},~\ref{fig:exp4:opa:P}, and~\ref{fig:exp4:diff:P}, indicating Tensorflow Privacy's advantage on continuous data.

\begin{figure}[!h]
	\centering
	\subfloat[]{\label{fig:exp4:ml:H:8000}\includegraphics[width=0.25\textwidth]{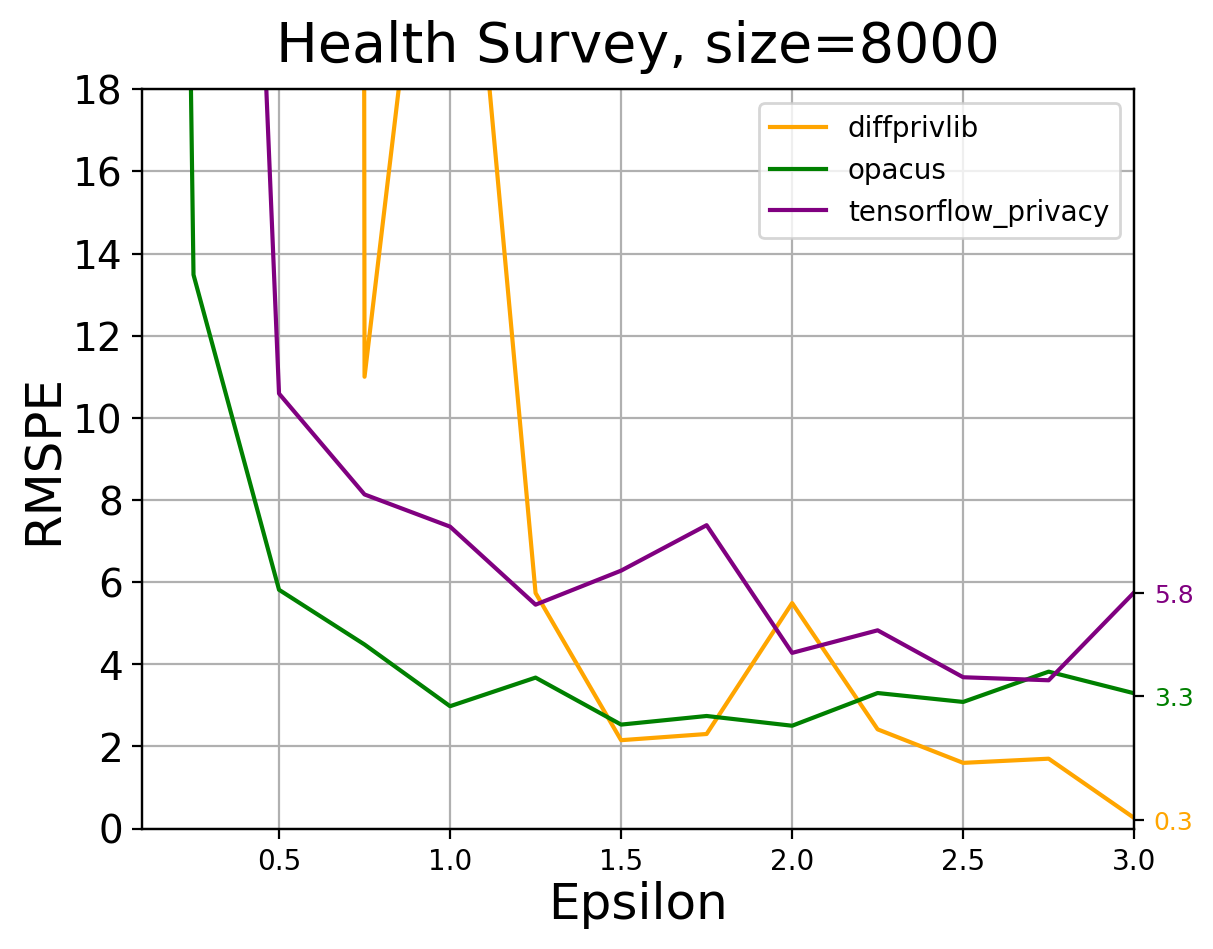}}
	\subfloat[]{\label{fig:exp4:ml:H:9000}\includegraphics[width=0.25\textwidth]{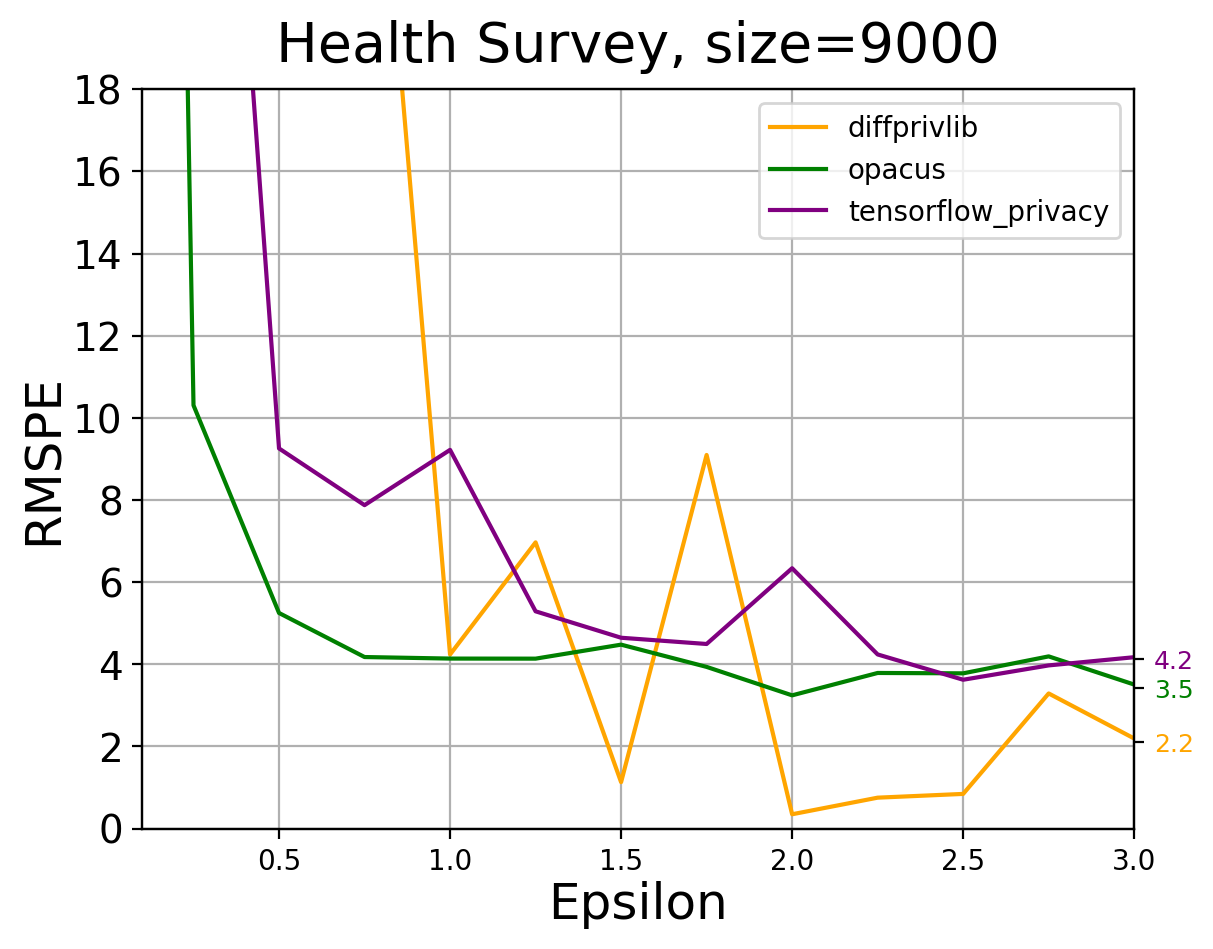}}
	\subfloat[]{\label{fig:exp4:ml:H:9358}\includegraphics[width=0.25\textwidth]{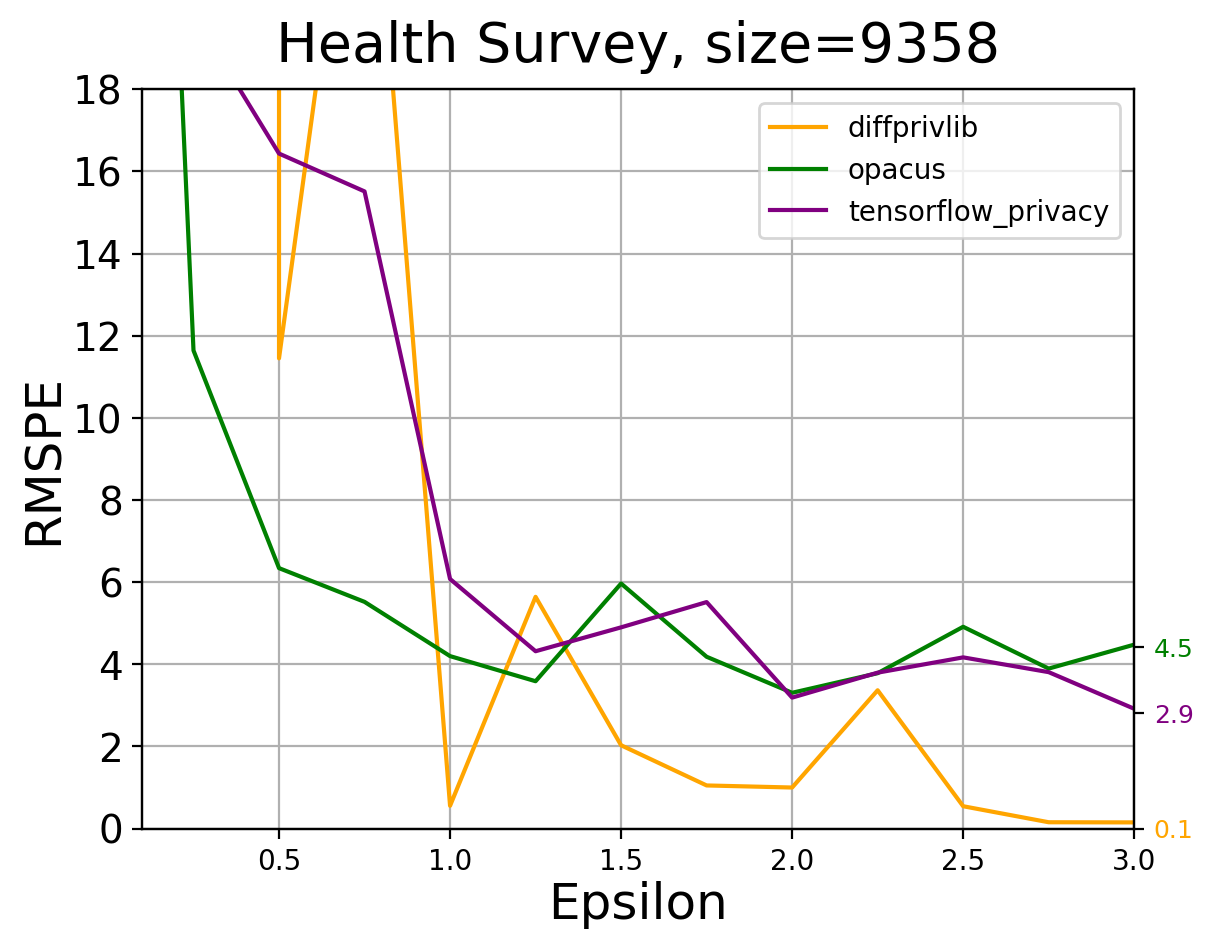}}
	\\
	\subfloat[]{\label{fig:exp4:ml:P:4000}\includegraphics[width=0.25\textwidth]{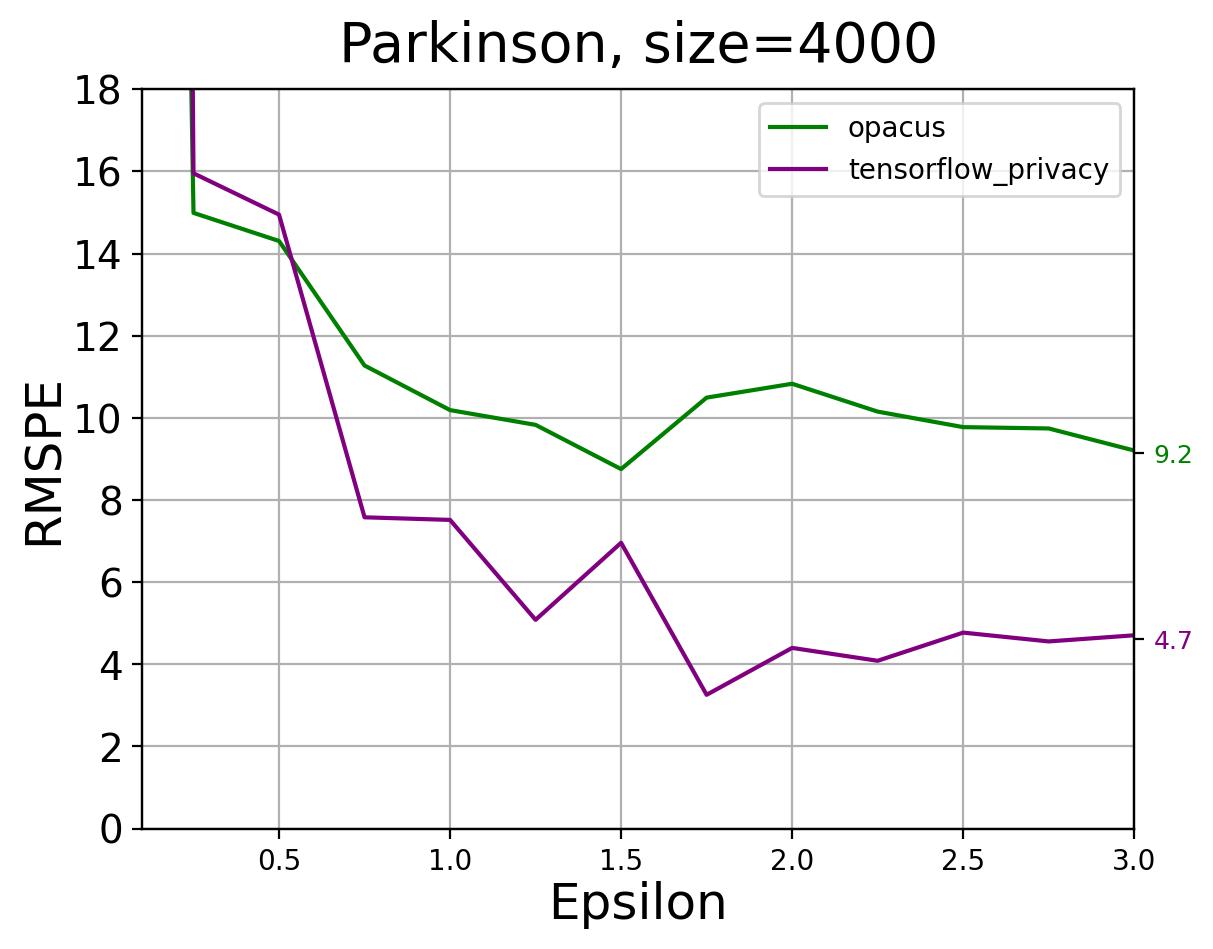}}
	\subfloat[]{\label{fig:exp4:ml:P:5000}\includegraphics[width=0.25\textwidth]{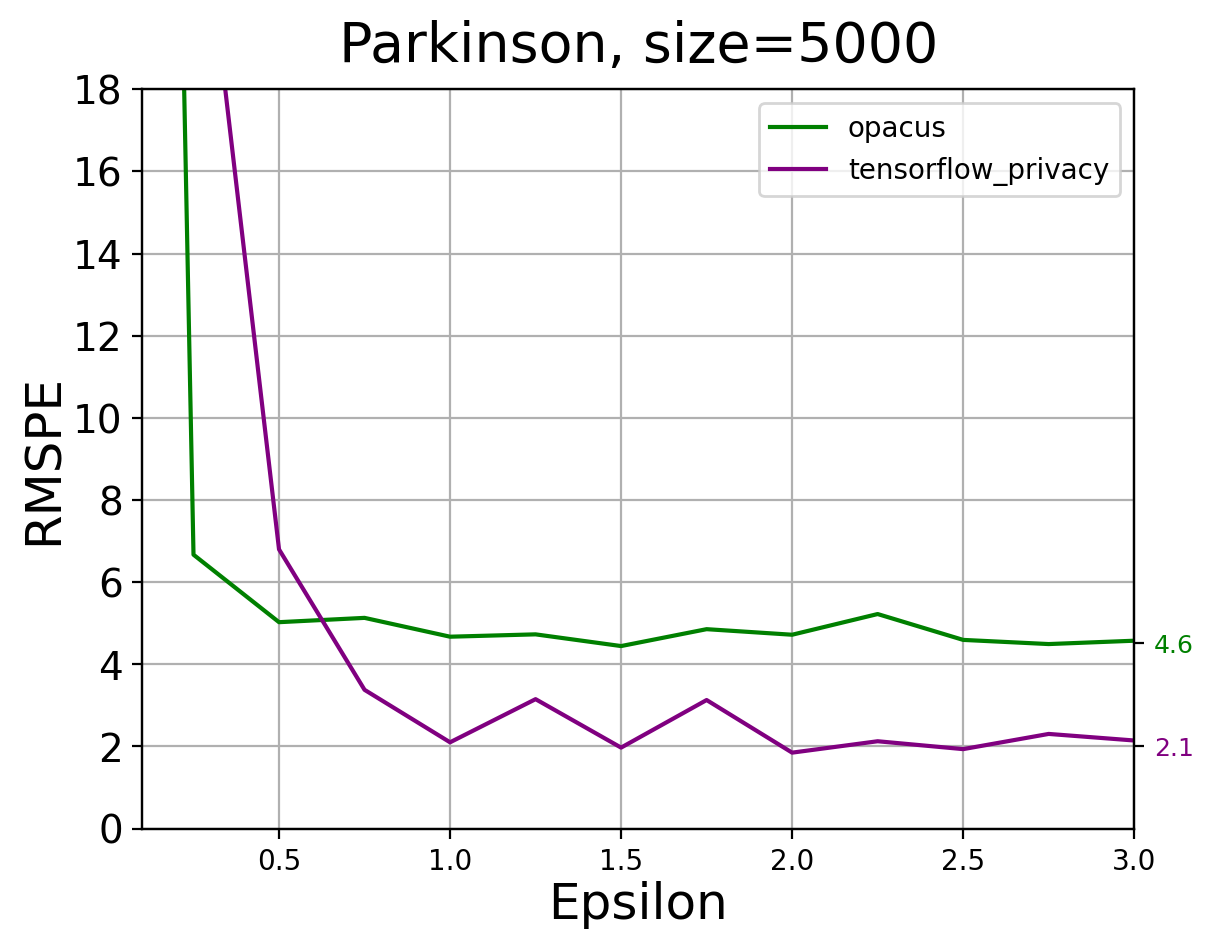}}
	\subfloat[]{\label{fig:exp4:ml:P:5499}\includegraphics[width=0.25\textwidth]{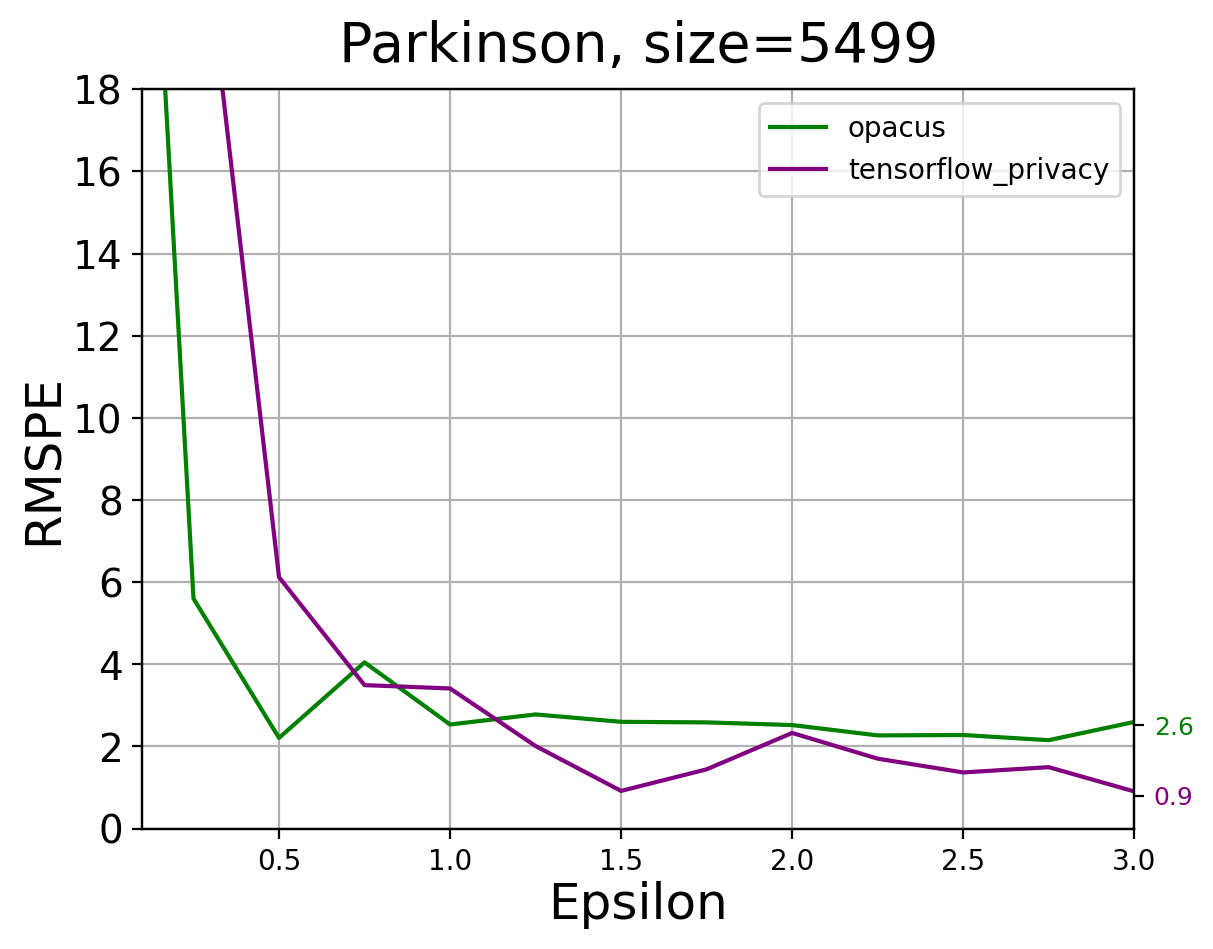}}
	\\
	\subfloat[]{\label{fig:exp4:diff:P:4000}\includegraphics[width=0.25\textwidth]{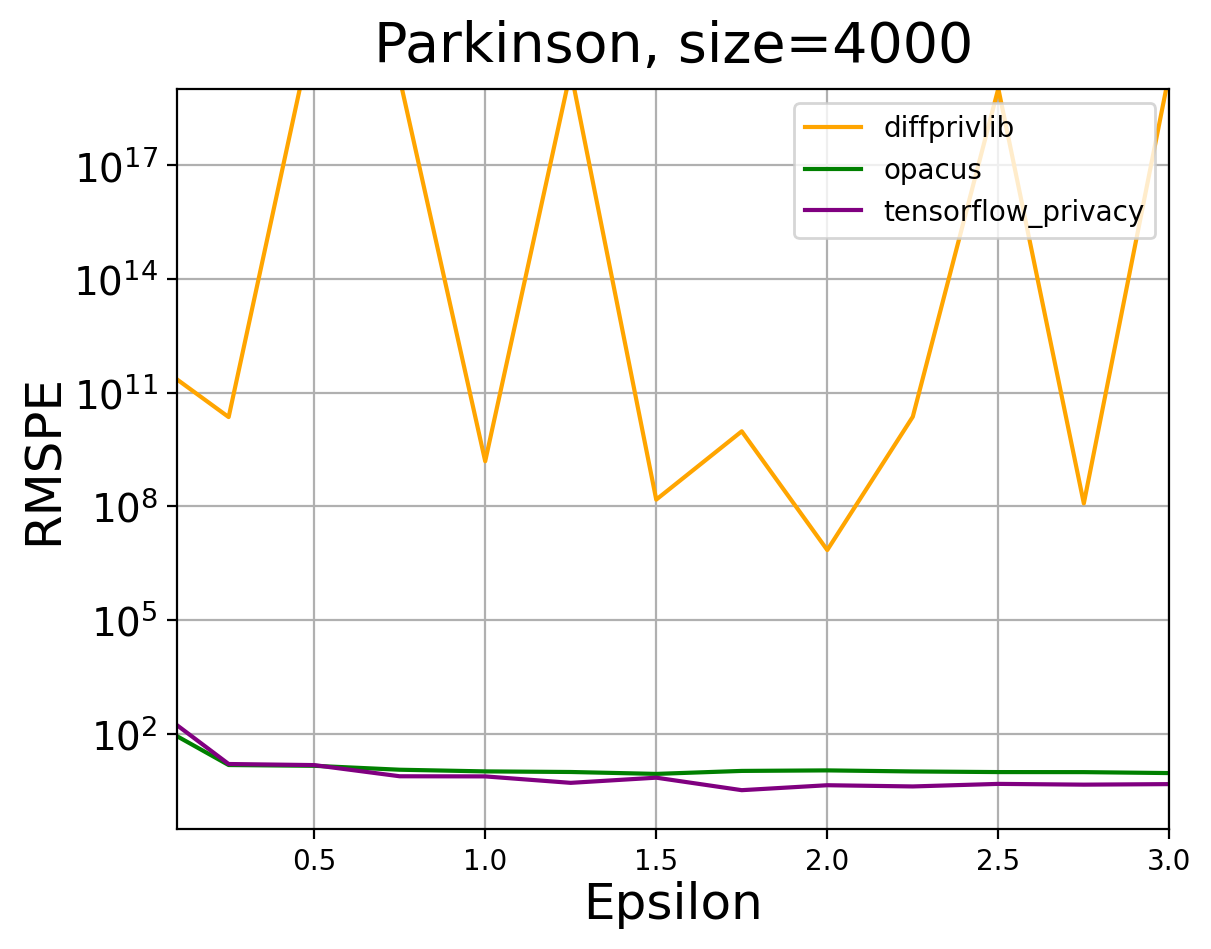}}
	\subfloat[]{\label{fig:exp4:diff:P:5000}\includegraphics[width=0.25\textwidth]{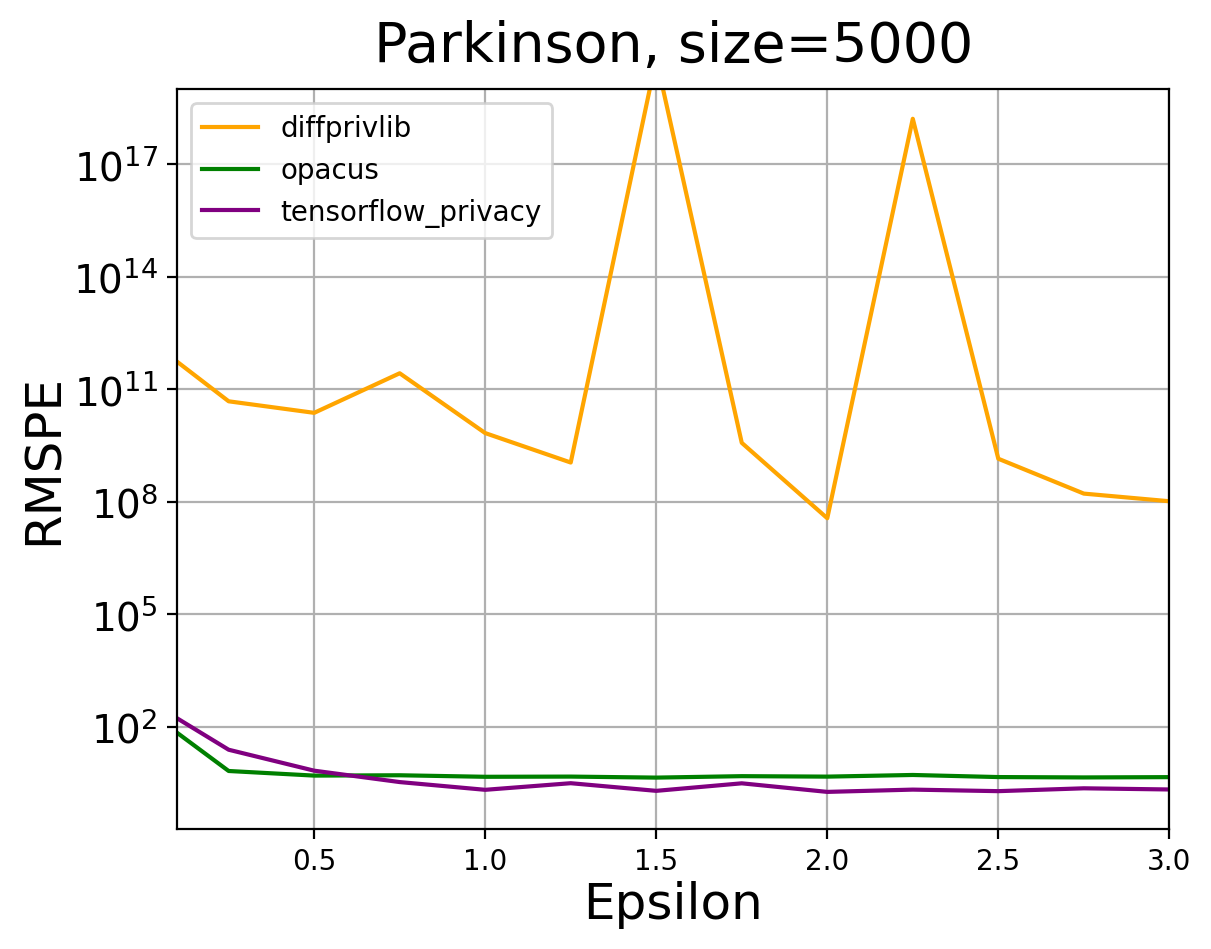}}
	\subfloat[]{\label{fig:exp4:diff:P:5499}\includegraphics[width=0.25\textwidth]{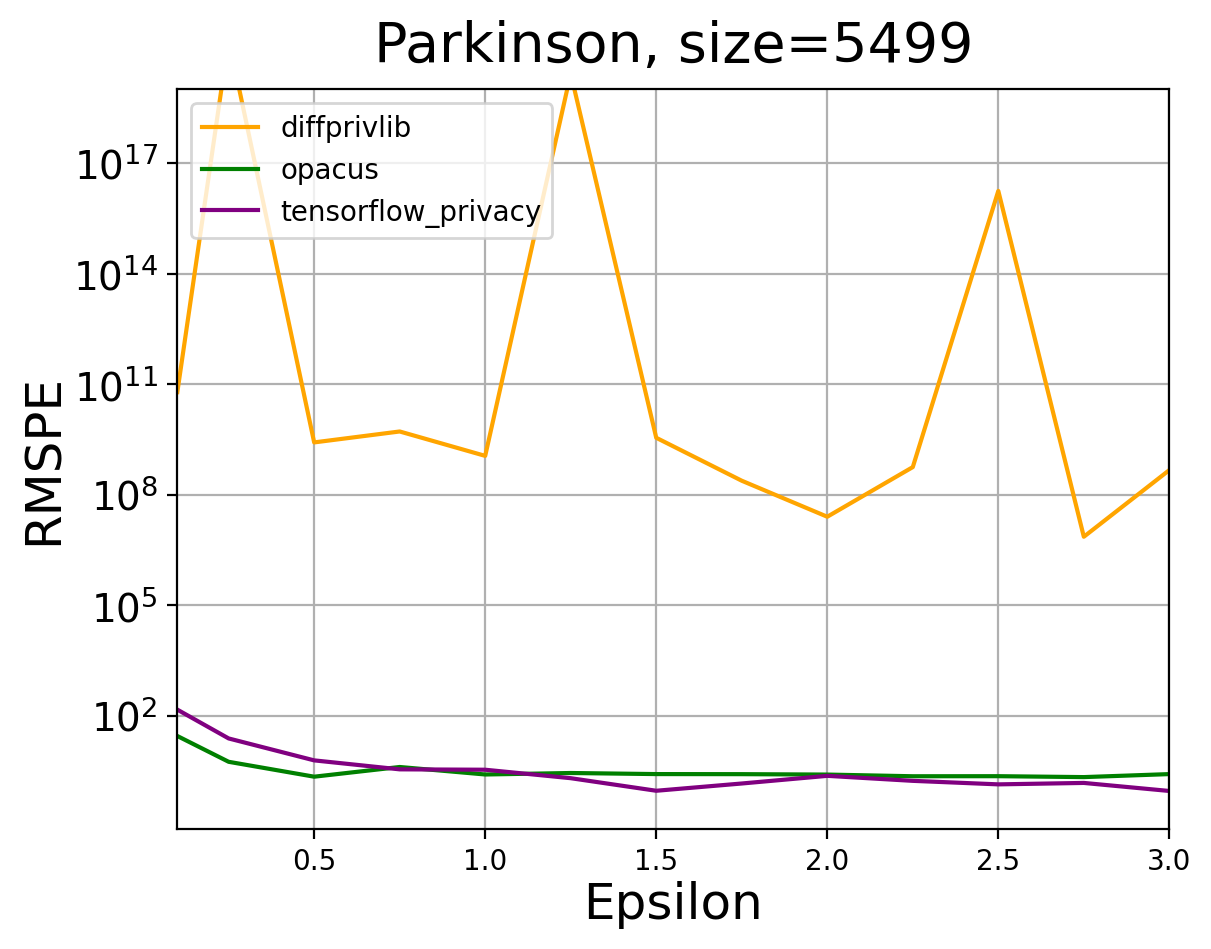}}
	\caption[Results of Experiment 4 for ML tools]{The evaluation results of DP ML tools on data utility for different data sizes (Table \ref{table_dataset_sizes}) and $\epsilon$ values (Table \ref{table_epsilons}). RMSPE is defined in Section~\ref{evaluation_criteria}.}
	\label{fig_res_ml_tools}
\end{figure}

Figure~\ref{fig_res_ml_tools} depicts model accuracy reduction under different $\epsilon$ values, where we observe that higher $\epsilon$ values decrease the RMSPE, implying less accuracy reduction of the DP-integrated-model compared with the benchmark. This trend holds for the two considered datasets by Tensorflow Privacy and Opacus, while Diffprivlib provides RMSPE of over $10^{8}$ on the \emph{Parkinson} data (figures~\ref{fig:exp4:diff:P:4000},~\ref{fig:exp4:diff:P:5000}, and~\ref{fig:exp4:diff:P:5499}), making it unacceptable under this scenario. Therefore, we neglect the experimental results of Diffprivlib on \emph{Parkinson} in the following analyses. Even though, it is worth noting that Diffprivlib generates comparably equal accuracy to Opacus and Tensorflow Privacy when no DP is applied on both datasets. In comparison, Opacus outperforms Tensorflow Privacy and Diffprivlib when $\epsilon\leq0.5$. However, since RMSPE rises abruptly when $\epsilon$ decreases away from 0.5, the advantage of Opacus here gets insignificant. In contrast, Tensorflow Privacy produces less accuracy reduction for \emph{Parkinson} data within a wide range of $\epsilon$ (0.5-3.0), implying its better performance on continuous data set than Opacus and Diffprivlib.

\subsubsection{Run-time overhead}\label{ml:exp5}

Evaluation of this part presents how the considered machine learning (ML) tools perform when combined with differential privacy (DP) regarding the run-time overhead induced due to DP. We vary the experimental settings to see how the results differ under various conditions.

The evaluation is anticipated to observe an increase in DP machine learning's run-time compared with the benchmark, and we also expect that the run-time overhead grows as the data size rises in the experiments since more data is processed. The results, as detailed below, show that Tensorflow Privacy poses the least run-time increase compared with Opacus and Diffprivlib, with the general RMSPE of $\leq40$ in run-time for Tensorflow Privacy versus $200-230$ for Opacus and $400-2000$ for Diffprivlib. Note that we did not display the results of Diffprivlib on \emph{Parkinson} data since there is no useful result generated, as elaborated in Section~\ref{ml:exp4}. Beyond that, we observe no clear trend between data size, $\epsilon$, and run-time overhead.

\begin{figure}[!htb]
	\centering
	\subfloat[]{\label{fig:exp5:tf:H}\includegraphics[width=0.25\textwidth]{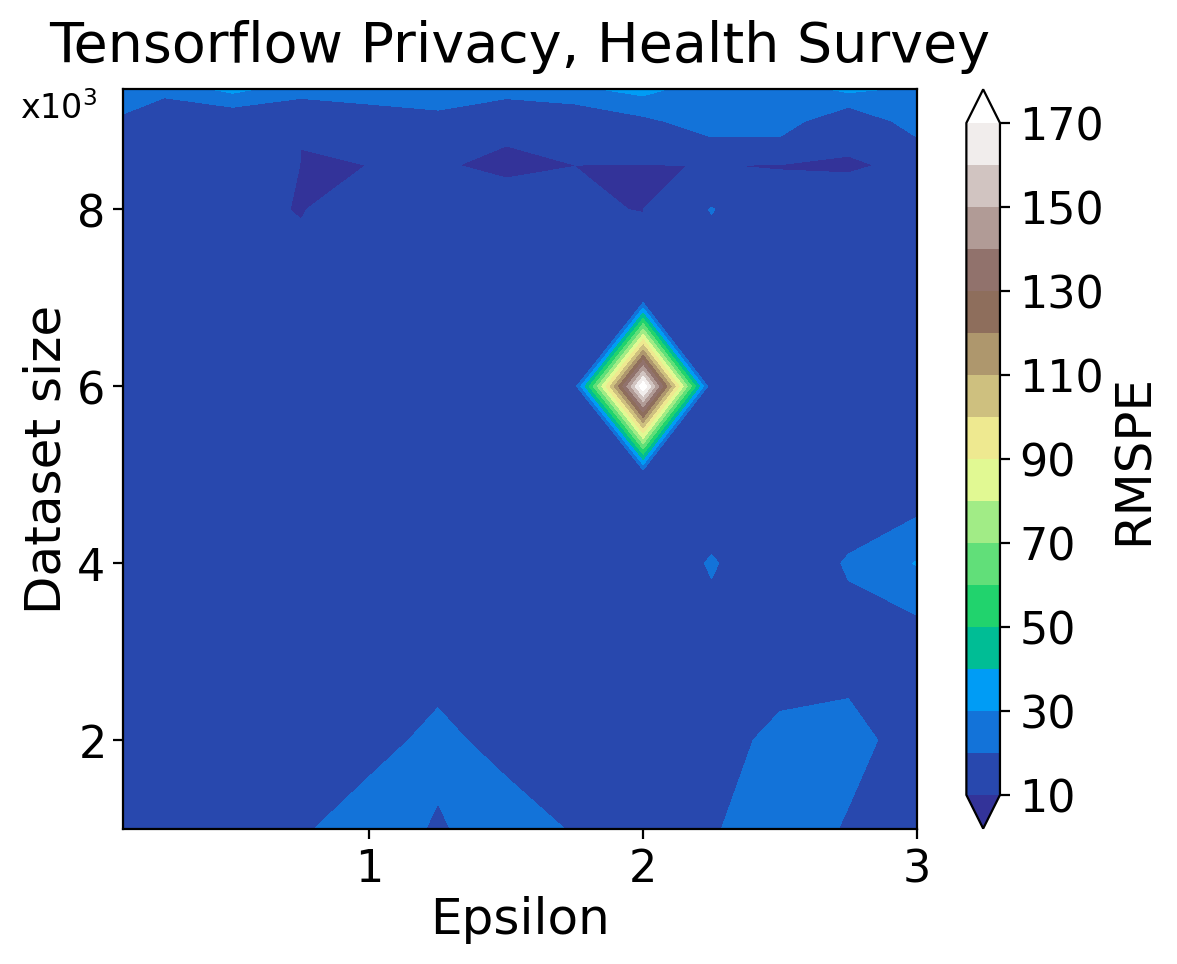}}
	\subfloat[]{\label{fig:exp5:opa:H}\includegraphics[width=0.25\textwidth]{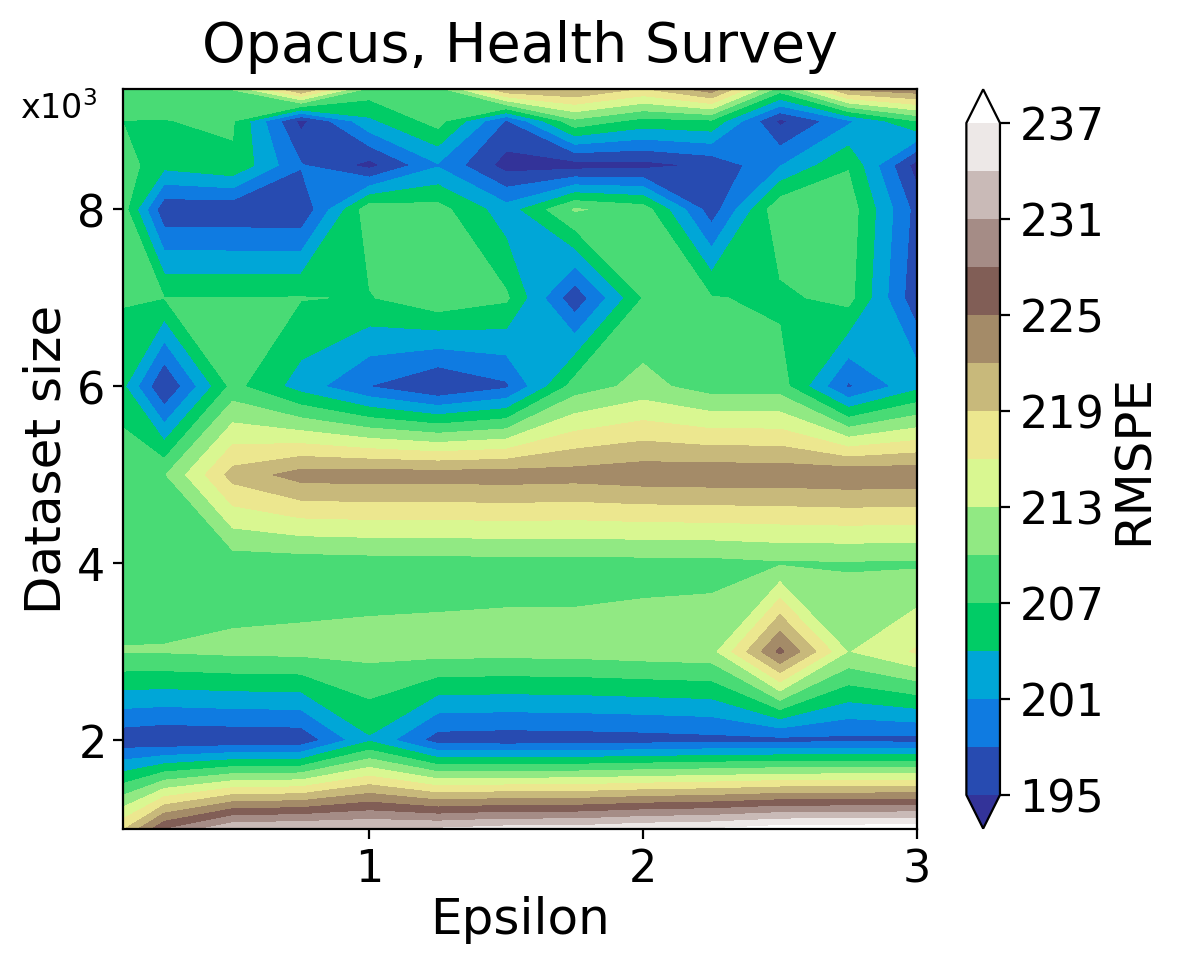}}
	\subfloat[]{\label{fig:exp5:diff:H}\includegraphics[width=0.25\textwidth]{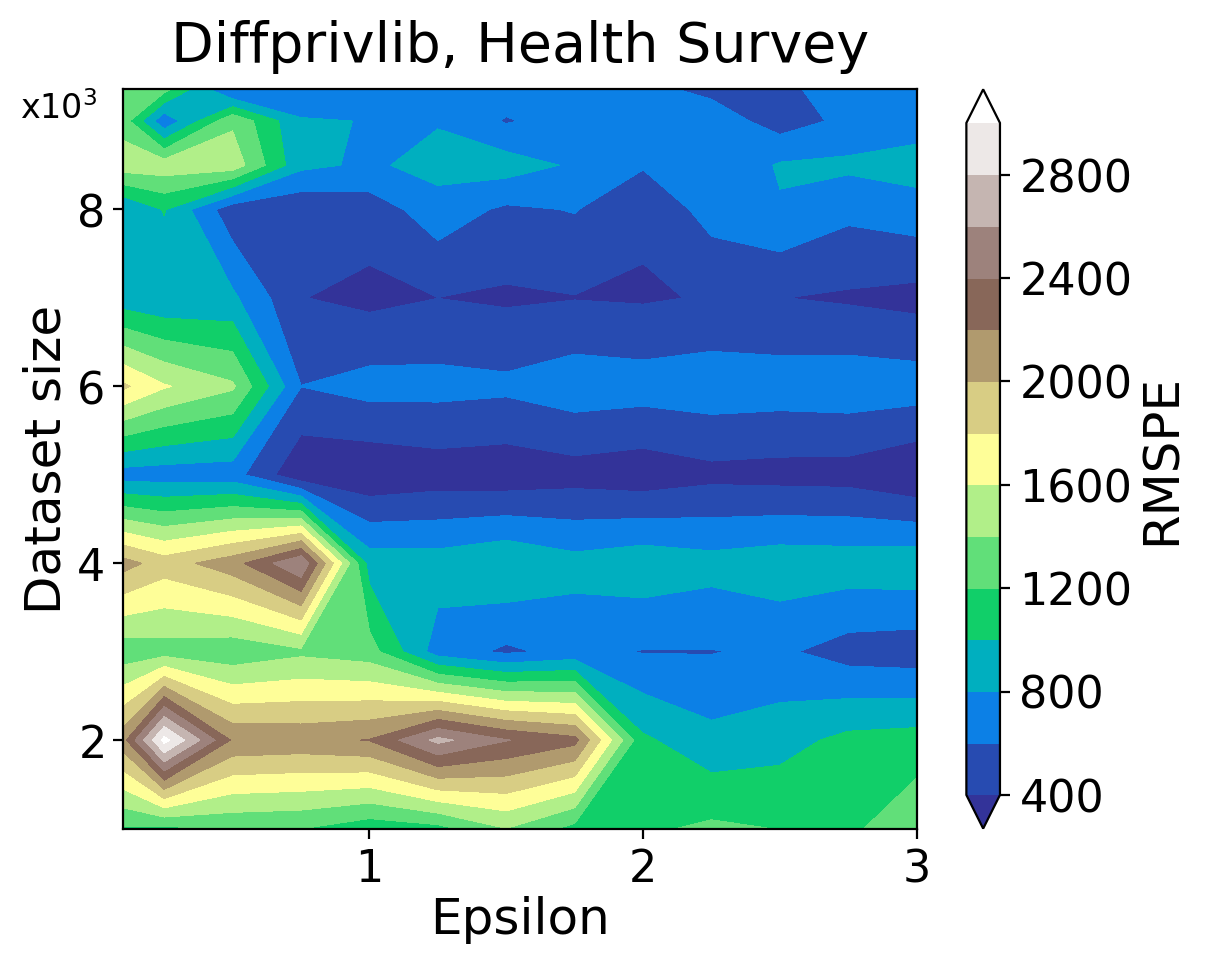}}
	\\
	\subfloat[]{\label{fig:exp5:tf:P}\includegraphics[width=0.25\textwidth]{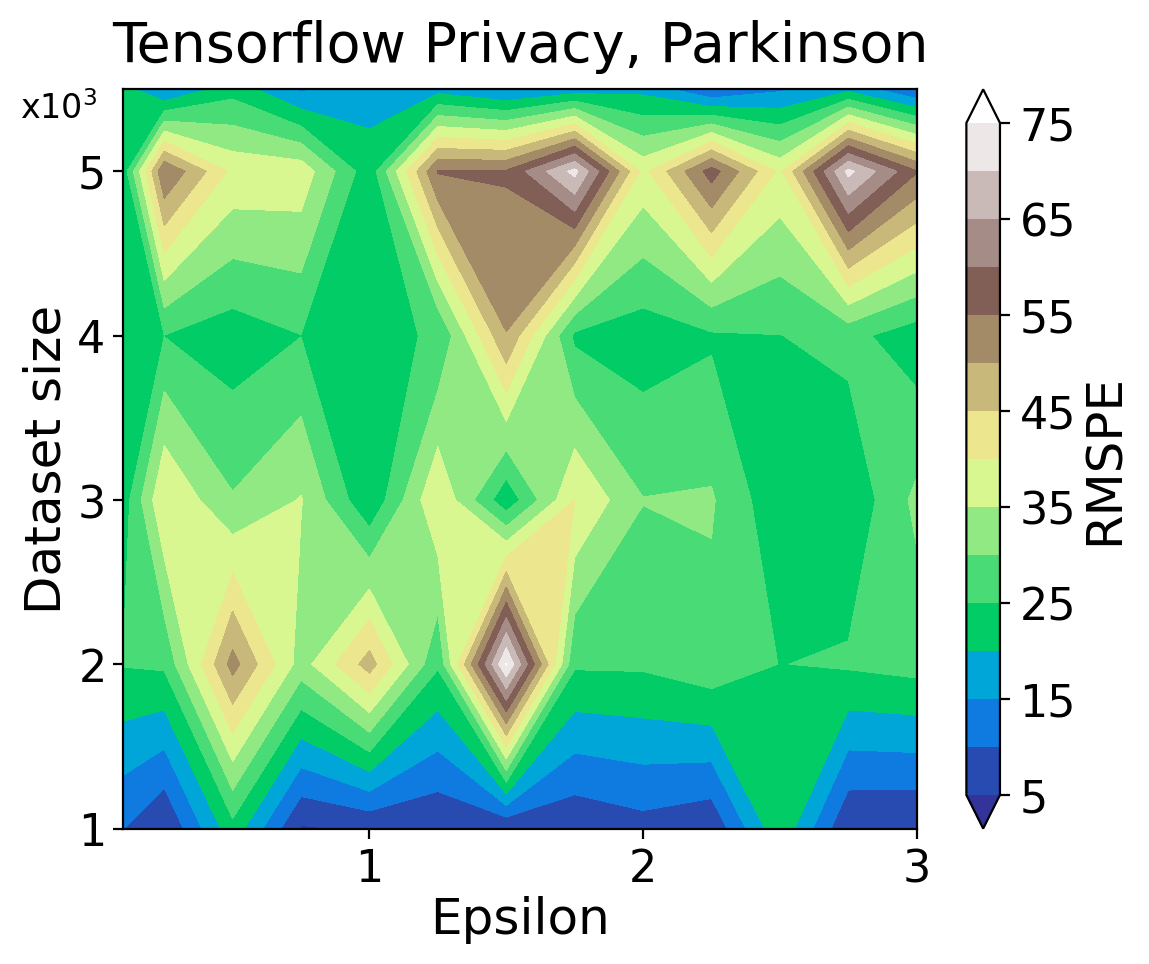}}
	\subfloat[]{\label{fig:exp5:diff:P}\includegraphics[width=0.25\textwidth]{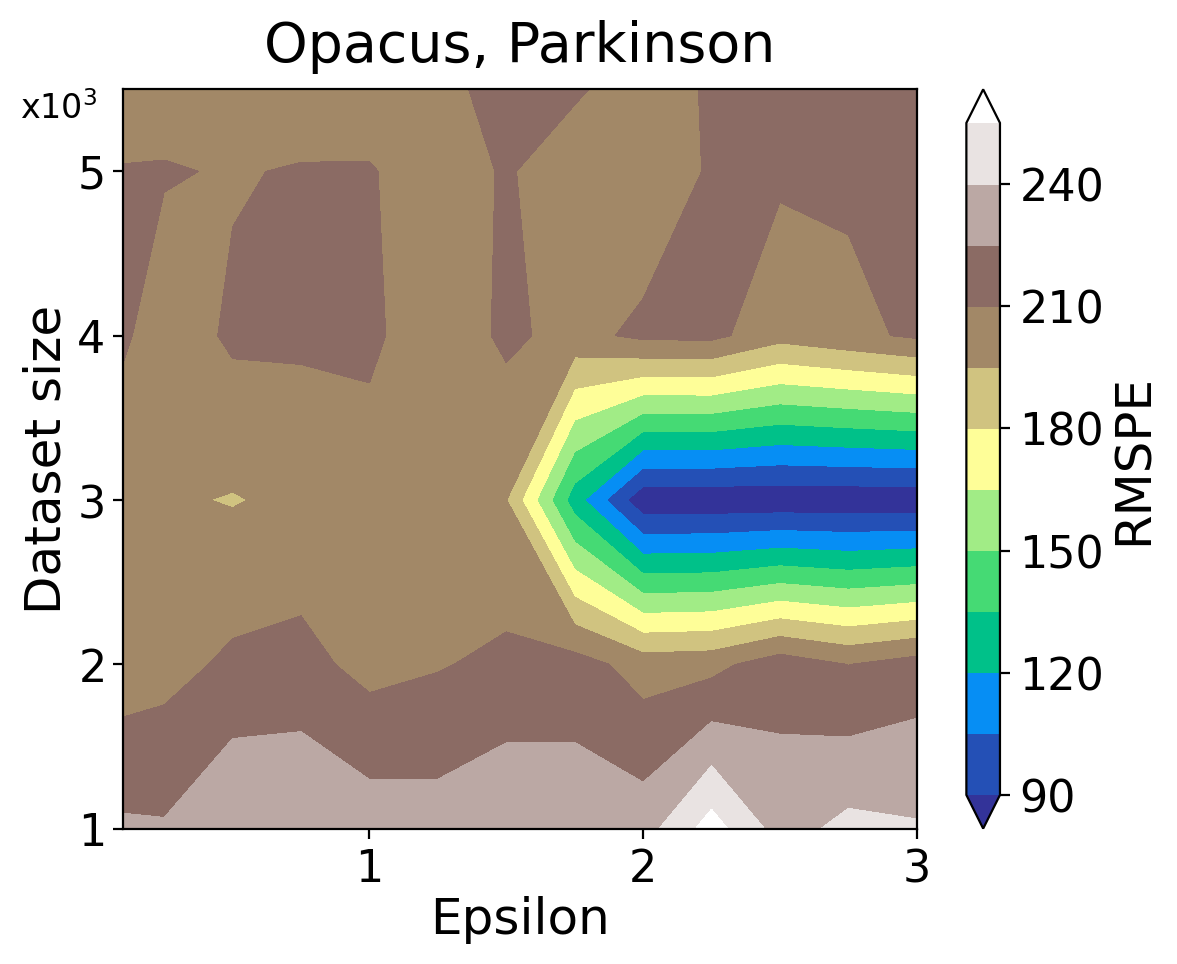}}\hspace{0.25\textwidth}
	\caption[Results of Experiment 5]{Contour plots for the evaluation of DP ML tools on run-time overhead for different data sizes (Table \ref{table_dataset_sizes}) and $\epsilon$ values (Table \ref{table_epsilons}). RMSPE is defined in Section~\ref{evaluation_criteria}.}
	\label{fig:exp5:contour}
\end{figure}

Contour plots in Figure~\ref{fig:exp5:contour} describe each tool's run-time overhead regarding different $\epsilon$ values and data sizes. Through the plots, we observe that both $\epsilon$ and data size affect the run-time overhead in an irregular manner, where no explicit patterns can be concluded. The results also present fluctuations and irregularities that are hard to explain, \eg~the abrupt increased RMSPE for Tensorflow Privacy on \emph{Health survey} when $\epsilon=2$ and $data\; size=6\times10^{3}$, and the sudden decreased RMSPE for Opacus on \emph{Parkinson} when $data\; size=3\times10^{3}$. Even though, it is clear that Tensorflow Privacy incurs less RMSPE due to DP in machine learning, compared with Opacus and Diffprivlib on both data sets.

\begin{figure}[!htb]
	\centering
	\subfloat[]{\label{fig:exp5:H:6000}\includegraphics[width=0.25\textwidth]{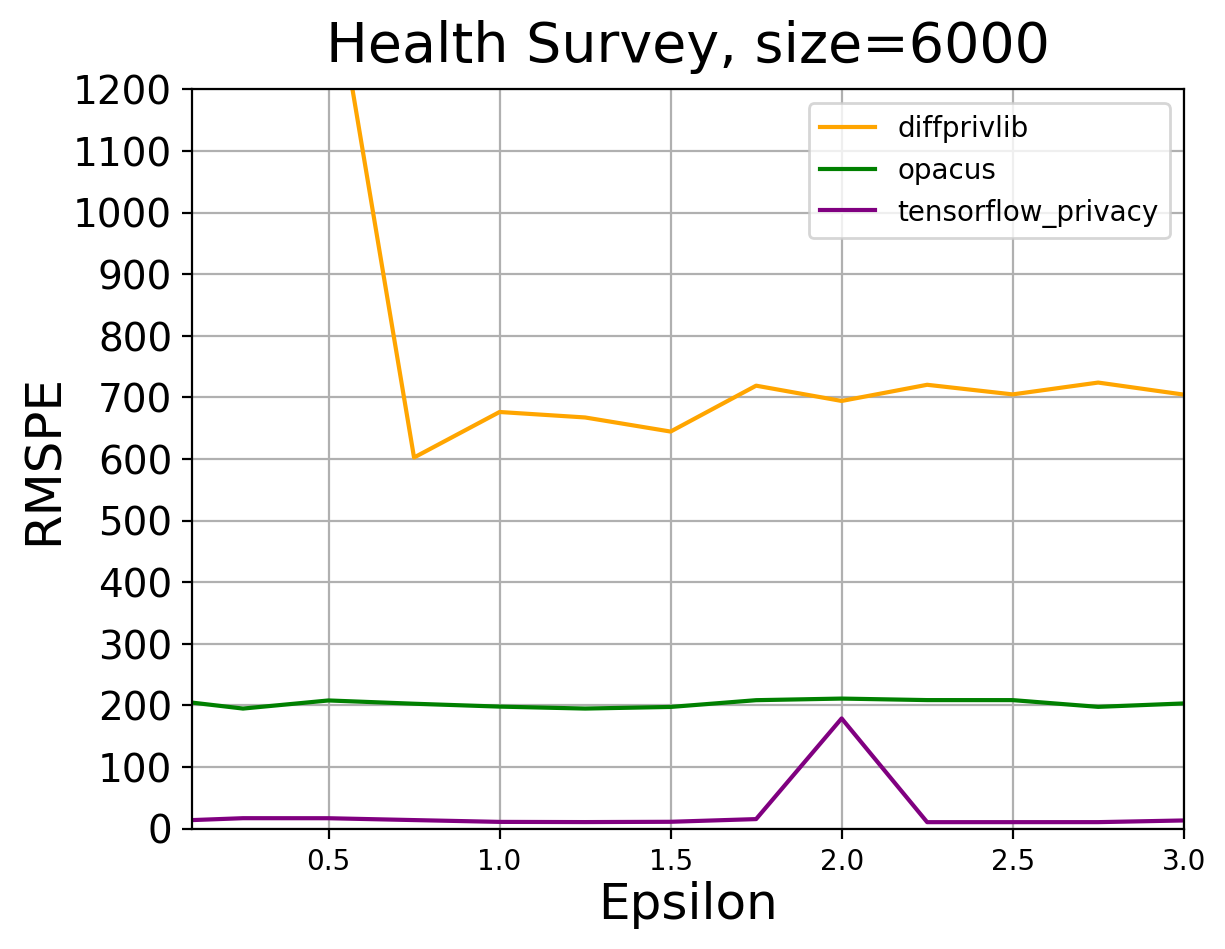}}
	\subfloat[]{\label{fig:exp5:H:8000}\includegraphics[width=0.25\textwidth]{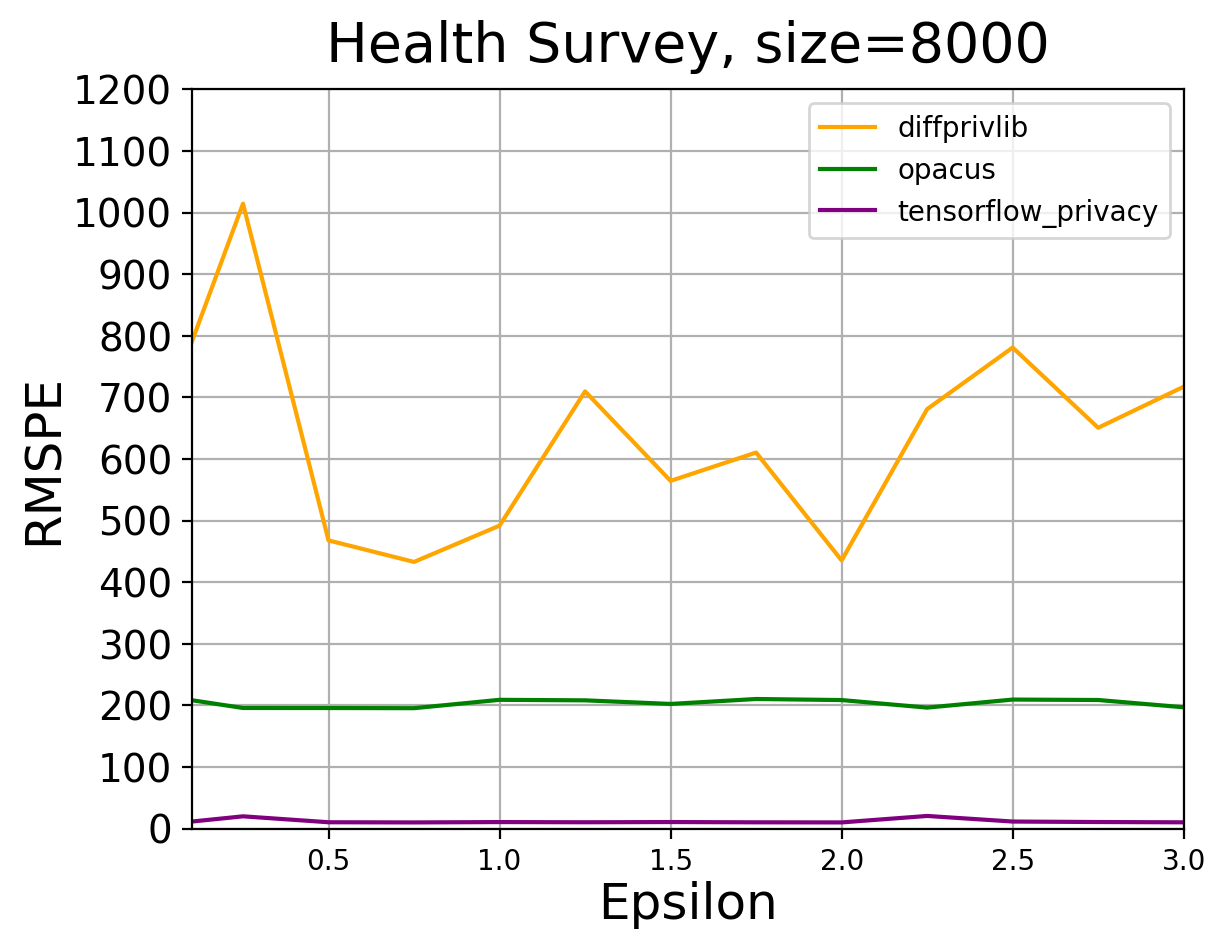}}
	\subfloat[]{\label{fig:exp5:H:9000}\includegraphics[width=0.25\textwidth]{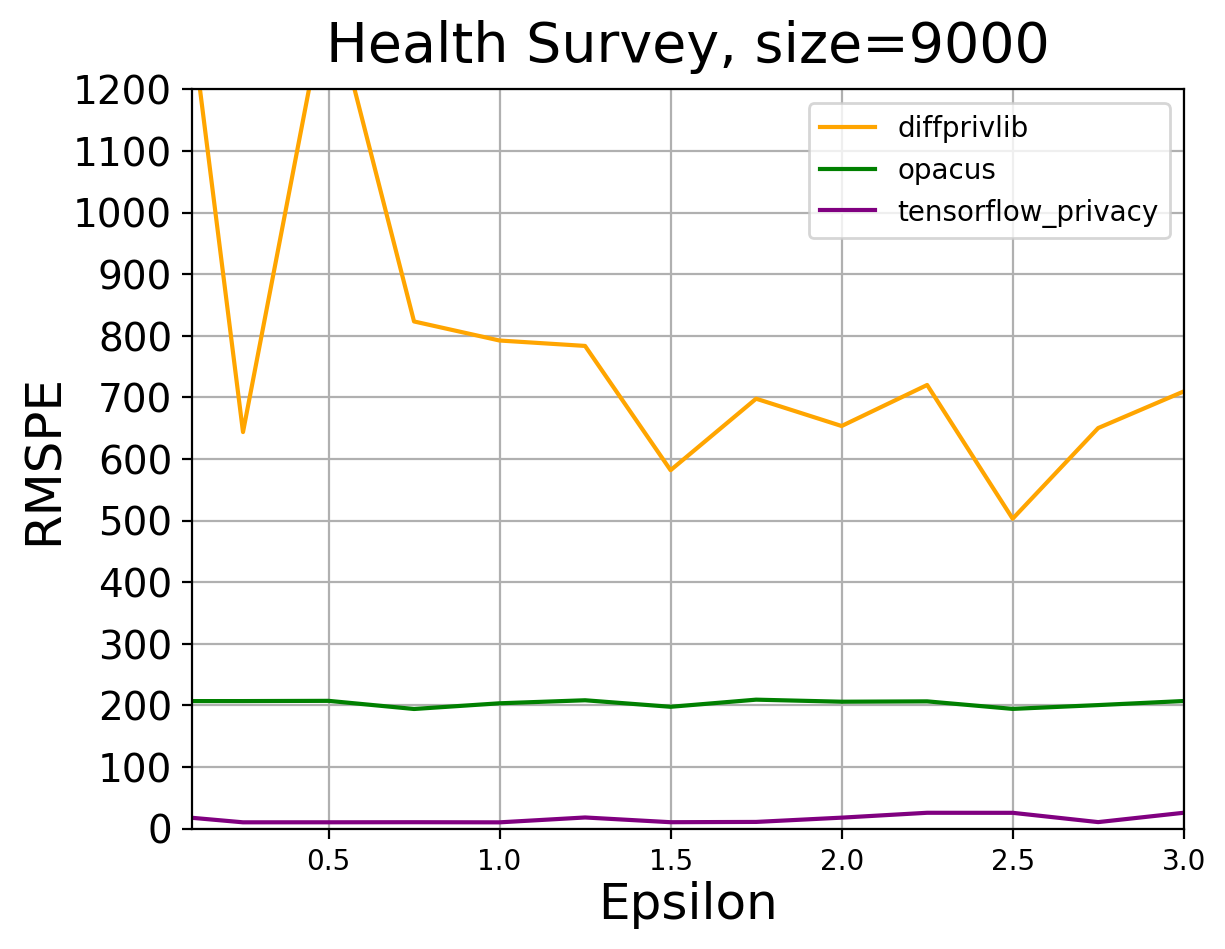}}
	\subfloat[]{\label{fig:exp5:H:9358}\includegraphics[width=0.25\textwidth]{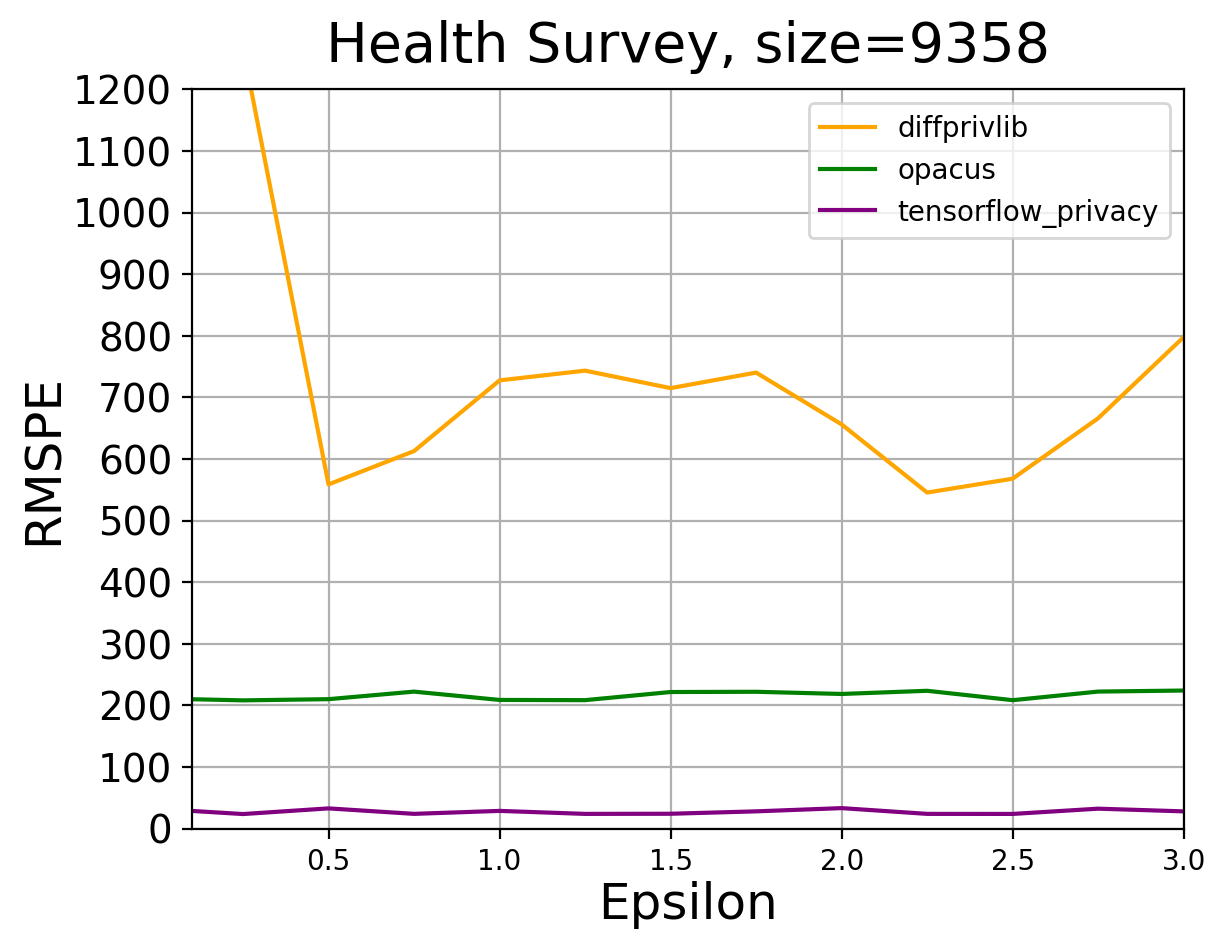}}
	\\
	\subfloat[]{\label{fig:exp5:P:3000}\includegraphics[width=0.25\textwidth]{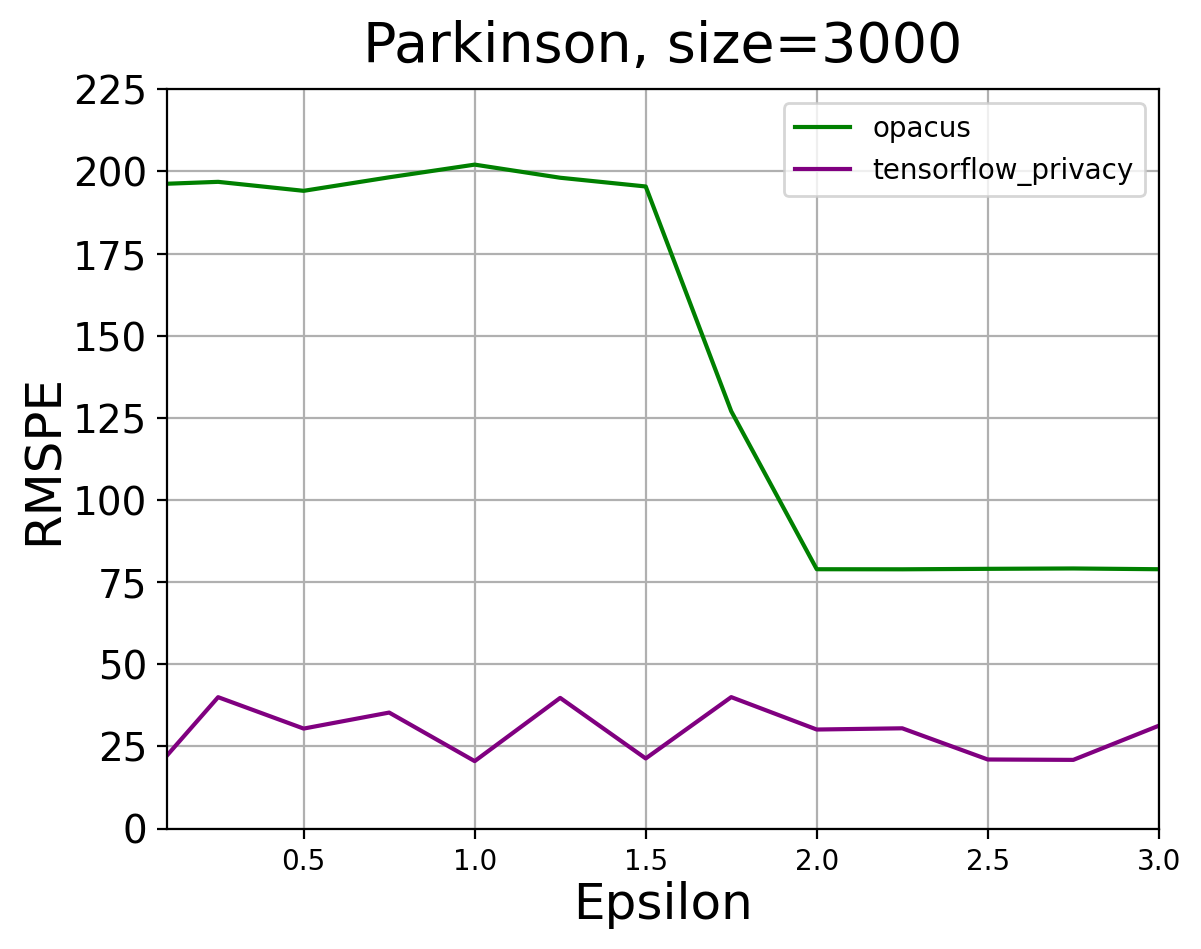}}
	\subfloat[]{\label{fig:exp5:P:4000}\includegraphics[width=0.25\textwidth]{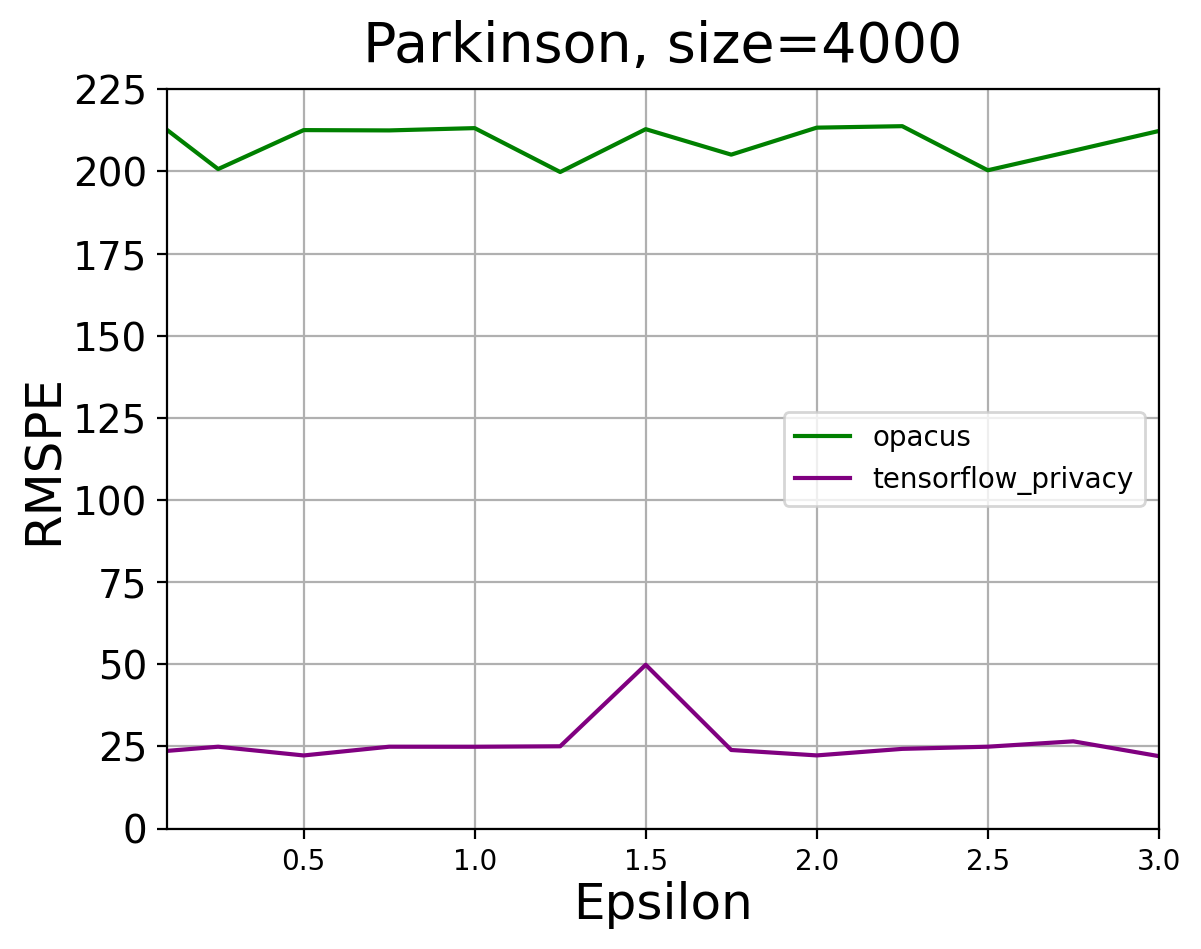}}
	\subfloat[]{\label{fig:exp5:P:5000}\includegraphics[width=0.25\textwidth]{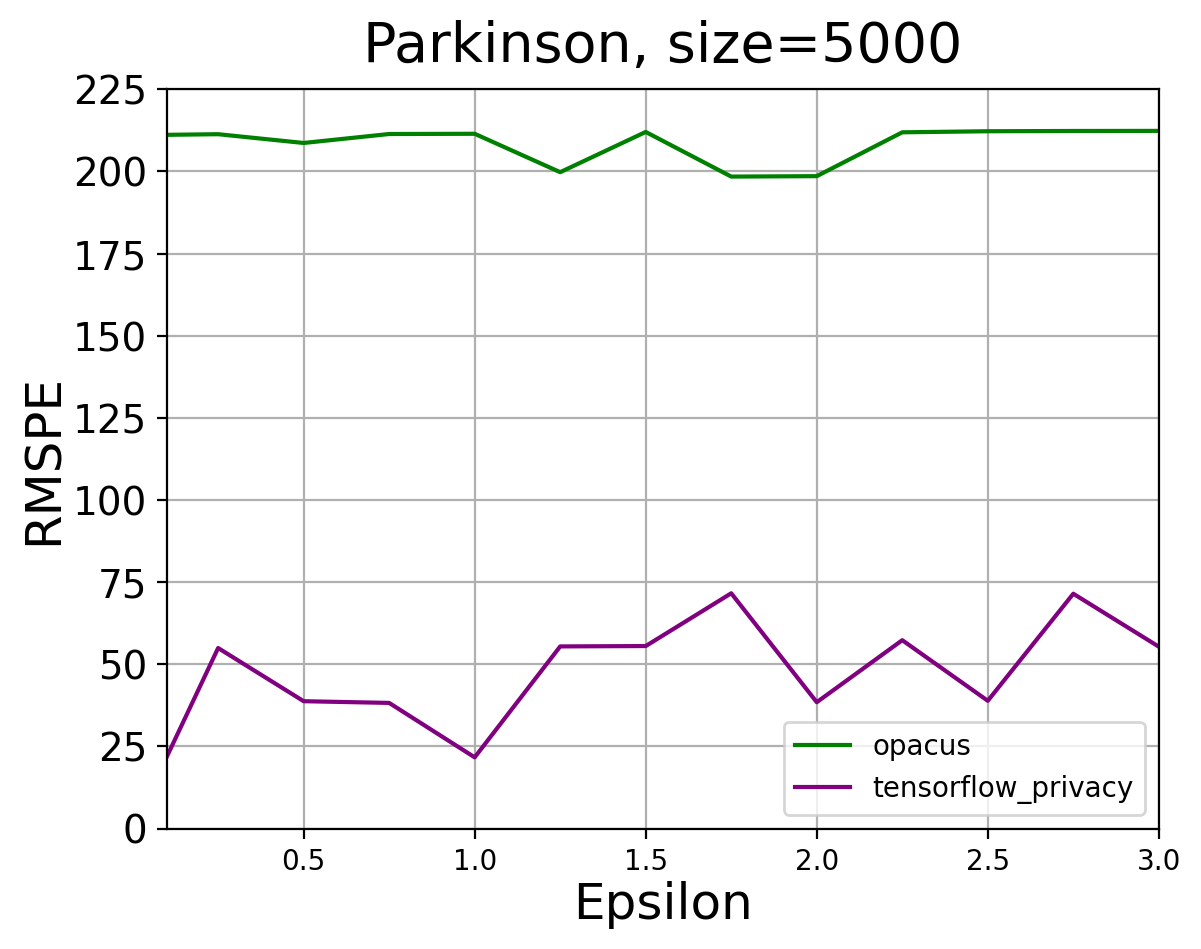}}
	\subfloat[]{\label{fig:exp5:P:5499}\includegraphics[width=0.25\textwidth]{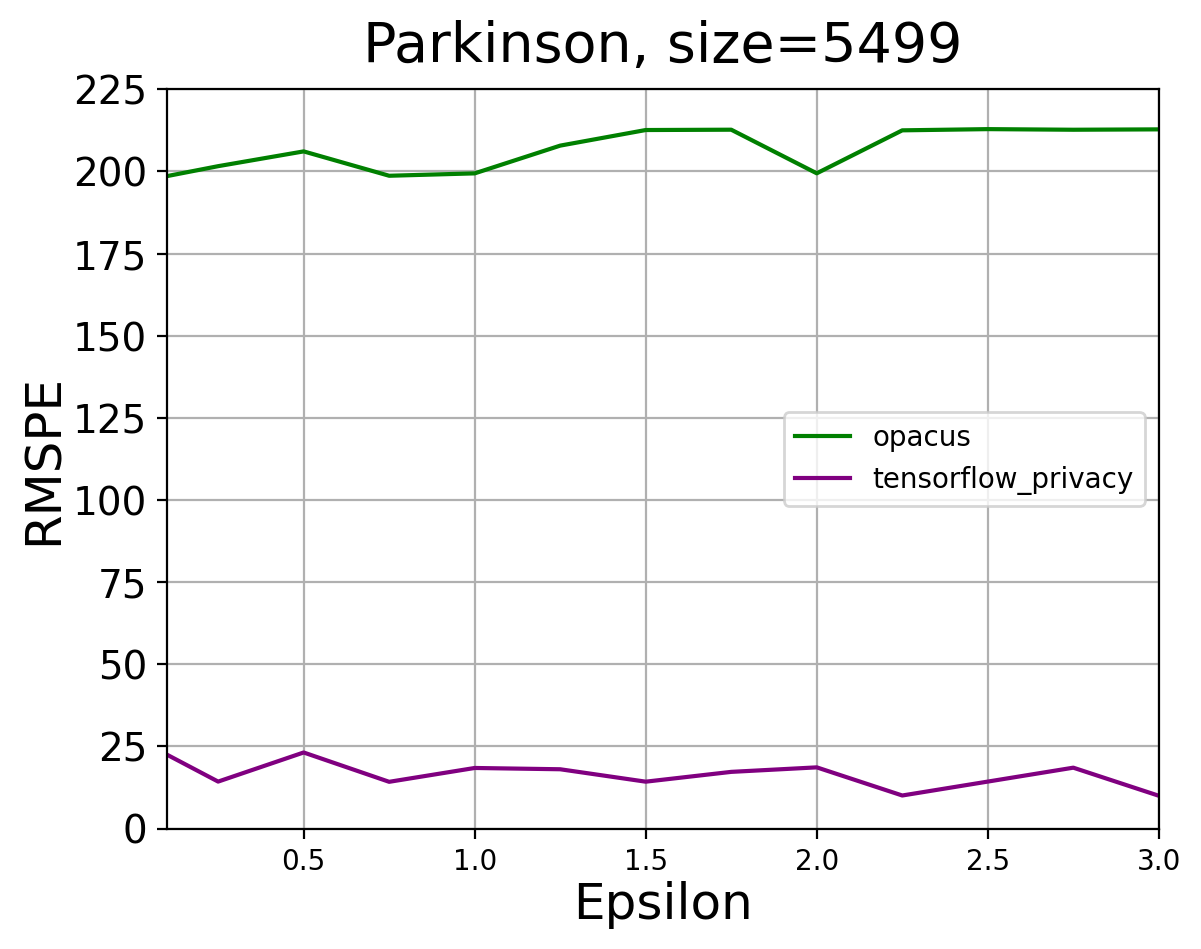}}
	\caption[Results of Experiment 5]{The evaluation results of DP ML tools on run-time overhead for different data sizes (Table \ref{table_dataset_sizes}) and $\epsilon$ values (Table \ref{table_epsilons}). RMSPE is defined in Section~\ref{evaluation_criteria}.}
	\label{fig:exp5:line}
\end{figure}

Figure~\ref{fig:exp5:line} illustrates the induced run-time under different $\epsilon$ values on the \emph{Health survey} and \emph{Parkinson} data for all the ML tools. The graphs reveal no relationship between $\epsilon$ and run-time during model training of all the considered ML tools, while the results of the \emph{Health survey} demonstrate a stable induced run-time by Tensorflow Privacy and Opacus. However, as shown in Figure~\ref{fig:exp5:H:6000}, local irregularity exists for Tensorflow Privacy. Overall, Tensorflow Privacy significantly outperforms Opacus and Diffprivlib regarding run-time on both the considered data sets.

\subsubsection{Memory overhead}\label{ml:exp6}

This section investigates the additional memory usage posed due to the integration of differential privacy (DP) in machine learning (ML). We conduct experiments using various settings to look into how the considered ML tools perform in DP ML models compared with non-private ones.

\begin{figure}[!ht]
	\centering
	\subfloat[]{\label{fig:exp6:tf:H}\includegraphics[width=0.25\textwidth]{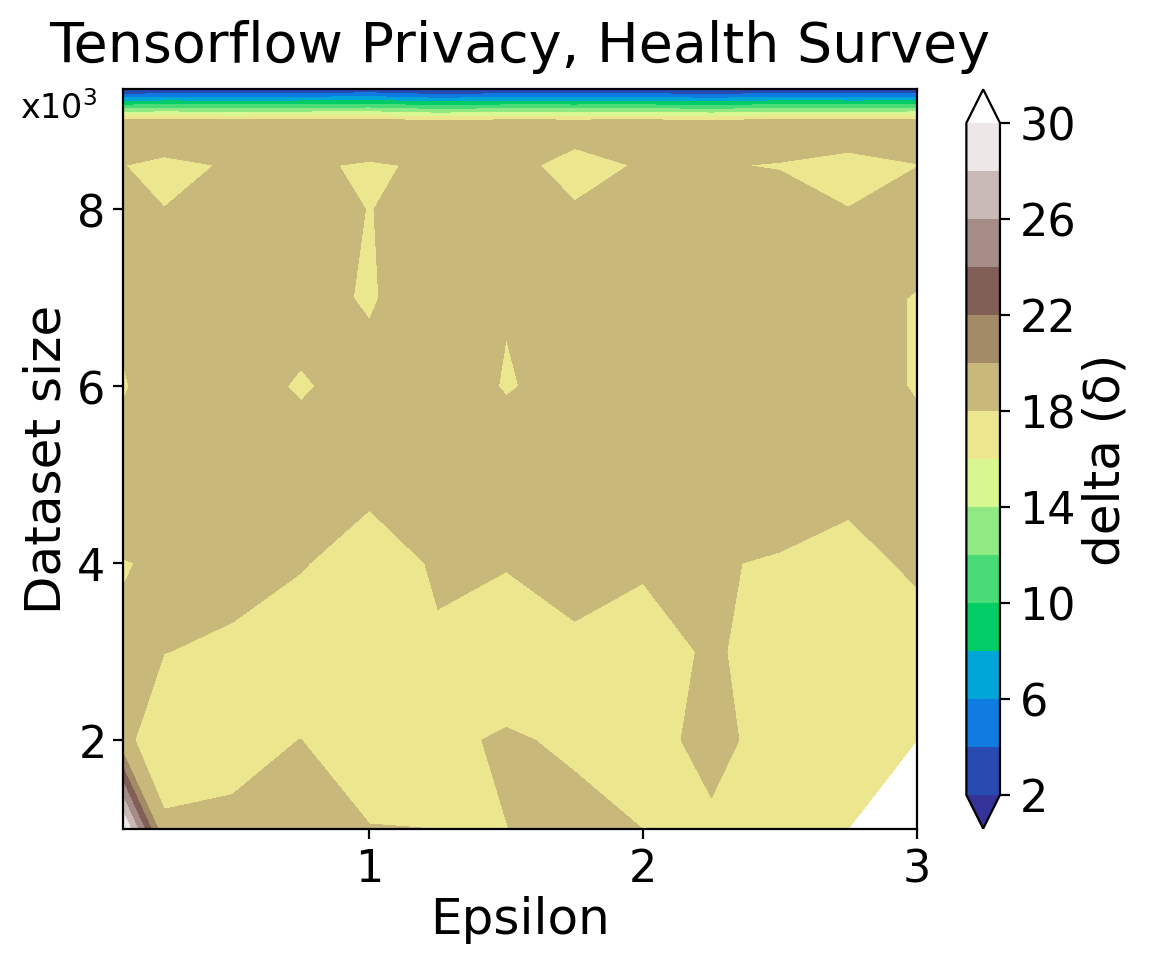}}
	\subfloat[]{\label{fig:exp6:opa:H}\includegraphics[width=0.25\textwidth]{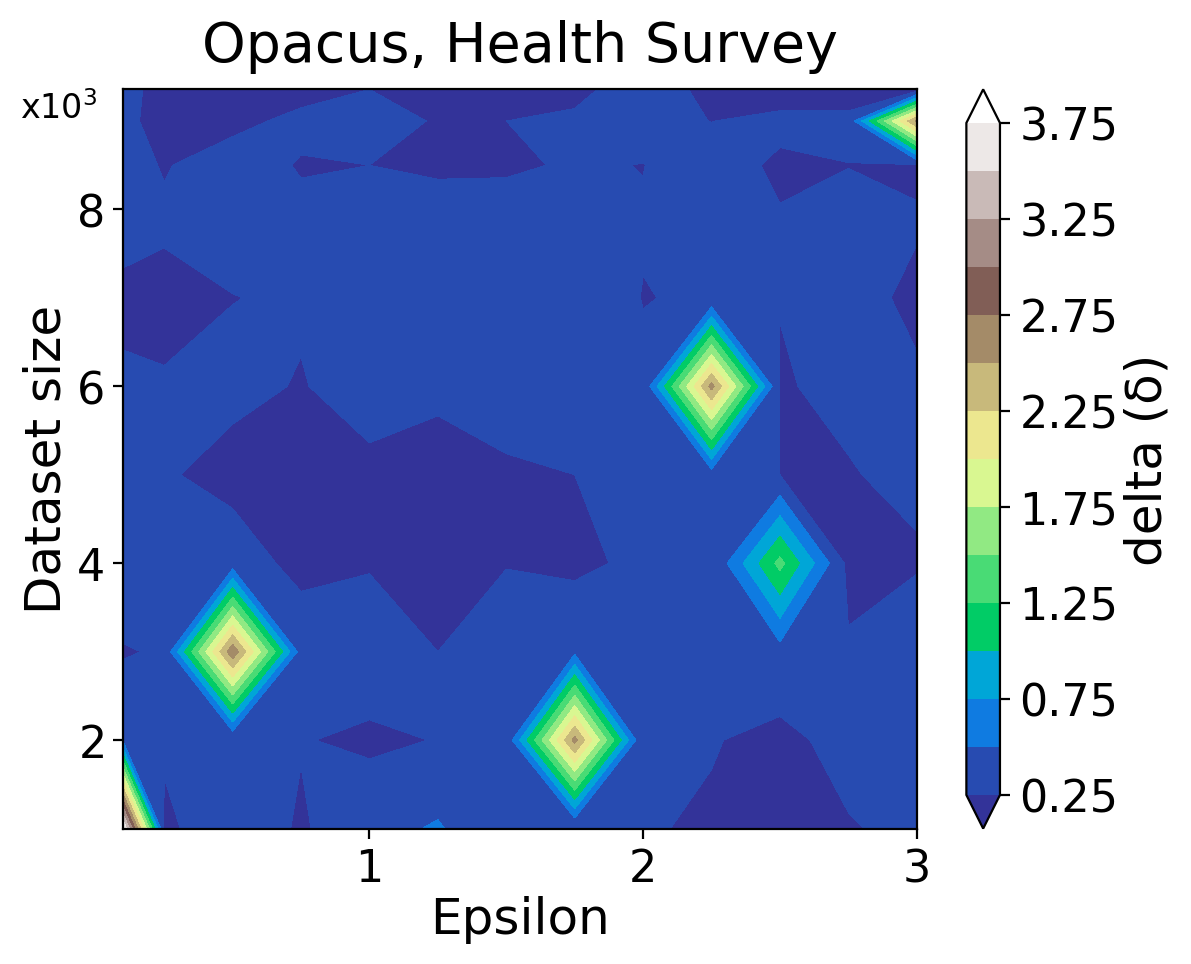}}
	\subfloat[]{\label{fig:exp6:diff:H}\includegraphics[width=0.25\textwidth]{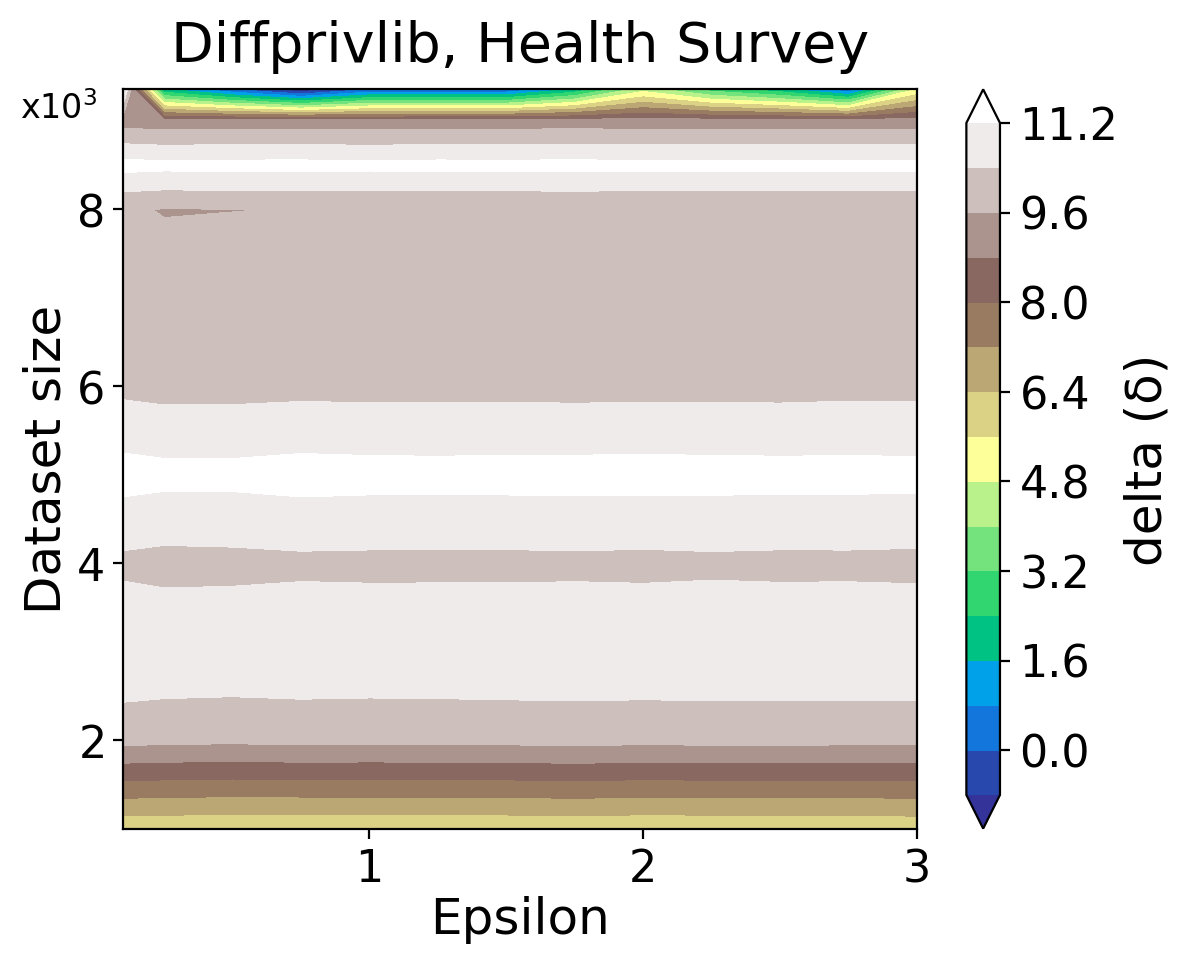}}
	\\
	\subfloat[]{\label{fig:exp6:opa:P}\includegraphics[width=0.25\textwidth]{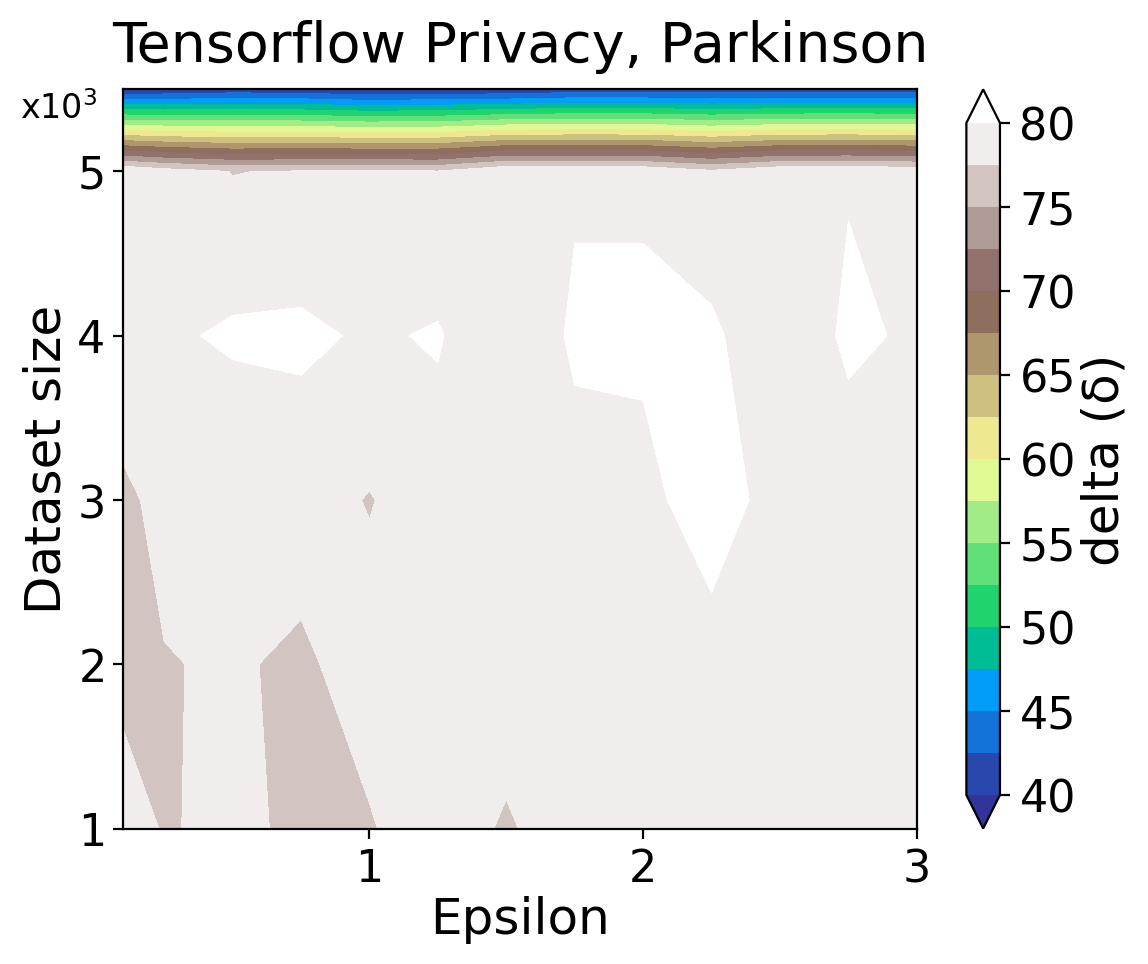}}
	\subfloat[]{\label{fig:exp6:tf:P}\includegraphics[width=0.25\textwidth]{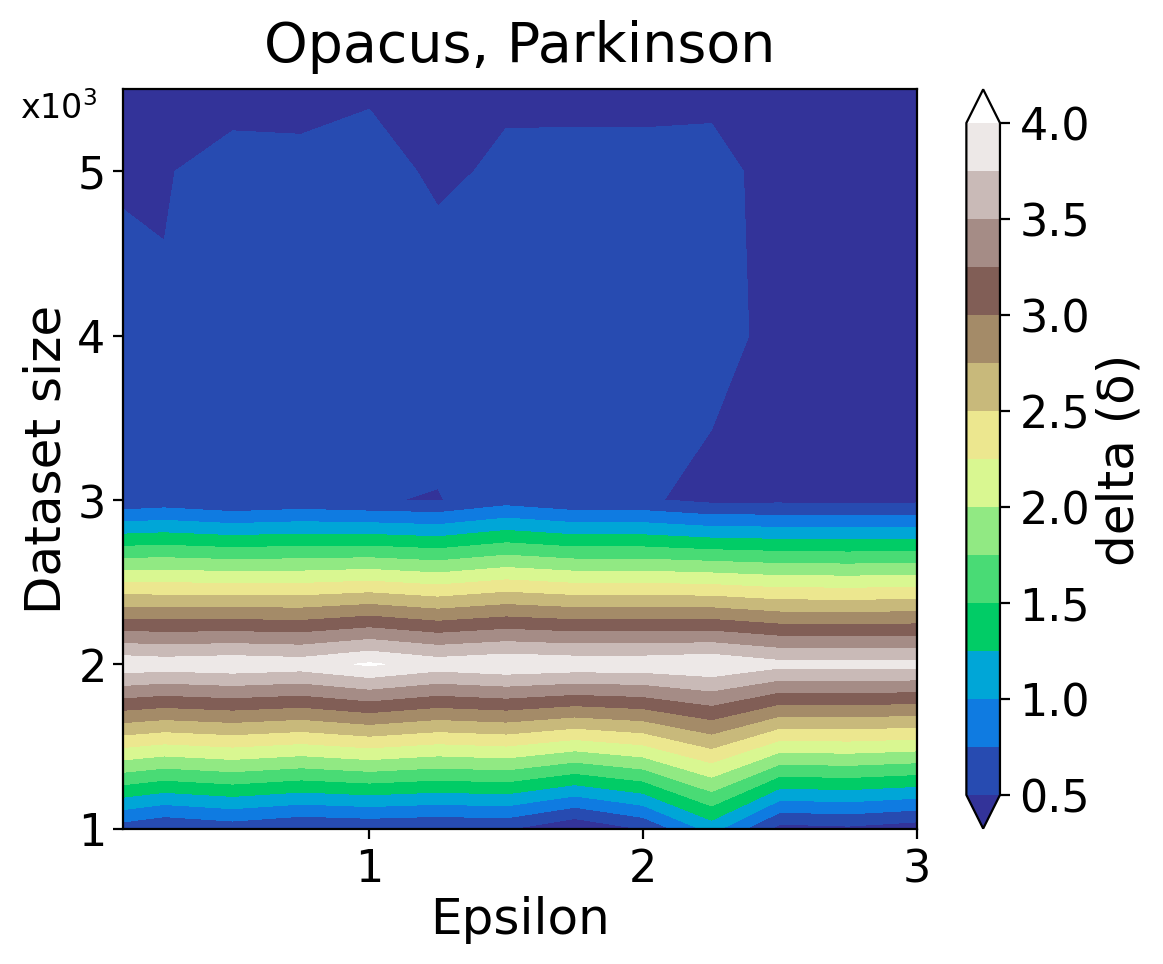}}\hspace{0.25\textwidth}
	\caption[Results of Experiment 6]{The results of the impact of ML tools on memory overhead for different data sizes (Table \ref{table_dataset_sizes}) and $\epsilon$ values (Table \ref{table_epsilons}). RMSPE is defined in Section~\ref{evaluation_criteria}.}
	\label{fig:exp6:contour}
\end{figure}

For the experiments, we anticipate an increase in memory usage in the training of DP ML models compared with non-private ones since more DP models involve more computation during the model training. As detailed below, the experimental results demonstrate extra memory usage of different levels for all the considered ML tools due to DP. Particularly, Tensorflow Privacy suffers the most memory usage (15\%-20\% for the \emph{Health Survey} and above 70\% for \emph{Parkinson}) compared with Opacus (below 3.75\% for the \emph{Health Survey} and below 4.0\% for \emph{Parkinson}). On the other hand, Diffprivlib generally brings an additional memory usage of 5\%-10\% for the \emph{Health Survey}, yet it does not provide useful results for \emph{Parkinson}. Overall, Opacus shows more advantage in memory overhead for DP machine learning.

\begin{figure}[!ht]
	\centering
	\subfloat[]{\label{fig:exp6:H:7000}\includegraphics[width=0.25\textwidth]{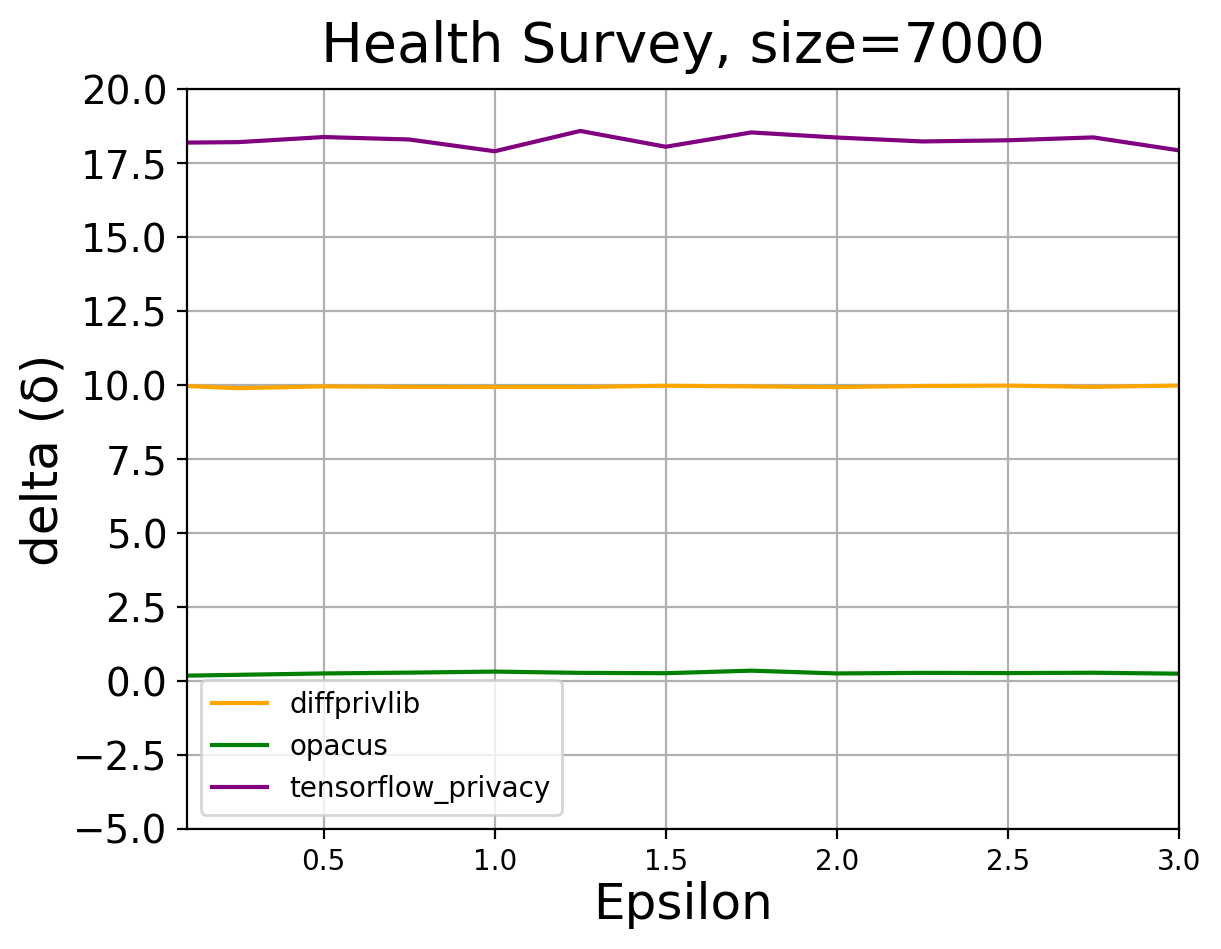}}
	\subfloat[]{\label{fig:exp6:H:8000}\includegraphics[width=0.25\textwidth]{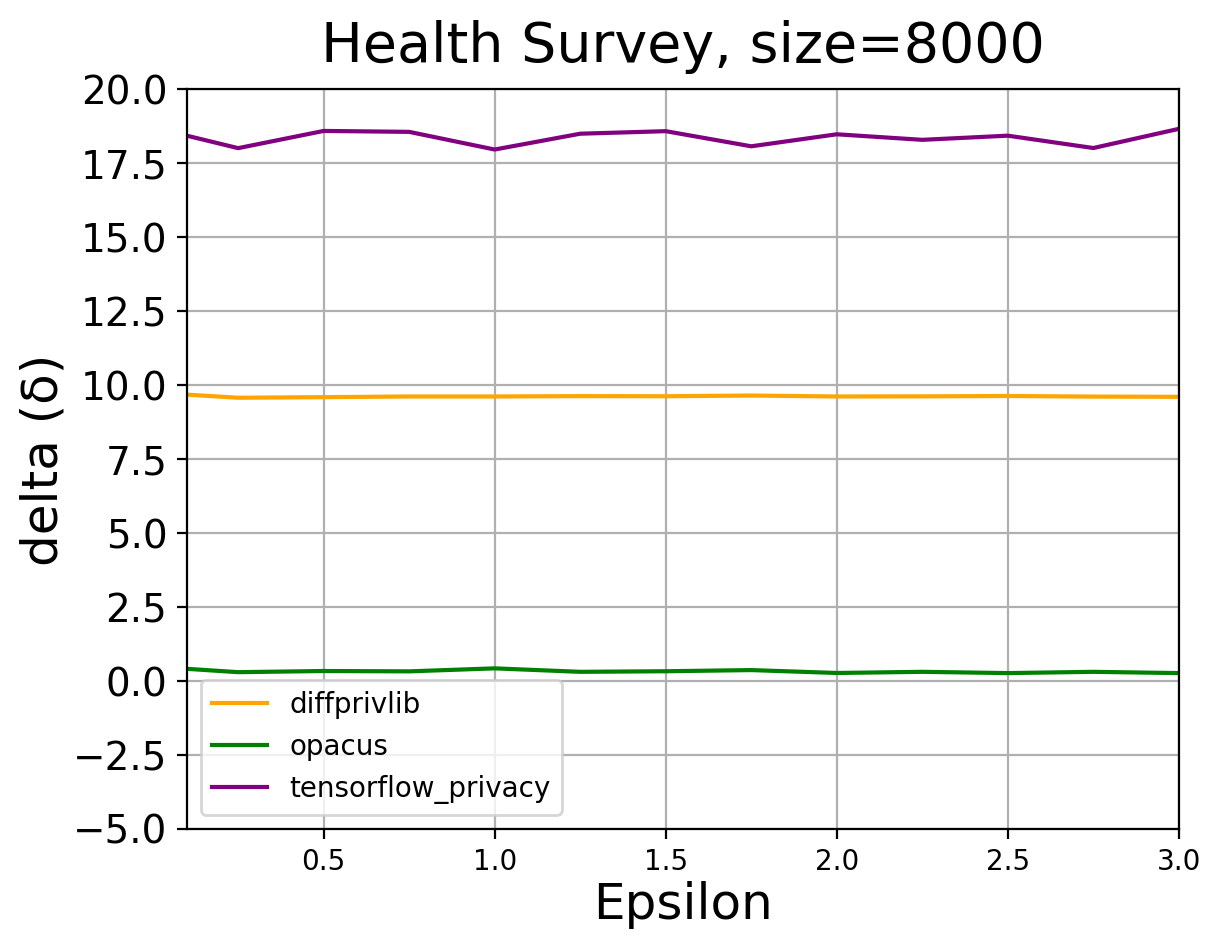}}
	\subfloat[]{\label{fig:exp6:H:9000}\includegraphics[width=0.25\textwidth]{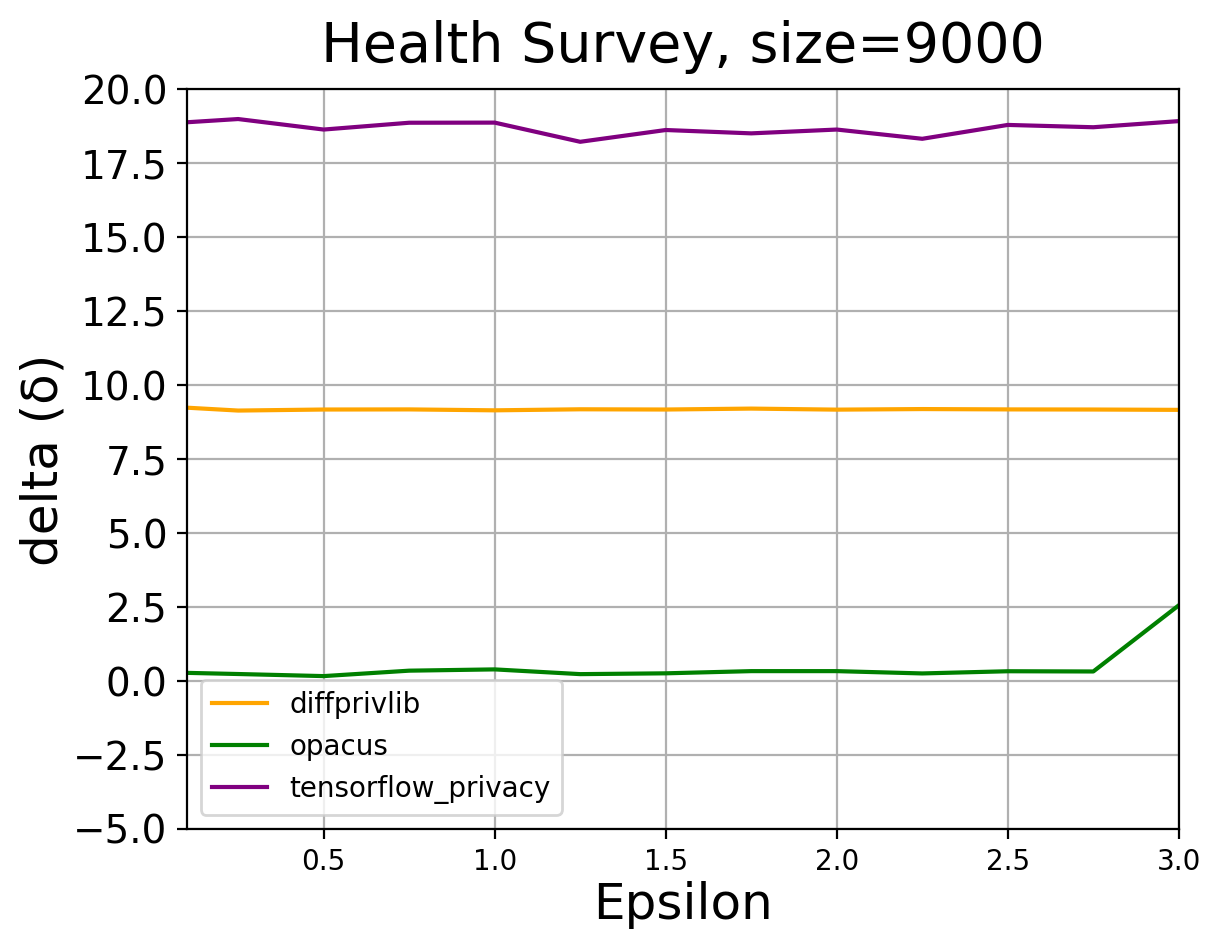}}
	\subfloat[]{\label{fig:exp6:H:9358}\includegraphics[width=0.25\textwidth]{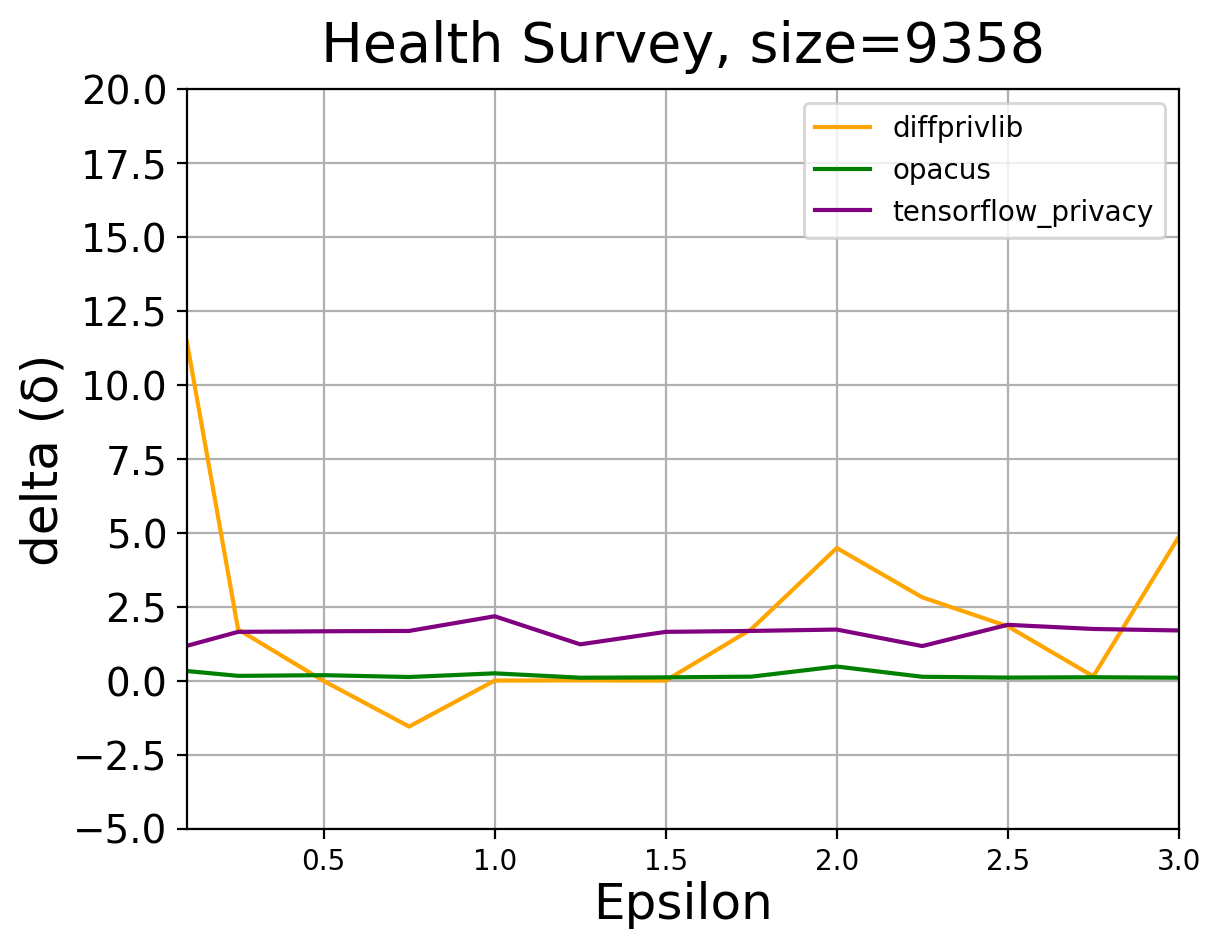}}
	\\
	\subfloat[]{\label{fig:exp6:P:3000}\includegraphics[width=0.25\textwidth]{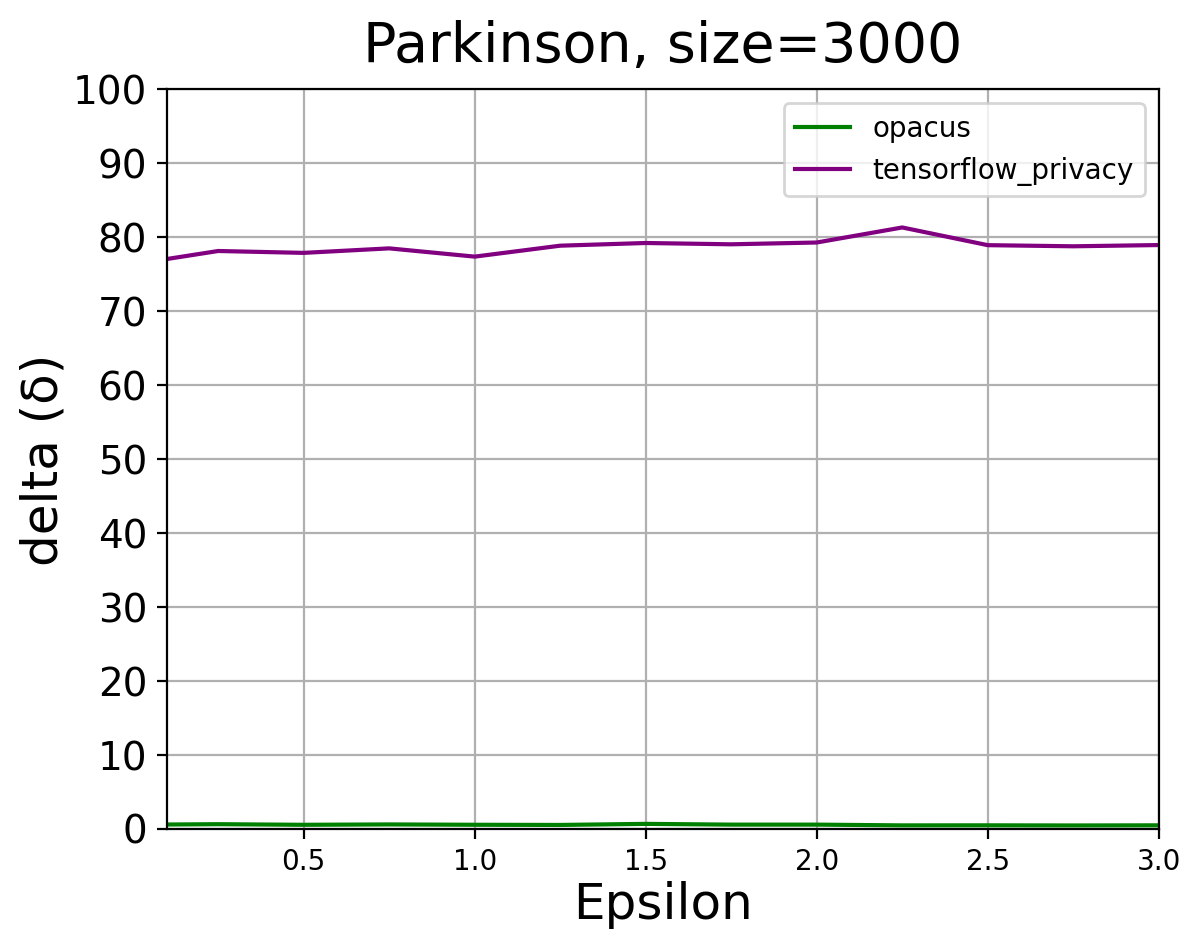}}
	\subfloat[]{\label{fig:exp6:P:4000}\includegraphics[width=0.25\textwidth]{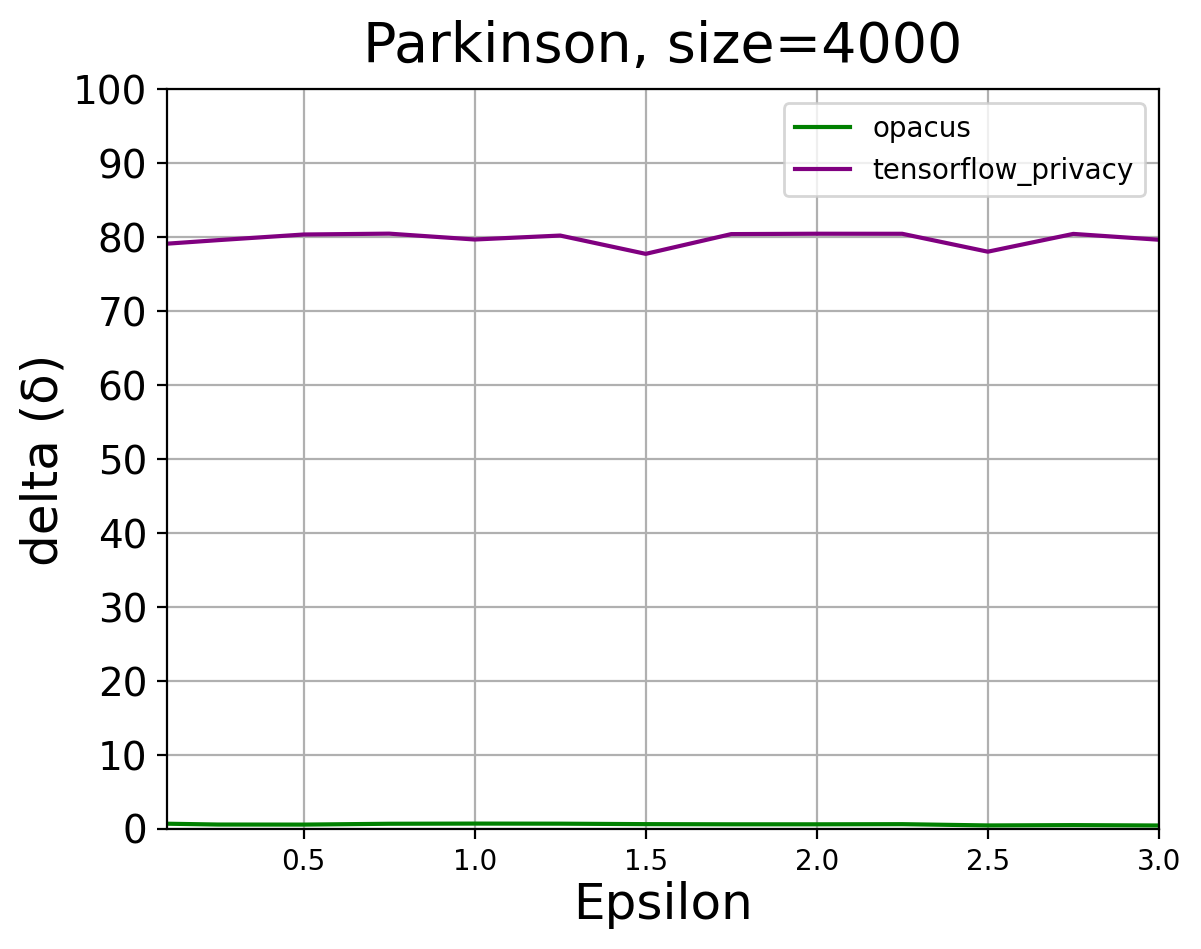}}
	\subfloat[]{\label{fig:exp6:P:5000}\includegraphics[width=0.25\textwidth]{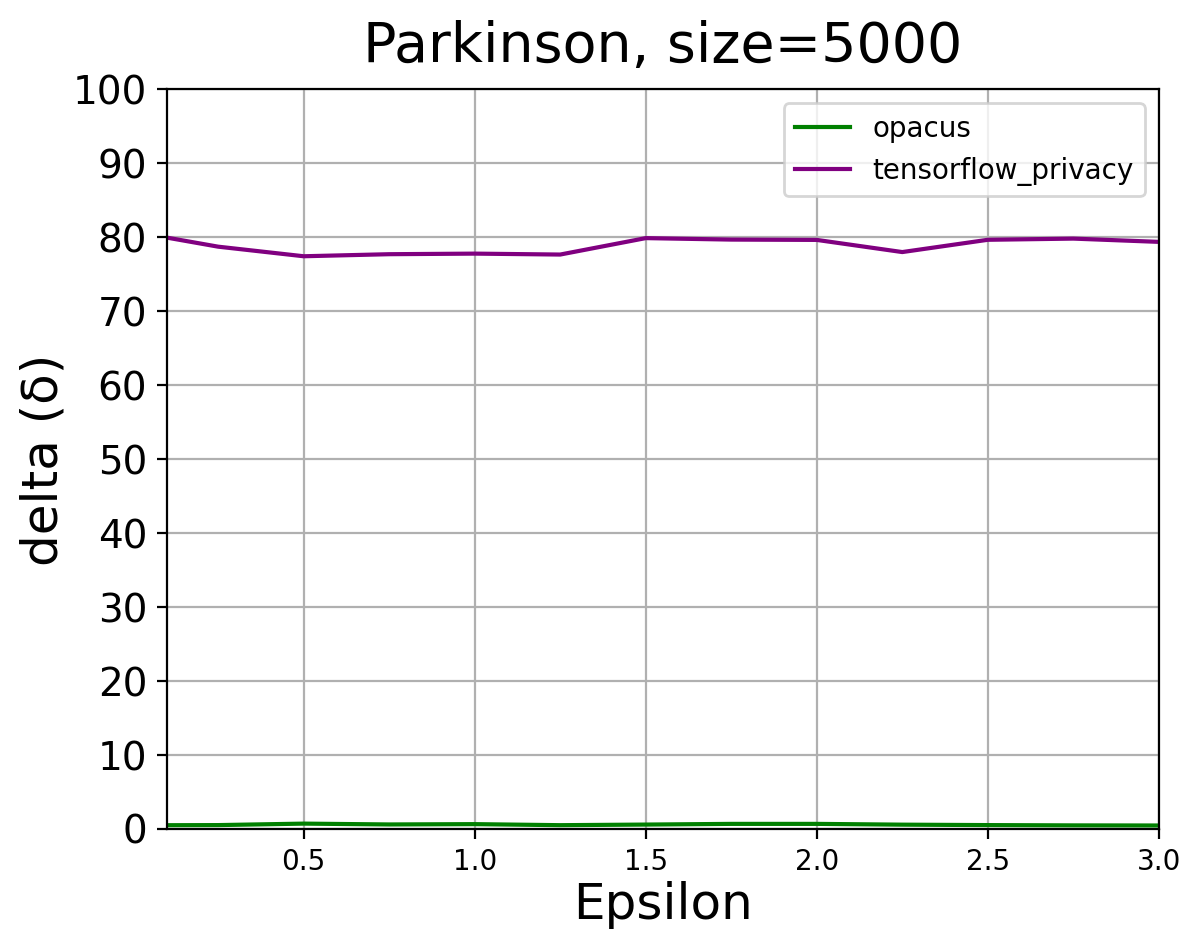}}
	\subfloat[]{\label{fig:exp6:P:5499}\includegraphics[width=0.25\textwidth]{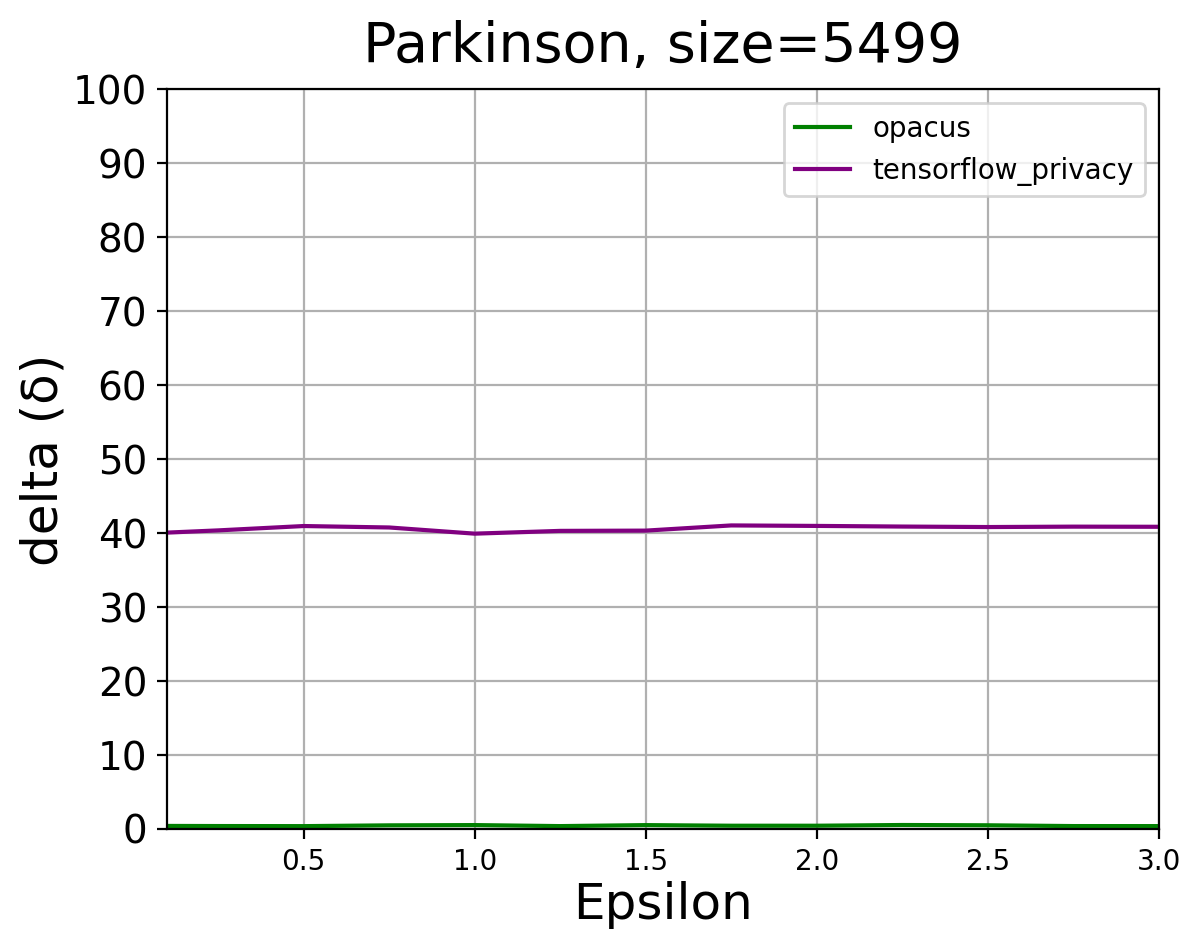}}
	\caption[Results of Experiment 6]{The results of the impact of ML tools on memory overhead for different data sizes (Table \ref{table_dataset_sizes}) and $\epsilon$ values (Table \ref{table_epsilons}).}
	\label{fig:exp6:line}
\end{figure}

Figure~\ref{fig:exp6:contour} shows contour plots of each tool's memory overhead regarding different $\epsilon$ values and data size. From the plots, we can observe that TFP has a memory overhead of about 90\% for the \emph{Parkinson} dataset for nearly all the data size and $\epsilon$, and about 20\% for the \emph{Health Survey} dataset (figures~\ref{fig:exp6:tf:H} and~\ref{fig:exp6:tf:P}), though there is an unexpected decreased memory usage for the highest data size of both \emph{Health Survey} and \emph{Parkinson}. Opacus (figures~\ref{fig:exp6:opa:H} and~\ref{fig:exp6:opa:P}), on the other hand, has less than 5\% memory overhead in both of the datasets, while irregularities exist for some data size and $\epsilon$ combinations for the \emph{Health Survey}, and an increased memory usage around the data size of 2000 for \emph{Parkinson}. Diffprivlib gains memory overhead mainly around 10\% and experiences a slightly decreased memory usage under the smallest data size and a significant decrease under the highest data size for the \emph{Health Survey}. We also note that the value of $\epsilon$ has no apparent influence on the results for all the three considered tools.

Figure~\ref{fig:exp6:line} shows representative results on how memory overhead varies over different combinations of data size and $\epsilon$. Beyond the conclusion from Figure~\ref{fig:exp6:contour}, we can also observe a higher memory overhead for the \emph{Parkinson} than that of the \emph{Health Survey} for Tensorflow Privacy, indicating its better memory usage for categoric data than continuous data. In contrast, Opacus performs stably for both the two data sets with the lowest memory overhead.


\subsection{Summary of results}\label{sec:summary}
We observed that experiment results depend on the studied tasks, the selected DP tools, and their configurations. We note that some patterns emerge from the results, see Table~\ref{table: pattern summary}. 
%
%
%
In order to facilitate the tool selection, we define the following performance criteria.

\begin{table*}[t!]
	\centering
	\resizebox{0.99\textwidth}{!}{
		\begin{tabular}{|c|l|l|}
			\hline
			\multicolumn{2}{|c|}{tool name} & \tabincell{c}{\qquad\qquad\qquad\qquad\qquad patterns revealed} \\
			\hline
			\multirow{5}{*}{\rotatebox{90}{\textbf{ML}}} & \tabincell{l}{Diffprivlib} & \tabincell{l}{$\bullet$~utility and run-time overhead grow with $\epsilon$ and data size on categorical data}  \\
			\cline{2-3}
			& \tabincell{l}{PyTorch Opacus} & \tabincell{l}{$\bullet$~utility grows with $\epsilon$ on categorical and continuous data\\$\bullet$~utility grows with data size on categorical data}  \\
			\cline{2-3}
			& \tabincell{l}{TensorFlow Privacy} & \tabincell{l}{$\bullet$~utility grows with $\epsilon$ and data size on categorical and continuous data\\$\bullet$~memory overhead grows with data size on categorical and continuous data}  \\
			\hline
			\multirow{4}{*}{\rotatebox{90}{\textbf{SQ}}} & \tabincell{l}{Google Differential Privacy} & \tabincell{l}{$\bullet$~utility grows with $\epsilon$ and data size on categorical and continuous data\\$\bullet$~higher run-time overhead for lower data size on categorical and continuous data}  \\
			\cline{2-3}
			& \tabincell{l}{OpenDP SmartNoise} & \tabincell{l}{$\bullet$~utility grows with $\epsilon$ and data size on categorical and continuous data\\$\bullet$~run-time overhead grows with data size on categorical and continuous data}  \\
			\hline
		\end{tabular}
	}
	\caption{\label{table: pattern summary}Summary of performance patterns emerging from the evaluation results of the studied tools for machine learning (ML) and statistical query (SQ).}
\end{table*}

\begin{table}[htp]
	\centering
	\resizebox{0.99\textwidth}{!}{
		\begin{tabular}{|l|l|l|l|l|l|l|l|l|l|l|l|l|l|l|l|l|l|l|l|}
			\cline{3-20}
			\multicolumn{2}{c|}{\multirow{3}{*}{}} & \multicolumn{18}{c|}{performance criteria (RMPSE)} \\
			\cline{3-20}
			\multicolumn{2}{c|}{}  & \multicolumn{9}{c|}{categorical data} & \multicolumn{9}{c|}{continuous data}\\
			\cline{3-20}
			\multicolumn{2}{c|}{}  & \rotatebox{90}{\textbf{utility FES}} & \rotatebox{90}{\textbf{run-time FES }} & \rotatebox{90}{\textbf{memory FES}} & \rotatebox{90}{\textbf{utility RES}} & \rotatebox{90}{\textbf{run-time RES}} & \rotatebox{90}{\textbf{memory RES}} & \rotatebox{90}{\textbf{utility DFC}} & \rotatebox{90}{\textbf{run-time DFC}} & \rotatebox{90}{\textbf{memory DFC}} & \rotatebox{90}{\textbf{utility FES}} & \rotatebox{90}{\textbf{run-time FES}} & \rotatebox{90}{\textbf{memory FES}} & \rotatebox{90}{\textbf{utility RES}} & \rotatebox{90}{\textbf{run-time RES}} & \rotatebox{90}{\textbf{memory RES}} & \rotatebox{90}{\textbf{utility DFC}} & \rotatebox{90}{\textbf{run-time DFC}} & \rotatebox{90}{\textbf{memory DFC}}\\
			\hline
			\multirow{12}{*}{\rotatebox{90}{\textbf{machine learning}}} & \multirow{4}{*}{\tabincell{l}{Diffprivlib}}  & \multirow{4}{*}{\tabincell{l}{1st\\{\tiny \textbf{0.15}}}} & \multirow{4}{*}{\tabincell{l}{3rd\\{\tiny \textbf{797.86}}}} & \multirow{4}{*}{\tabincell{l}{3rd\\{\tiny \textbf{6.35}}}} & \multirow{4}{*}{\tabincell{l}{3rd\\{\tiny \textbf{6.4e+8}}}} & \multirow{4}{*}{\tabincell{l}{3rd\\{\tiny \textbf{1176.20}}}} & \multirow{4}{*}{\tabincell{l}{3rd\\{\tiny \textbf{46.21}}}} & \multirow{4}{*}{\tabincell{l}{3rd\\{\tiny \textbf{2.76}}\\{\tiny \textbf{to}}\\{\tiny \textbf{1.21e+8}}}} & \multirow{4}{*}{\tabincell{l}{3rd\\{\tiny \textbf{603.17}}\\{\tiny \textbf{to}}\\{\tiny \textbf{1.15e+3}}}} & \multirow{4}{*}{\tabincell{l}{3rd\\{\tiny \textbf{9.11}}\\{\tiny \textbf{to}}\\{\tiny \textbf{16.64}}}} & \multicolumn{9}{c|}{Results far away from usable} \\
			\cline{12-20}
			&  &  &  &  &  &  &  &  &  &   & \multirow{3}{*}{\tabincell{l}{{\tiny \textbf{4.6e+8}}}} & \multirow{3}{*}{\tabincell{l}{{\tiny \textbf{2323.46}}}} & \multirow{3}{*}{\tabincell{l}{{\tiny \textbf{35.64}}}} & \multirow{3}{*}{\tabincell{l}{{\tiny \textbf{1.96e+11}}}} & \multirow{3}{*}{\tabincell{l}{{\tiny \textbf{856.53}}}} & \multirow{3}{*}{\tabincell{l}{{\tiny \textbf{31.54}}}} & \multirow{3}{*}{\tabincell{l}{{\tiny \textbf{1.53e+9}}\\{\tiny \textbf{to}}\\{\tiny \textbf{1.89e+18}}}} & \multirow{3}{*}{\tabincell{l}{{\tiny \textbf{603.17}}\\{\tiny \textbf{to}}\\{\tiny \textbf{1.15e+3}}}} & \multirow{3}{*}{\tabincell{l}{{\tiny \textbf{40.88}}\\{\tiny \textbf{to}}\\{\tiny \textbf{45.38}}}} \\
			&  &  &  &  &  &  &  &  &  &   &  &  &  &  &  &  &  &  &  \\
			&  &  &  &  &  &  &  &  &  &   &  &  &  &  &  &  &  &  &  \\
			\cline{2-20}
			& \tabincell{l}{PyTorch\\ Opacus} & \tabincell{l}{3rd\\{\tiny \textbf{4.47}}} & \tabincell{l}{2nd\\{\tiny \textbf{221.13}}} & \tabincell{l}{2nd\\{\tiny \textbf{0.72}}} & \tabincell{l}{1st\\{\tiny \textbf{1283.47}}} & \tabincell{l}{2nd\\{\tiny \textbf{217.85}}} & \tabincell{l}{1st\\{\tiny \textbf{2.35}}} & \tabincell{l}{1st\\{\tiny \textbf{3.44}}\\{\tiny \textbf{to}}\\{\tiny \textbf{7.10}}} & \tabincell{l}{2nd\\{\tiny \textbf{197.67}}\\{\tiny \textbf{to}}\\{\tiny \textbf{211.96}}} & \tabincell{l}{1st\\{\tiny \textbf{0.11}}\\{\tiny \textbf{to}}\\{\tiny \textbf{0.16}}}  & \tabincell{l}{2nd\\{\tiny \textbf{2.59}}} & \tabincell{l}{2nd\\{\tiny \textbf{213.23}}} & \tabincell{l}{1st\\{\tiny \textbf{1.77}}} & \tabincell{l}{1st\\{\tiny \textbf{5122.05}}} & \tabincell{l}{2nd\\{\tiny \textbf{224.95}}} & \tabincell{l}{1st\\{\tiny \textbf{1.81}}} & \tabincell{l}{2nd\\{\tiny \textbf{3.99}}\\{\tiny \textbf{to}}\\{\tiny \textbf{38.21}}} & \tabincell{l}{2nd\\{\tiny \textbf{199.77}}\\{\tiny \textbf{to}}\\{\tiny \textbf{213.90}}} & \tabincell{l}{1st\\{\tiny \textbf{0.55}}\\{\tiny \textbf{to}}\\{\tiny \textbf{1.68}}} \\
			\cline{2-20}
			& \tabincell{l}{TensorFlow\\ Privacy} & \tabincell{l}{2nd\\{\tiny \textbf{2.91}}} & \tabincell{l}{1st\\{\tiny \textbf{22.52}}} & \tabincell{l}{1st\\{\tiny \textbf{6.1e-4}}} & \tabincell{l}{2nd\\{\tiny \textbf{6960.15}}} & \tabincell{l}{1st\\{\tiny \textbf{17.25}}} & \tabincell{l}{2rd\\{\tiny \textbf{9.74}}} & \tabincell{l}{2rd\\{\tiny \textbf{4.65}}\\{\tiny \textbf{to}}\\{\tiny \textbf{12.73}}} & \tabincell{l}{1st\\{\tiny \textbf{10.30}}\\{\tiny \textbf{to}}\\{\tiny \textbf{14.23}}} & \tabincell{l}{2nd\\{\tiny \textbf{0.64}}\\{\tiny \textbf{to}}\\{\tiny \textbf{7.06}}}  & \tabincell{l}{1st\\{\tiny \textbf{0.90}}} & \tabincell{l}{1st\\{\tiny \textbf{35.69}}} & \tabincell{l}{2nd\\{\tiny \textbf{15.77}}} & \tabincell{l}{2nd\\{\tiny \textbf{6177.35}}} & \tabincell{l}{1st\\{\tiny \textbf{1.57}}} & \tabincell{l}{2nd\\{\tiny \textbf{28.32}}} & \tabincell{l}{1st\\{\tiny \textbf{2.32}}\\{\tiny \textbf{to}}\\{\tiny \textbf{26.37}}} & \tabincell{l}{1st\\{\tiny \textbf{21.01}}\\{\tiny \textbf{to}}\\{\tiny \textbf{35.25}}} & \tabincell{l}{2nd\\{\tiny \textbf{13.48}}\\{\tiny \textbf{to}}\\{\tiny \textbf{26.88}}} \\
			\hline
			
			\hline
			\multirow{4}{*}{\rotatebox{90}{\textbf{statistical query}}} & \tabincell{l}{Google\\ Differential\\ Privacy} & \tabincell{l}{1st\\{\tiny \textbf{0.61}}} & \tabincell{l}{1st\\{\tiny \textbf{115.63}}} & \tabincell{l}{1st\\{\tiny \textbf{8.12}}} & \tabincell{l}{1st\\{\tiny \textbf{74.3}}} & \tabincell{l}{1st\\{\tiny \textbf{137.37}}} & \tabincell{l}{1st\\{\tiny \textbf{25.19}}} & \tabincell{l}{1st\\{\tiny \textbf{0.016}}\\{\tiny \textbf{to}}\\{\tiny \textbf{0.31}}} & \tabincell{l}{1st\\{\tiny \textbf{108.25}}\\{\tiny \textbf{to}}\\{\tiny \textbf{129.35}}} & \tabincell{l}{1st\\{\tiny \textbf{0.05}}\\{\tiny \textbf{to}}\\{\tiny \textbf{1.88}}} & \tabincell{l}{1st\\{\tiny \textbf{0.5}}} & \tabincell{l}{1st\\{\tiny \textbf{120.32}}} & \tabincell{l}{1st\\{\tiny \textbf{4.6e-4}}} & \tabincell{l}{1st\\{\tiny \textbf{75.9}}} & \tabincell{l}{1st\\{\tiny \textbf{126.63}}} & \tabincell{l}{1st\\{\tiny \textbf{7.6e-5}}} & \tabincell{l}{1st\\{\tiny \textbf{0.02}}\\{\tiny \textbf{to}}\\{\tiny \textbf{0.36}}} & \tabincell{l}{1st\\{\tiny \textbf{115.11}}\\{\tiny \textbf{to}}\\{\tiny \textbf{137.43}}} & \tabincell{l}{1st\\{\tiny \textbf{0}}\\{\tiny \textbf{to}}\\{\tiny \textbf{2.16}}} \\
			\cline{2-20}
			& \tabincell{l}{OpenDP\\ SmartNoise} & \tabincell{l}{2rd\\{\tiny \textbf{2.9}}} & \tabincell{l}{2rd\\{\tiny \textbf{481.11}}} & \tabincell{l}{2rd\\{\tiny \textbf{14.08}}} & \tabincell{l}{2nd\\{\tiny \textbf{801}}} & \tabincell{l}{2rd\\{\tiny \textbf{413.96}}} & \tabincell{l}{2rd\\{\tiny \textbf{29.27}}} & \tabincell{l}{2rd\\{\tiny \textbf{0.11}}\\{\tiny \textbf{to}}\\{\tiny \textbf{2.08}}} & \tabincell{l}{2rd\\{\tiny \textbf{408.74}}\\{\tiny \textbf{to}}\\{\tiny \textbf{466.23}}} & \tabincell{l}{2rd\\{\tiny \textbf{0.50}}\\{\tiny \textbf{to}}\\{\tiny \textbf{2.75}}}  & \tabincell{l}{2rd\\{\tiny \textbf{3.9}}} & \tabincell{l}{2rd\\{\tiny \textbf{473.63}}} & \tabincell{l}{2rd\\{\tiny \textbf{4.56}}} & \tabincell{l}{2nd\\{\tiny \textbf{746}}} & \tabincell{l}{2rd\\{\tiny \textbf{558.16}}} & \tabincell{l}{2rd\\{\tiny \textbf{4.67}}} & \tabincell{l}{2rd\\{\tiny \textbf{0.19}}\\{\tiny \textbf{to}}\\{\tiny \textbf{2.24}}} & \tabincell{l}{2rd\\{\tiny \textbf{427.19}}\\{\tiny \textbf{to}}\\{\tiny \textbf{465.05}}} & \tabincell{l}{2rd\\{\tiny \textbf{5.06}}\\{\tiny \textbf{to}}\\{\tiny \textbf{6.47}}}  \\
			\hline
		\end{tabular}
	}
	\caption{\label{table: tool comparison}Tool performance comparison for the tasks of machine learning and statistical query. For each criterion, we provide the ranking (in the upper part of the table cells) of different tools associated with the numerical performance (in the lower part of the table cells) under the considered criterion.}
\end{table}


Any DP tool inherently follows some basic tradeoffs. For example, when processing a large dataset, the system costs, \eg processing time and memory usage, increase. Another well-known trade-off exists between the privacy budget and the utility of the privacy-protected data. Naturally, the evaluated tools might include additional dependencies and tradeoffs. In order to highlight such basic boundaries, Table~\ref{table: tool comparison} provides the values of the utility and system costs under  \emph{Favorable Experiment Settings} (FES) and \emph{Restrictive Experiment Settings} (RES) w.r.t. the size of the dataset and privacy budget. Specifically for FES, the size of the dataset is 9358 records and $\epsilon$=3.0 for the categorical data. The size is 5499 records and $\epsilon$=3.0 for the continuous data. For the case of RES, the size is 1000 records and $\epsilon$=0.1 both for  categorical and continuous data. For a more \emph{Detailed version of Performance Comparison} (DPC), we provide the utility and system costs that the tools have when considering an exhaustive set of pairs of dataset size (Table~\ref{table_dataset_sizes}) and privacy budget (Table~\ref{table_epsilons}).

With the defined criteria and the performance measurement under each criterion, we facilitate a need-based selection of DP tools, \eg the highest utility, the lightest computing resource requirement, and the least running time. Table~\ref{table: tool comparison} suggests that TensorFlow Privacy and PyTorch Opacus perform equivalently well for machine learning tasks. Also, Google DP works well for statistical queries.


\section{Conclusions}
\label{sec:con}
We propose an evaluation framework for differential privacy (DP) tools and offer evaluation for state-of-the-art open-source DP tools. We define criteria to quantify how different DP tools perform so that they can be selected. Specifically, we evaluate and measure the impact of DP on different functionality that the studied tools provide. We use two data sources of different types to obtain a nuanced picture of how well the studied tools perform when DP is applied. The evaluation results demonstrate the degree to which the use of DP tools impacts data utility and system overhead. Our results can support practitioners that consider using these tools.



\section*{Acknowledgement}

The work of S. Zhang and E. M. Schiller was partially supported by the project `Privacy-Protected Machine Learning for Transport Systems' of Area of Advance Transport and Chalmers AI Research Centre (CHAIR) as well as by AutoSPADA (Automotive Stream Processing and Distributed Analytics) OODIDA Phase 2 by Vinnova's FFI framework (reference number 2019-05884). The work of A. Hagermalm and S. Slavnic was partially supported by AstraZeneca AB. The computations were enabled by resources provided by the Swedish National Infrastructure for Computing (SNIC) at C3SE partially funded by the Swedish Research Council through grant agreement no. 2018-05973.

\bibliographystyle{IEEEtran}
\bibliography{IEEE}

\end{document}